\documentclass[showpacs,pra,onecolumn]{revtex4-1}
\usepackage{amssymb}
\usepackage{amsmath}
\usepackage{graphicx}
\usepackage{subfigure}
\usepackage{epstopdf}
\usepackage{color}

\begin{document}

\title{Solitons and vortices in nonlinear potential wells}
\author{Nir Dror and Boris A. Malomed}
\address{Department of Physical Electronics, School of Electrical
Engineering, Faculty of Engineering,Tel Aviv University, Tel Aviv
69978, Israel}

\begin{abstract}
We consider self-trapping of topological modes governed by the one- and
two-dimensional (1D and 2D) nonlinear-Schr\"{o}dinger/Gross-Pitaevskii
equation with effective single- and double-well (DW) nonlinear potentials
induced by spatial modulation of the local strength of the self-defocusing
nonlinearity. This setting, which may be implemented in optics and
Bose-Einstein condensates, aims to extend previous studies, which dealt with
single-well nonlinear potentials. In the 1D setting, we find several types
of symmetric, asymmetric and antisymmetric states, focusing on scenarios of
the spontaneous symmetry breaking. The single-well model is extended by
including rocking motion of the well, which gives rise to Rabi oscillations
between the fundamental and dipole modes. Analysis of the 2D single-well
setting gives rise to stable modes in the form of ordinary dipoles,
vortex-antivortex dipoles (VADs), and vortex triangles (VTs), which may be
considered as produced by spontaneous breaking of the axial symmetry. The
consideration of the DW configuration in 2D reveals diverse types of modes
built of components trapped in the two wells, which may be fundamental
states and vortices with topological charges $m=1$ and $2$, as well as VADs
(with $m=0$) and VTs (with $m=2$).
\end{abstract}

\maketitle

\section{Introduction and the model}

\label{sec:Introduction} It is commonly known that in spaces of different
dimension, $D$, bright solitons of scalar complex field $\phi $ may be
supported by balance between the focusing nonlinearity and diffraction \cite%
{Agrawal}. The solitons created by focusing term $|\phi |^{2m}\phi $ in the
underlying nonlinear Schr\"{o}dinger/Gross-Pitaevskii equation (NLSE/GPE)
are unstable in the case when the same setting gives rise to the collapse,
i.e., at $mD\geq 2$, according to the Talanov's criterion \cite{Talanov}. In
particular, the Townes' solitons \cite{Townes}, which form degenerate
families with the norm that does not depend on the propagation constant, are
destabilized by the critical collapse (corresponding to $mD=2$) in the 2D
space with the cubic nonlinearity, $m=1$ \cite{Berge'}, and in the 1D space
with the quintic self-focusing, $m=2$ \cite{Mario}.

In the absence of linear trapping potentials, an effective nonlinear
potential (alias \textit{pseudopotential} \cite{pseudo}) for optical waves
in photonic media, and for matter waves in a Bose-Einstein condensate (BEC)
\cite{BEC} can be induced by the spatial modulation of the local strength of
the cubic nonlinearity, accounted for by coefficient $\sigma (\mathbf{r})$
\cite{RMP}:
\begin{equation}
i\psi _{z}=-\frac{1}{2}\nabla ^{2}\psi +\sigma (\mathbf{r})|\psi |^{2}\psi ,
\label{NLSE}
\end{equation}%
the respective Hamiltonian being%
\begin{equation}
H=\frac{1}{2}\int \int \left[ \left\vert \nabla \psi \right\vert ^{2}+\sigma
(x)|\phi (x)|^{4}\right] dx.  \label{Ham}
\end{equation}%
In optics, $\psi $ is the scaled amplitude of the guided electromagnetic
field, $z$ is the propagation distance and the set of transverse coordinates
is $\mathbf{r}=(x,y)$ or $x$, in the 1D and 2D settings, respectively. In
terms of the GPE, $z$ is the scaled time, and the scattering length, which
is proportional to $\sigma (\mathbf{r})$ in Eq. (\ref{NLSE}), can be
modified by means of the Feshbach resonance in external magnetic or optical
fields \cite{FR-Randy}-\cite{FR-Tom}. The spatial modulation of the
scattering length in atomic condensates was experimentally demonstrated on
the submicron scale \cite{experiment-inhom-Feshbach}. The necessary spatial
profile may be also induced as an averaged spatial pattern ``painted" by a
fast moving laser beam \cite{painting}, or by an optical-flux lattice \cite%
{Cooper}. Another approach makes use of an appropriate magnetic lattice,
into which the condensate is loaded \cite{magn-latt}, or of magnetic-field
concentrators \cite{concentrator}. In optics, the modulation of the
nonlinearity strength can be achieved by means of inhomogeneous doping of
the waveguide with nonlinearity-enhancing impurities \cite{Kip}.
Alternatively, one can use a uniform dopant density, onto which an external
field imposes an inhomogeneous distribution of detuning from the respective
two-photon resonance.

The nonlinear potential induced by the modulation of the self-focusing
nonlinearity, which corresponds to $\sigma (\mathbf{r})<0$ in Eq. (\ref{NLSE}%
), can readily support stable solitons in the 1D geometry, as has been
demonstrated in various settings \cite{1D-attr,Thawatchai}. On the other
hand, in the 2D geometry, stable fundamental solitons can be maintained only
by modulation profiles with sharp edges, all vortex solitons being unstable
\cite{2D-attr,2D-Thawatchai,RMP}.

Recently, an alternative scheme was theoretically elaborated, based on the
defocusing-nonlinearity strength growing at $r\rightarrow \infty $ at any
rate exceeding $r^{D}$ \cite{Barcelona}-\cite{SR}. This scheme secures
stable self-trapping of a great variety of fundamental and higher-order
solitons, including solitary vortices \cite{Barcelona,Barcelona2,SR}, and
complex 3D modes, such as soliton gyroscopes \cite{gyroscope},
vortex-antivortex hybrids \cite{hybrid}, and \textquotedblleft hopfions"
(vortex tori with intrinsic twist) \cite{Yasha}. Moreover, the scheme was
extended for discrete solitons \cite{discrete}, quantum solitons in the
Bose-Hubbard model \cite{LucaLuca}, and 1D and 2SD settings with the
spatially modulated long-range dipole-dipole repulsive interactions \cite%
{Raymond}.

Stationary solutions to Eq. (\ref{NLSE}) are looked for as
\begin{equation}
\psi (\mathbf{r},z)=\phi (\mathbf{r})e^{i\mu z},  \label{psiphi}
\end{equation}
where $\mu $ determines the propagation constant, in terms of photonic
models, or the chemical potential, $-\mu $, in BEC, and stationary wave
function $\phi $ obeys equation

\begin{equation}
\mu \phi =\frac{1}{2}\nabla ^{2}\phi +\sigma (\mathbf{r})|\phi |^{2}\phi .
\label{StatNLSE}
\end{equation}%
Self-trapped solutions supported by the system are characterized by the norm
(energy flow), defined as
\begin{equation}
N=\int \left\vert \phi (\mathbf{r})\right\vert ^{2}d\mathbf{r}.  \label{Norm}
\end{equation}

We aim to consider two types of nonlinearity-modulation profiles, in the 1D
and 2D geometries alike. The first corresponds to the isotropic single-well
setting, with the steep anti-Gaussian shape, which was introduced in Ref.
\cite{Barcelona2}:

\begin{equation}
\sigma (\mathbf{r})=\exp \left( \alpha r^{2}\right) ,
\label{BasicSingleWellProfile}
\end{equation}%
with constant $\alpha >0$ which determines the width of the well, $\sim
\alpha ^{-1/2}$. Another version of the single-well profile, including a
pre-exponential factor, was also considered in Ref. \cite{Barcelona2}:%
\begin{equation}
\sigma (\mathbf{r})=\left( \sigma _{0}+\frac{\sigma _{2}}{2}r^{2}\right)
\exp \left( \alpha r^{2}\right) .  \label{sophisticated}
\end{equation}%
The main subject of the present work is an anisotropic double-well (DW)
profile, whose 2D form can be defined as a natural extension of its
single-well counterpart, possibly with the addition of a pre-exponential
factor:
\begin{equation}
\sigma (\mathbf{r})=\left\{ \sigma _{0}+\frac{\sigma _{2}}{2}\left[
(|x|-x_{0})^{2}+y^{2}\right] \right\} \exp \left\{ \alpha \left[ (|x|-\beta
x_{0})^{2}+y^{2}\right] \right\}  \label{FullDoubleWellProfile}
\end{equation}%
[cf. Eq. (\ref{sophisticated})], with constants $\sigma _{0,2}\geq 0$, and $%
x_{0}\geq 0$, where $x=\pm x_{0}$ are positions of centers of the two wells.
Different types of the DW structure correspond to $\beta =1$ and $\beta =0$
in Eq. (\ref{FullDoubleWellProfile}) (in the latter case, the bottom of each
well is shifted from $x=\pm x_{0}$ towards $x=0$). A particular DW profile
corresponds to $\sigma _{0}=1,\sigma _{2}=0,\beta =1$ in Eq. (\ref%
{FullDoubleWellProfile}) [cf. Eq. (\ref{BasicSingleWellProfile}) for the
single well]:%
\begin{equation}
\sigma (\mathbf{r})=\exp \left\{ \alpha \left[ (|x|-\beta x_{0})^{2}+y^{2}%
\right] \right\} .  \label{simple}
\end{equation}%
The 1D counterpart of the DW setting based on Eq. (\ref{simple}) is
considered below too.

It is commonly known that the ground state generated by the linear Schr\"{o}%
dinger equation with a DW potential is always symmetric, with respect to the
two potential wells \cite{LL}. The interplay of the linear DW potentials
with uniform self-focusing nonlinearities in models based on the NLSE/GPE
gives rise to the fundamental effect of the \textit{spontaneous symmetry
breaking} (SSB) \cite{book,NewsViews}. In its simplest manifestation, the
SSB implies that the probability to find the particle in one well of the DW
potential is larger than in the other. This also means that another
principle of quantum mechanics, according to which the ground state cannot
be degenerate, is no longer valid in the nonlinear systems, as the SSB
creates a degenerate pair of two mutually symmetric ground states, with the
maximum of the wave function trapped in either potential well. While the
same system admits a symmetric state coexisting with the asymmetric ones, it
no longer represents the ground state above the SSB point, being, unstable
against symmetry-breaking perturbations. In systems with the defocusing
nonlinearity, the ground state remains symmetric and stable, while the SSB
manifests itself in the form of the spontaneous breaking of the spatial
\textit{antisymmetry} of the first excited state.

The SSB was introduced in early works \cite{Davies}, and then developed in
detail in the system modeling the propagation of continuous-wave optical
beams in dual-core nonlinear optical fibers \cite{Snyder}. Depending on the
form of the nonlinearity, it gives rise to symmetry-breaking bifurcations of
the supercritical (alias forward) or subcritical (backward) type \cite{bif}.
The next step in the studies of the SSB phenomenology in dual-core systems
was the detailed consideration of this effect for solitons, described by a
system of linearly coupled partial differential NLSEs \cite{Wabnitz}-\cite%
{Pak}. The transition to asymmetric solitons in this system was predicted by
means of the variational approximation \cite{Pare,Pak} and investigated in a
numerical form \cite{Akhmed,Pak}. The analysis of the SSB in BEC and other
models based on the GPE with the DW potential was initiated in Ref. \cite%
{Milburn}, and later extended for bosonic Josephson junctions \cite{junction}%
-\cite{junction-reviews} and matter-wave solitons \cite{Warsaw}.

Experimentally, the self-trapping of asymmetric states in the BEC of $^{87}$%
Rb atoms loaded into the DW potential, as well as Josephson oscillations in
the same setting, were reported in Ref. \cite{Markus}. The SSB of laser
beams coupled into an effective transverse DW potential created in a
photorefractive medium was demonstrated in Ref. \cite{photo}. A
spontaneously established asymmetric regime of the operation of a
symmetrically coupled pair of lasers was reported too \cite{lasers}.

Here, our main objective is to study the SSB of self-trapped modes supported
by effective nonlinear (pseudo)potentials. Previously, some results for such
settings were reported in Refs. \cite{Thawatchai} and \cite{China-DW}, but
the systematic analysis based on the model with the spatially modulated
strength of the self-defocusing nonlinearity was not developed. Because we
consider the models with the defocusing sign of the nonlinearity, the
respective symmetric ground state is always stable and is not subject to the

SSB, as mentioned above. Therefore, we focus on the SSB featured by
antisymmetric (dipole) modes, in the form of the spontaneous breaking of
their spatial antisymmetry. Another manifestation of the SSB that we address
in this work is spontaneous formation of anisotropic patterns in the 2D
isotropic single-well configuration, a known example of that in usual models
with the uniform nonlinearity being the creation of azimuthons \cite{azi}.

Localized solutions to the 1D version of the stationary equation (\ref%
{StatNLSE}) are obtained in the numerical form by means of the
Newton-Raphson method \cite{Yang}. Symmetric 1D and 2D modes are also
produced in an approximate analytical form by means of the Thomas-Fermi
approximation (TFA). The stability was then studied by adding small
perturbations to the stationary solutions, $e^{i\mu z}\phi _{s}(x)$, in the
form of

\begin{equation}
\psi (x,z)=e^{i\mu z}\left[ \phi _{s}(x)+g(x)e^{-i\lambda t}+f^{\ast
}(x)e^{i\lambda ^{\ast }z}\right] ,  \label{Perturbed_solution}
\end{equation}%
cf. Eq. (\ref{psiphi}), where $g(x)$ and $f(x)$ are eigenmodes of the
infinitesimal perturbation, $\lambda $ is the corresponding eigenfrequency,
which is complex (in particular, imaginary) in the case of instability, and
the asterisk stands for the complex conjugation. Substituting the perturbed
solution, (\ref{Perturbed_solution}), in Eq. (\ref{StatNLSE}) and the
subsequent linearization results in the following eigenvalue problem,

\begin{equation}
\left(
\begin{array}{cc}
\hat{L} & \sigma (x)\left( \phi _{s}(x)\right) ^{2} \\
-\sigma (x)\left( \phi _{s}(x)\right) ^{2} & -\hat{L}%
\end{array}%
\right) \left(
\begin{array}{c}
g(x) \\
f(x)%
\end{array}%
\right) =\lambda \left(
\begin{array}{c}
g(x) \\
f(x)%
\end{array}%
\right) ,  \label{Eigenvalue_problem}
\end{equation}%
with $\hat{L}=\mu -(1/2)d^{2}/dx^{2}+2\sigma (x)\left( \phi _{s}(x)\right)
^{2}$. This problem can be solved using the basic finite-difference scheme,
thus finding the set of eigenfrequencies $\lambda $ and determining the
stability of the underlying solution. In addition, direct numerical
simulations of initially perturbed solutions are performed by means of the
pseudospectral split-step Fourier method, to verify the predicted stability,
as well as to explore the evolution of unstable states. The direct
simulations are run with absorbers installed at edges of the computation
domain. In the 2D setting, stationary solutions are obtained by dint of the
modified squared-operator method introduced in Ref. \cite{Lakoba} (see also
book \cite{Yang}), and the stability is then investigated primarily through
direct simulations.

The rest of the paper is organized as follows. The 1D DW setting and the SSB
effect in it are considered in Section II. The dynamics of trapped modes in
a single \textit{rocking} (periodically moving) 1D well is addressed in
Section III. Various stable and unstable states trapped in the single- and
double-well structures in 2D are considered in Sections IV and V,
respectively. The paper is concluded by Section VI.

\section{The one-dimensional setting}

\label{sec:1DResults}

\subsection{Basic 1D states: Symmetric, anti-symmetric, and asymmetric
solitons}

\label{sec:Basic1DSolutions}

\begin{figure}[tbp]
\subfigure[]{\includegraphics[width=2.2in]{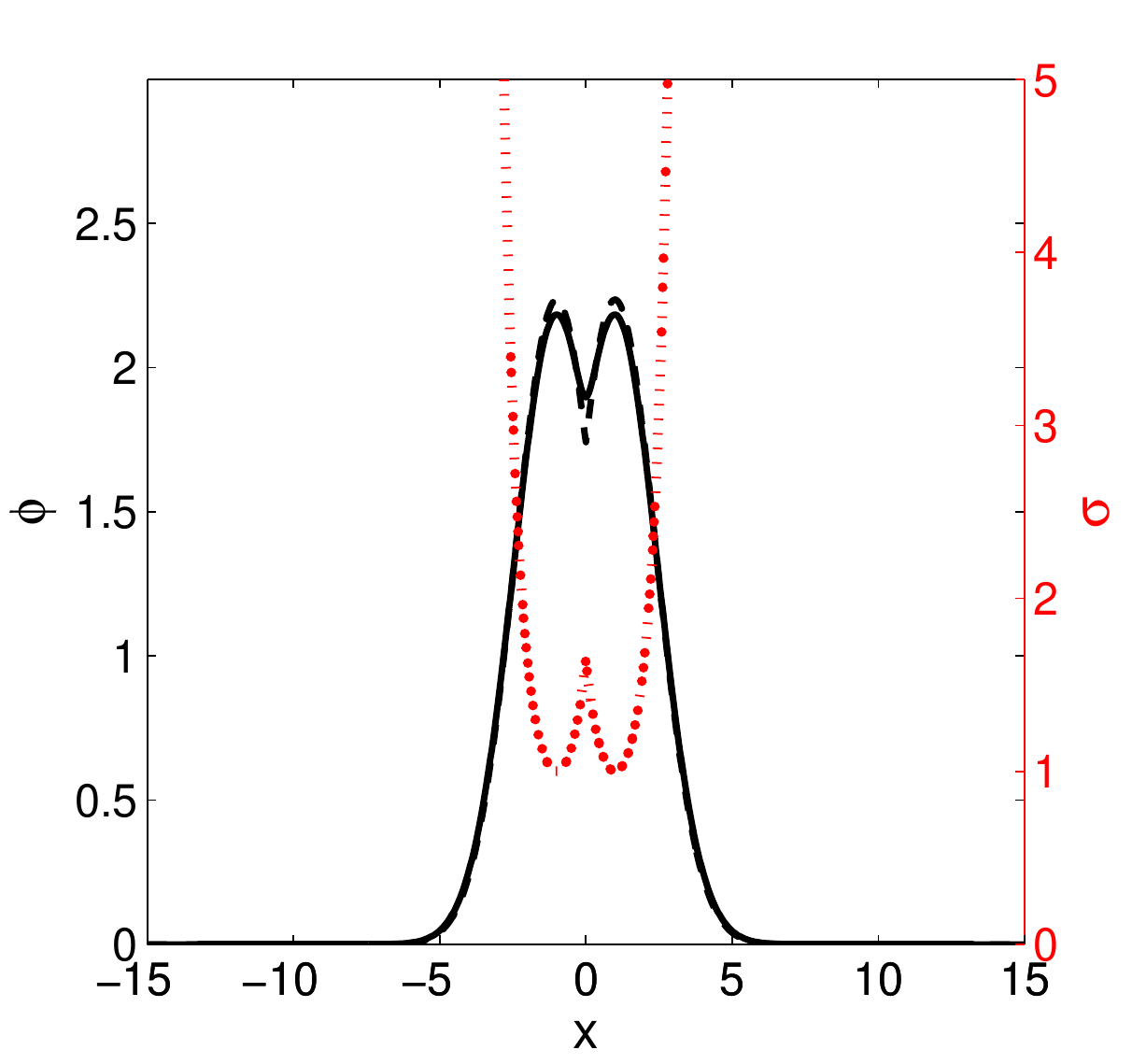}
\label{1D2WellSymmetricProfileMu5}}
\subfigure[]{\includegraphics[width=2.2in]{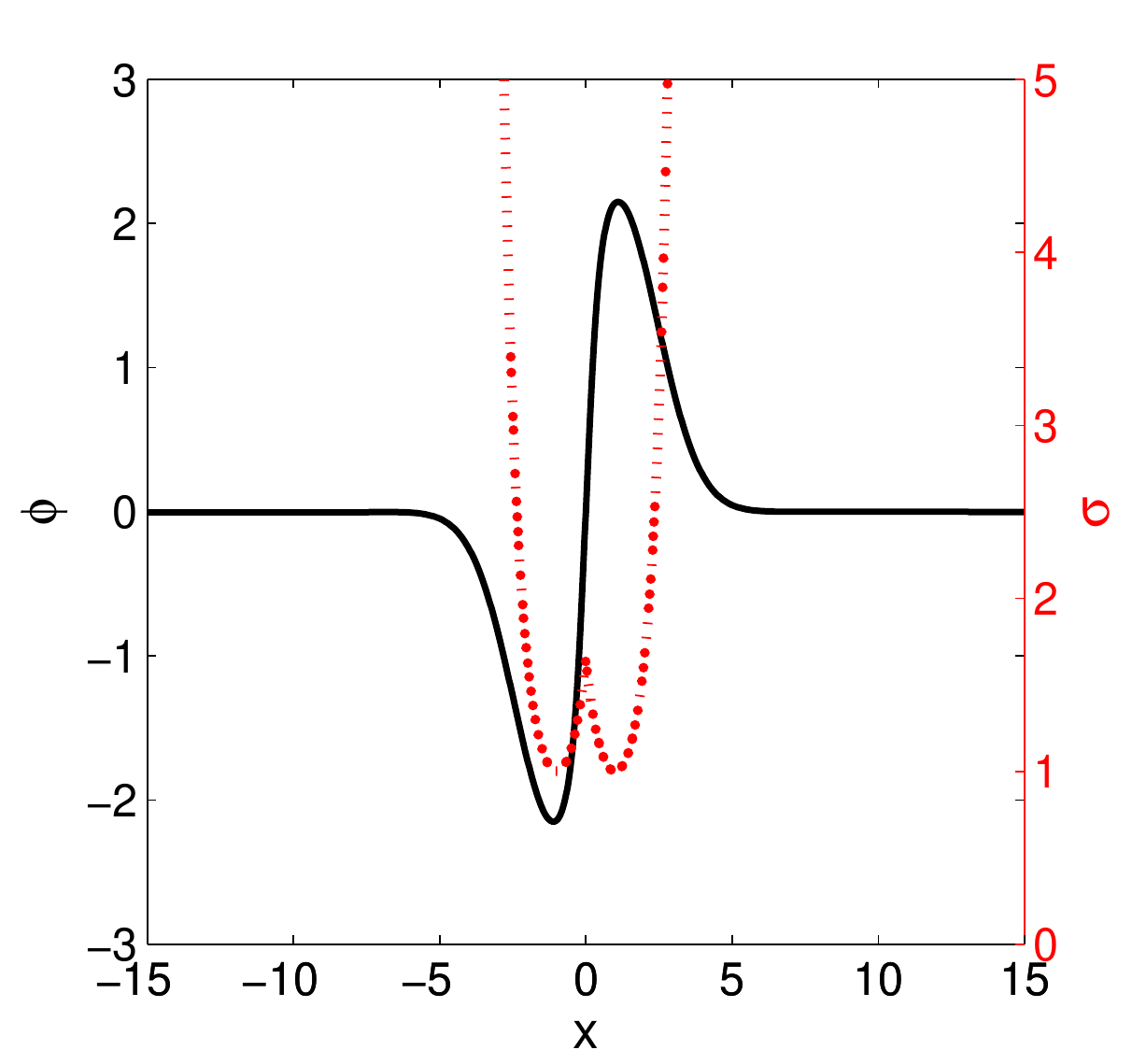}
\label{1D2WellAntisymmetricProfileMu5}}
\subfigure[]{\includegraphics[width=2.2in]{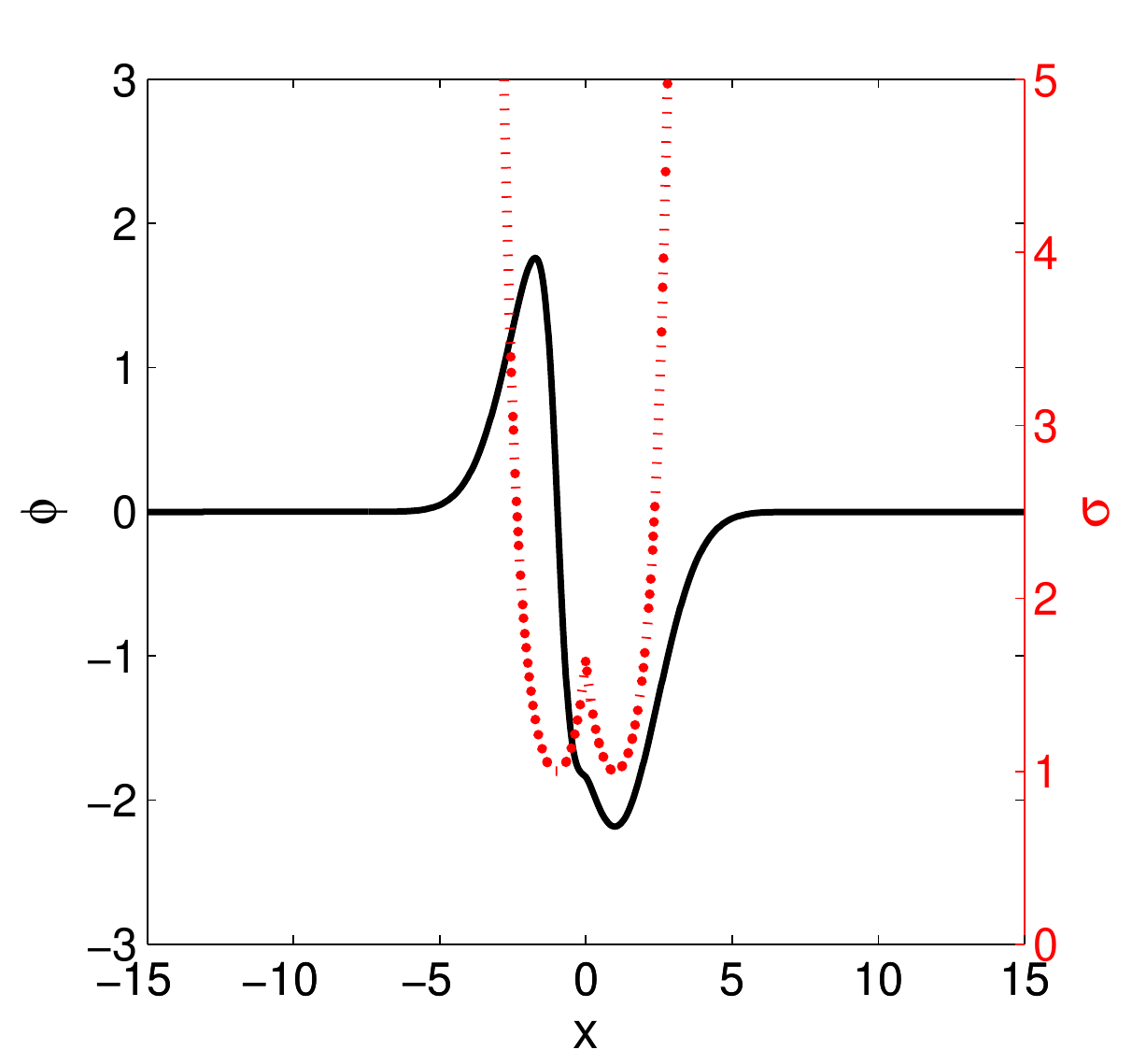}
\label{1D2WellAsymmetricProfileMu5}}
\caption{(Color online) Examples of numerically found 1D symmetric (a),
antisymmetric (b) and asymmetric (c) states, for $\protect\mu =5$ and the
double-well nonlinearity-modulation profile given by Eq. (\protect\ref%
{1Dsimple}) (depicted by the red dotted line), with $\protect\alpha =0.5$
and $x_{0}=1$. The dashed line in (a), which almost overlaps with the
numerically generated continuous profile, displays the respective TFA\
(Thomas-Fermi approximation), as given by Eq. (\protect\ref{1DTFA}).}
\label{1dDoubleWellBasicSolutions}
\end{figure}

The 1D version of model (\ref{StatNLSE}), with the simplest DW modulation
profile taken as the 1D variant of Eq. (\ref{simple}),%
\begin{equation}
\sigma (x)=\exp \left[ \alpha \left( \left( |x|-x_{0}\right) ^{2}\right) %
\right] ,  \label{1Dsimple}
\end{equation}%
gives rise to basic families of symmetric, antisymmetric and asymmetric
(antisymmetry-breaking) states. Typical examples of shapes of these three
types are presented in Fig. \ref{1dDoubleWellBasicSolutions}, for $\alpha
=0.5$, $x_{0}=1$ and $\mu =5$.

The symmetric solutions, which represent the ground state of the model (see
below), can be analytically approximated \ by means of the TFA, which was
efficiently applied to the description of ground states in 1D, 2D, and 3D
versions of the model with the single-well structure \cite{Barcelona}-\cite%
{Yasha}, \cite{Raymond}. The TFA neglects the kinetic-energy term (the
second-order derivative) in the 1D version of Eq. (\ref{StatNLSE}), with $%
\sigma (x)$ substituted by expression (\ref{1Dsimple}):%
\begin{equation}
\phi _{\mathrm{TFA}}(x)=\sqrt{\mu }\exp \left[ -\left( \alpha /2\right)
\left( |x|-x_{0}\right) ^{2}\right] .  \label{1DTFA}
\end{equation}%
The respective approximation for the norm of the symmetric modes is%
\begin{equation}
N_{\mathrm{TFA}}(\mu )=\sqrt{\pi /\alpha }\mu \left[ 1+\mathrm{erf}\left(
\sqrt{\alpha }x_{0}\right) \right] ,  \label{1DNTFA}
\end{equation}%
where $\mathrm{erf}$ is the standard error function. The comparison of the
TFA profile with its numerically generated counterpart is displayed in Fig. %
\ref{1dDoubleWellBasicSolutions}.

The stability investigation was conducted for all the three basic families,
by numerically solving the eigenvalue problem based on Eq. (\ref%
{Eigenvalue_problem}), at different values of $\mu $ (i.e., different norms $%
N$), and different values of parameters $x_{0}$ and $\alpha $. It has been
found that the symmetric family, which does not undergo any bifurcation, is
completely stable.

The antisymmetric solutions are stable for low values of $\mu $
(sufficiently small $N$). Increasing $\mu $, one hits a bifurcation point,
above which the antisymmetric state loses its stability and a new asymmetric
(antisymmetry-breaking) branch emerges. This asymmetric branch may be
stable, at least partially, depending on values of $x_{0}$ and $\alpha $.

\begin{figure}[tbp]
\includegraphics[width=3.2in]{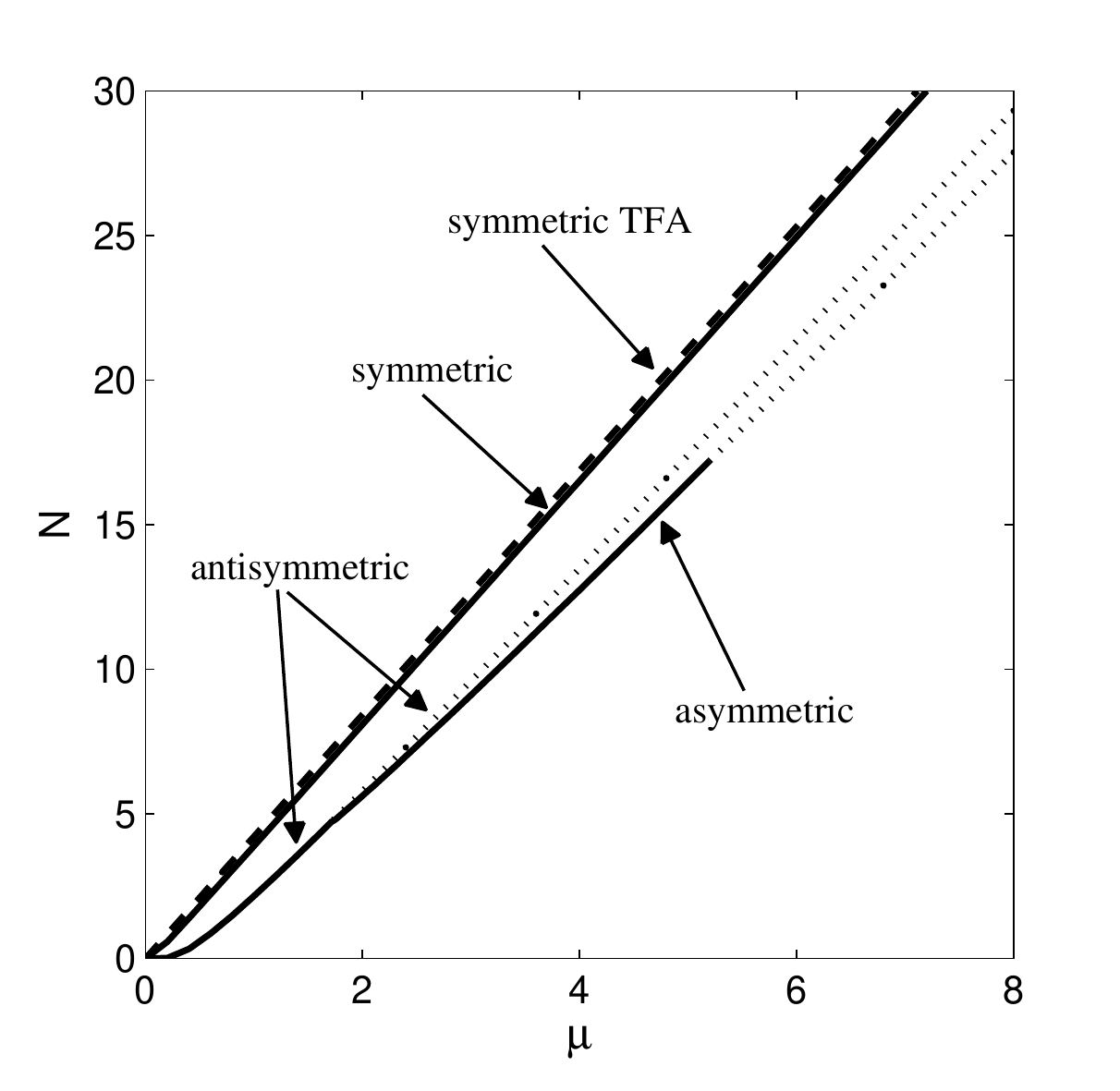}
\caption{Branches $N(\protect\mu )$ for symmetric, antisymmetric and
asymmetric modes trapped in the 1D double-well nonlinear potential, for $%
x_{0}=1$ and $\protect\alpha =0.5$. Here and in similar figures following
below, stable and unstable solutions are indicated by continuous and dotted
lines, respectively. The dashed line shows the TFA, as produced by Eq. (%
\protect\ref{1DNTFA}) for the symmetric modes. This analytical approximation
readily explains the nearly linear form of $N(\protect\mu )$.}
\label{1D2WellBasicNVsMu}
\end{figure}

Figure \ref{1D2WellBasicNVsMu} demonstrates the stable symmetric branch, as
well as the antisymmetric/asymmetric bifurcation scenario, for $x_{0}=1$ and
$\alpha =0.5$. As shown in this example, and is true in the general case
too, for all values of $x_{0}$ and $\alpha $ examined, the bifurcation is of
the \textit{supercritical} type, which means that the asymmetric solutions
are stable immediately after the bifurcation point. Figure \ref%
{1D2WellBasicNVsMu} shows that bistability exists between the symmetric and
antisymmetric modes, and between the symmetric and asymmetric ones, below
and above the bifurcation, respectively.

\begin{figure}[tbp]
\subfigure[]{\includegraphics[width=3in]{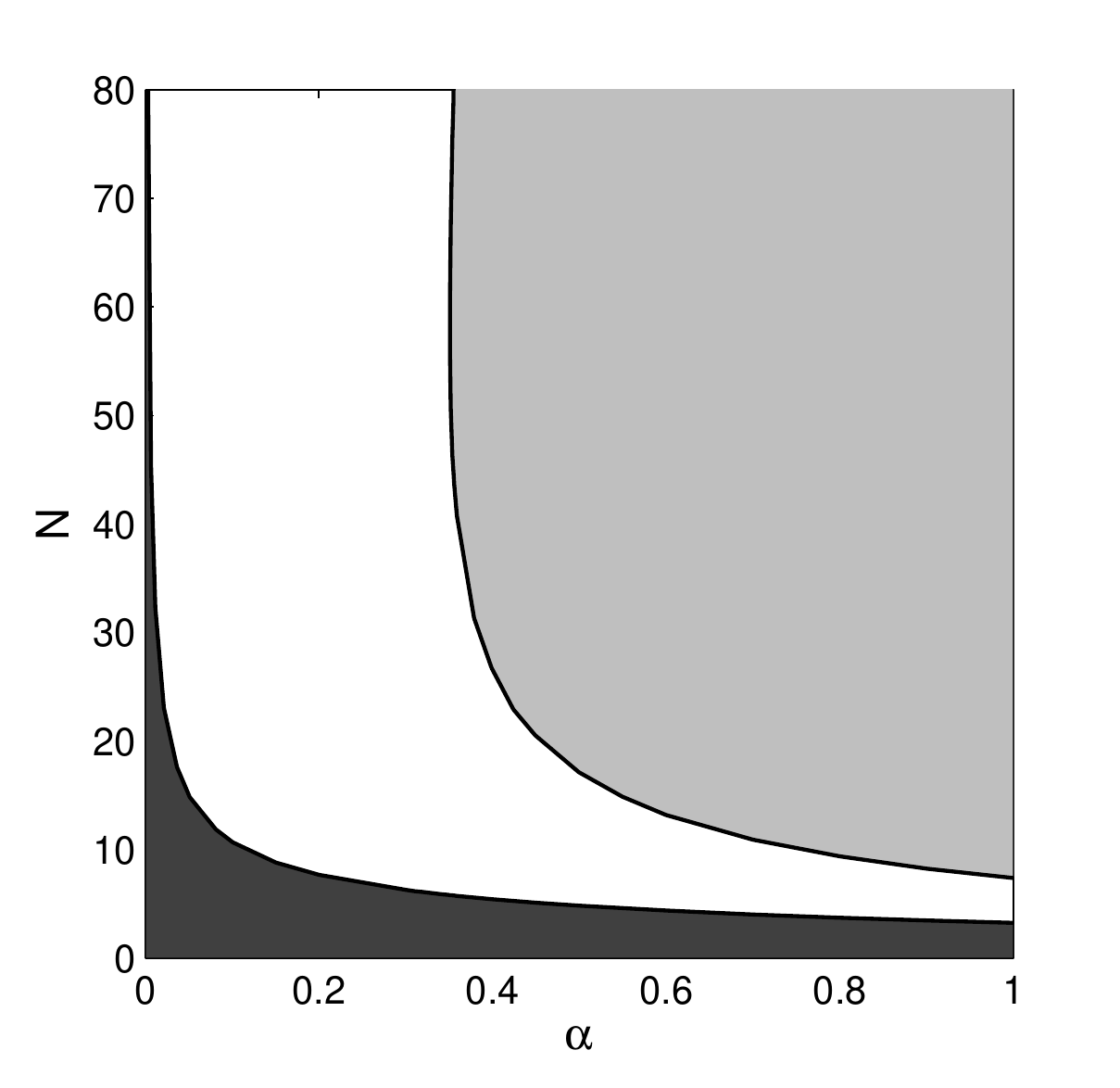}
\label{1D2WellAsymmetricNVsAlpha}}
\caption{The stability diagram for the 1D asymmetric modes, in the $\left(
\protect\alpha ,N\right) $ plane, at fixed $x_{0}=1$. Stable solutions exist
in the white region. Asymmetric solutions do not exist in the dark area, at
the bottom and left side of the diagram (the border between the white and
dark regions is the line of the antisymmetry-breaking bifurcation for the
antisymmetric branch). Unstable asymmetric solutions exist in the light-gray
region. }
\label{1D2WellAsymmetric}
\end{figure}

In the case shown in Fig. \ref{1D2WellBasicNVsMu}, the asymmetric branch is
stable in a limited region, starting from the bifurcation point (at $\mu
=1.719$, $N=4.748$) and up to $\mu =5.206$, $N=17.21$. The stability domain
strongly depends on $\alpha $, as shown in Fig. \ref{1D2WellAsymmetric},
which displays stability/instability domains for the asymmetric states in
the $\left( \alpha ,N\right) $ plane for $x_{0}=1$ [this may be fixed in Eq.
(\ref{StatNLSE}) by means of obvious rescaling]. In particular, for $\alpha
<0.351$ the asymmetric states are stable in their \emph{entire existence
region}.

\begin{figure}[tbp]
\subfigure[]{\includegraphics[width=2.8in]{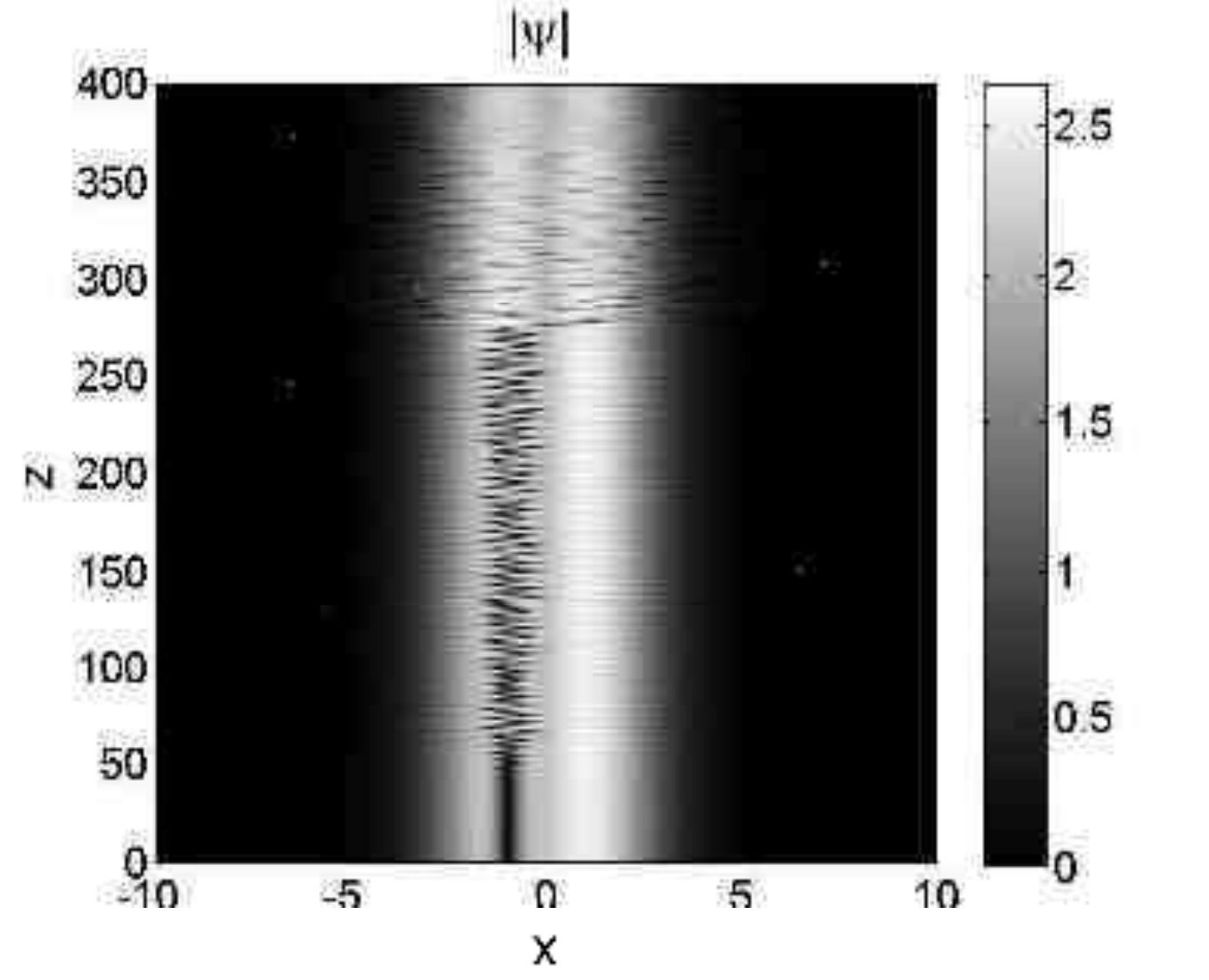}
\label{1D2WellAsymmetricEvolutionMu6p5}}
\subfigure[]{\includegraphics[width=2.8in]{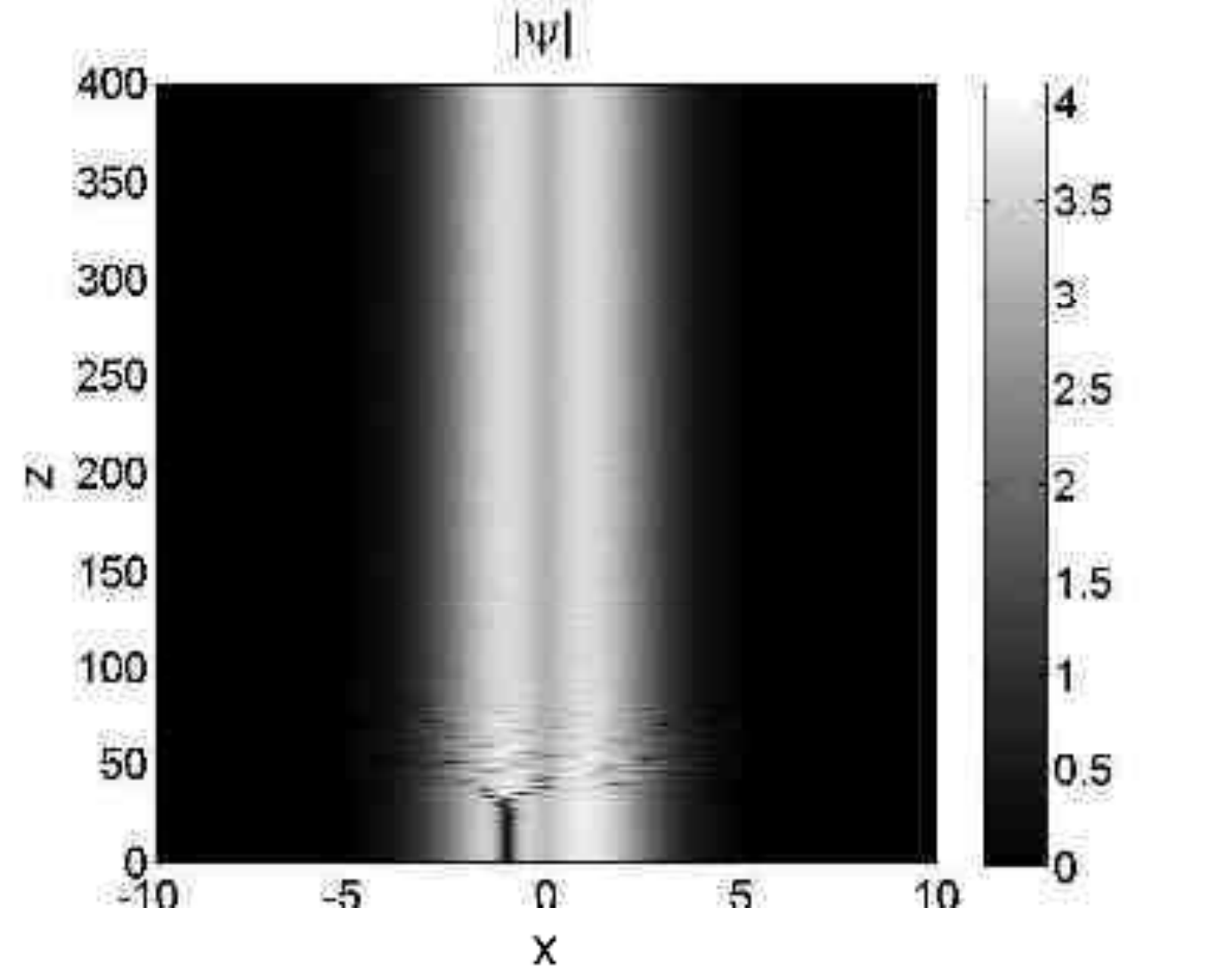}
\label{1D2WellAsymmetricEvolutionMu15}}
\caption{(Color online) The evolution of unstable 1D asymmetric states, at $x_{0}=1$, $%
\protect\alpha =0.5$, into stable symmetric modes. (a) A conspicuous
interval of the oscillatory instability is observed at relatively low values
of $\protect\mu $ (or $N$), here for $\protect\mu =6.5$. (b) For large $%
\protect\mu $ (here, for $\protect\mu =15$). the transition to the symmetric
state is much faster. }
\label{1D2WellAsymmetricEvolution}
\end{figure}

Direct simulations, for unstable asymmetric states and unstable
antisymmetric ones, demonstrate that they develop an oscillatory instability
\ and eventually converge to stable symmetric modes. The linear stability
analysis shows that the instability of asymmetric states (when they are
unstable) are accounted for by complex eigenvalues, while the unstable
antisymmetric states are characterized by imaginary eigenvalues. With the
increase of $\mu $ (i.e., the increase of the norm), the duration of the
intermediate oscillatory evolution shrinks, which is related to the fact
that the imaginary part of the stability eigenvalues (i.e., the instability
growth rate) increases with the norm. Examples of the evolution of unstable
asymmetric states are shown in Fig.~\ref{1D2WellAsymmetricEvolution}, for $%
x_{0}=1$, $\alpha =0.5$, and $\mu =6.5$ or $15$. In these examples, the
interval of the oscillatory behavior shrinks from $50<z<275$ at $\mu =6.5$ ($%
N=22.12$), to virtually no oscillations at $\mu =15$ ($N=55.26$). Similar
results were obtained for unstable antisymmetric solutions, see Fig.~\ref%
{1D2WellAntisymmetricEvolution}. In this case, the interval of the
oscillations shrinks from $20<z<215$ at $\mu =6.5$ ($N=23.34$) to $10<z<75$
at $\mu =15$ ($N=57.57$).

\begin{figure}[tbp]
\subfigure[]{\includegraphics[width=2.8in]{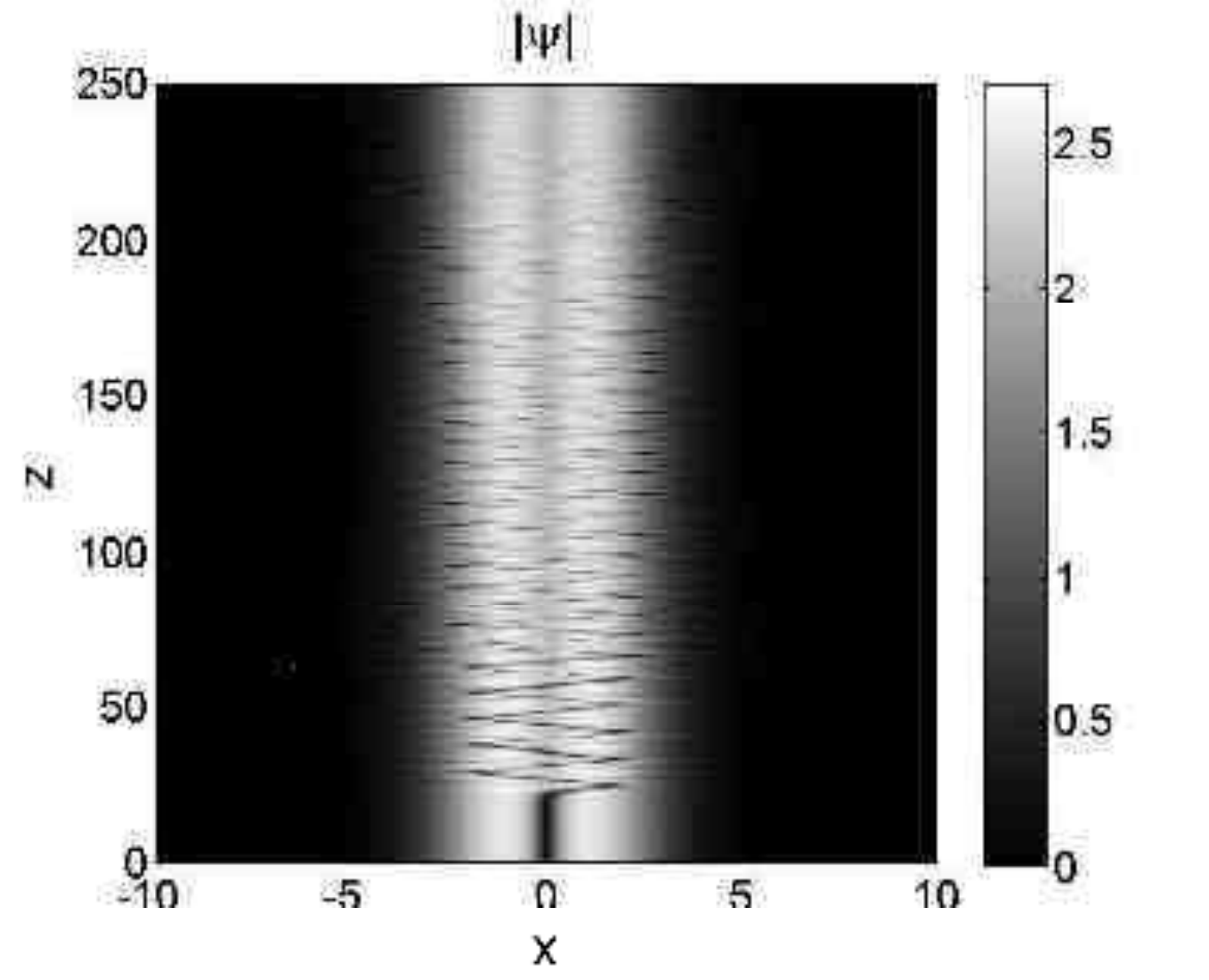}
\label{1D2WellAntisymmetricEvolutionMu6p5}}
\subfigure[]{\includegraphics[width=2.8in]{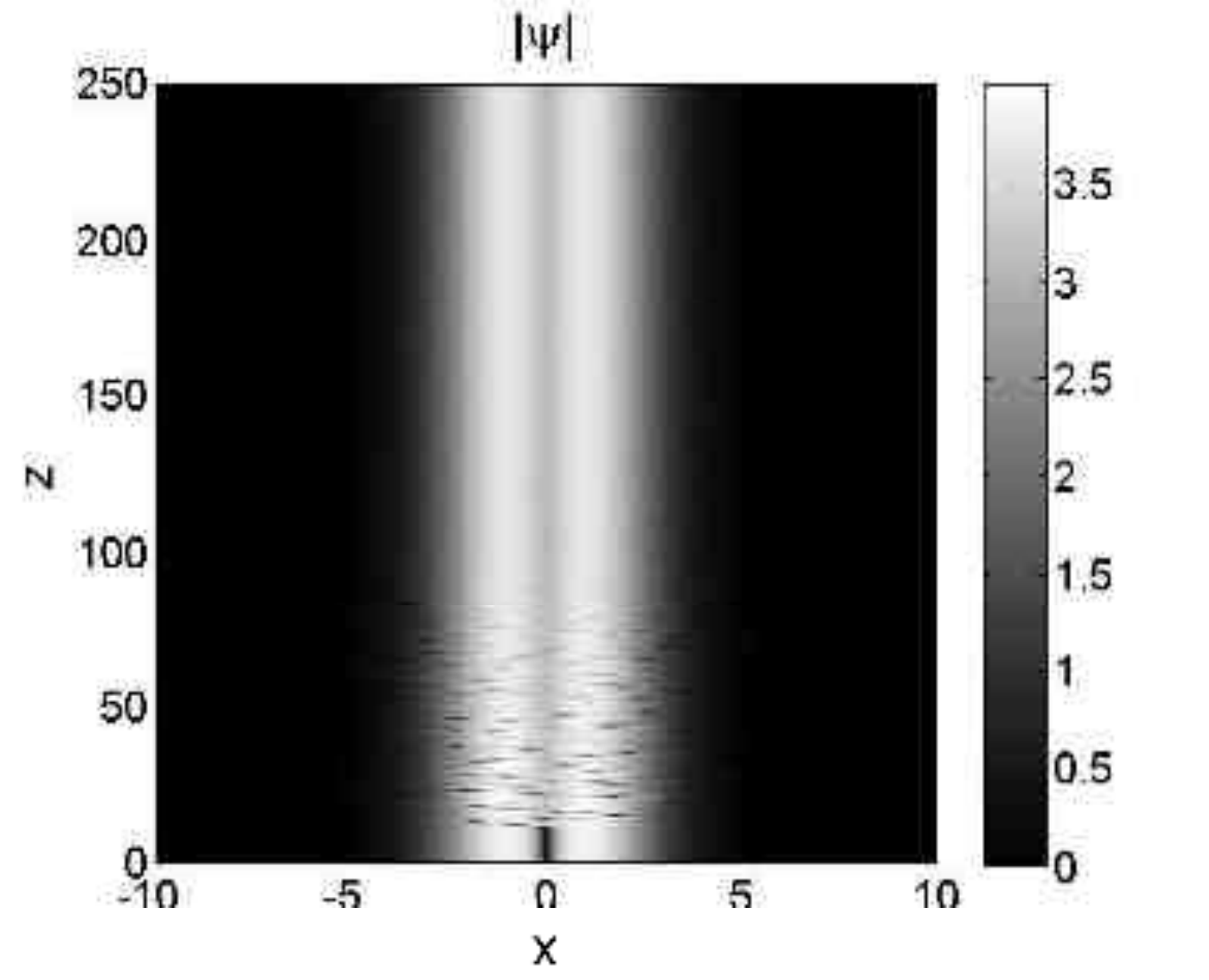}
\label{1D2WellAntisymmetricEvolutionMu15}}
\caption{(Color online) The evolution of unstable 1D antisymmetric states into stable
symmetric ones, at $x_{0}=1$ and $\protect\alpha =0.5$: (a) $\protect\mu %
=6.5 $; (b) $\protect\mu =15$.}
\label{1D2WellAntisymmetricEvolution}
\end{figure}

In fact, the dynamics observed in Figs. \ref{1D2WellAsymmetricEvolution}(a)
and \ref{1D2WellAntisymmetricEvolution}(a) may be regarded as a
manifestation of Josephson oscillations in the underlying bosonic coupler
\cite{junction,Markus,junction2,junction-reviews}.

\subsection{Novel 1D modes: excited symmetric and composite asymmetric modes}

\label{sec:Novel1DSolutions} 
\begin{figure}[tbp]
\subfigure[]{\includegraphics[width=2.8in]{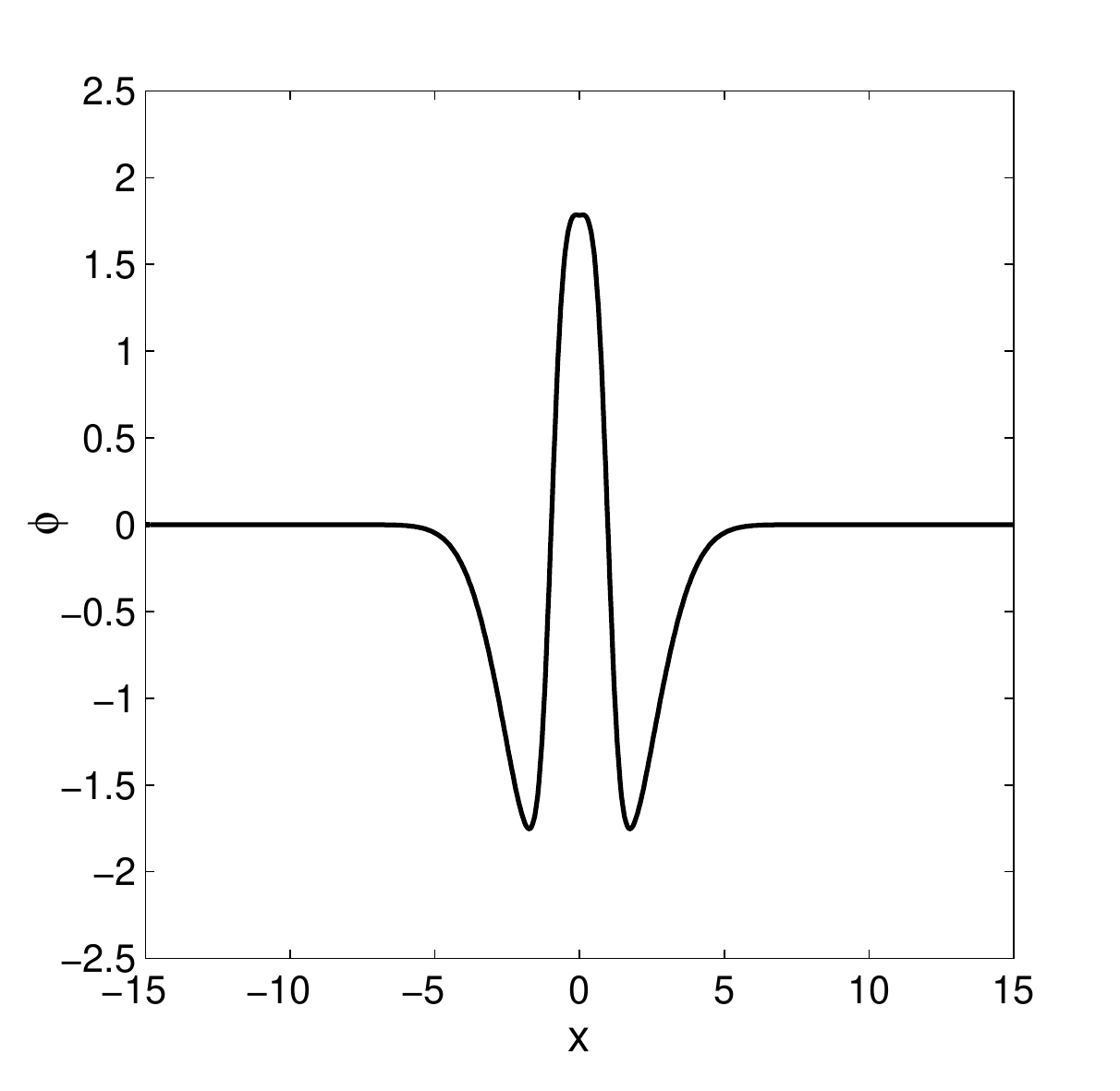}
\label{1D2WellFirstExcitedStateMu5}}
\subfigure[]{\includegraphics[width=2.8in]{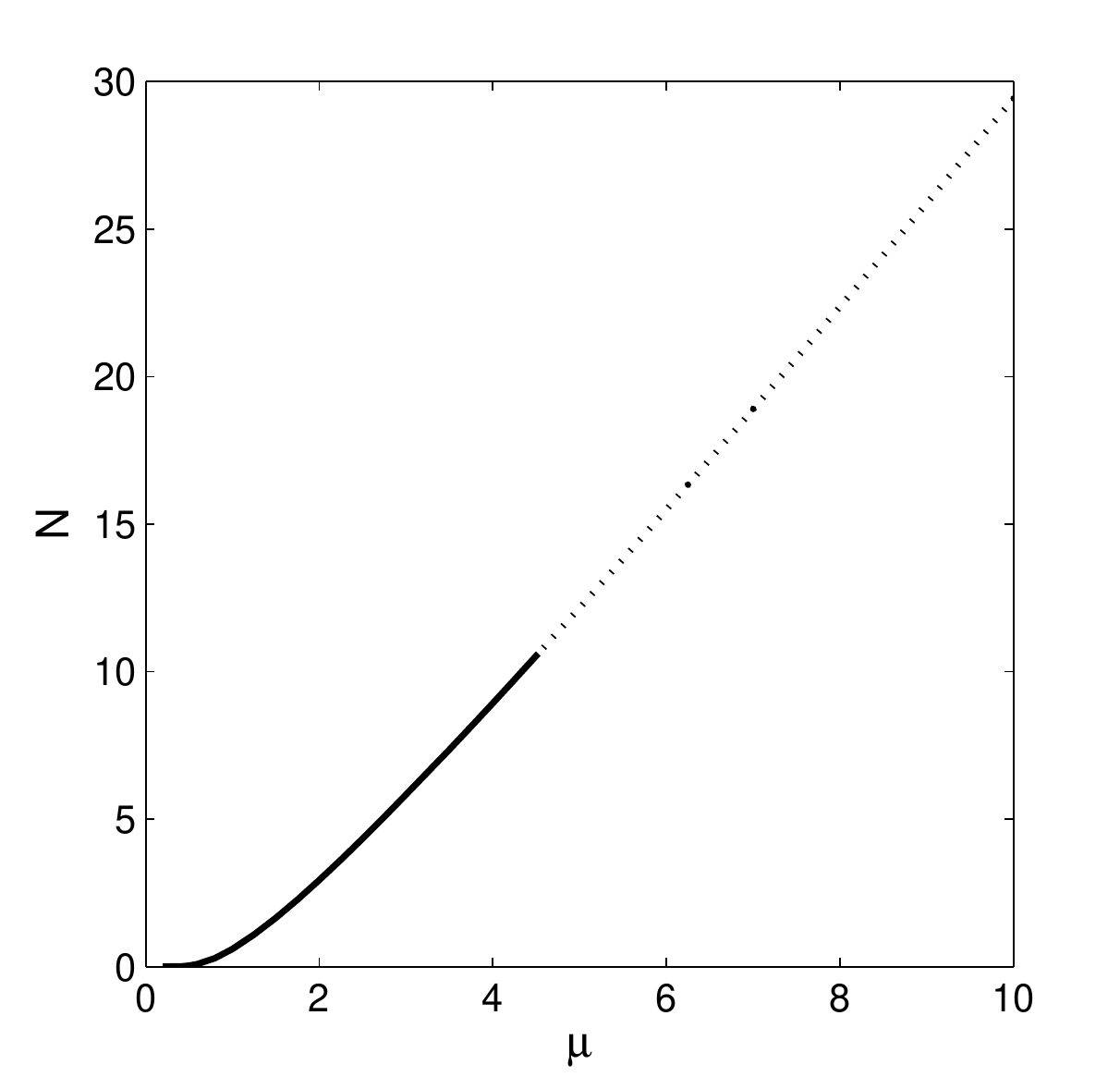}
\label{1D2WellFirstExcitedStateNVsMu}}
\caption{(a) A typical example of the first excited 1D symmetric state, for $%
x_{0}=1$, $\protect\mu =5$ and $\protect\alpha =0.5$. (b) The respective
solution branch $N(\protect\mu )$, for $x_{0}=1$ and $\protect\alpha =0.5$.
It is stable at $\protect\mu <4.523$, i.e., $N<10.6059$.}
\label{1D2WellFirstExcitedState}
\end{figure}

The numerical analysis has revealed several new
types of 1D states. One of them is the first excited symmetric state, which
is shown in Fig. \ref{1D2WellFirstExcitedStateMu5} for $x_{0}=1$, $\mu =5$
and $\alpha =0.5$. The classification of this state is based on the
consideration of its energy, see below. This solution is actually a
composition of two mutually reversed dipole states, centered in each well
(this is better seen when the wells are set farther apart). The stability
diagram for this type of the solutions is presented in Fig. \ref%
{1D2WellFirstExcitedStateNVsMu}. Note that (completely stable) excited
symmetric states with two nodes ($k=2$) were also found in the model of the
same type, but with a single-well shape of the local-nonlinearity modulation
\cite{Barcelona2}.

Similar to the asymmetric mode, the stability of the first excited symmetric
state strongly depends on $x_{0}$ and $\alpha $, see Fig. \ref%
{1D2WellFirstExcitedStateNVsAlpha}, the instability being always accounted
for by complex eigenvalues. An example of the evolution of an unstable
solution of this type is shown in Fig. \ref%
{1D2WellFirstExcitedStateUnstableMu7}. Like the unstable asymmetric and
antisymmetric states presented above, it converges to a stable symmetric
solution.

\begin{figure}[tbp]
\subfigure[]{\includegraphics[width=2.8in]{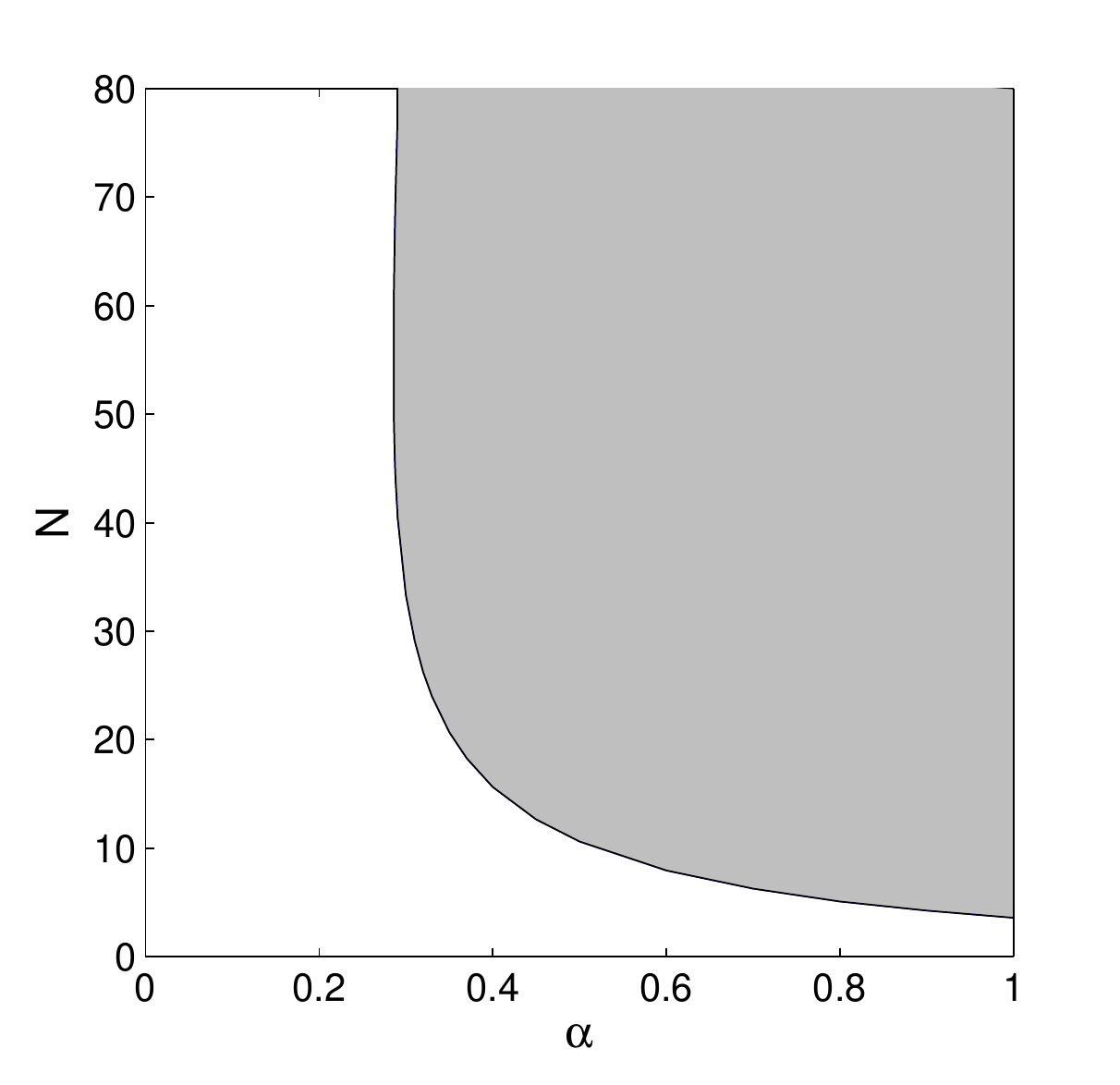}
\label{1D2WellFirstExcitedStateNVsAlpha}}
\subfigure[]{\includegraphics[width=2.8in]{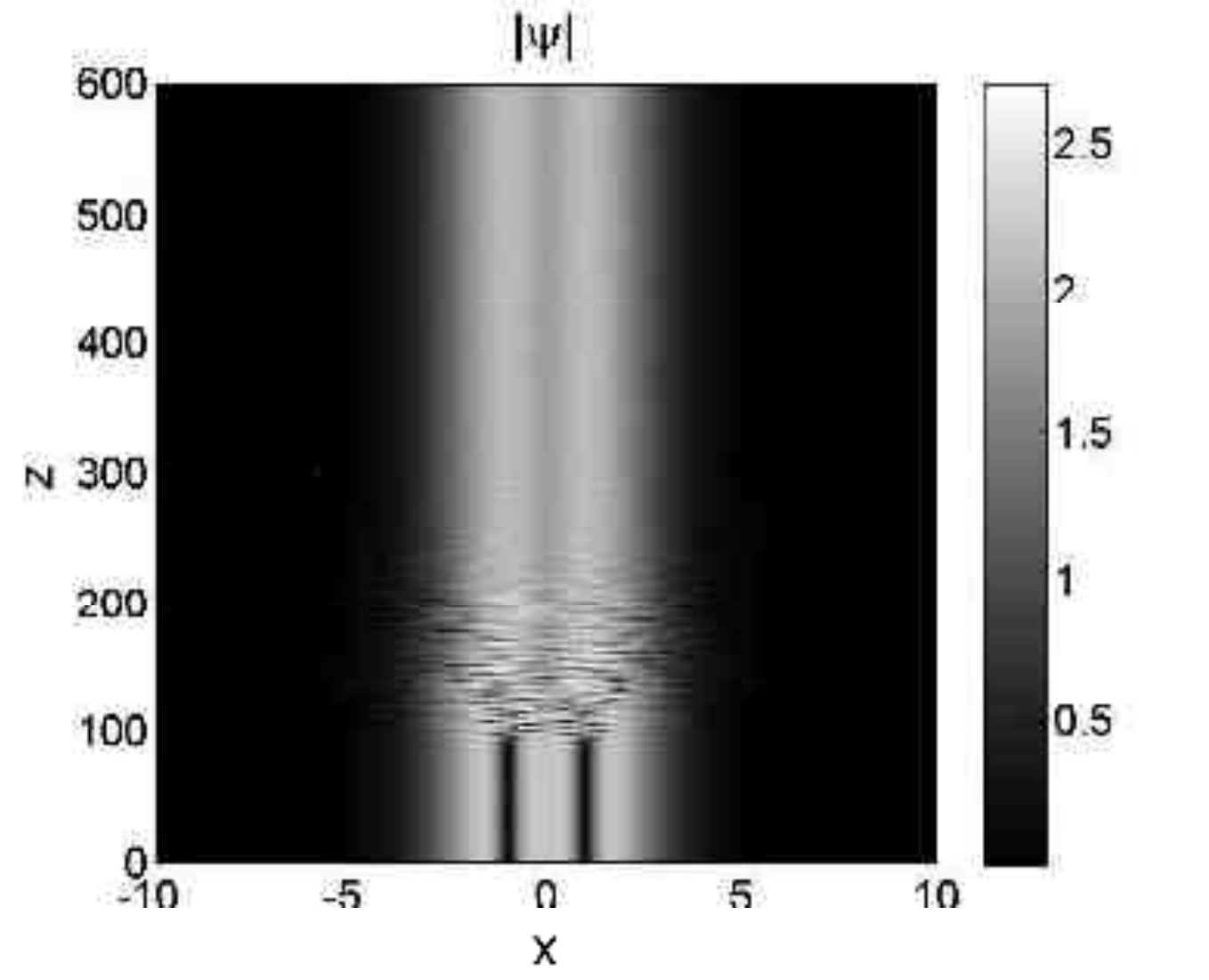}
\label{1D2WellFirstExcitedStateUnstableMu7}}
\caption{(Color online) (a) The stability diagram for the first excited 1D symmetric state
(an example of which is shown in Fig. \protect\ref{1D2WellFirstExcitedState}%
), in the $\left( \protect\alpha ,N\right) $ plane, for $x_{0}=1$. As above,
the white and gray regions represent stable and unstable solutions,
respectively. (b) The evolution of an unstable first excited symmetric
state, for $x_{0}=1$, $\protect\alpha =0.5$ and $\protect\mu =7$.}
\label{1D2WellFirstExcitedStateStability}
\end{figure}

\begin{figure}[tbp]
\subfigure[]{\includegraphics[width=2.8in]{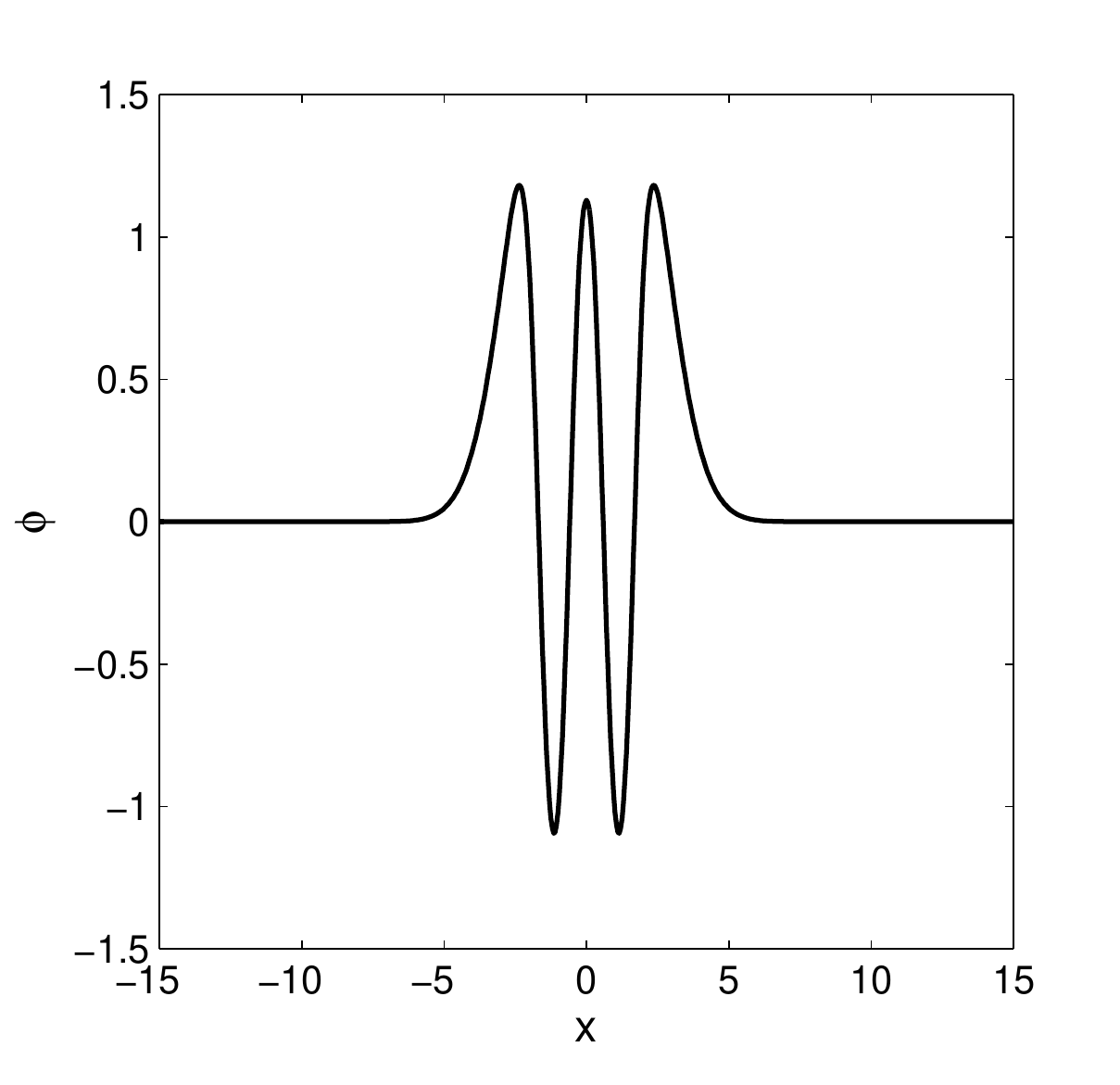}
\label{1D2WellSecondExcitedStateMu5}}
\subfigure[]{\includegraphics[width=2.8in]{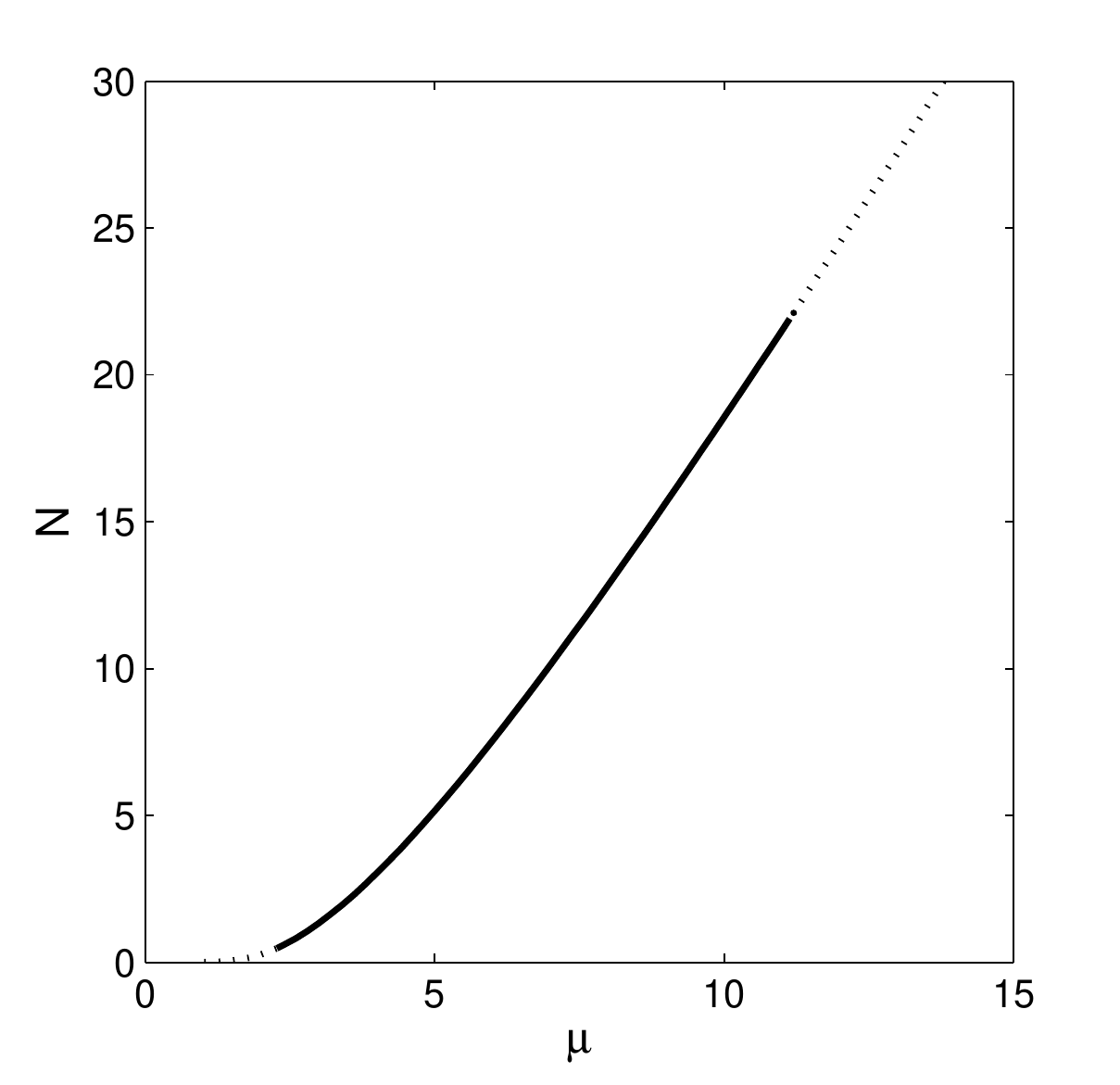}
\label{1D2WellSecondExcitedStateNVsMu}}
\caption{(a) A typical example of the second excited 1D symmetric state, for
$x_{0}=1$, $\protect\mu =5$ and $\protect\alpha =0.5$. (b) The $N(\protect%
\mu )$ curve for these solutions. The stability segment is $2.26<\protect\mu %
<11.13$, i.e., $0.48<N<21.93$.}
\label{1D2WellSecondExcitedState}
\end{figure}

\begin{figure}[tbp]
{\ \includegraphics[width=3.0in]{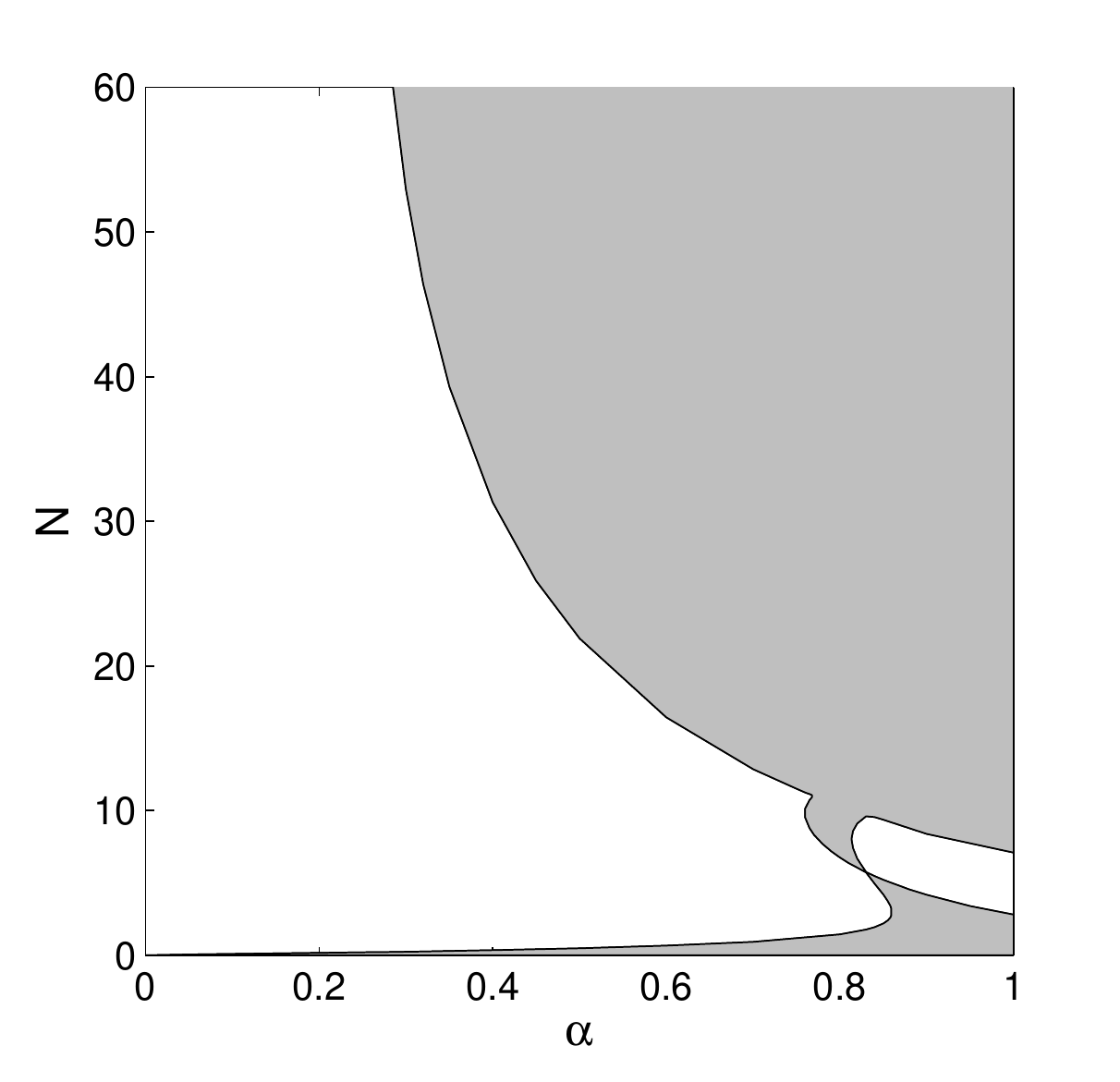}}
\caption{The stability diagram for the second excited 1D symmetric state, in
the $\left( \protect\alpha ,N\right) $ plane, for $x_{0}=1$. The solutions
are stable in white regions.}
\label{1D2WellSecondExcitedStateNVsAlpha}
\end{figure}

Also found was a second excited symmetric state, an example of which is
shown in Fig. \ref{1D2WellSecondExcitedStateMu5} for $x_{0}=1$, $\mu =5$ and
$\alpha =0.5$. This mode is composed of two $k=2$ single-well solutions,
where, as mentioned above, $k$ is the number of nodes (zeroes) of the 1D
mode trapped in the single nonlinear pseudopotential well \cite{Barcelona2}.
Results for this solution family and its stability, for $x_{0}=1$ and $%
\alpha =0.5$, are presented in Fig. \ref{1D2WellSecondExcitedStateNVsMu},
and the dependence of the solution's stability on $\alpha $ is shown in Fig. %
\ref{1D2WellSecondExcitedStateNVsAlpha}

Similar to the unstable antisymmetric and asymmetric states considered
above, the instability of the second excited symmetric states is accounted
for by complex eigenvalues. The development of the instability transforms
them into stable fundamental symmetric states, via a stage of oscillatory
behavior. The latter stage is very long for low values of $\mu $ (or $N$),
shrinking at larger $\mu $ (not shown here in detail).

\begin{figure}[tbp]
\subfigure[]{\includegraphics[width=2.8in]{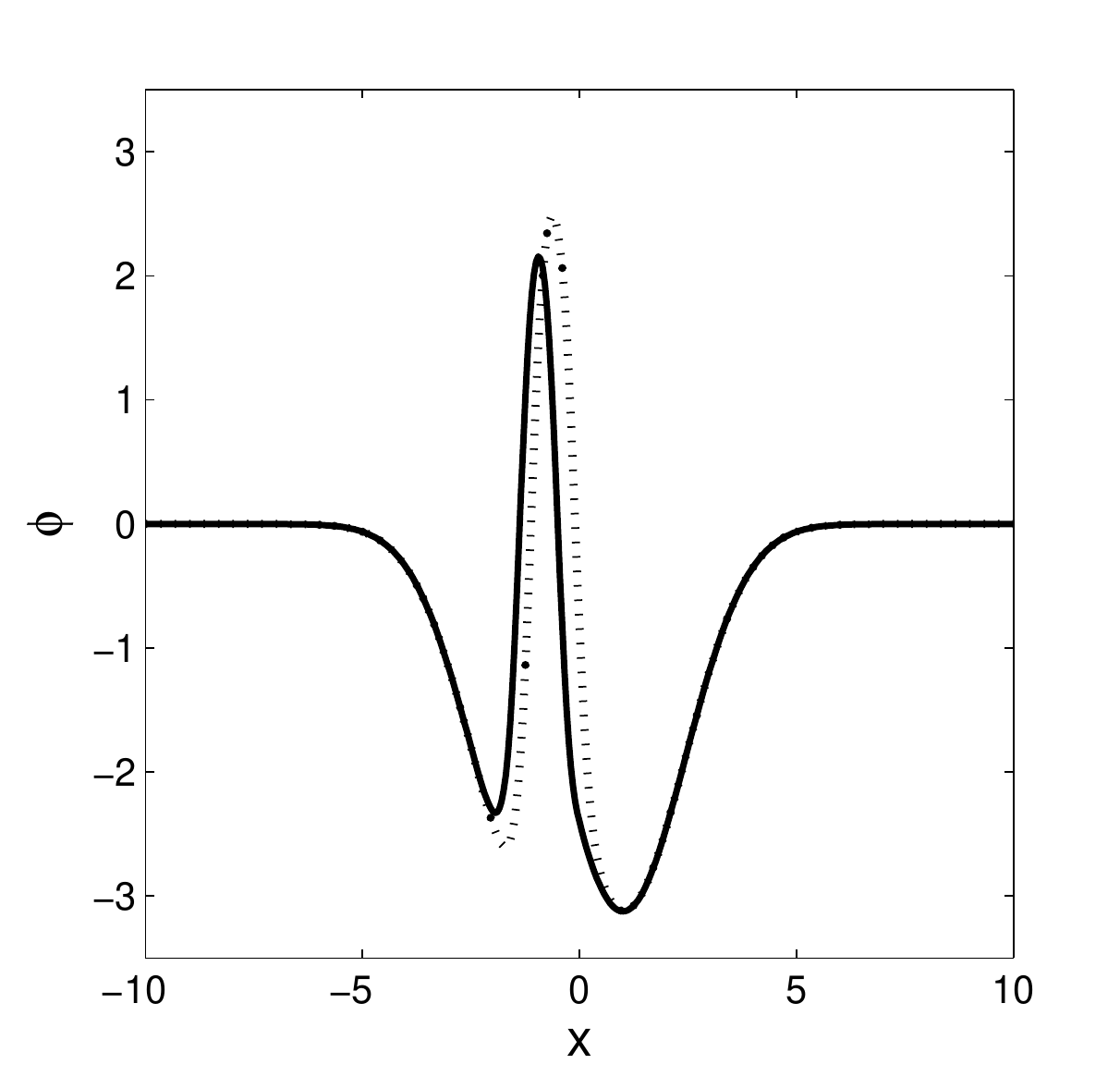}
\label{1D2WellCompositeAsymmetricMu10}}
\subfigure[]{\includegraphics[width=2.8in]{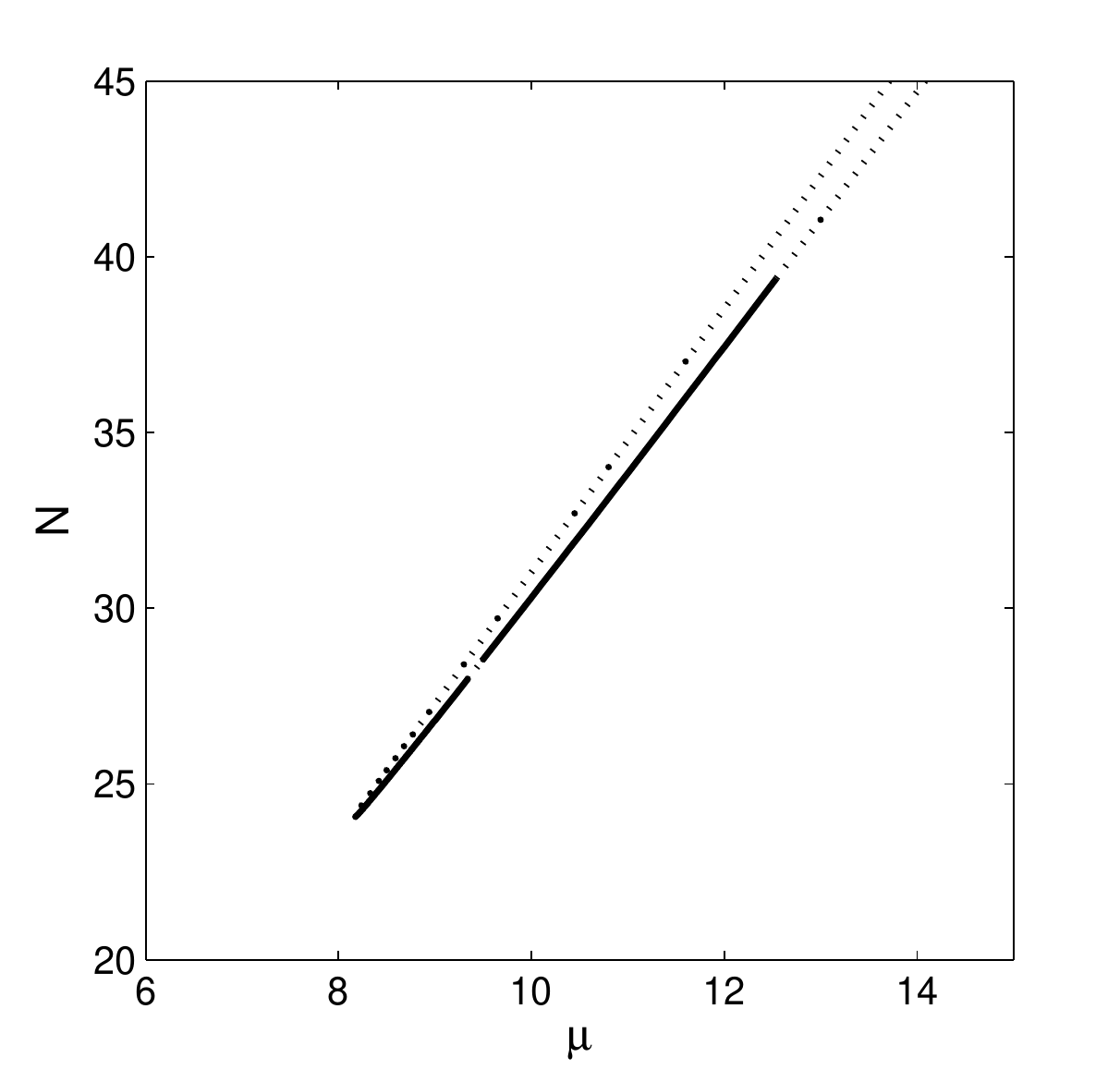}
\label{1D2WellCompositeAsymmetricNVsMu}}
\caption{(a) The solid and dotted profiles show, severally, examples of
stable and unstable composite 1D asymmetric solutions of the [$k=0,k=2$]
type, for $x_{0}=1$, $\protect\mu =10$ and $\protect\alpha =0.5$. (b) $N(%
\protect\mu )$ curves of the composite asymmetric states of these types,
also for $x_{0}=1$ and $\protect\alpha =0.5$. Two stable(solid) segments
were found on the lower branch: at $8.17<\protect\mu <9.34$, i.e., $%
24.05<N<27.98$, and at $9.49<\protect\mu <12.54$, i.e., $28.51<N<39.38$.}
\label{1D2WellCompositeAsymmetric}
\end{figure}

Alongside the higher-order symmetric states introduced above, higher-order
asymmetric solutions were found too. An example is a family of composite
asymmetric states of the $[k=0,k=2]$ type, which are introduced in Fig. \ref%
{1D2WellCompositeAsymmetricMu10}, for $x_{0}=1$, $\mu =5$ and $\alpha =0.5$.
This mode is a combination of two single-well-based constituents: a
fundamental solution on one side ($k=0$), and a solution with two nodes ($%
k=2 $) on the other. Figure \ref{1D2WellCompositeAsymmetricNVsMu} exhibits a
typical $N(\mu )$ branch for this type of composite modes. It is seen that
the branch consists of two, almost coinciding, curves that merge at a
certain point (in the present case, at $\mu =8.17$, $N=24.07$), with only
the lower curve having a stability segment.

The stability map for this type of the asymmetric composite modes in the
plane of $\left( \alpha ,N\right) $ is shown in Fig. \ref%
{1D2WellCompositeAsymmetricStateNVsAlpha}. It exhibits not a single
stability region, but a more complex map, with internal instability strips.
The dark region at low values of $N$ (or $\mu $) refers to the region where
the solutions do not exist, above the merger point of the two branches.
Similar to what was shown before for symmetric unstable modes [see Fig. (\ref%
{1D2WellFirstExcitedStateStability})], the evolution of unstable asymmetric
composite modes originally exhibits an oscillatory behavior, leading to
transformation into the stable fundamental symmetric mode (not shown here in
detail).

\begin{figure}[tbp]
{\ \includegraphics[width=3.0in]{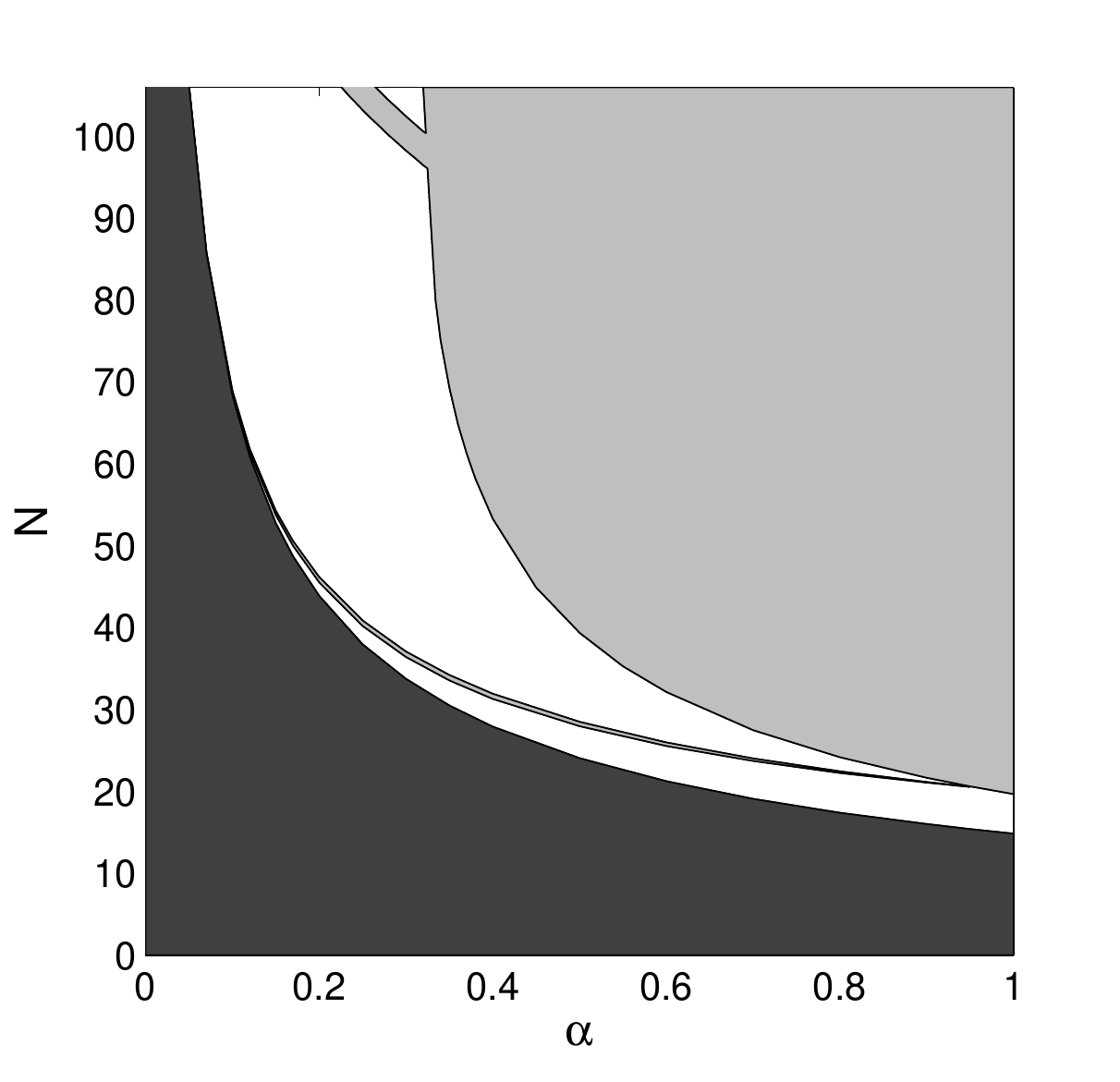}}
\caption{ The stability diagram of the composite 1D modes of the $[k=0,k=0]$
type in the $\left( \protect\alpha ,N\right) $ plane, for $x_{0}=1$.
Solutions do not exist in the dark region at the bottom of the plot. As
above, the solutions are unstable in gray areas.}
\label{1D2WellCompositeAsymmetricStateNVsAlpha}
\end{figure}

\begin{figure}[tbp]
\subfigure[]{\includegraphics[width=2.8in]{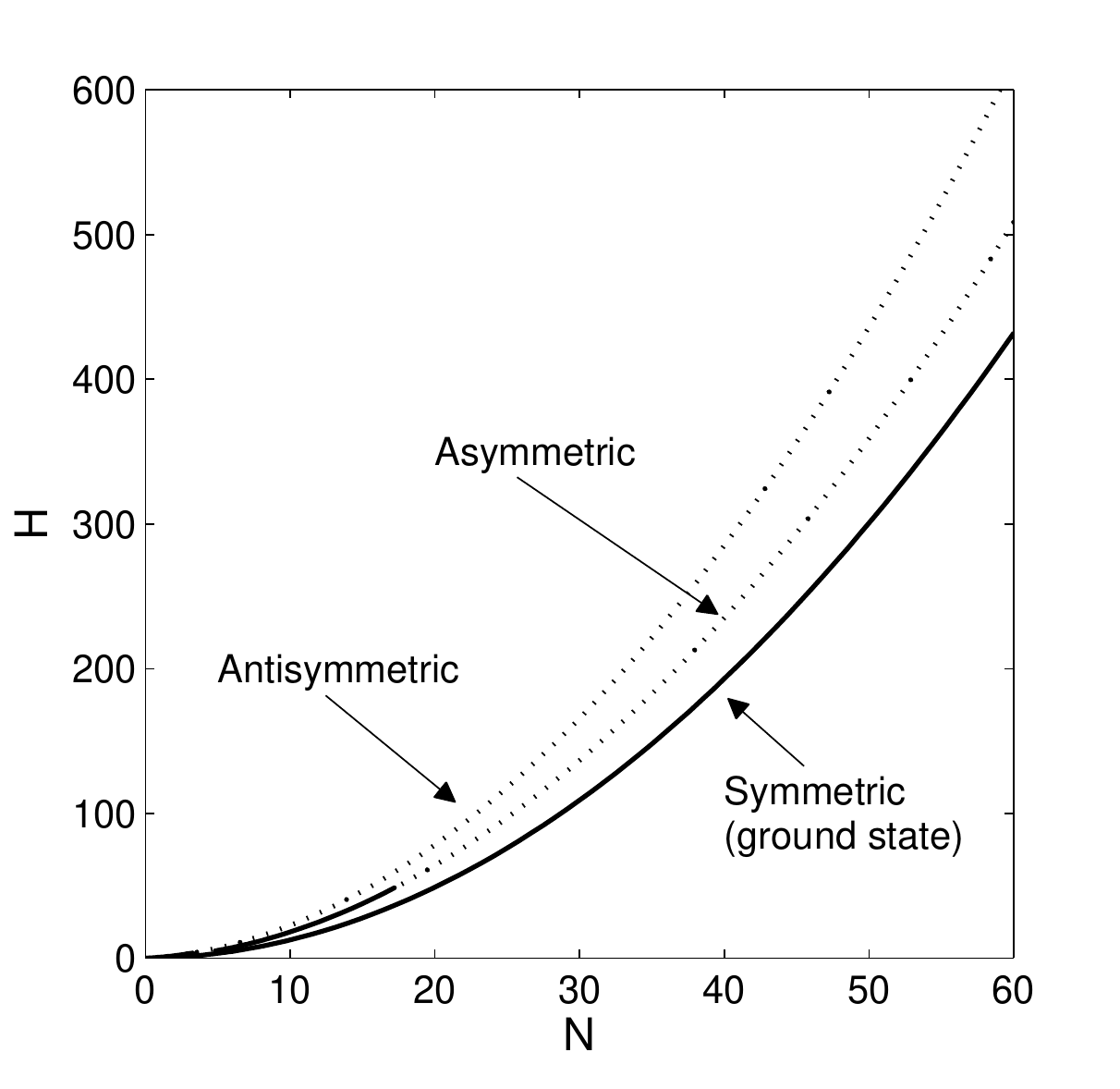}
\label{1D2WellHamiltonianBasic}}
\subfigure[]{\includegraphics[width=2.8in]{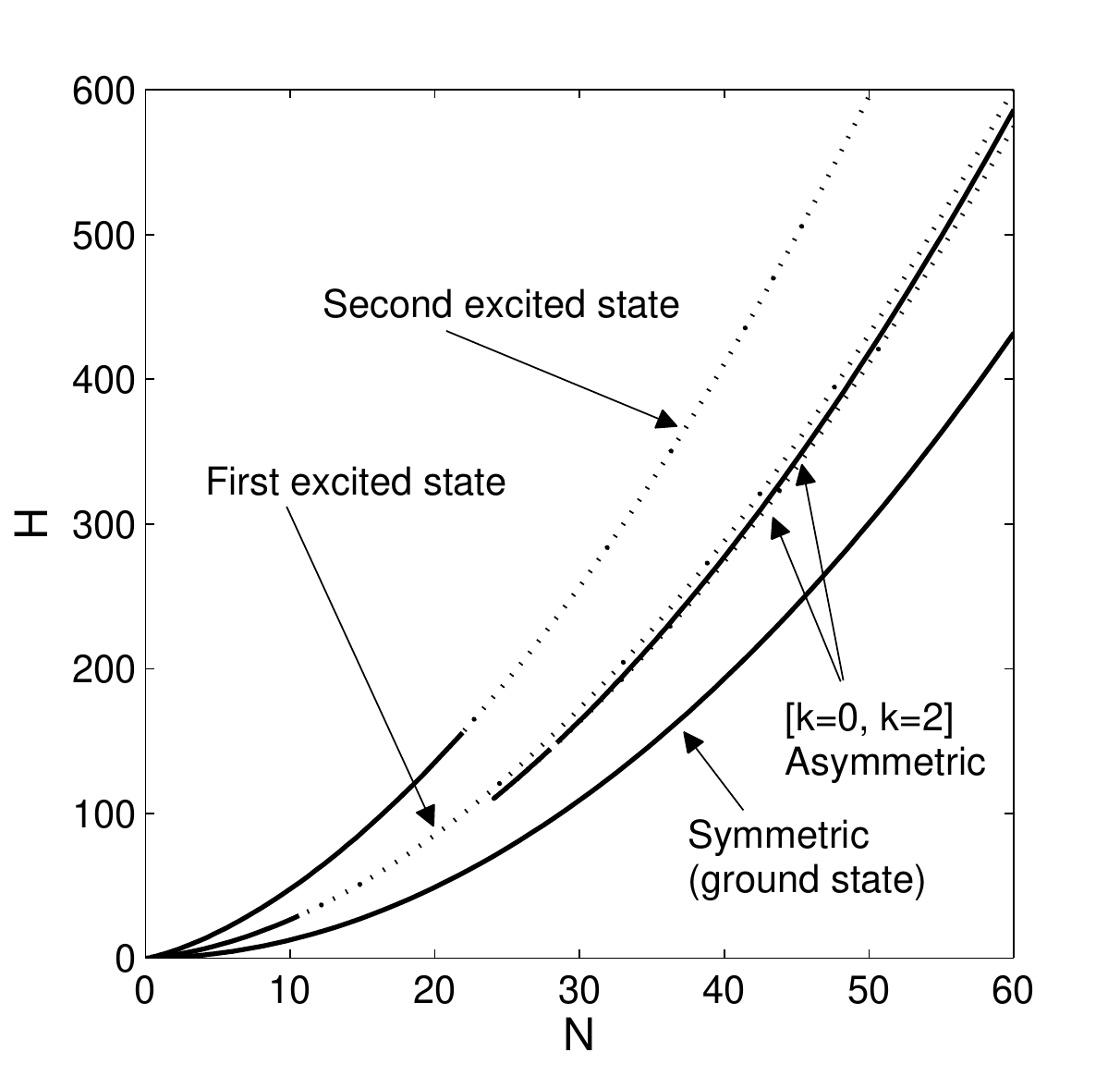}
\label{1D2WellHamiltonianNovel}}
\caption{The Hamiltonian-versus-norm curves for families of 1D states, at $%
x_{0}=1$ and $\protect\alpha =0.5$. Continuous and dotted lines refer to
stable and unstable solutions, respectively. Panel (a) exhibits the curves
for the basic symmetric, antisymmetric and asymmetric states. The composite
asymmetric state of the $[k=0,k=2]$ type, as well as the first and second
excited symmetric modes, are presented in panel (b) (the curve for the basic
symmetric solutions, which represent the ground state, is included in both
panels, for the sake of comparison).}
\label{1D2WellHamiltonians}
\end{figure}

The relative stability of the different coexisting modes with equal norms is
determined by the comparison of their Hamiltonians, given by the 1D version
of Eq. (\ref{Ham}). The respective $H(N)$ curves for all the above-mentioned
1D solutions are displayed in Fig. \ref{1D2WellHamiltonians}, for $x_{0}=1$,
$\alpha =0.5$. As expected, the ground state, with the lowest value of the
Hamiltonian, corresponds to the basic symmetric solution, while the first
and second excited symmetric states have, respectively, higher values of $H$.

\subsection{The general 1D model: asymmetric modes}

\label{sec:General1DModel} 
\begin{figure}[tbp]
\subfigure[]{\includegraphics[width=2.8in]{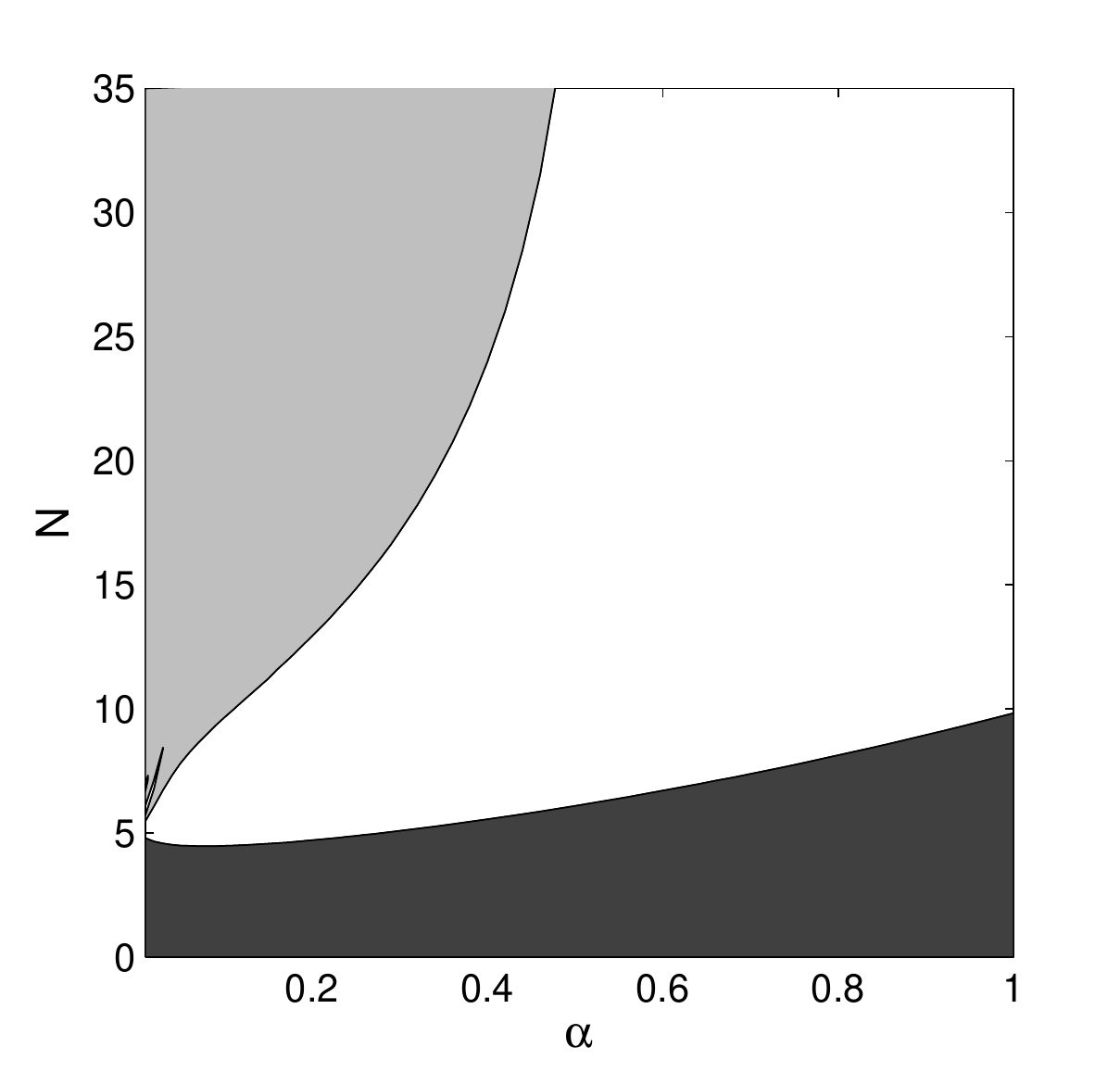}
\label{1D2WellFullProfileG01G22}}
\subfigure[]{\includegraphics[width=2.8in]{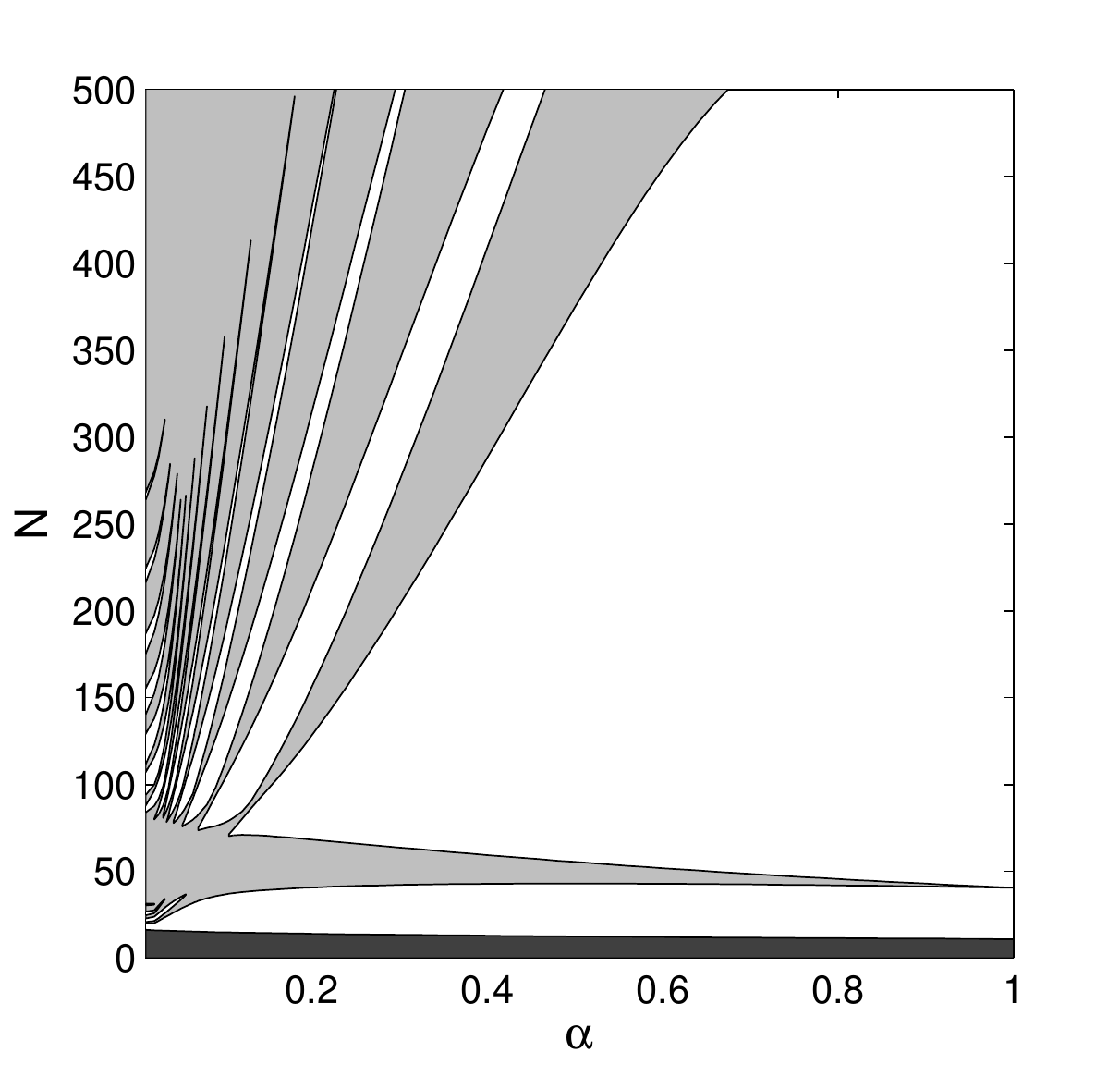}
\label{1D2WellFullProfileG00G22}}
\caption{Stability diagrams for 1D asymmetric modes, obtained with the
generalized nonlinearity-modulation profile (\protect\ref%
{FullDoubleWellProfile}), for $\protect\sigma _{2}=2$, $x_{0}=1$, $\protect%
\beta =1$, and (a) $\protect\sigma _{0}=1$ or (b) $\protect\sigma _{0}=0$.
White and gray regions represent, as before, stable and unstable solutions,
respectively. The asymmetric solutions do not exist in the dark region at
the bottom of the plots, below the bifurcation of the antisymmetric mode.}
\label{1D2WellFullProfile}
\end{figure}

In addition to the detailed investigation
reported above for the simplest version of the DW profile, based on Eq. (\ref%
{simple}), a partial analysis has been performed for the more general
profile corresponding to Eq. (\ref{FullDoubleWellProfile}). Specifically, we
have examined the influence of parameters $\sigma _{0}$ and $\sigma _{2}$ on
the stability of the asymmetric state. Figure \ref{1D2WellFullProfileG01G22}
displays the respective stability diagram in the $\left( \alpha ,N\right) $
plane, with $\sigma _{0}=1$, $\sigma _{2}=2$, $x_{0}=1$, and $\beta =0$.
Similar results were also obtained for $\beta =1$ (not shown here). As seen
in Fig. \ref{1D2WellFullProfileG01G22}, asymmetric solutions were found for
all values of $\alpha $, provided that the soliton's norm is high enough (an
expected outcome, as the effective DW potential depends on $N$). The
stability region expands with the increase of $\alpha $, opposite to what is
reported above for the simplified profile (\ref{simple}), cf. Fig. \ref%
{1D2WellAsymmetricNVsAlpha}. When the constant term is absent, $\sigma
_{0}=0 $, the stability map features alternating stability and instability
strips, as seen in Fig. \ref{1D2WellFullProfile}(b). For the same
parameters, but with $\sigma _{0}=1$ [Fig. \ref{1D2WellFullProfile}(a)], the
striped pattern, barely observed at small values of $\alpha $, is much less
salient.

\section{A rocking one-dimensional potential well}

\label{sec:OscillatingSolutions} It is natural to extend the consideration
of modes trapped in the single (pseudo)potential well to the case when the
well is subject to rocking motion, which corresponds to
\begin{equation}
\sigma (z)=\exp \left[ \alpha (x-A_{0}\cos (W_{0}z)^{2})\right]
\label{OscillatingSigma}
\end{equation}%
in the 1D version of Eq. (\ref{NLSE}), where the $A_{0}$ and $W_{0}$ are the
rocking amplitude and frequency. In the optical realization, this
corresponds to to the planar waveguide with an undular guiding channel [in
the plane of $\left( z,x\right) $], written by the local modulation of the
defocusing nonlinearity \cite{rocking-optical}. In the case of BEC, the
rocking implies oscillatory motion of the nonlinearity-modulation profile,
which can be readily implemented if the modulation is induced by the
optically-controlled Feshbach resonance \cite{FR-Tom}, as the controlling
laser beam may be made moving \cite{painting}.

Here, we focus on the fundamental and first-order (dipole) modes, both
well-known to be stable in the static model \cite{Barcelona}. In all the
examples shown below, we fix $\alpha =0.5$, while the parameters of the
rocking well were taken in ranges $0<A_{0}<4$ and $2\pi /40<W_{0}<2\pi $.

\begin{figure}[tbp]
{\ \includegraphics[width=3.2in]{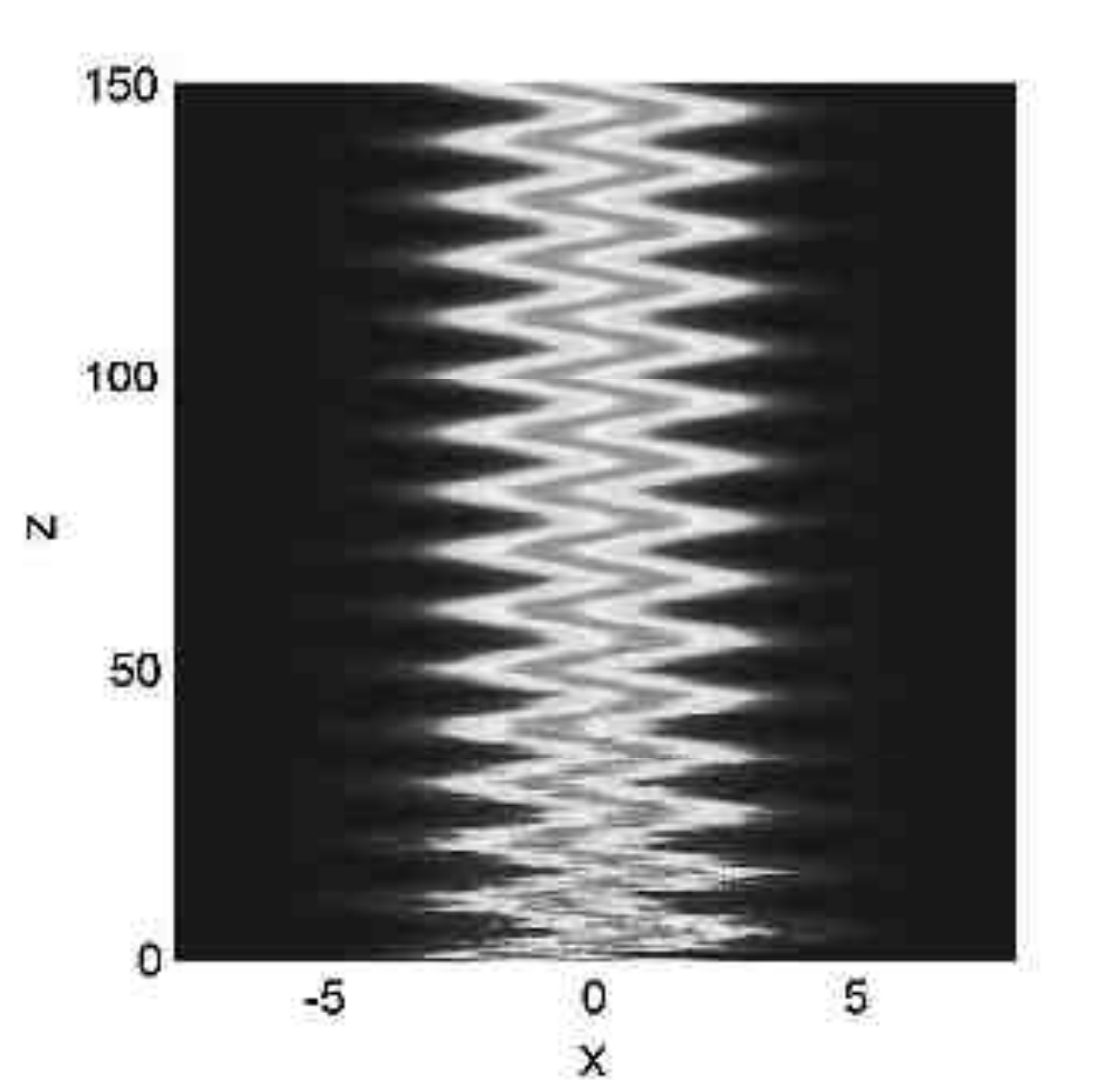}}
\caption{(Color online) An example of the stable evolution of a 1D
fundamental mode, with $\protect\mu =5$ ($N=12.22$), which adiabatically
follows the single-well rocking with a large period, $Z_{0}=10$, and
amplitude $A_{0}=1$.}
\label{EvolutionOfSmoothFund}
\end{figure}

Starting with the fundamental modes, two evolution scenarios can be
identified, depending on the initial value of the norm, $N$ (or $\mu $), and
parameters $A_{0}$ and $W_{0}$. Namely, for the rocking period, $Z_{0}=2\pi
/W_{0}$, exceeding a certain threshold value, the mode adiabatically follows
the slowly rocking nonlinear well, maintaining its original shape. An
example is shown in Fig. \ref{EvolutionOfSmoothFund}.

\begin{figure}[tbp]
\subfigure[]{\includegraphics[width=2.8in]{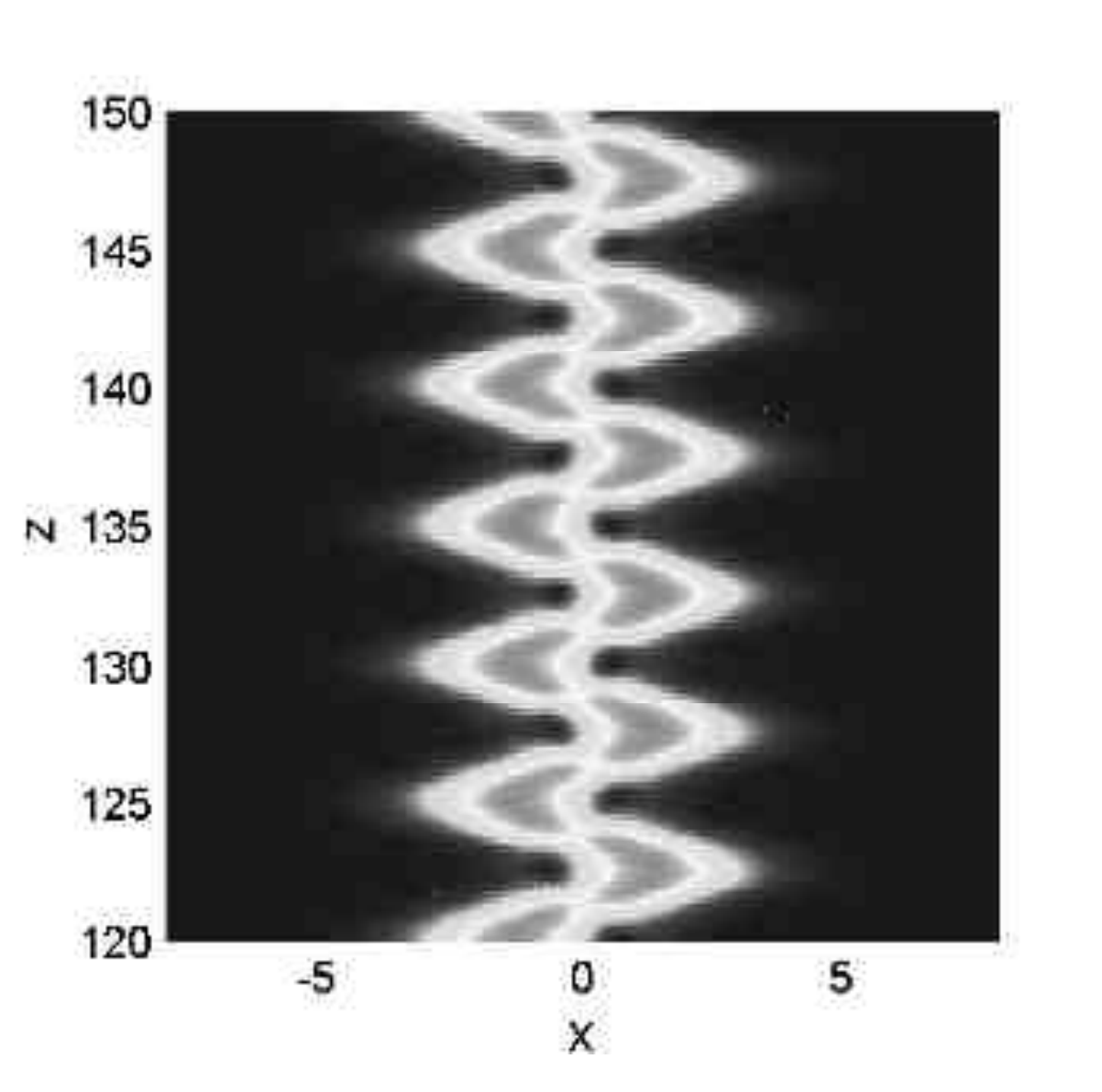}
\label{EvolutionOfOsc1FundMu0p5}}
\subfigure[]{\includegraphics[width=2.8in]{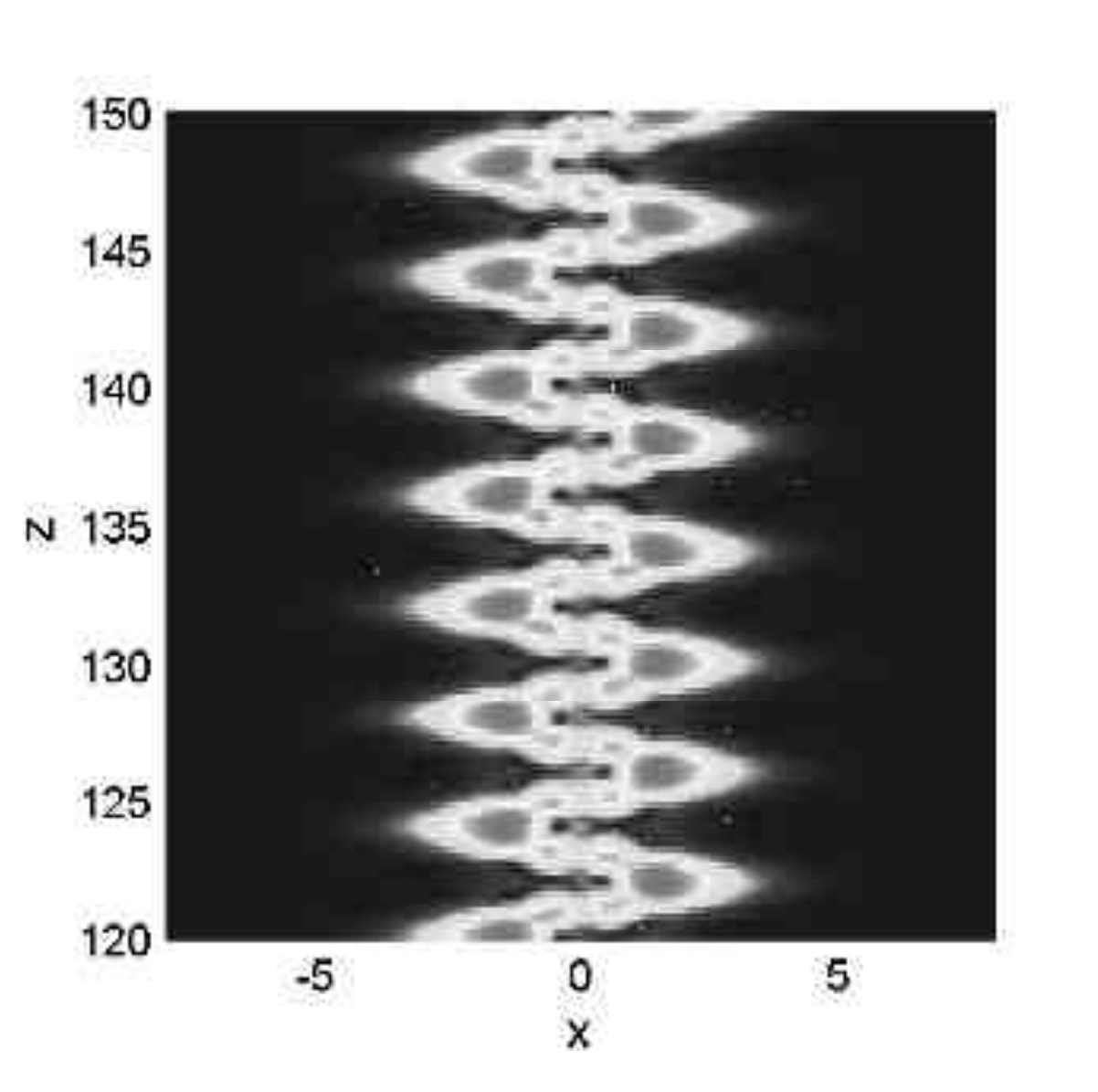}
\label{EvolutionOfOsc2FundMu0p5}}
\caption{(Color online) Typical examples of the non-adiabatic evolution of
the 1D fundamental mode, in the case of relatively fast rocking motion of
the underlying nonlinear well. For $\protect\mu =5$, $A_{0}=1$ and $Z_{0}=5$
(a), the wave pattern keeps a single-peak shape. With the rocking period
further (slightly) decreasing to $Z_{0}=4$, the pattern breaks into
fragments.}
\label{EvolutionOfOscFund}
\end{figure}

On the other hand, for period $Z_{0}$ below this threshold, the shape of the
initial fundamental soliton is not kept. In this case, the evolution of the
solution is not smooth, and may exhibit different oscillatory patterns.
Examples are shown in Fig. \ref{EvolutionOfOscFund} for $\mu =5$, $A_{0}=1$
and $Z_{0}=4$ and $5$. Further, the dependence of the threshold value of the
rocking period on the rocking amplitude, $A_{0}$, is shown in Fig. \ref%
{AvsT_FundThreshold}, for two fixed values of the norm.

\begin{figure}[tbp]
{\ \includegraphics[width=3.5in]{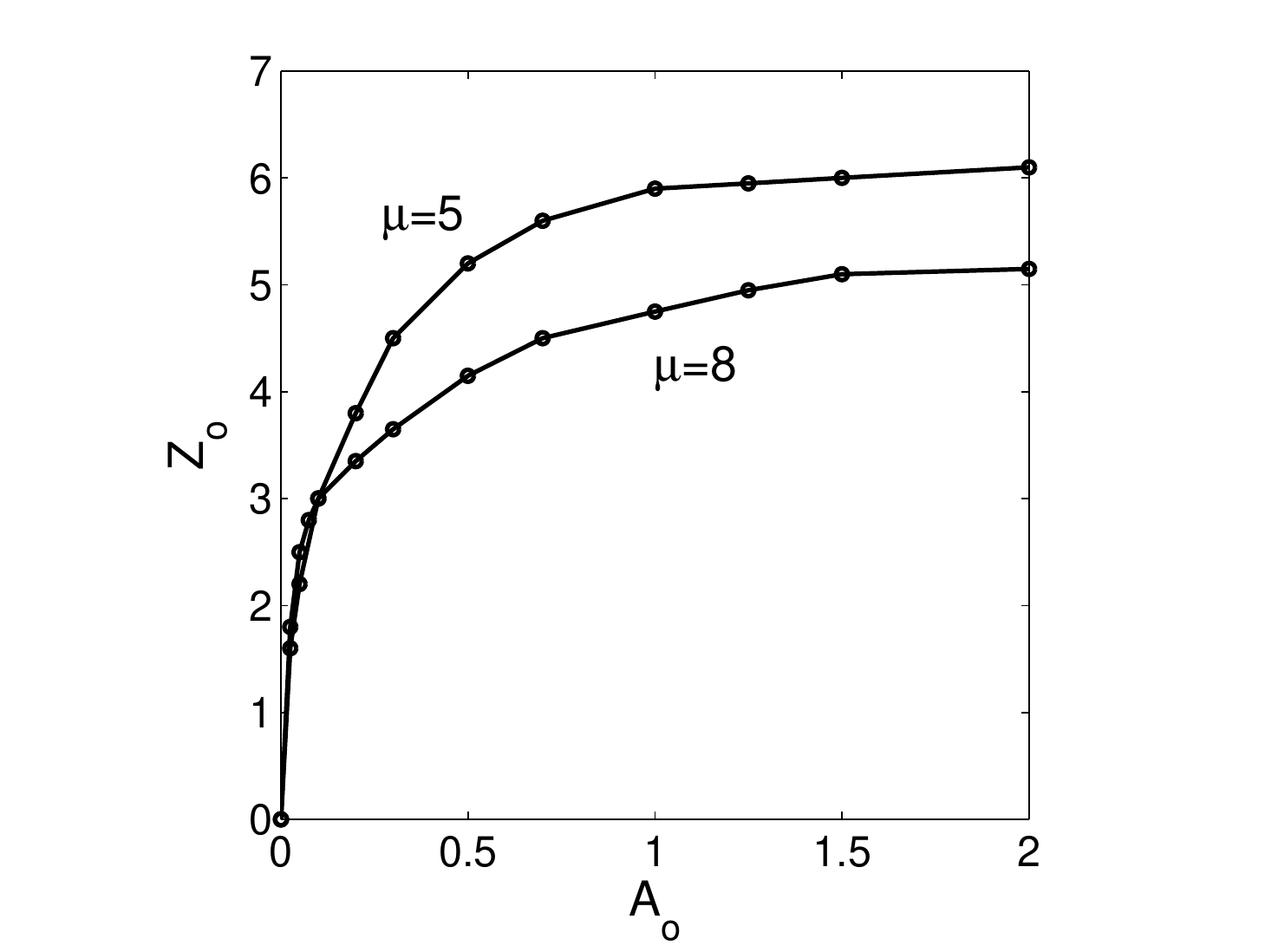}}
\caption{The threshold value of the rocking period, below which the
evolution of the trapped 1D fundamental mode is non-adiabatic, see Fig.
\protect\ref{EvolutionOfOscFund}. Two curves pertain to fixed $N=12.22$ (the
respective propagation constant in the absence of the rocking is $\protect%
\mu _{0}=5$) and $N=19.74$ ($\protect\mu _{0}=8$).}
\label{AvsT_FundThreshold}
\end{figure}

\begin{figure}[tbp]
\subfigure[]{\includegraphics[width=2.8in]{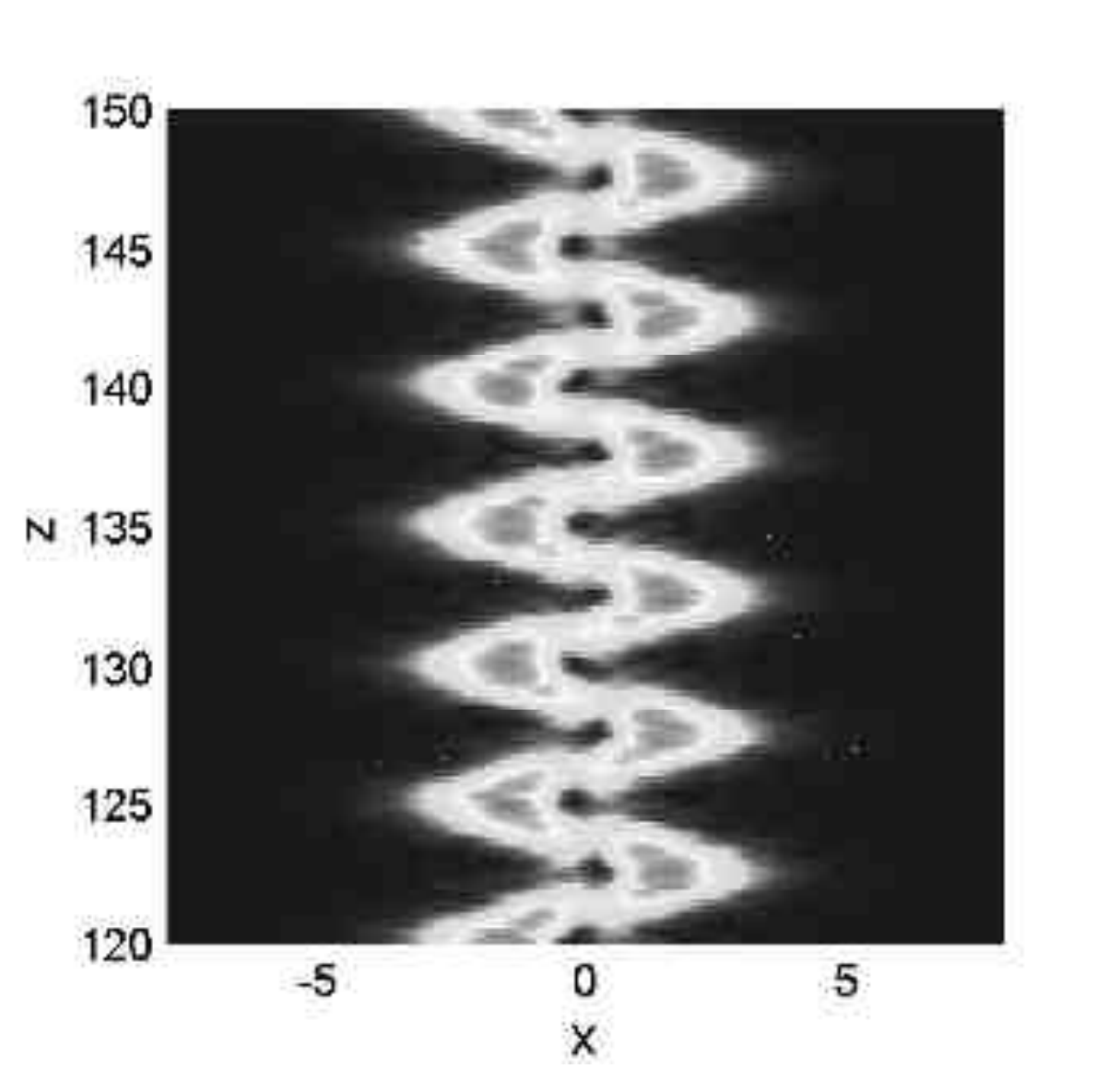}
\label{EvolutionOfOsc1DipoleMu0p5}}
\subfigure[]{\includegraphics[width=2.8in]{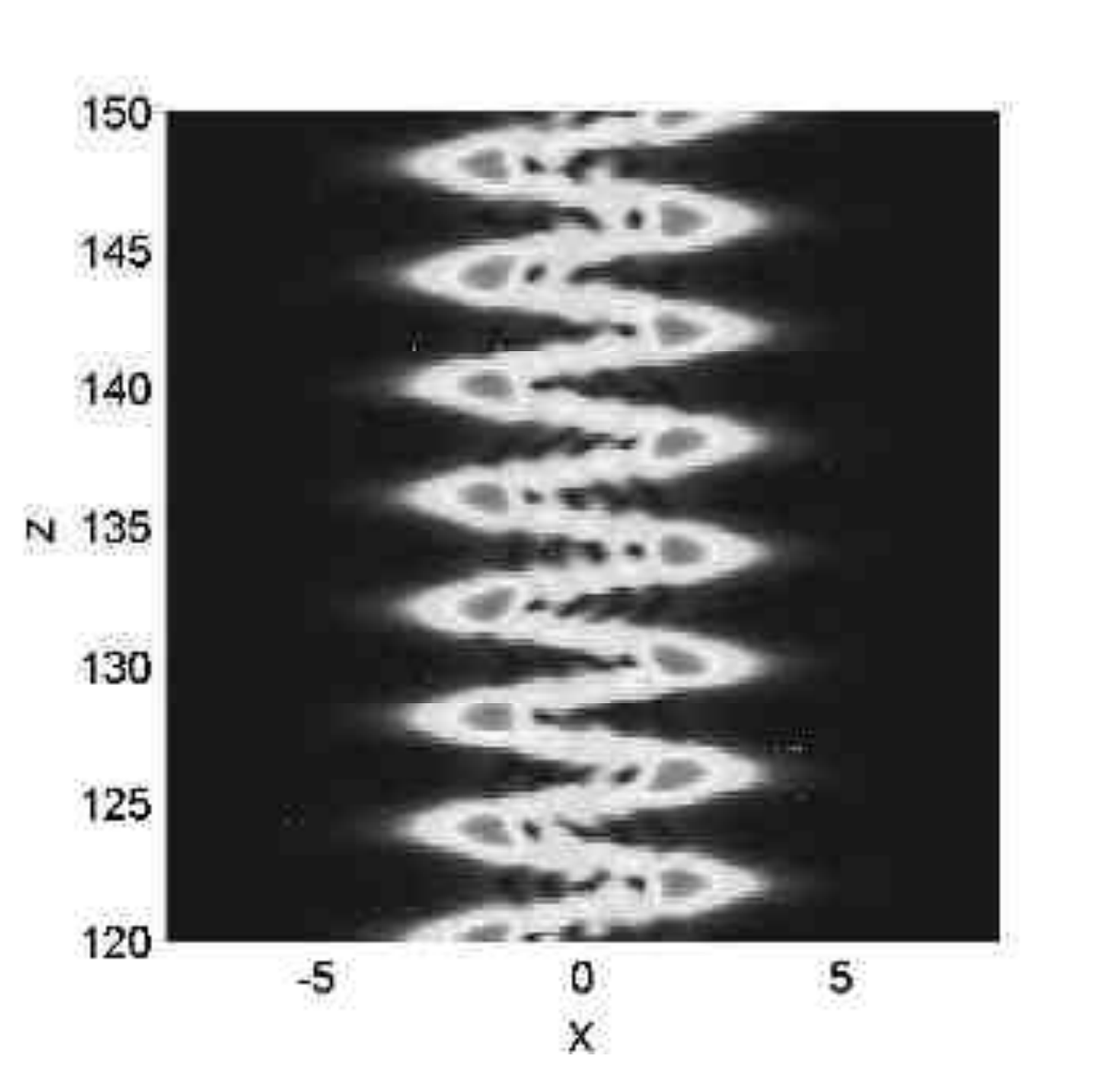}
\label{EvolutionOfOsc2DipoleMu0p5}}
\caption{(Color online) The evolution of the 1D dipole mode trapped in the
single rocking nonlinear well below the threshold, when the initial profile
is destroyed. Similar to the situation with the fundamental mode (see Fig.
\protect\ref{EvolutionOfOscFund}), the dipole transforms into an oscillating
single-peak pattern, at $\protect\mu =5$, $A_{0}=1$ and $Z_{0}=5$ (a). At a
still smaller rocking period, $Z_{0}=4$, the solution breaks into fragments
(b).}
\label{EvolutionOfOscDipole}
\end{figure}

The evolution of the trapped dipole mode also turns out to be different
above and below the threshold value of the rocking period, which is found to
be virtually indistinguishable from the one shown in Fig. \ref%
{AvsT_FundThreshold} for the fundamental solution. Below the threshold, the
dipoles quickly transform into single-peak or fragmented patterns, which are
quite similar to those observed in the case of the fundamental mode, see
typical examples in Fig. \ref{EvolutionOfOscDipole} for $\mu =5$, $A_{0}=1$
and $Z_{0}=4$ or $5$.

\begin{figure}[tbp]
\subfigure[]{\includegraphics[width=2.2in]{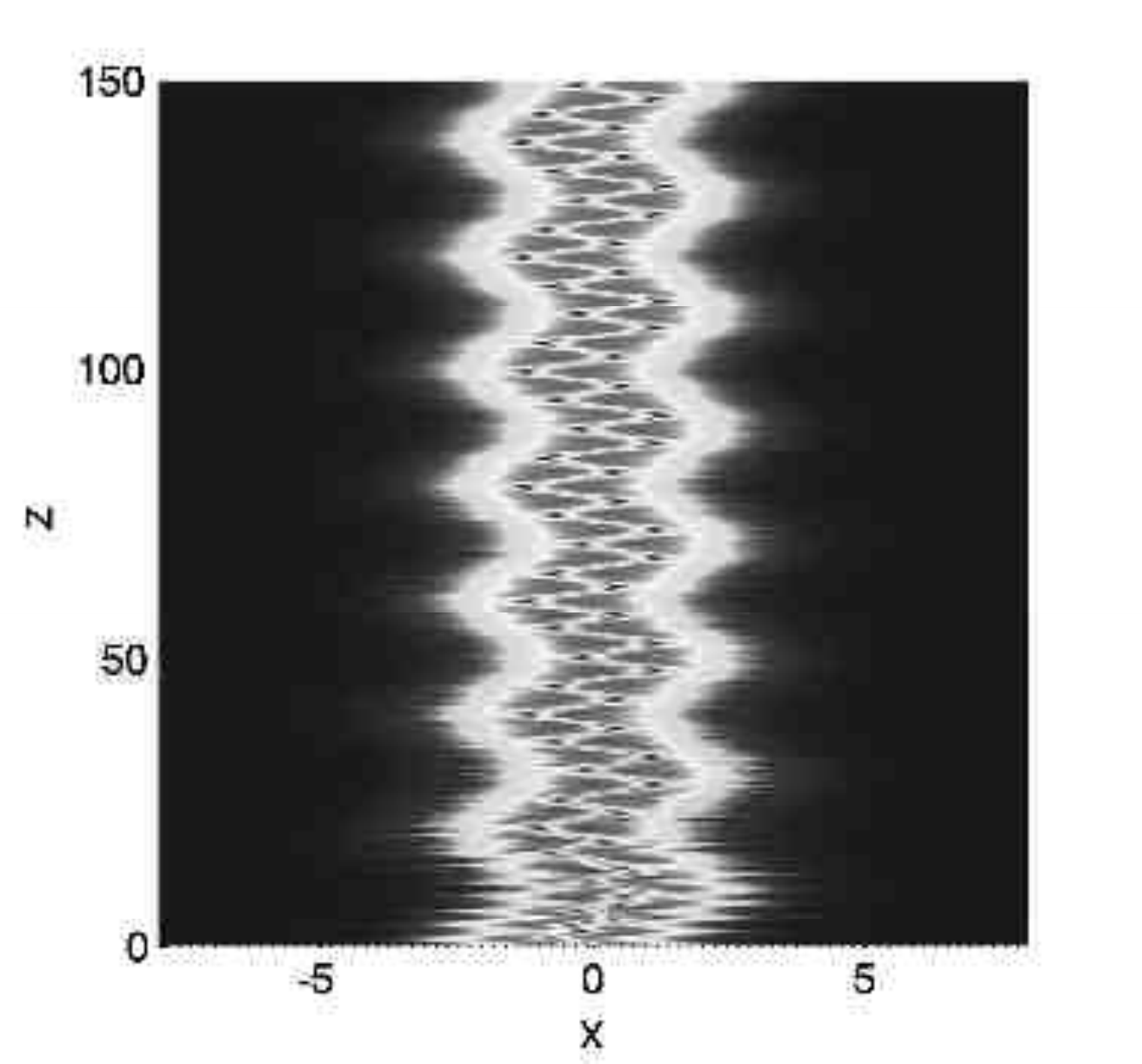}
\label{EvolutionOfDipoleMu0p5A0p5Zo20}}
\subfigure[]{\includegraphics[width=2.3in]{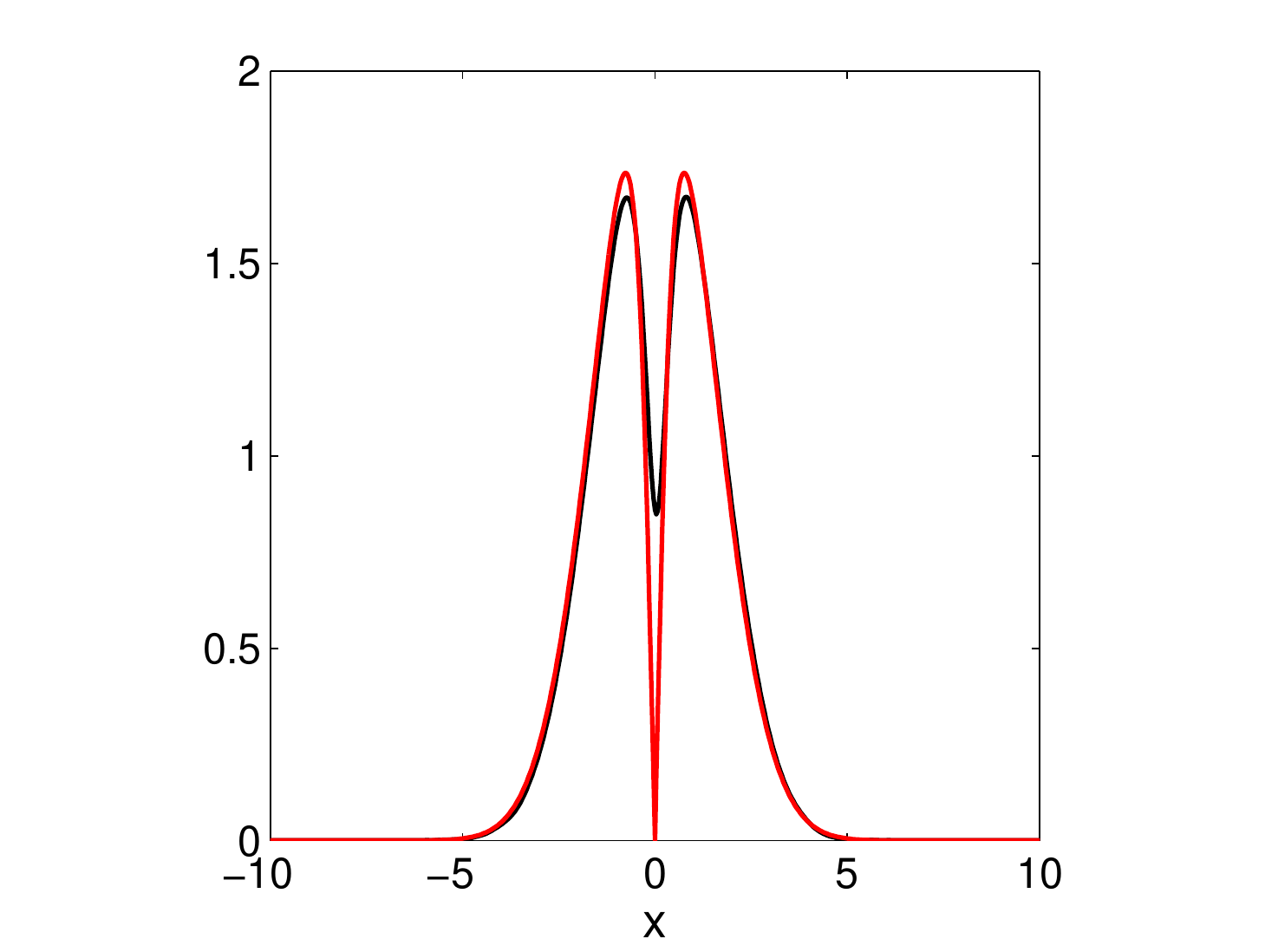}
\label{DipoleProfileCompare_Mu0p5A0p5Zo20}}
\subfigure[]{\includegraphics[width=2.3in]{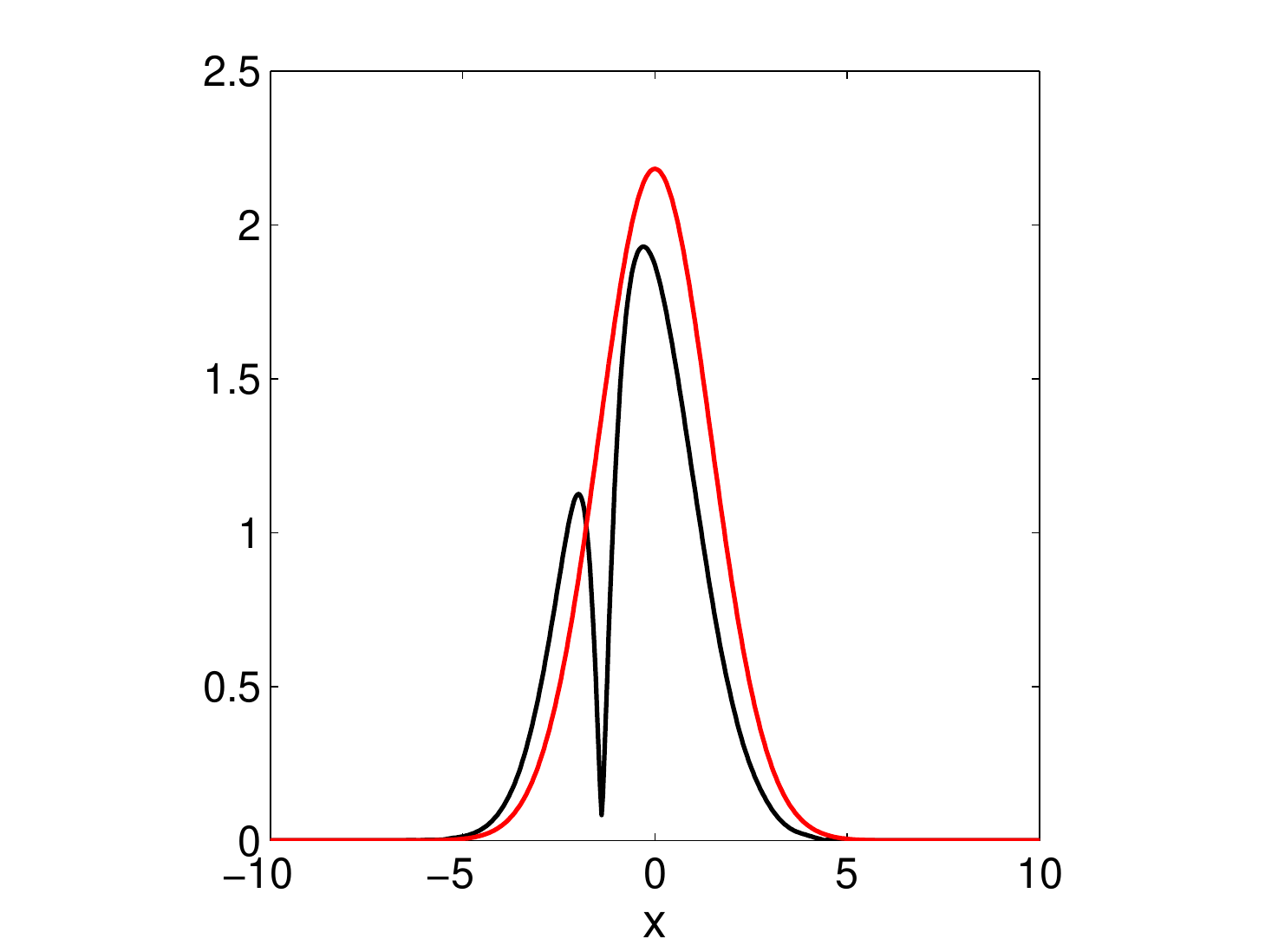}
\label{FundProfileCompare_Mu0p5A0p5Zo20}}
\caption{(Color online) (a) A typical example of the evolution of the1D
dipole solution trapped in the slowly rocking nonlinear well (above the
threshold value of the period), which results in Rabi oscillations between
the dipole (b) and single-peak fundamental (c) modes. The parameters are $%
N=7.92$, $A_{0}=0.5$ and $Z_{0}=20$. Panels (b) and (c) compare the shapes
of the periodically appearing fundamental and dipole modes (shown by black
curves), and their stationary counterparts with the same norm (plotted by
red curves). }
\label{RabiOscillations}
\end{figure}

On the other hand, when the rocking period is taken above the threshold, the
mode periodically switches between the initial dipole shape and the
fundamental soliton. This scenario, which may be considered as \textit{Rabi
oscillations }between the two soliton species (dipole and fundamental ones)
in the rocking (pseudo)potential well \cite{Rabi}, is demonstrated in Fig. %
\ref{RabiOscillations} for $N=7.92$ (the respective propagation constant,
corresponding to the dipole mode in the absence of the rocking, is $\mu =5$).

\begin{figure}[tbp]
\subfigure[]{\includegraphics[width=2.2in]{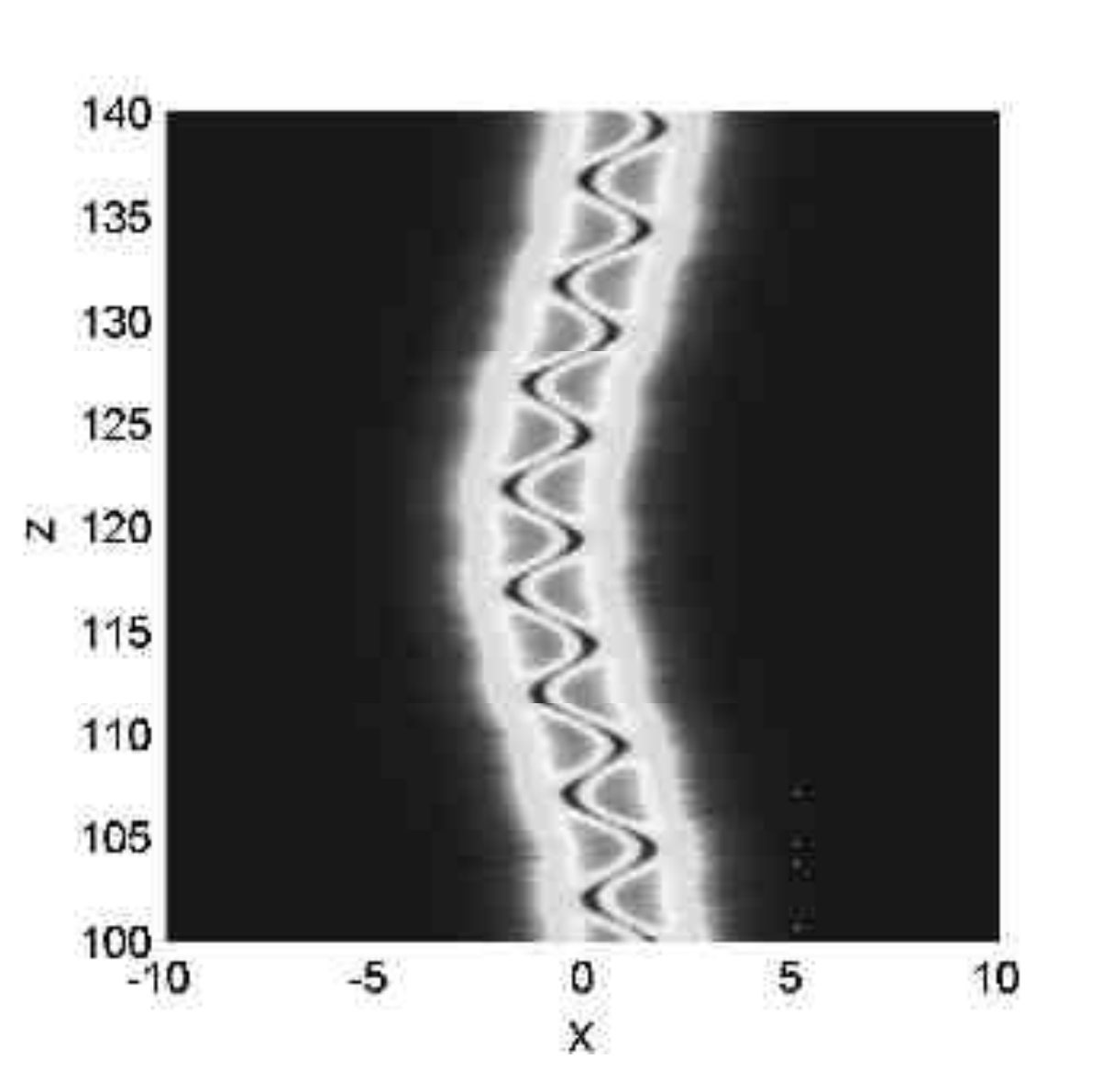}
\label{EvolutionOfDipoleMu0p5A1p0Zo40}}
\subfigure[]{\includegraphics[width=2.2in]{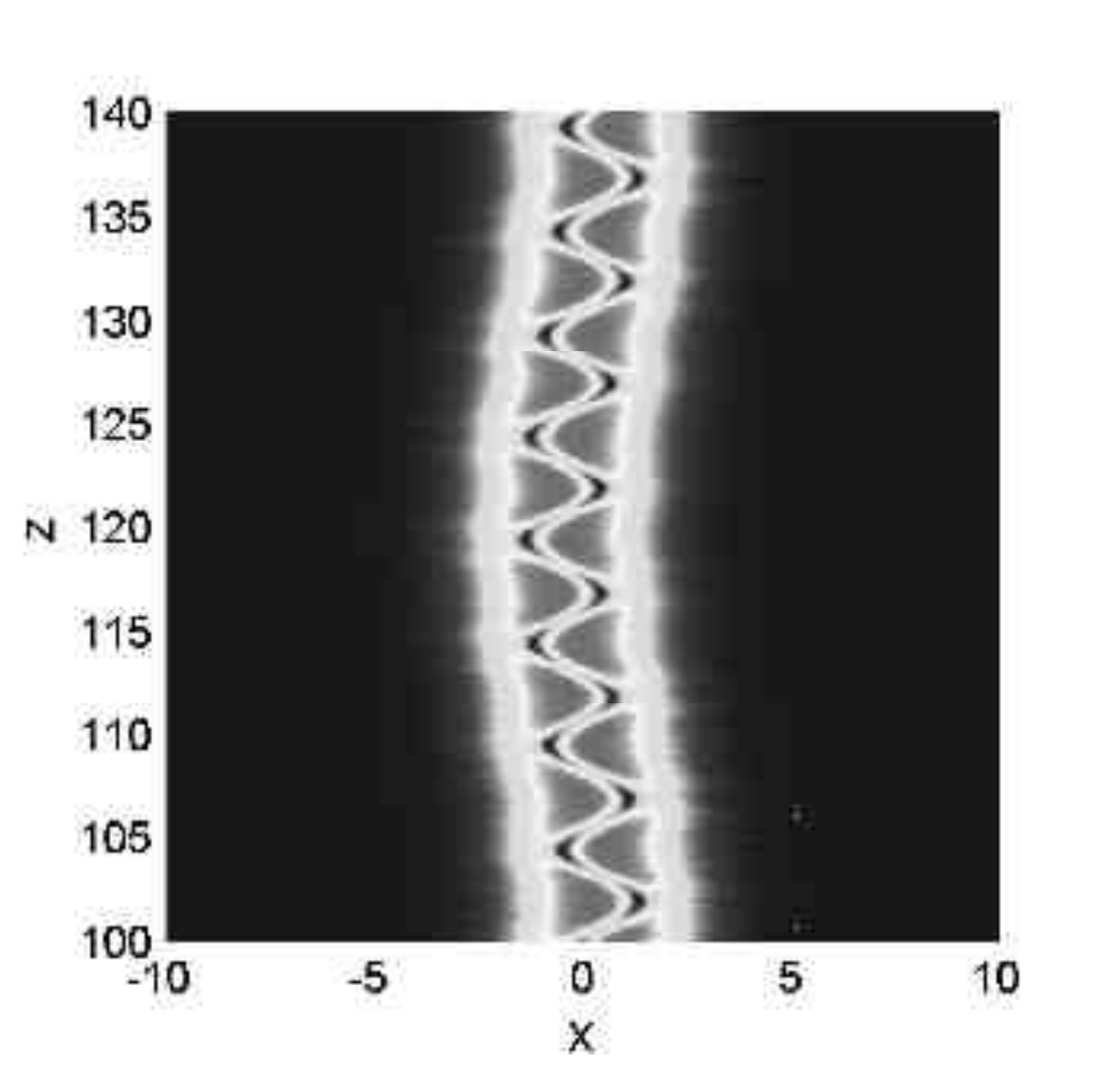}
\label{EvolutionOfDipoleMu0p5A0p5Zo40}}
\subfigure[]{\includegraphics[width=2.2in]{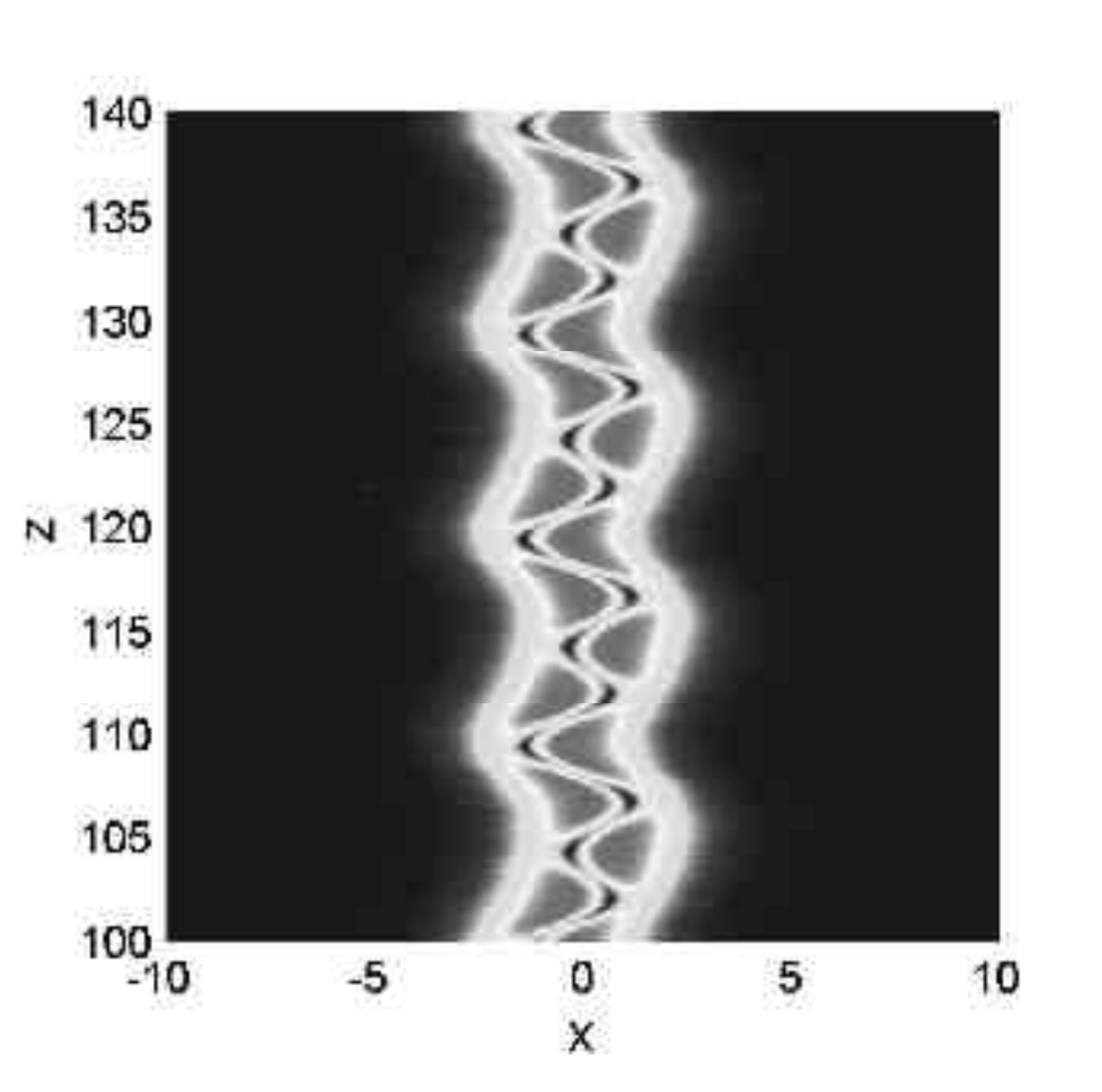}
\label{EvolutionOfDipoleMu0p5A0p5Zo10}}
\caption{(Color online) Examples of the Rabi oscillations of the 1D trapped
mode: (a) $A_{0}=1$, $Z_{0}=40$; (b) $A_{0}=0.5$, $Z_{0}=40$; (c) $A_{0}=0.5$%
, $Z_{0}=10$. In all the cases, $N=7.92$. }
\label{EvolutionOfDipoleMu0p5}
\end{figure}

Unlike the destructive oscillations below the threshold (see Fig. \ref%
{EvolutionOfOscDipole}), the frequency of the Rabi oscillations above the
threshold is different from the rocking frequency, $W_{0}$. Figure \ref%
{Zi_InnerOscillations} shows the period of the intrinsic Rabi oscillations
of the trapped mode, $Z_{\mathrm{i}}$, as a function of $N$. Detailed
numerical analysis demonstrates that $Z_{\mathrm{i}}$ is insensitive to the
rocking parameters, $Z_{0}$ and $A_{0}$. This feature is demonstrated in
Fig. \ref{EvolutionOfDipoleMu0p5}, where the frequency of the Rabi
oscillations remains constant, while the rocking frequency varies. Thus, the
Rabi oscillations between the fundamental and dipole states are actually a
dynamical feature of the system based on the stationary profile of the
nonlinearity modulation, while the rocking motion is a drive which helps to
excite the oscillations.

\begin{figure}[tbp]
\includegraphics[width=2.8in]{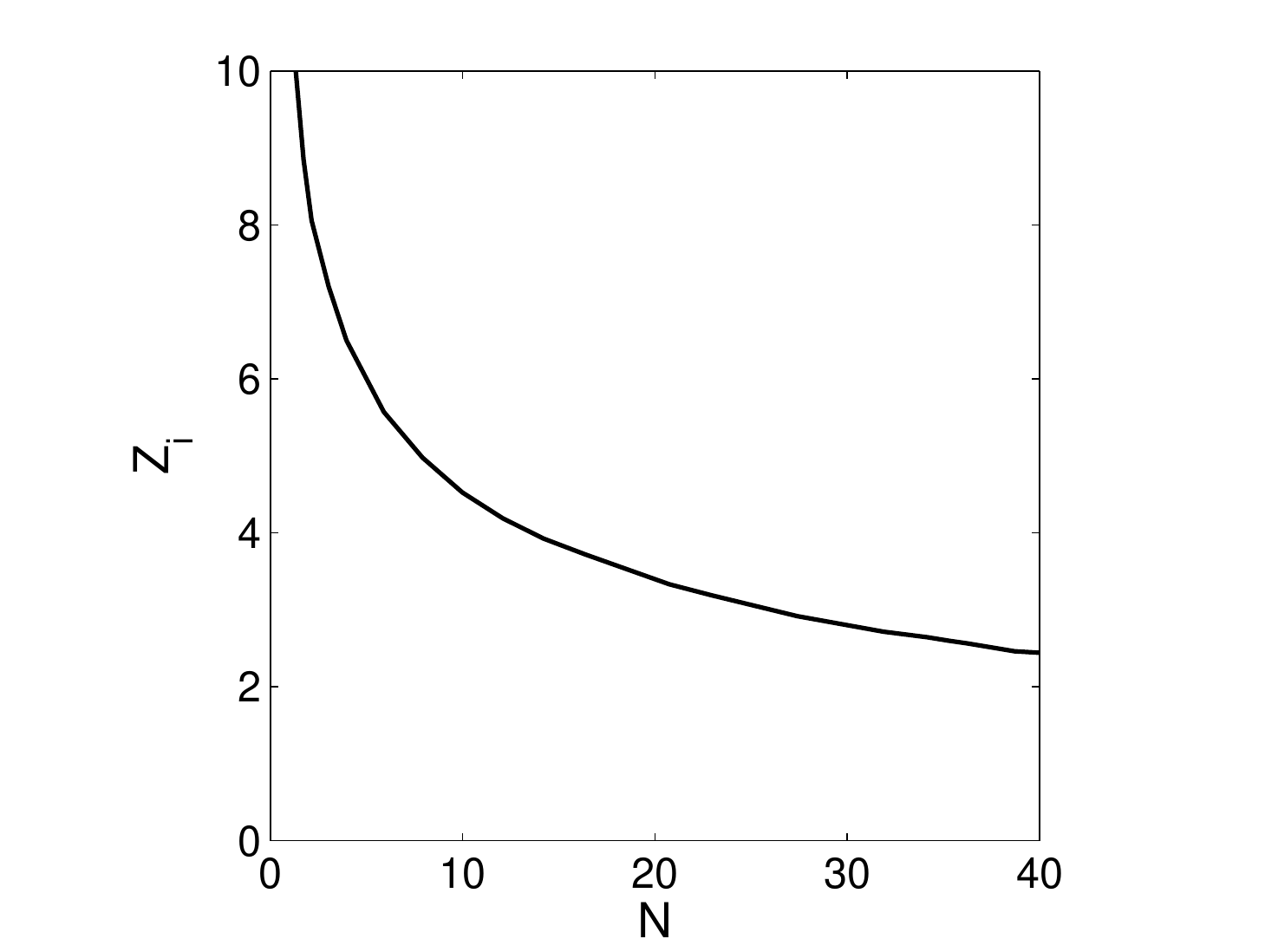} \label%
{ZivsN_InnerOscillations}
\caption{(Color online) The period of the Rabi (intrinsic) oscillations of
the trapped 1D mode between the dipole and fundamental shapes, $Z_{\mathrm{i}%
}$, as a function of $N$.}
\label{Zi_InnerOscillations}
\end{figure}

Actually, the Rabi oscillations are not completely robust. Gradually, the
oscillations fade away, and the trapped mode slowly relaxes into the
fundamental solution. The evolution period in the course of which the Rabi
oscillations remain conspicuous reduces with the increase of the underlying
rocking amplitude, $A_{0}$, and frequency, $W_{0}$.

\section{The two-dimensional setting: the single-well nonlinear potential}

The existence and stability of the axially-symmetric states, both
fundamental (zero-vorticity) ones and vortices with a nonzero topological
charge, in the 2D model with the isotropic single-well modulation profile,
based on Eq.~(\ref{sophisticated}) with $\mathbf{r}_{0}=0$, including the
simplified version (\ref{BasicSingleWellProfile}), were reported in Ref.~%
\cite{Barcelona2}. In this section we introduce and briefly consider
additional higher-order modes, both nontopological and topological ones,
which demonstrate the SSB effect in the form of spontaneous breaking of the
axial symmetry. This setting, which is fundamental by itself, deserves a
detailed study, results of which will be reported elsewhere \cite{Radik}.

\begin{figure}[tbp]
\subfigure[]{\includegraphics[width=2.25in]{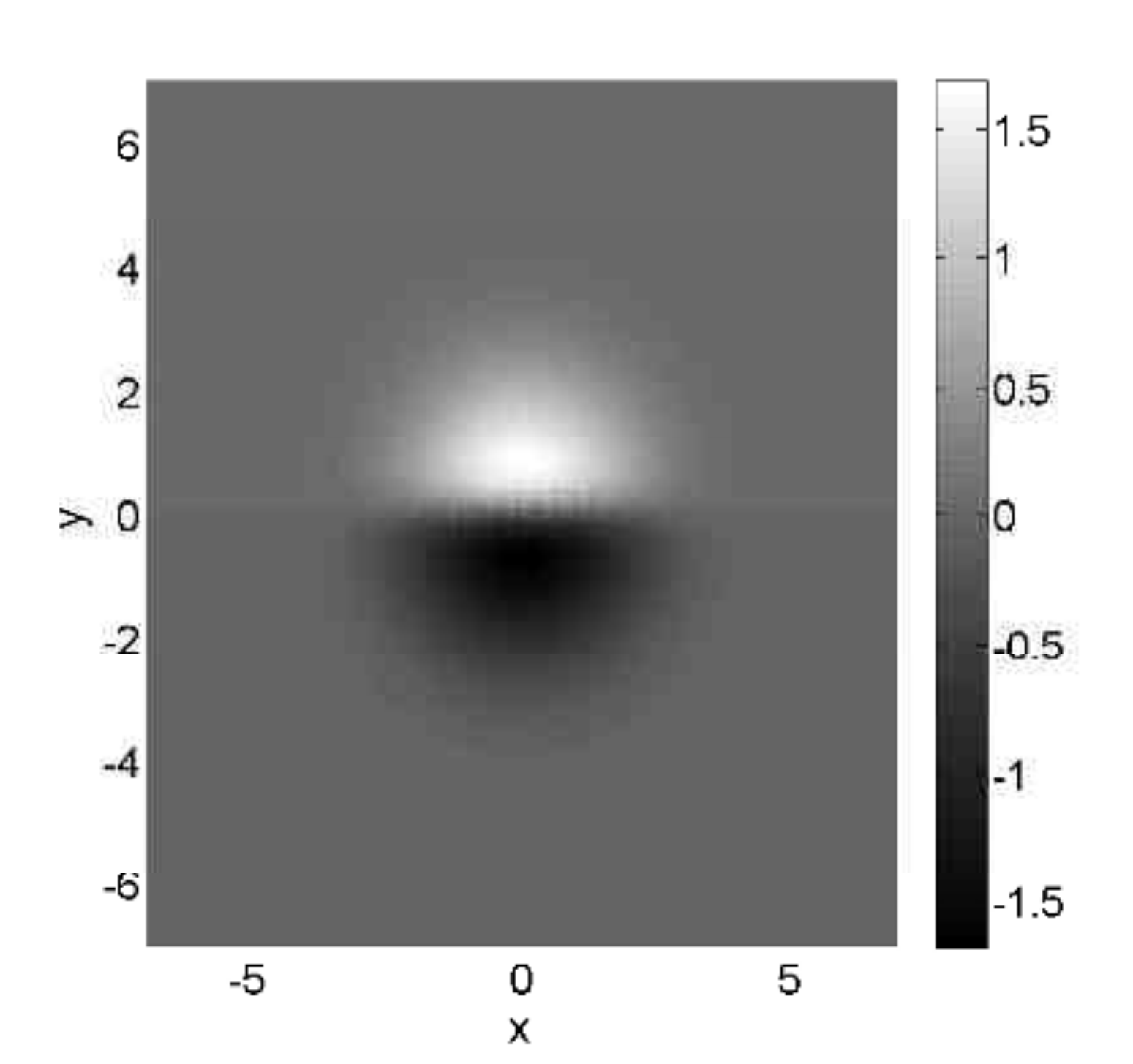}
\label{1WellDipoleSolutionMu5}}
\subfigure[]{\includegraphics[width=2.25in]{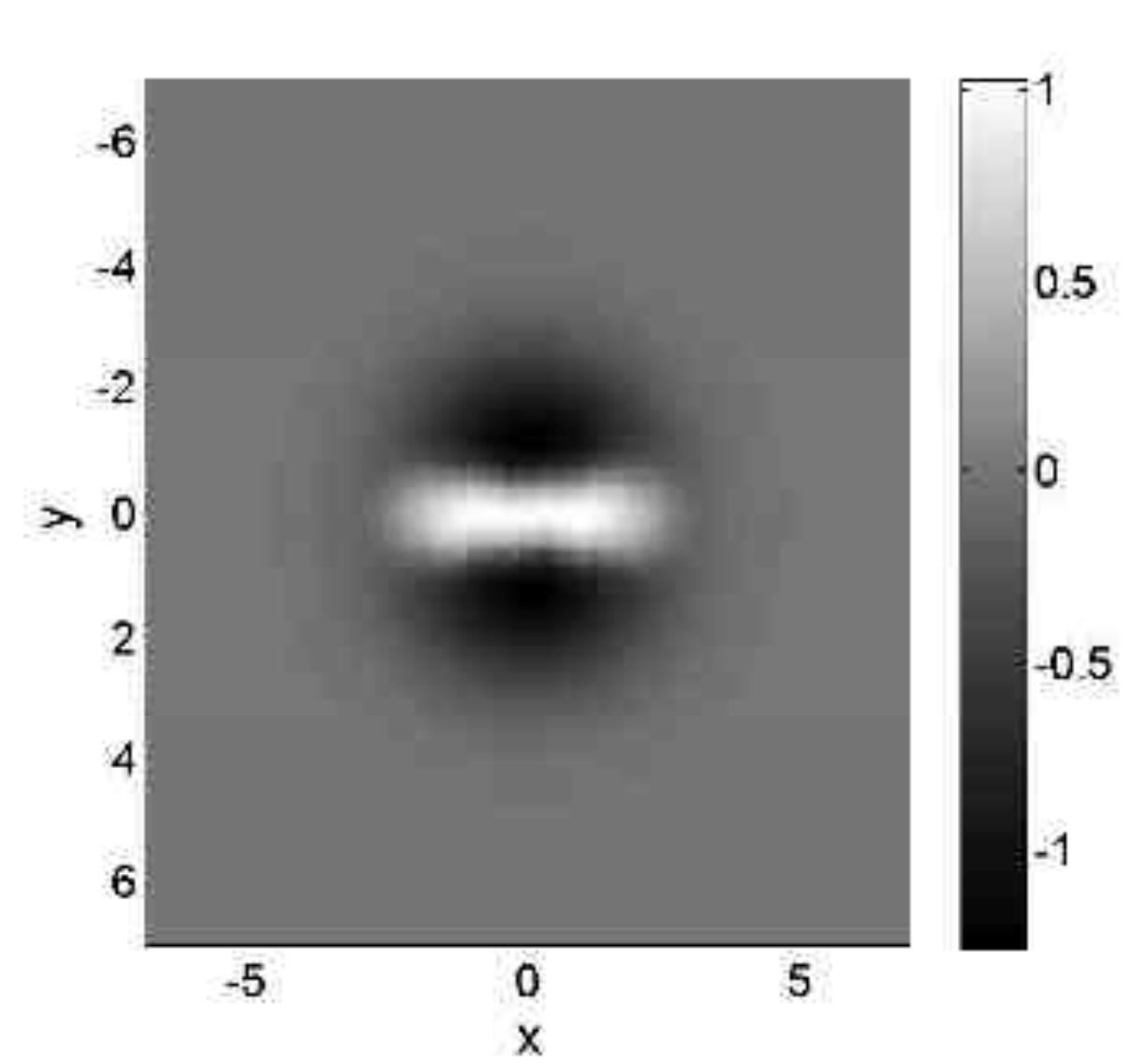}
\label{1Well2ndOrderSolutionMu5}}
\subfigure[]{\includegraphics[width=2.25in]{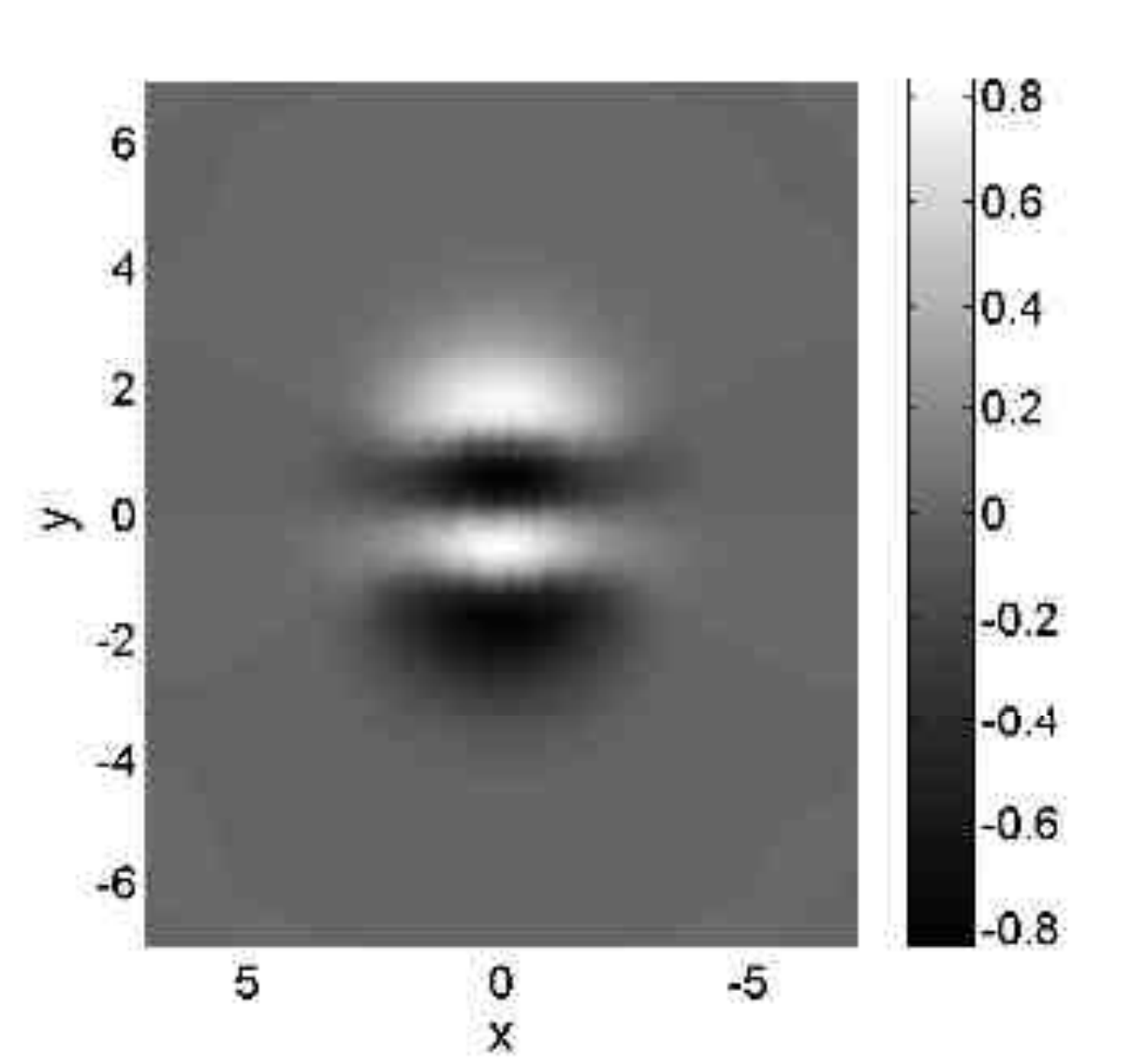}
\label{1Well3rdOrderSolutionMu5}}
\caption{(Color online) Examples of unstable anisotropic higher-order 2D solutions,
produced by stationary equation (\protect\ref{StatNLSE}), with the
single-well nonlinearity-modulation profile (\protect\ref%
{BasicSingleWellProfile}). Panels (a), (b) and (c) display examples of
dipoles, \textquotedblleft string tripoles", and \textquotedblleft string
quadrupoles", respectively, for $\protect\mu =5$ and $\protect\alpha =0.5.$}
\label{1WellHighOrderSolutions}
\end{figure}

\begin{figure}[tbp]
{\ \includegraphics[width=2.8in]{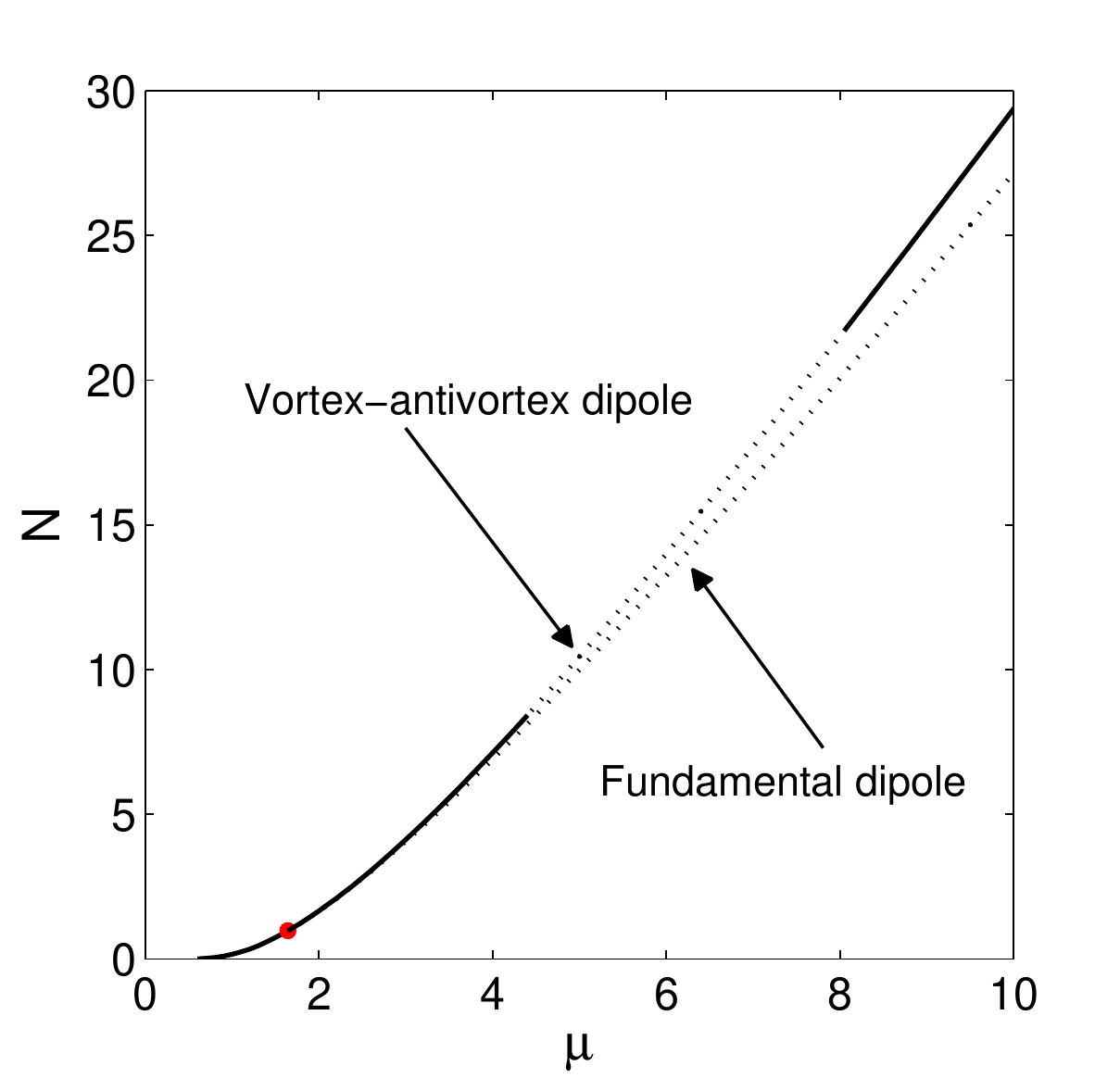}}
\caption{(Color online) The red dot (at $\protect\mu =1.64,N=0.98$) marks
the bifurcation of the family of VADs (vortex-antivortex dipoles) from the
fundamental-dipole-mode branch in the single nonlinear 2D potential well,
for $\protect\alpha =0.5$. As before, solid and dashed curves refer to
stable and unstable states, respectively. The fundamental-dipole-mode branch
is destabilized at the bifurcation point.}
\label{2HoleSolutionMuVsN}
\end{figure}

Examples of first-, second- and third-order states, which may be
interpreted, respectively, as dipoles, ``string tripoles" and ``string
quadrupoles", are displayed in Fig. \ref{1WellHighOrderSolutions}, for $\mu
=5$ and $\alpha =0.5,$ the respective $N(\mu )$ curves being shown, for $%
\alpha =0.5$, in panel (d) of the figure. Direct simulations have
demonstrated that families of the second- and third-order solutions are
entirely unstable, and it is reasonable to conclude that their counterparts
of still higher orders are unstable too. The instability-induced evolution
transform the unstable modes into the ground state, as shown in\ Fig. \ref%
{1Well2DUnstableDipoleEvolution} for an unstable dipole. On the other hand,
as shown in Fig. \ref{2HoleSolutionMuVsN}, dipoles have a narrow stability
interval at small values of $\mu $ (or $N$) (to the left of the red dot in
the figure), which will be reported in detail in another work \cite{Radik}.

\begin{figure}[tbp]
\subfigure[]{\includegraphics[width=2.25in]{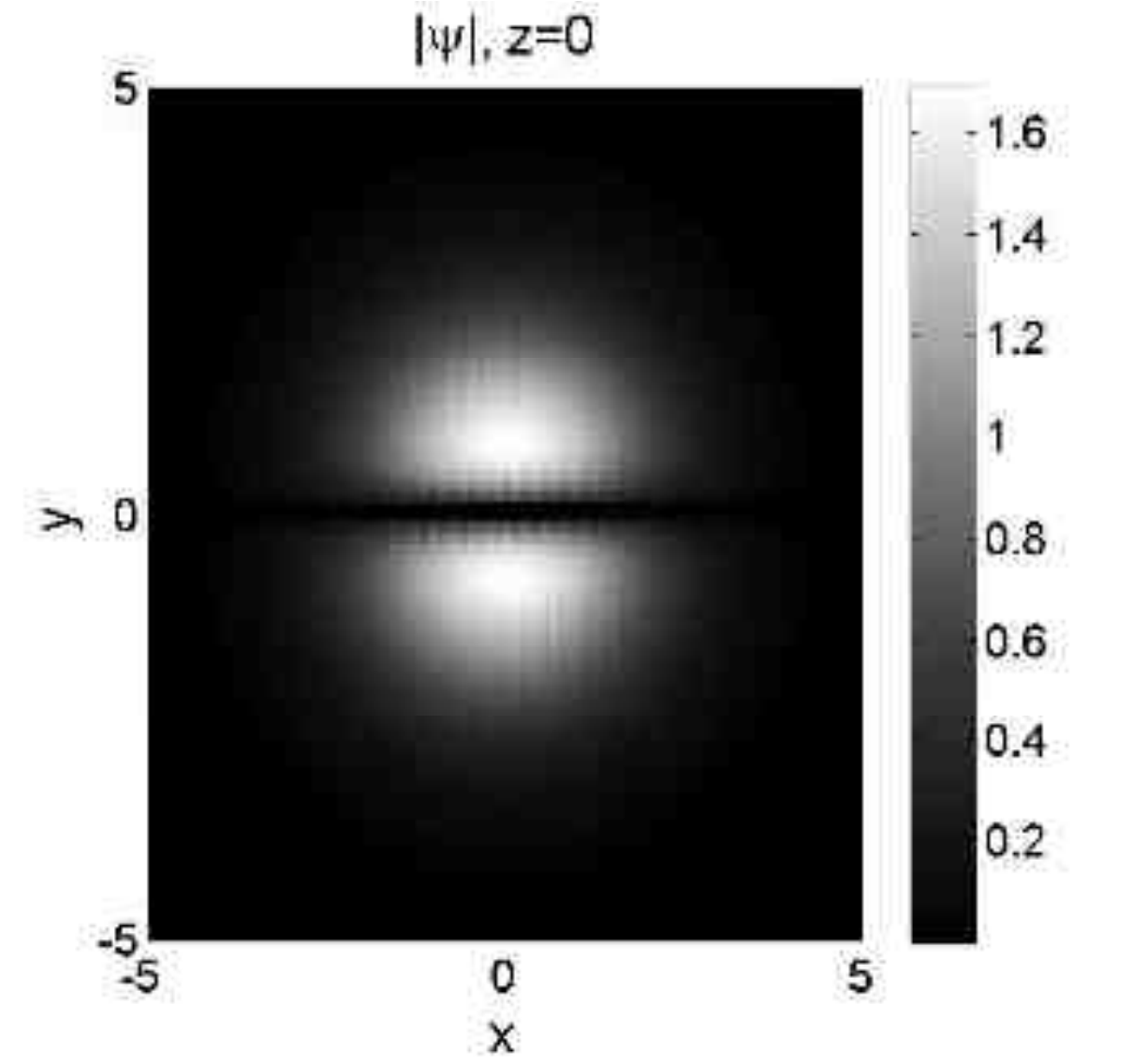}
\label{1Well2DDipoleEvolutiont0}}
\subfigure[]{\includegraphics[width=2.25in]{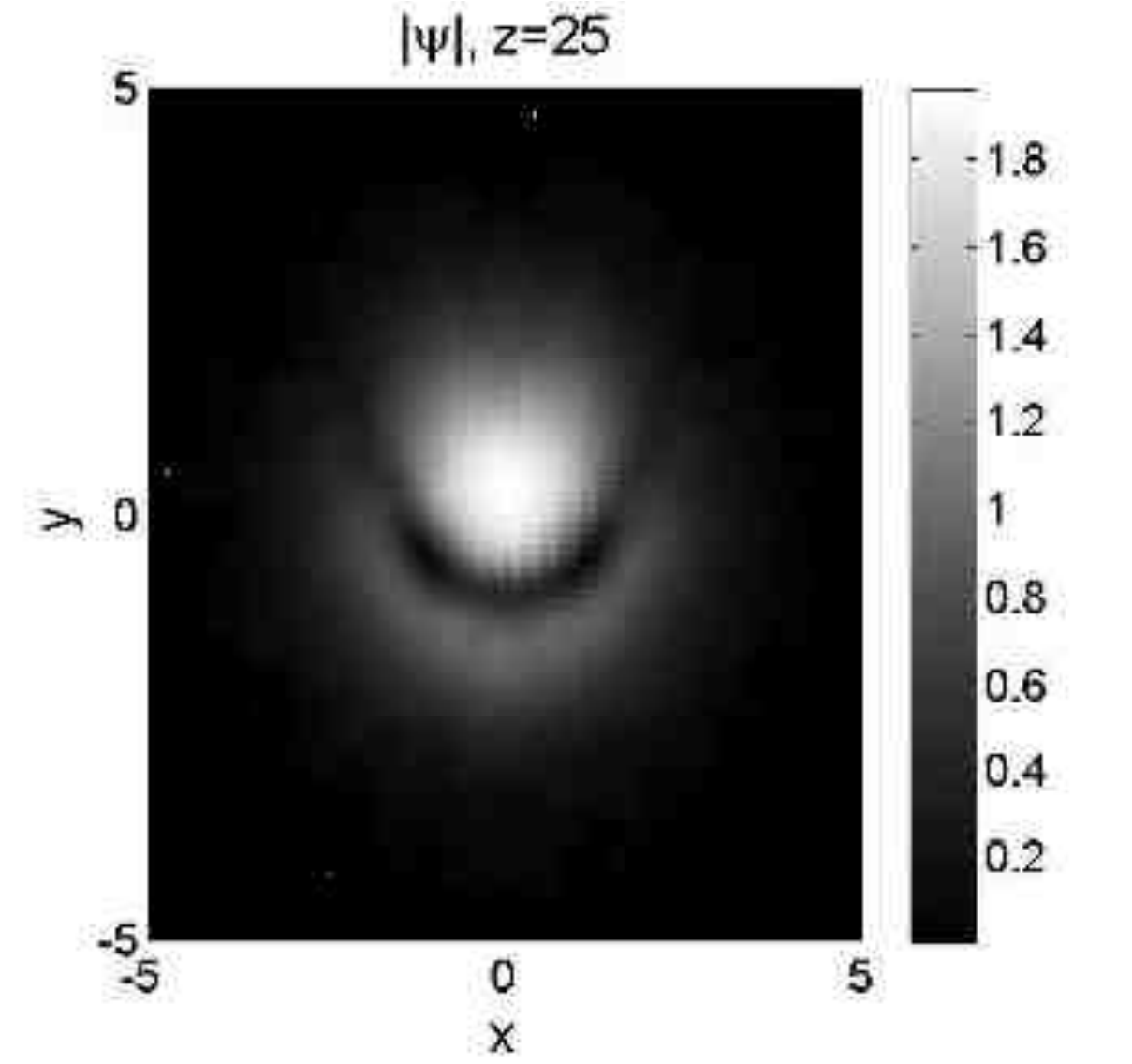}
\label{1Well2DDipoleEvolutiont25}}
\subfigure[]{\includegraphics[width=2.25in]{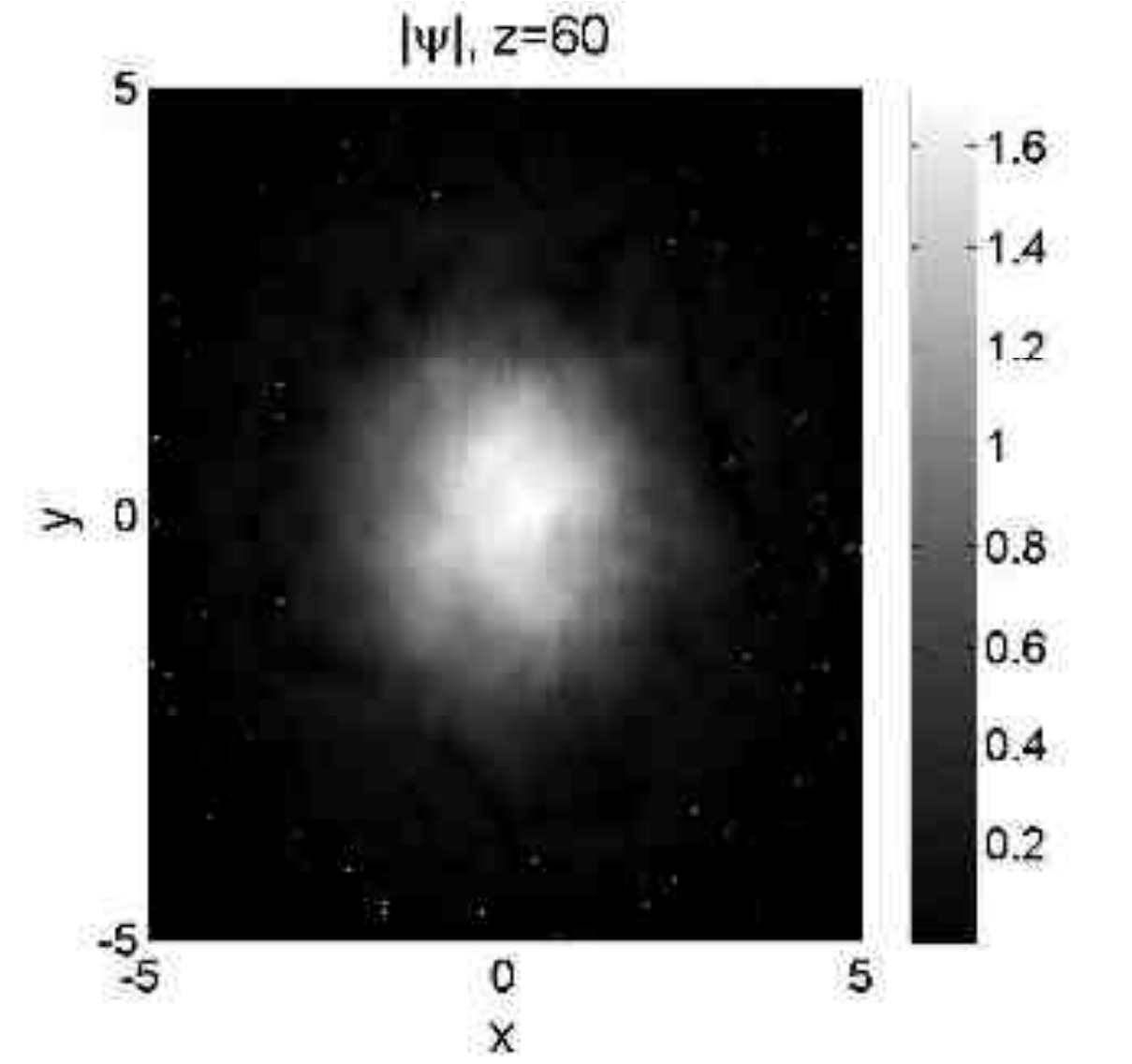}
\label{1Well2DDipoleEvolutiont60}}
\caption{(Color online) The evolution of an unstable 2D dipole mode (the \textquotedblleft
first order" state from Fig. \protect\ref{1WellHighOrderSolutions}), at $%
\protect\mu =5$ and $\protect\alpha =0.5$.}
\label{1Well2DUnstableDipoleEvolution}
\end{figure}

\begin{figure}[tbp]
\subfigure[]{\includegraphics[width=2.6in]{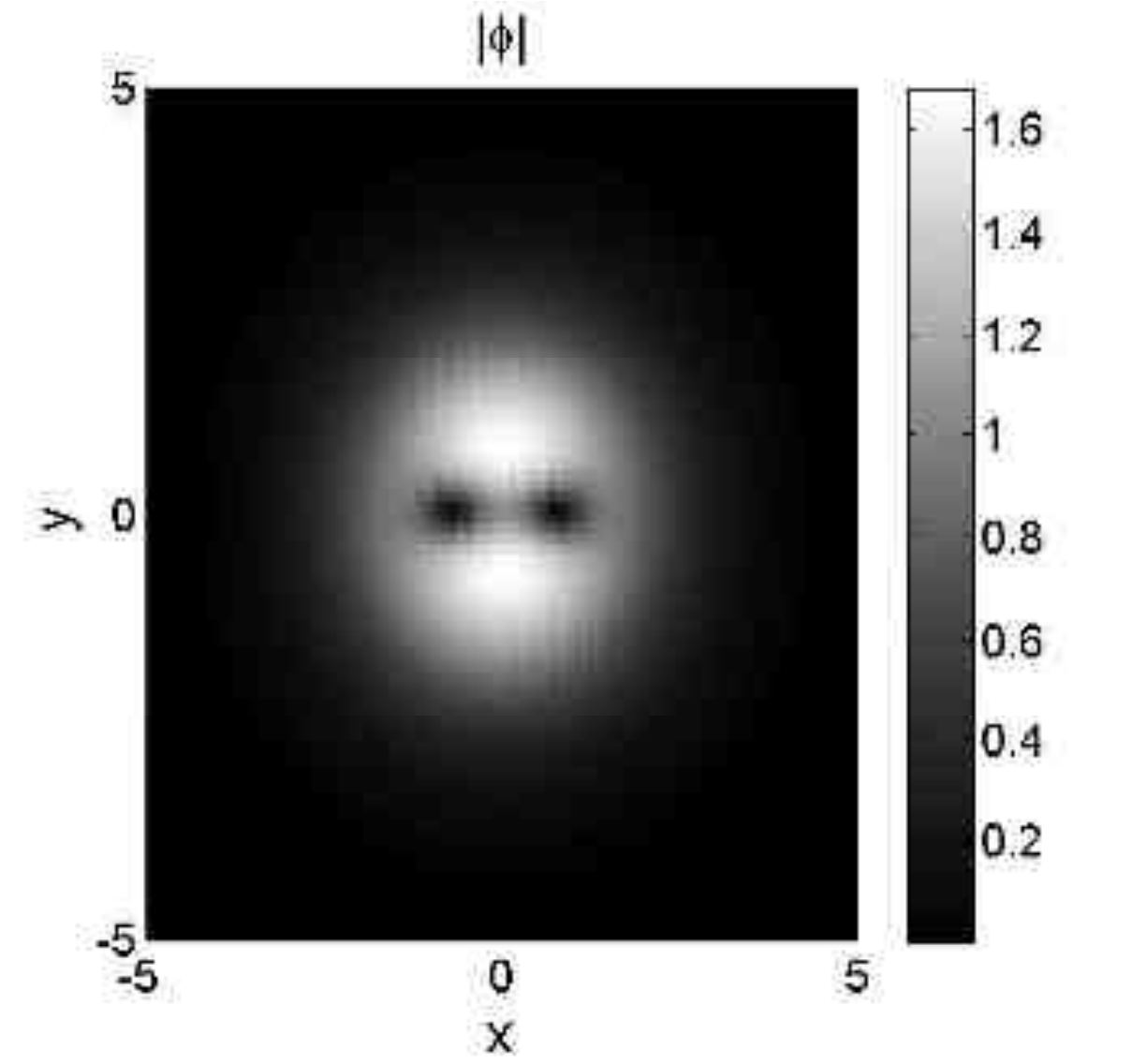}
\label{2HoleSolutionMu5Abs}}
\subfigure[]{\includegraphics[width=2.6in]{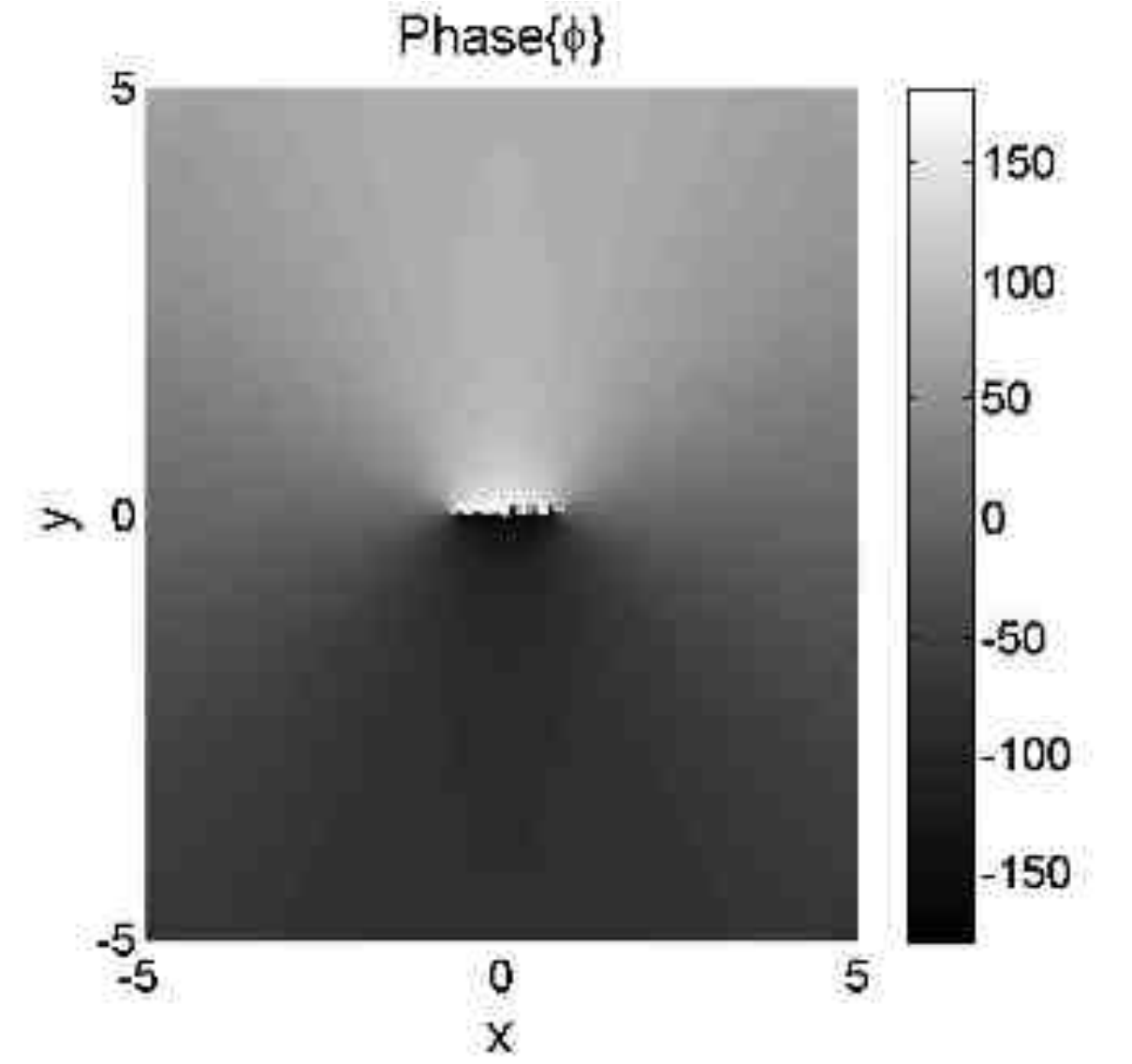}
\label{2HoleSolutionMu5Phase}}
\caption{(Color online) An example of the amplitude (a) and phase (b) structure of a \emph{%
stable} vortex-antivortex dipole trapped in the single 2D nonlinear well,
for $\protect\mu =5$ and $\protect\alpha =0.5$. Here and in similar patterns
displayed below, values of the phase are given in degrees (not in radians).}
\label{2HoleSolutionMu5}
\end{figure}

More interesting 2D topological states trapped in the single nonlinear
potential well are vortex-antivortex dipoles (VADs), shaped as fundamental
solutions with an embedded pair of vortical holes carrying opposite
topological charges, $\pm 1$, so that the total charge is zero, see an
example in Fig. \ref{2HoleSolutionMu5} for $\mu =5$ and $\alpha =0.5$.
Systematic results for this family will be presented in Ref. \cite{Radik}.
In particular, it bifurcates, as a stable branch, from the family of the
above-mentioned ordinary (fundamental) dipoles, as shown in Fig. \ref%
{2HoleSolutionMuVsN}. With the increase of $\mu $ and $N$, the VAD gets
destabilized by oscillatory perturbations, and then \emph{restabilizes}.

\begin{figure}[tbp]
\subfigure[]{\includegraphics[width=2.6in]{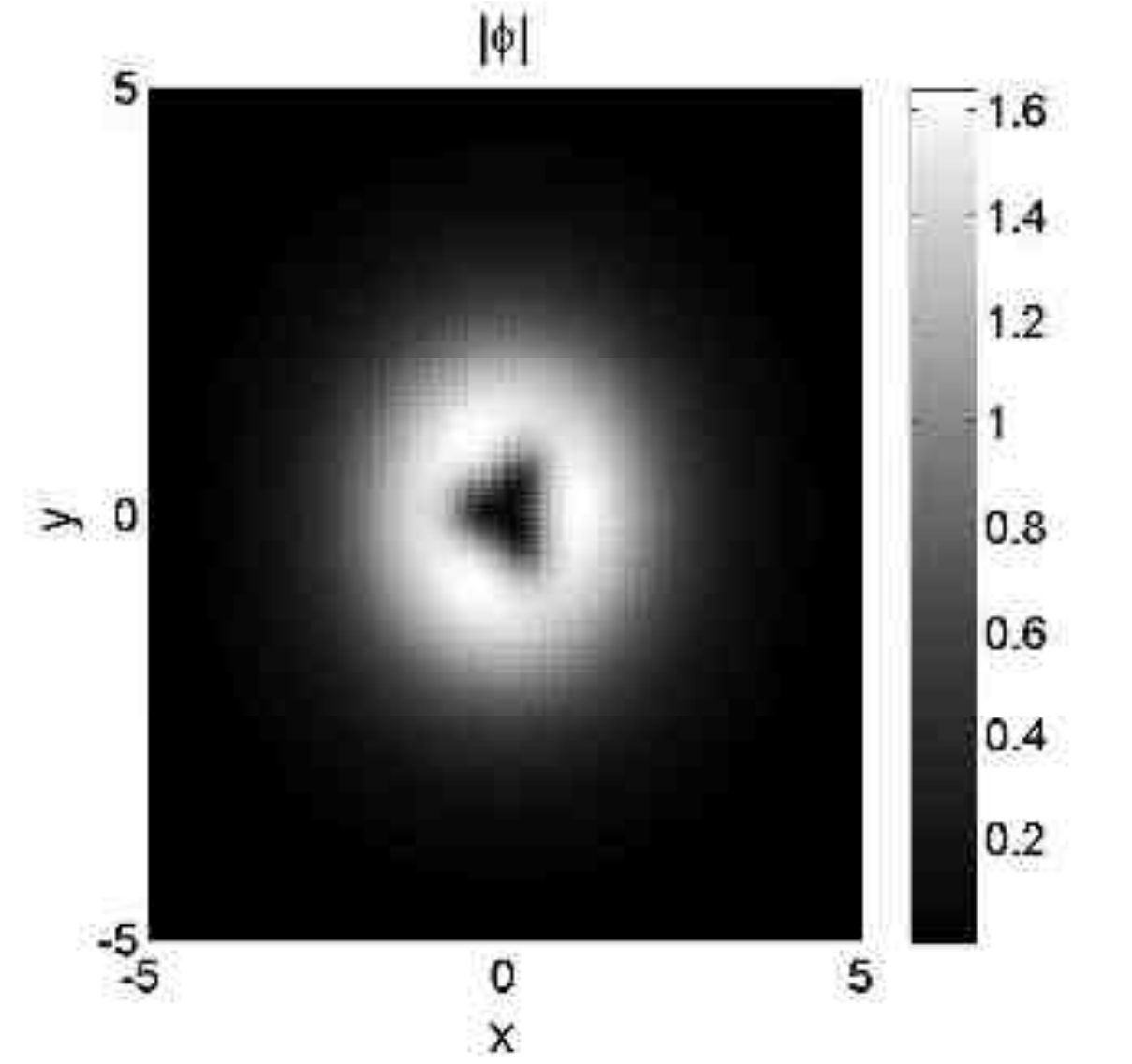}
\label{3HoleSolutionMu5Abs}}
\subfigure[]{\includegraphics[width=2.6in]{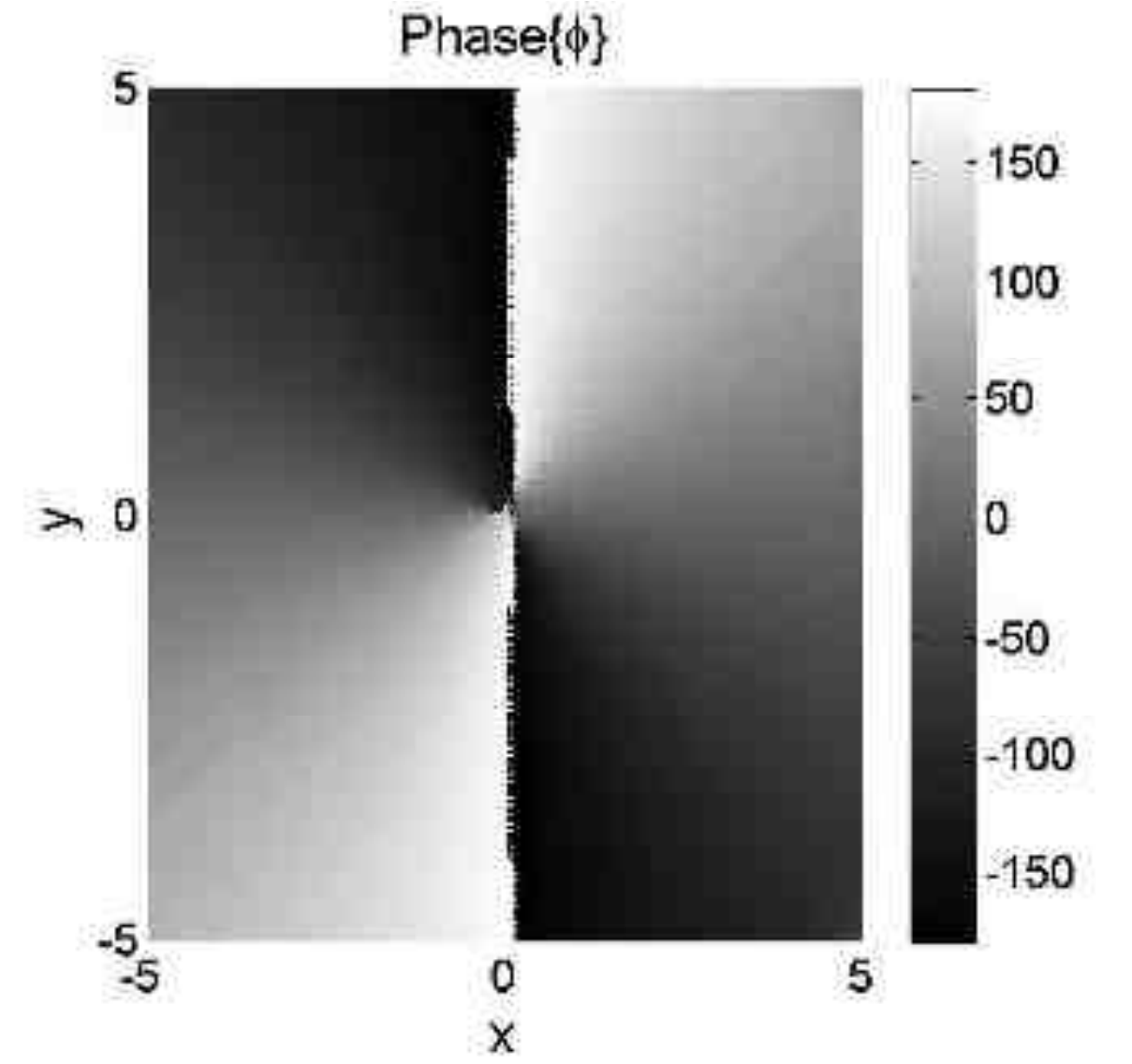}
\label{3HoleSolutionMu5Phase}}
\caption{(Color online) The amplitude (a) and phase (b) structure of a vortex triangle
supported by the single 2D nonlinear potential well, at $\protect\alpha =0.5$%
, $\protect\mu =7$ ($N=25.86$). This mode is weakly unstable, see Fig.
\protect\ref{3HoleSolutionMuVsN} below.}
\label{3HoleSolutionMu5}
\end{figure}

\begin{figure}[tbp]
{\ \includegraphics[width=2.8in]{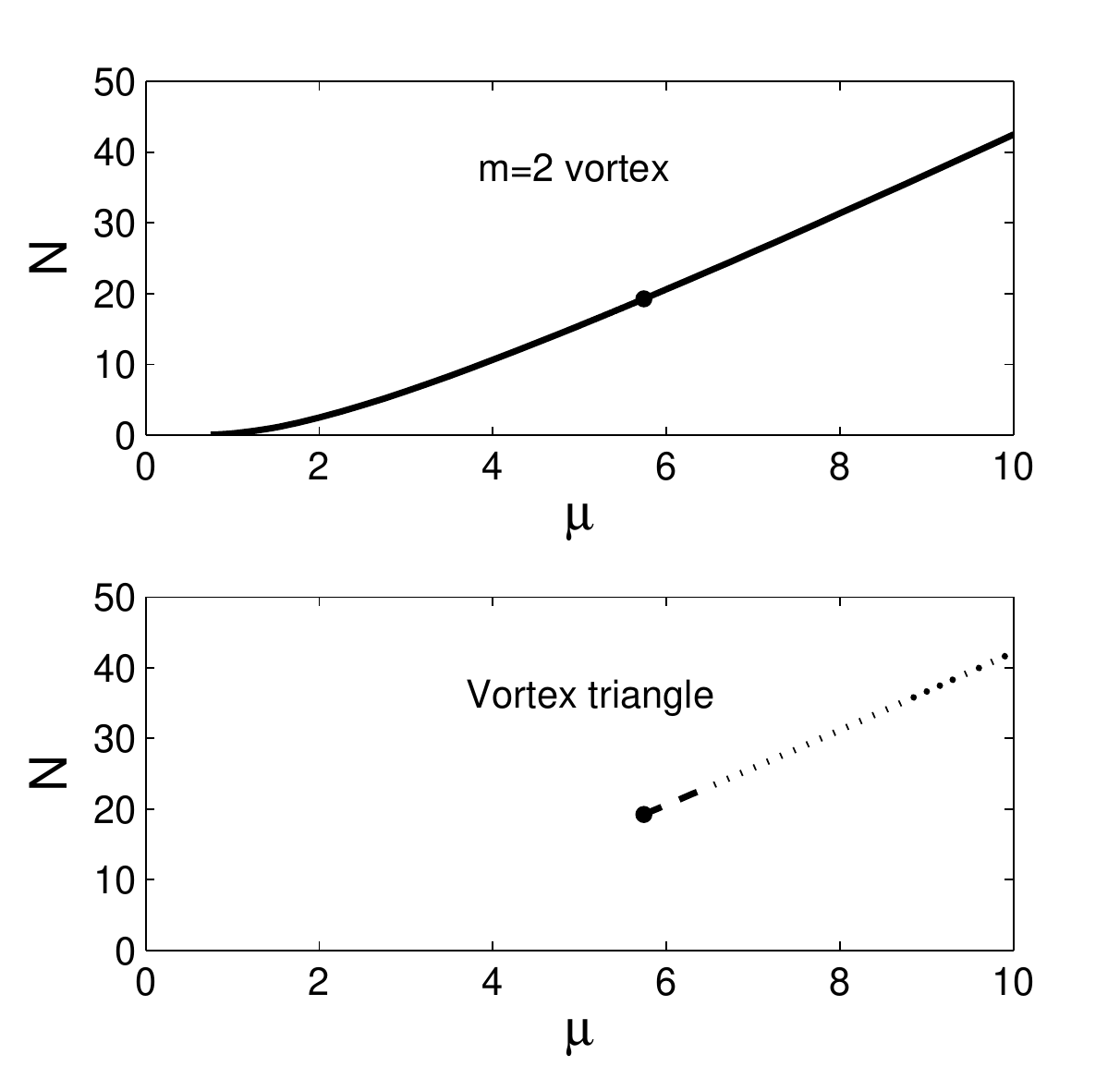}}
\caption{The $N(\protect\mu )$ curves, demonstrating, for $\protect\alpha %
=0.5$, the bifurcation of the the VT (vortex-triangle) branch (the bottom
panel), from the family of ordinary stable vortices with topological charge $%
m=2$ (the upper panel; the curves are displayed in the different panels, as
they would otherwise completely overlap). The circles mark, in both panels,
the bifurcation point, at $\protect\mu =5.73,N=19.2$. As before, solid and
dotted lines refer to stable and unstable solutions, respectively. The short
dashed bold segment refers to spinning VTs.}
\label{3HoleSolutionMuVsN}
\end{figure}

The branch of $m=2$ vortices gives rise, via a bifurcation, to another
anisotropic topologically charged mode trapped in the single nonlinear
potential well, namely, \textit{vortex triangles }(VTs), see an example in
Fig. \ref{3HoleSolutionMu5} for $\mu =7$, $\alpha =0.5$. As shown in Fig. %
\ref{3HoleSolutionMuVsN}, this family indeed bifurcates, at
\begin{equation}
\mu =5.73,N=19.2  \label{VT-bif}
\end{equation}
(if $\alpha =0.5$ is fixed), from the ordinary stable vortex branch with $%
m=2 $, which was constructed, in the framework of the present model, in Ref.
\cite{Barcelona2}. Figure \ref{3HoleSolutionMu5}(b) confirms that, following
its parent vortex mode, the VT carries topological charge $m=2$. The fact
that the \emph{triangular} vortical state emerges from the \emph{%
double-charged} vortex is a counter-intuitive manifestation of the SSB in
the present setting, which, however, does not contradict general principles.
Note that the present VT modes are different from rotating triangles built
of \emph{well-separated} unitary vortices, which were observed, as a
dynamical regime initiated by splitting of unstable vortices with
topological charge $m=3$ (rather than $m=2$), in the same model \cite%
{Barcelona2}.

The stability analysis of the VTs demonstrates that, strictly speaking, the
entire family is unstable. Further, direct simulations of the perturbed
evolution demonstrate that, from bifurcation point (\ref{VT-bif}), at which
this branch appears, and up to $\mu =6.55$ ($N=23.43$), the VT exhibits
rotation, as a robust object, at an angular velocity whose value depends on
the initial perturbation. The rotation interval is designated by the (short)
dashed bold segment in the bottom panel of Fig. \ref{3HoleSolutionMuVsN}).
An example for such a spinning VT is shown in Fig. \ref%
{1Well2DRotatingVortexTrioEvolution} for $\mu =6$.

At $\mu >6.55$, the VTs develop real instability, evolving into the
above-mentioned stable VADs, see an example in Fig. \ref%
{1Well2DUnstableVortexTrioEvolution}, or into the ground state, as observed
above for other species of unstable modes.

\begin{figure}[tbp]
\subfigure[]{\includegraphics[width=1.60in]{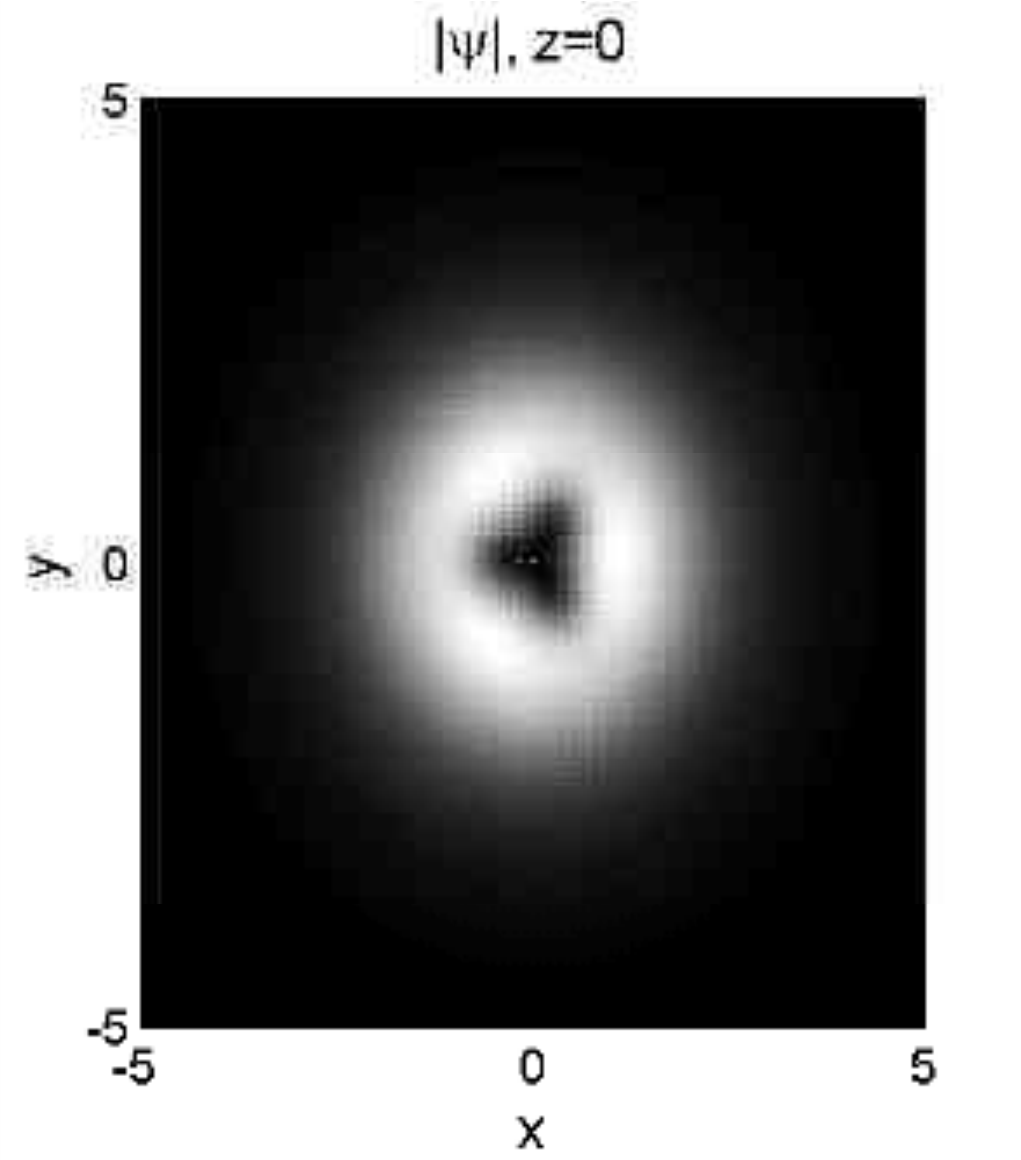}
\label{1Well2DVortexTrioEvolutionMu6z0s}}
\subfigure[]{\includegraphics[width=1.60in]{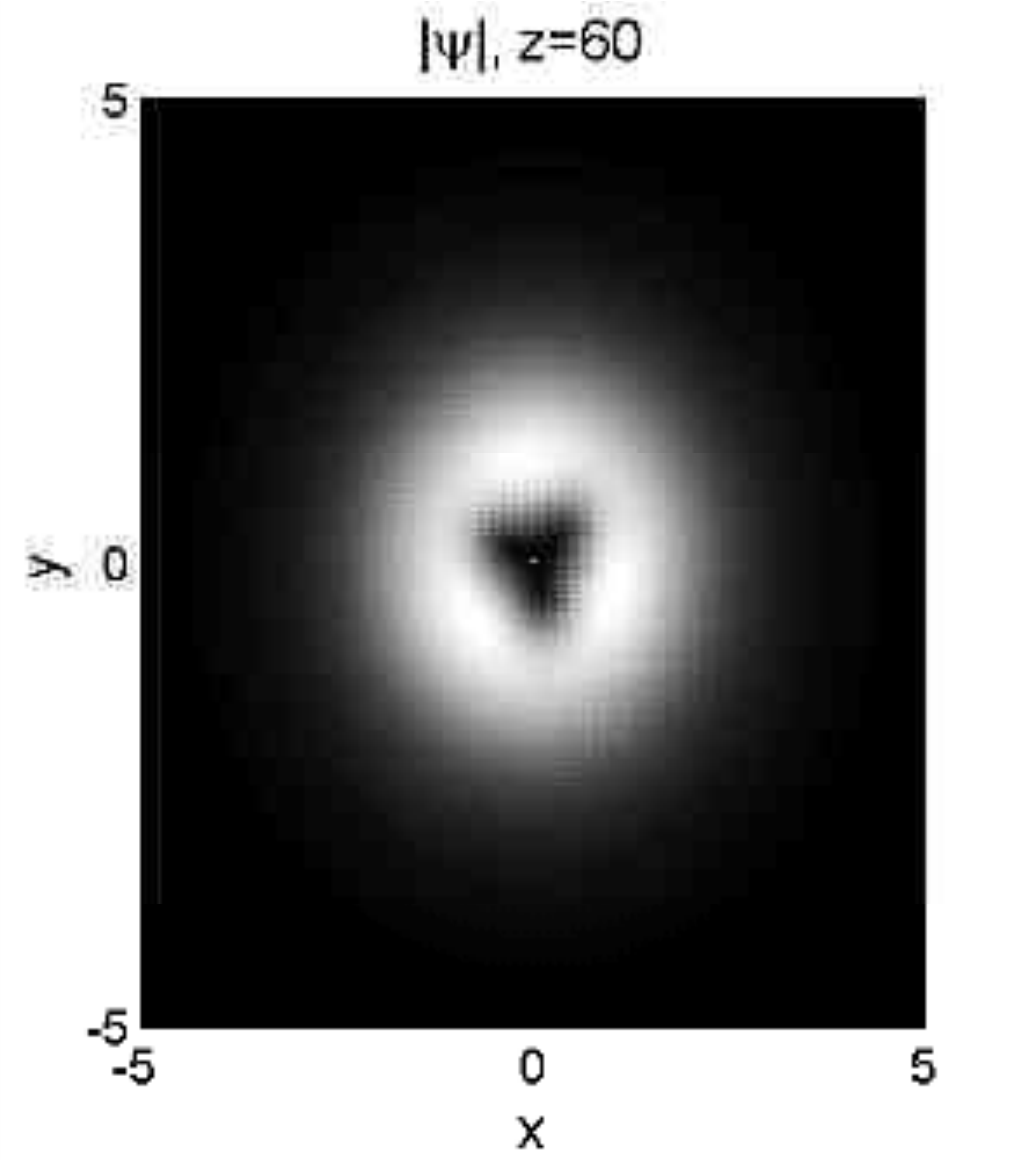}
\label{1Well2DVortexTrioEvolutionMu6z60s}}
\subfigure[]{\includegraphics[width=1.60in]{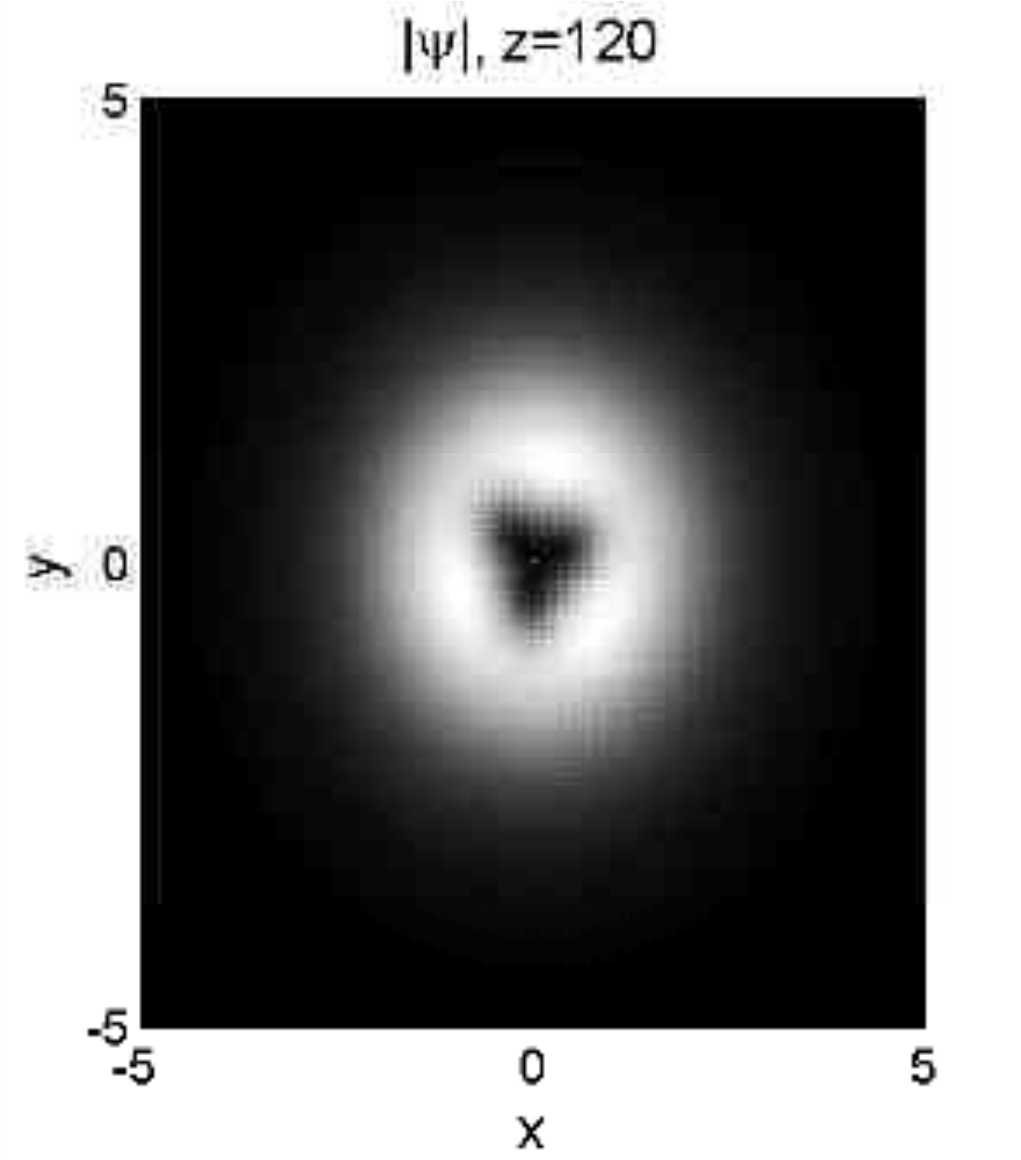}
\label{1Well2DVortexTrioEvolutionMu6z120s}}
\subfigure[]{\includegraphics[width=1.93in]{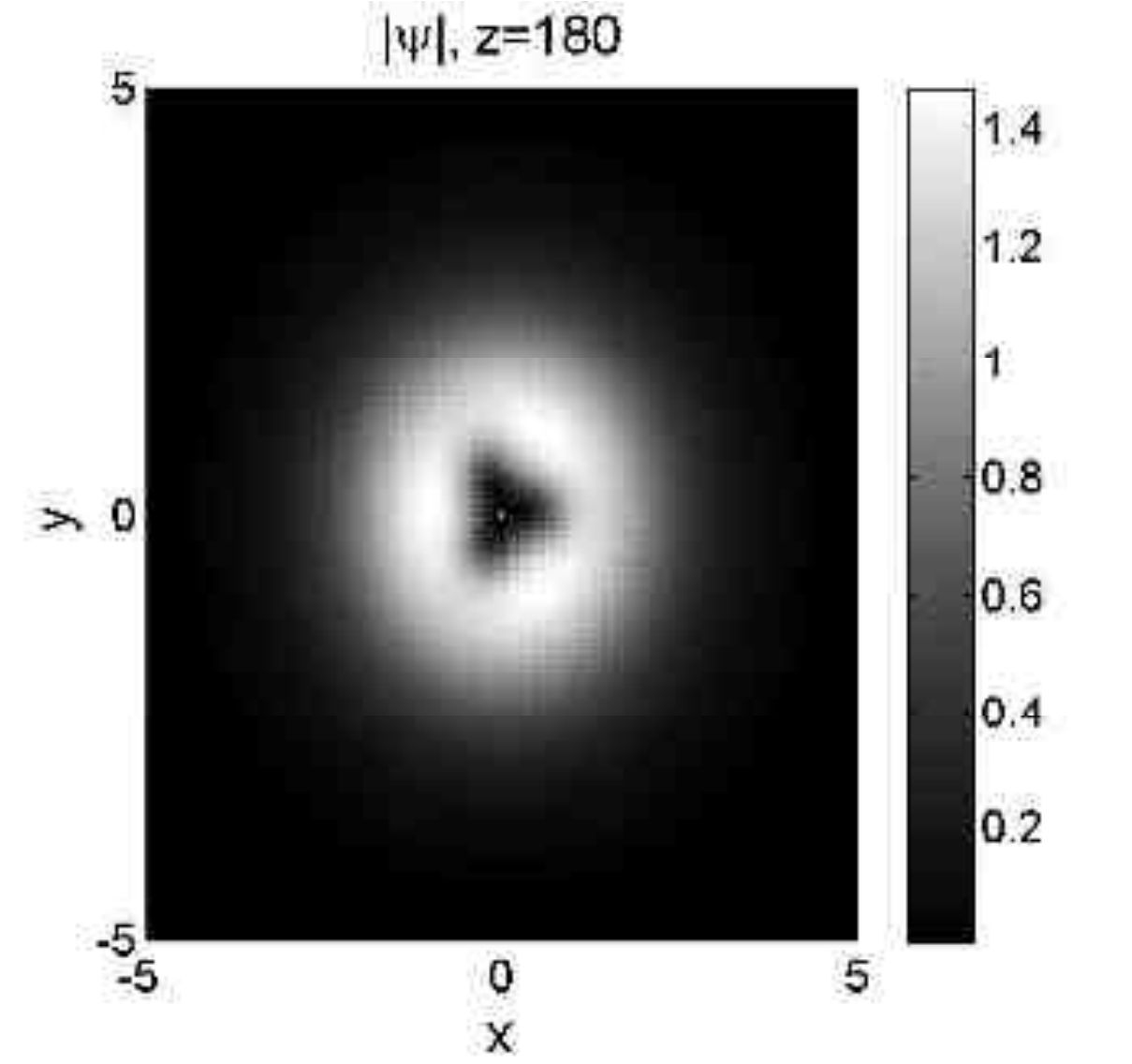}
\label{1Well2DVortexTrioEvolutionMu6z180s}}
\caption{(Color online) The evolution of a robust vortex triangle, trapped in the single
nonlinear potential well, at $\protect\alpha =0.5$ and $\protect\mu =6$ ($%
N=20.58$), with an initial perturbation that slightly increases its norm.
The result is a triangle steadily rotating in the clockwise direction.}
\label{1Well2DRotatingVortexTrioEvolution}
\end{figure}

\begin{figure}[tbp]
\subfigure[]{\includegraphics[width=1.60in]{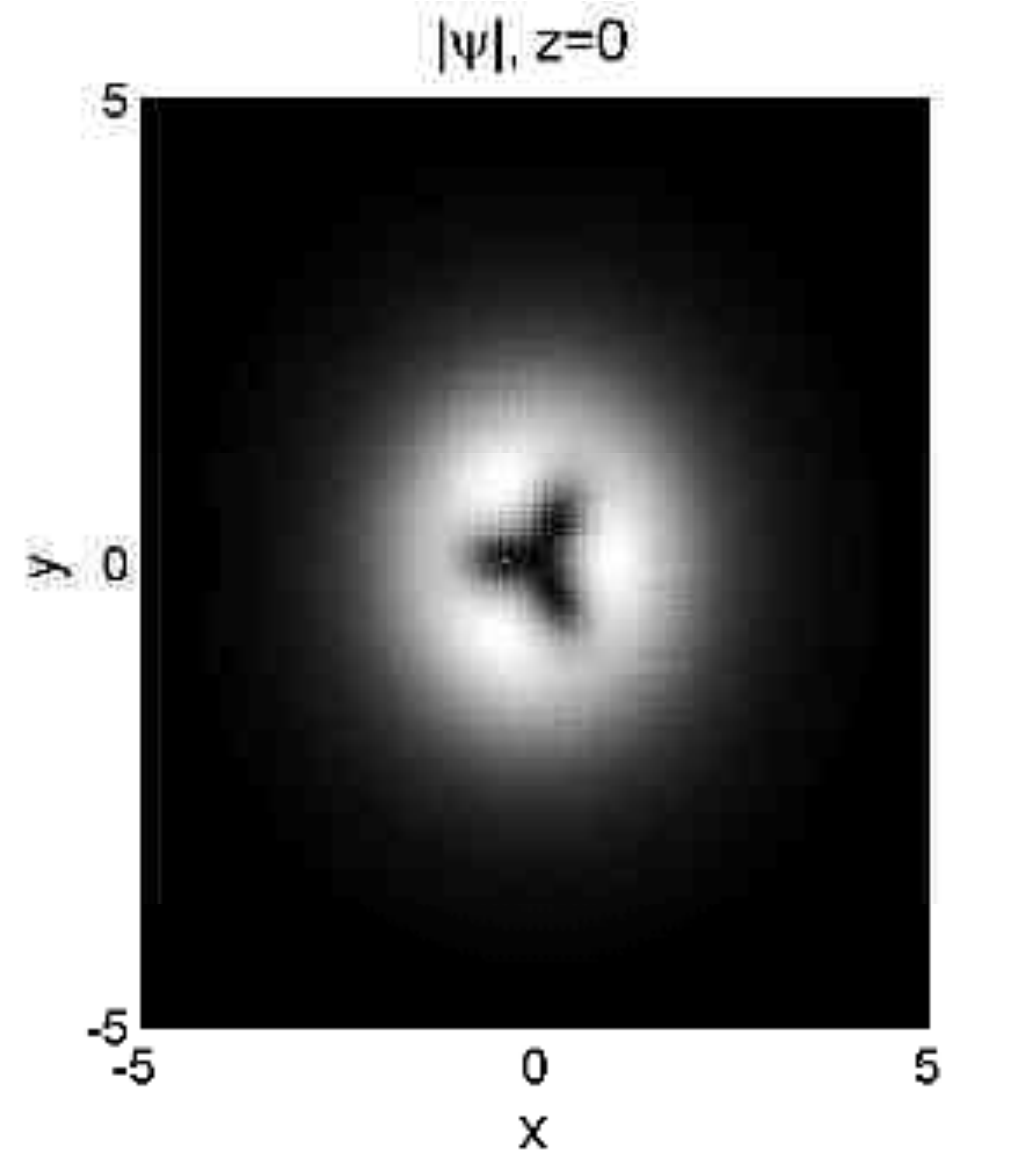}
\label{1Well2DVortexTrioEvolutionMu7t0}}
\subfigure[]{\includegraphics[width=1.60in]{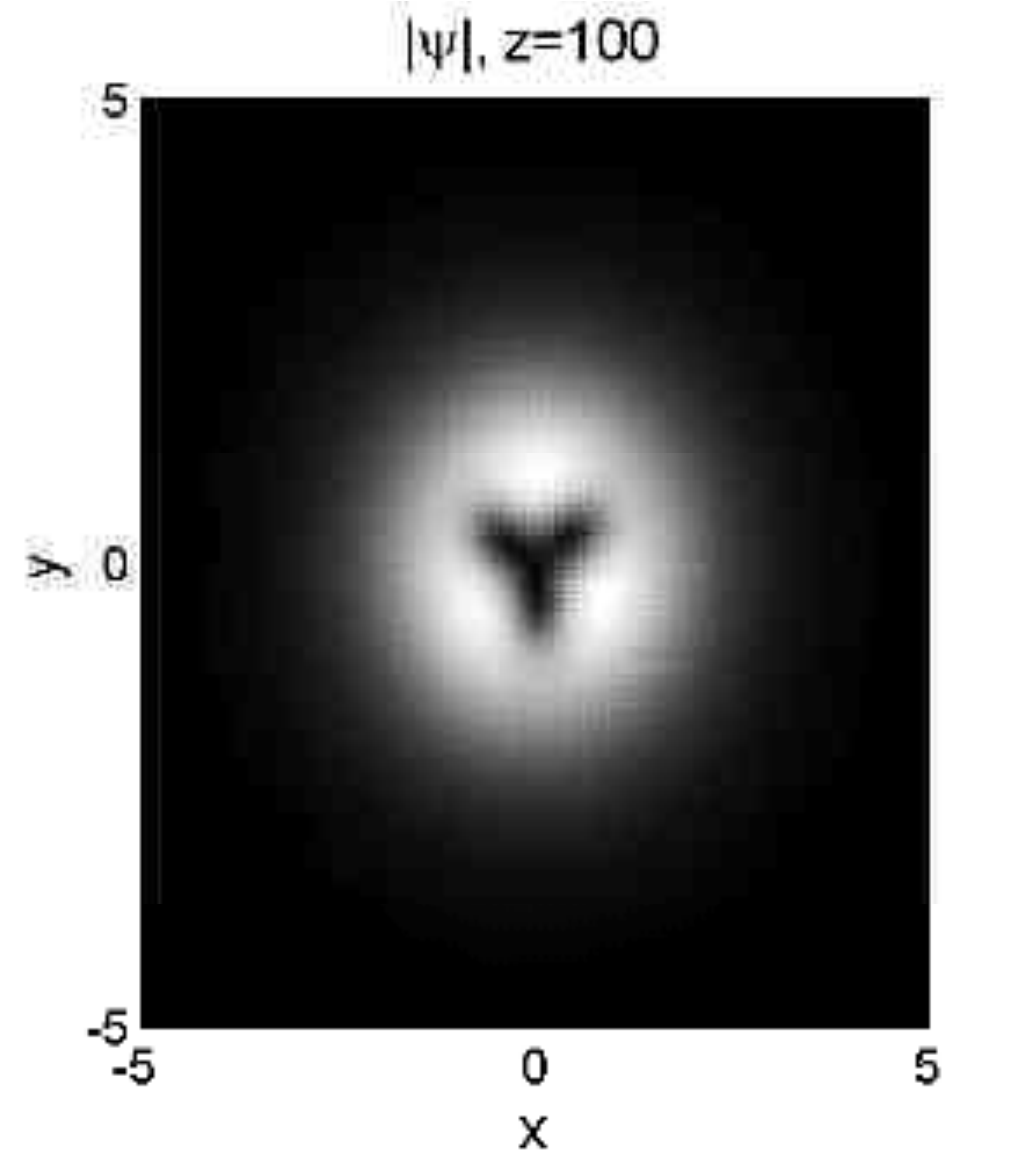}
\label{1Well2DVortexTrioEvolutionMu7t100}}
\subfigure[]{\includegraphics[width=1.60in]{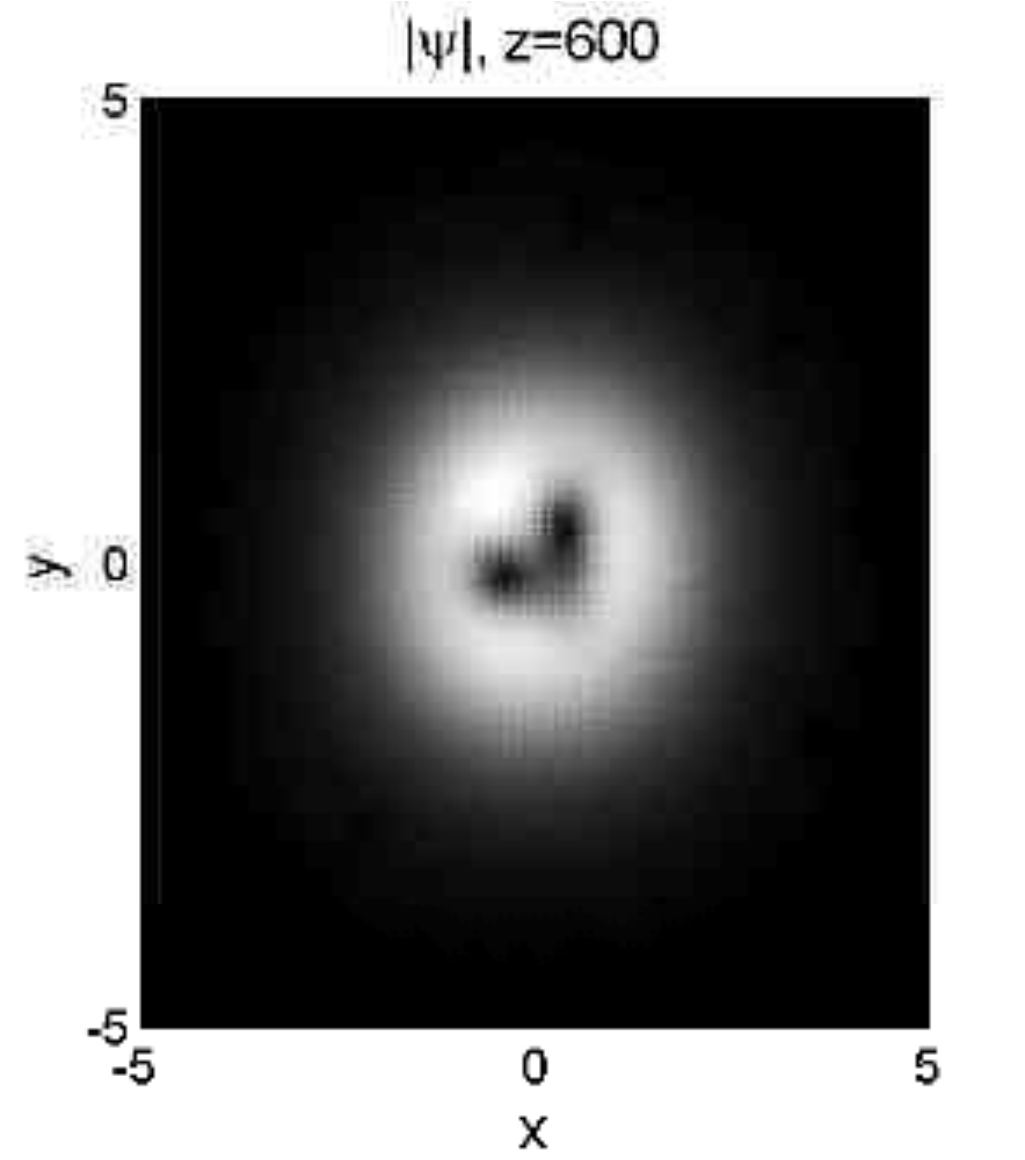}
\label{1Well2DVortexTrioEvolutionMu7t600}}
\subfigure[]{\includegraphics[width=1.93in]{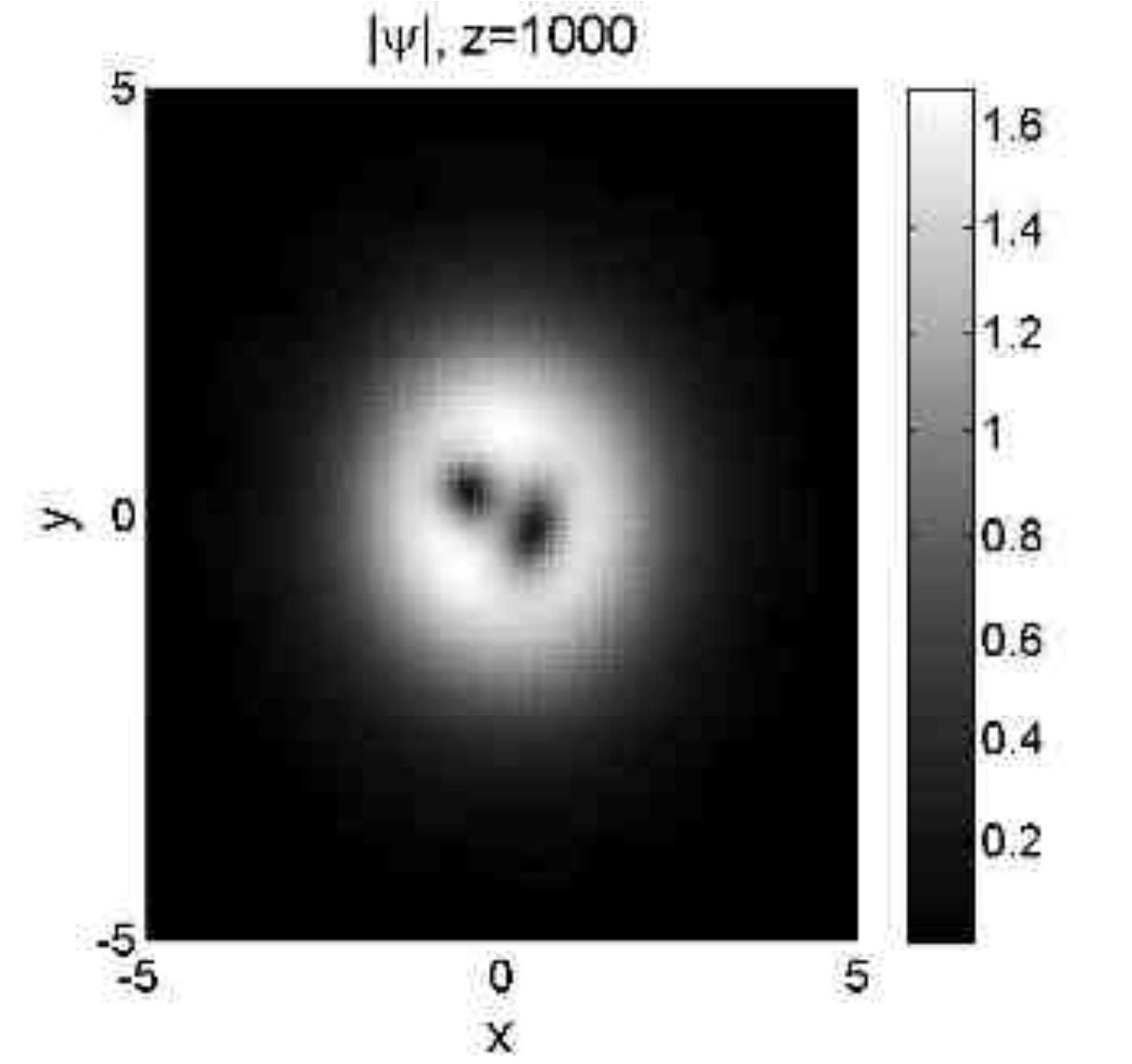}
\label{1Well2DVortexTrioEvolutionMu7t1000}}
\caption{(Color online) The evolution of an unstable vortex triangle at $\protect\alpha %
=0.5 $ and $\protect\mu =7$ ($N=25.86$).}
\label{1Well2DUnstableVortexTrioEvolution}
\end{figure}

\begin{figure}[tbp]
\subfigure[]{\includegraphics[width=2.6in]{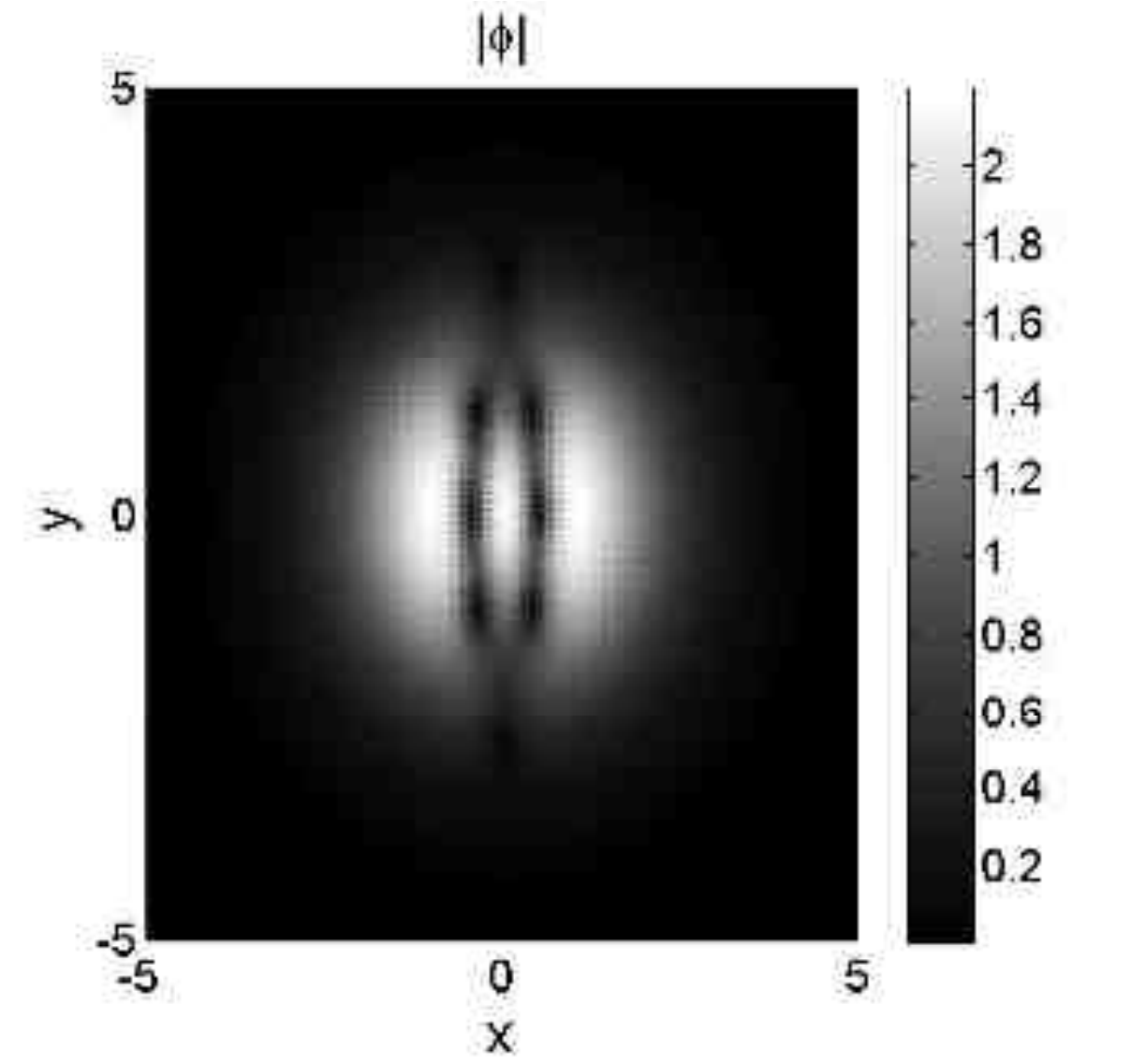}
\label{InternalVortexHexagonMu5Abs}}
\subfigure[]{\includegraphics[width=2.6in]{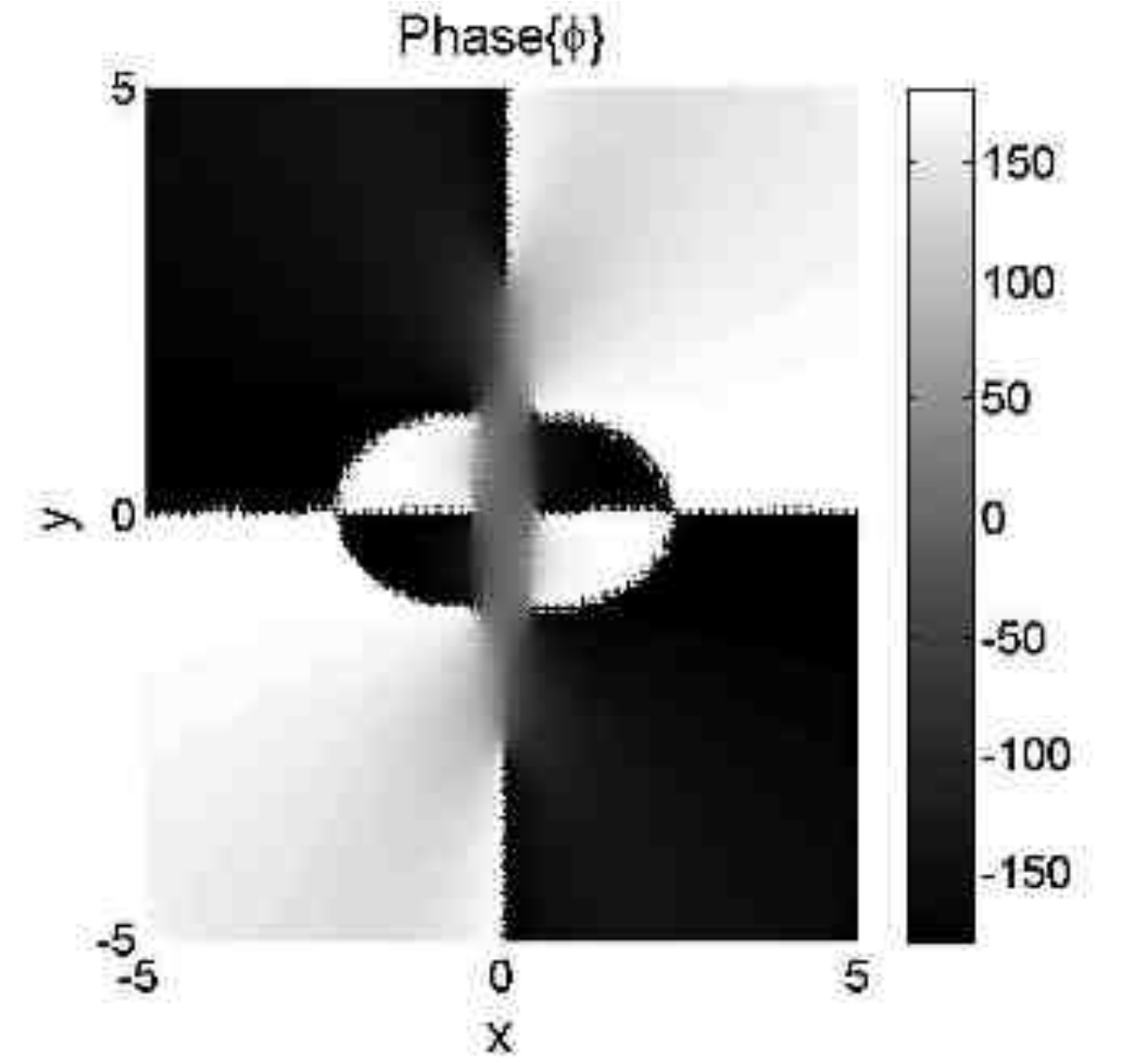}
\label{InternalVortexHexagonMu5Phase}}
\caption{(Color online) An example of the amplitude (a) and phase (b) structure of an
unstable vortex hexagon trapped in a single nonlinear potential well, for $%
\protect\mu =10$, $\protect\alpha =0.5$.}
\label{AdditionalInternalVortexSolutions}
\end{figure}

The systematic numerical analysis has revealed other species of anisotropic
modes trapped in the single isotropic nonlinear potential well. In
particular, bound states of multiple (more than three) vortices were
observed. One such species, a ``vortex hexagon" with zero total topological
charge, which is shown in Fig. \ref{AdditionalInternalVortexSolutions},
bifurcates from the unstable branch of tripoles from Fig. \ref%
{1WellHighOrderSolutions}(b). This state, as well all other bound states of
many vortices, are completely unstable.

Another type of solutions, in the form of isotropic higher-order (alias,
excited) radial states of vortices, were found too, for the first time in
the model of the present type. An example of an excited radial state of the
vortex solution, with $m=1$ and one radial node (zero), is shown in Fig. \ref%
{HighOrderVortexes}. These solutions are unstable too, relaxing into the
stable basic vortices (with the node-free radial structure).

\begin{figure}[tbp]
\subfigure[]{\includegraphics[width=2.6in]{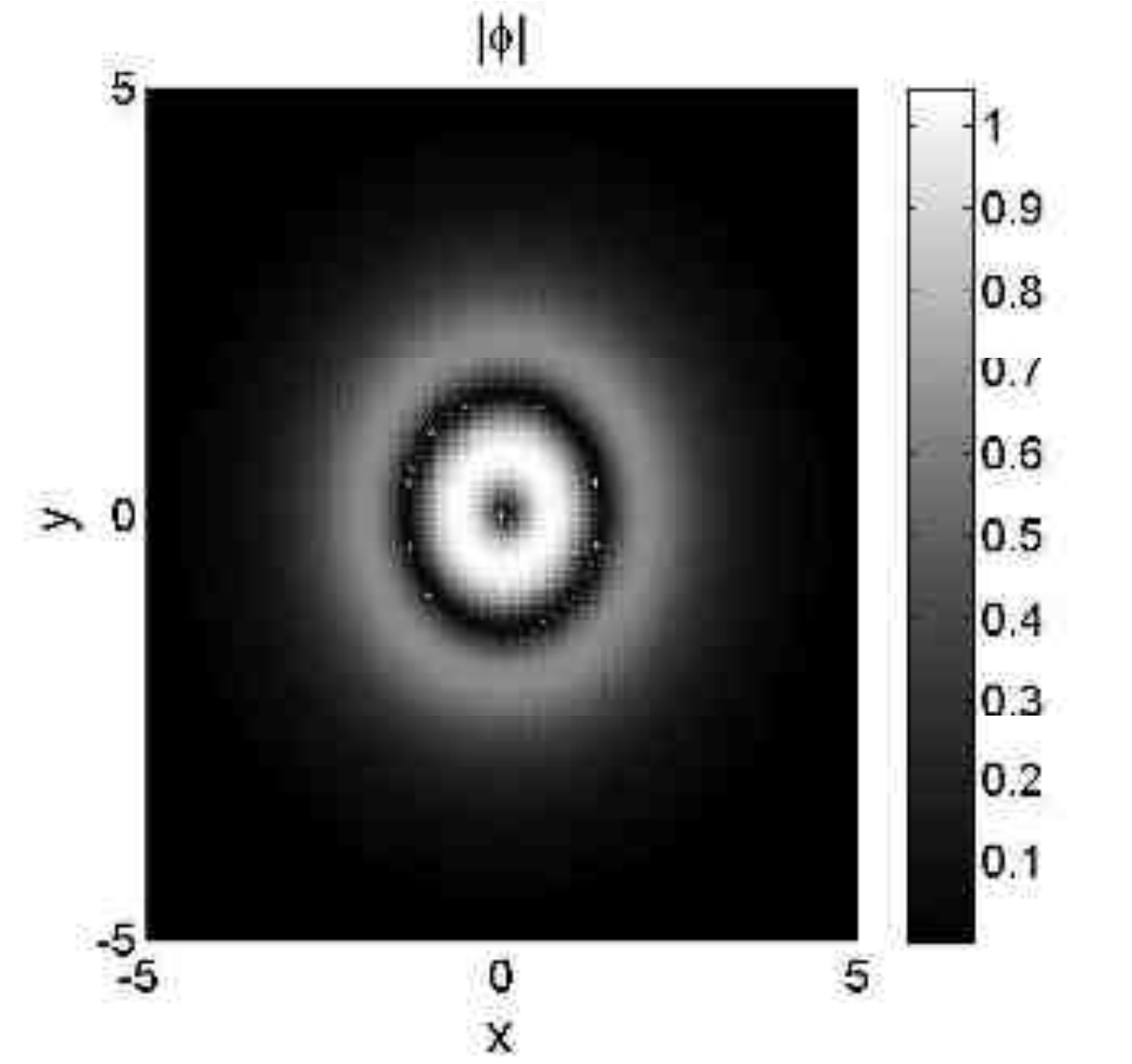}
\label{SecondOrderVortexMu5Abs}}
\subfigure[]{\includegraphics[width=2.6in]{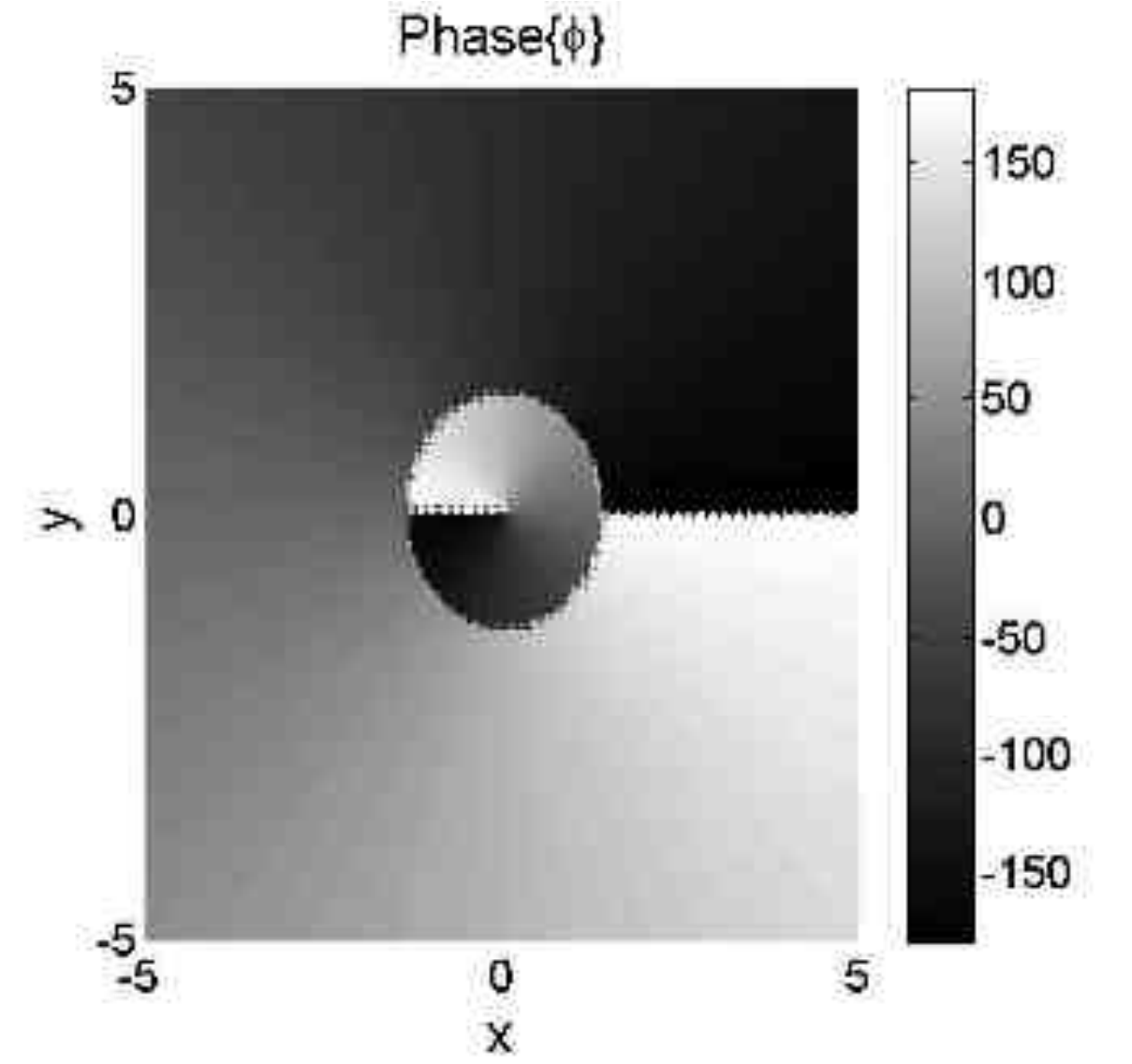}
\label{SecondOrderVortexMu5Phase}}
\caption{(Color online) An example of the amplitude (a) and phase (b) structure of an
unstable excited radial vortex state, with $m=1$, $\protect\mu =5$ and $%
\protect\alpha =0.5$.}
\label{HighOrderVortexes}
\end{figure}

\section{The two-dimensional setting: the double-well (DW) nonlinear
potential}

\subsection{Zero-vorticity states}

\begin{figure}[tbp]
\subfigure[]{\includegraphics[width=2.2in]{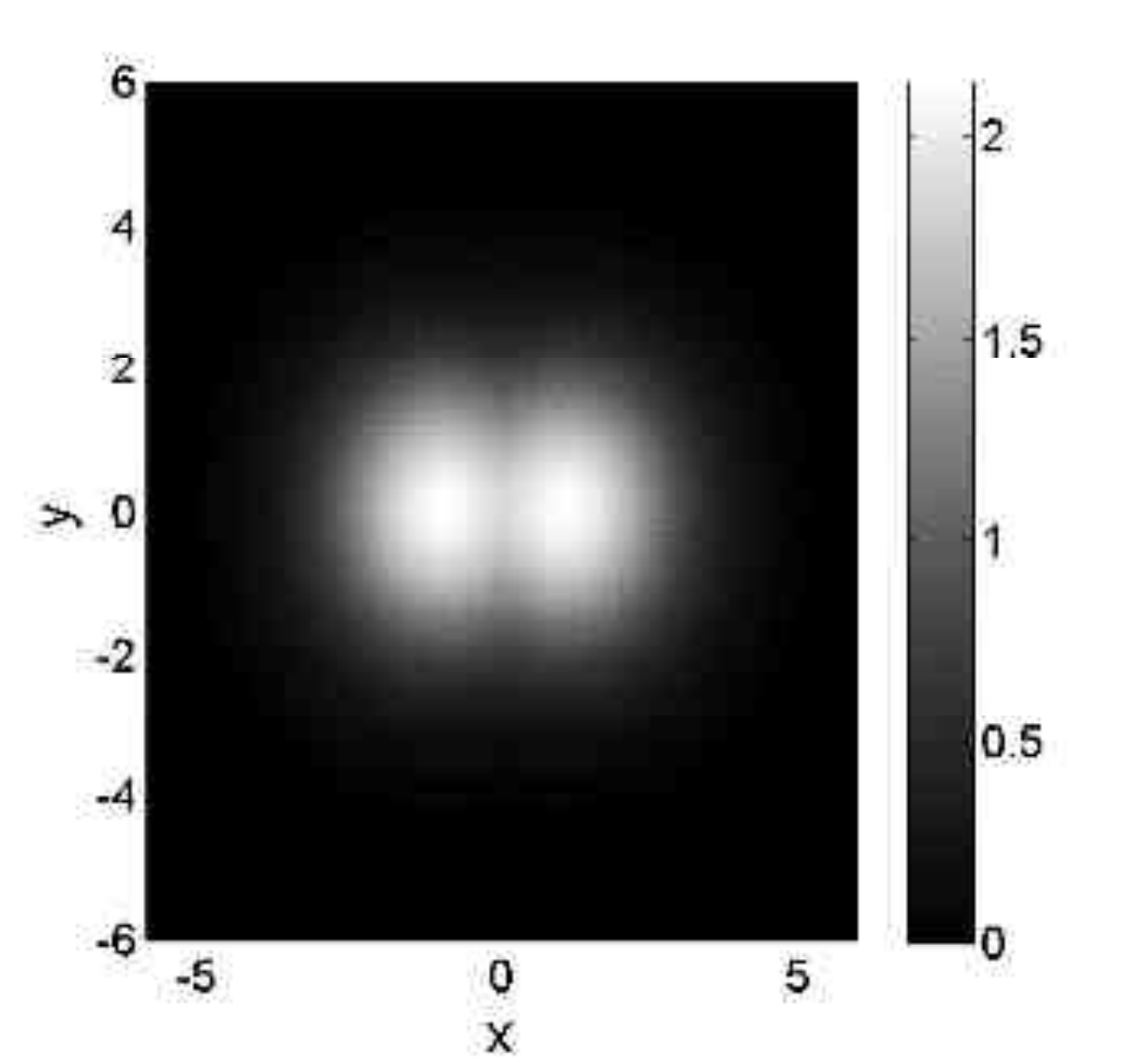}
\label{2WellSymmetricSolutionMu5}}
\subfigure[]{\includegraphics[width=2.2in]{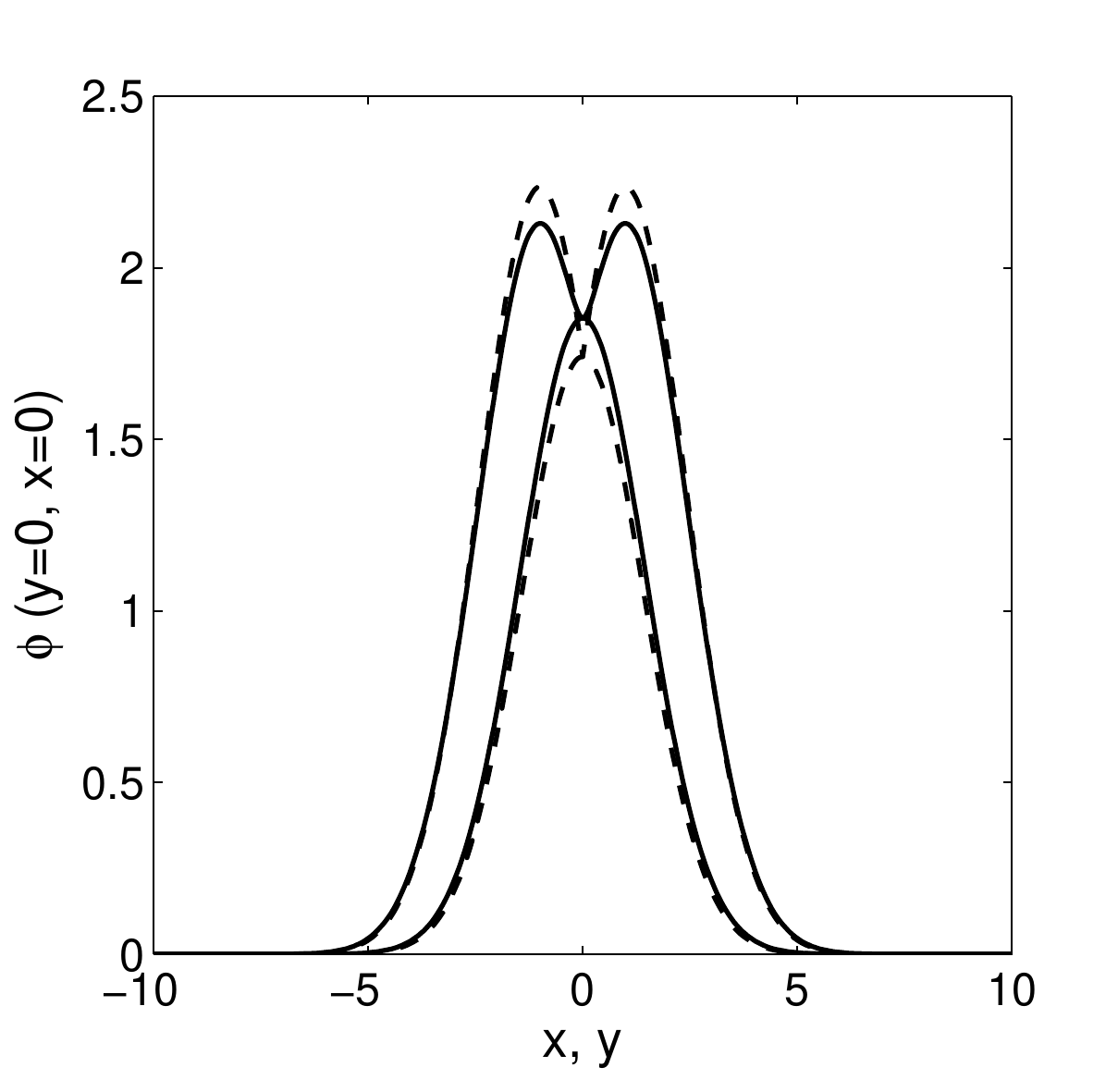}
\label{2WellSymmetricSolutionMu5TFA}}
\subfigure[]{\includegraphics[width=2.2in]{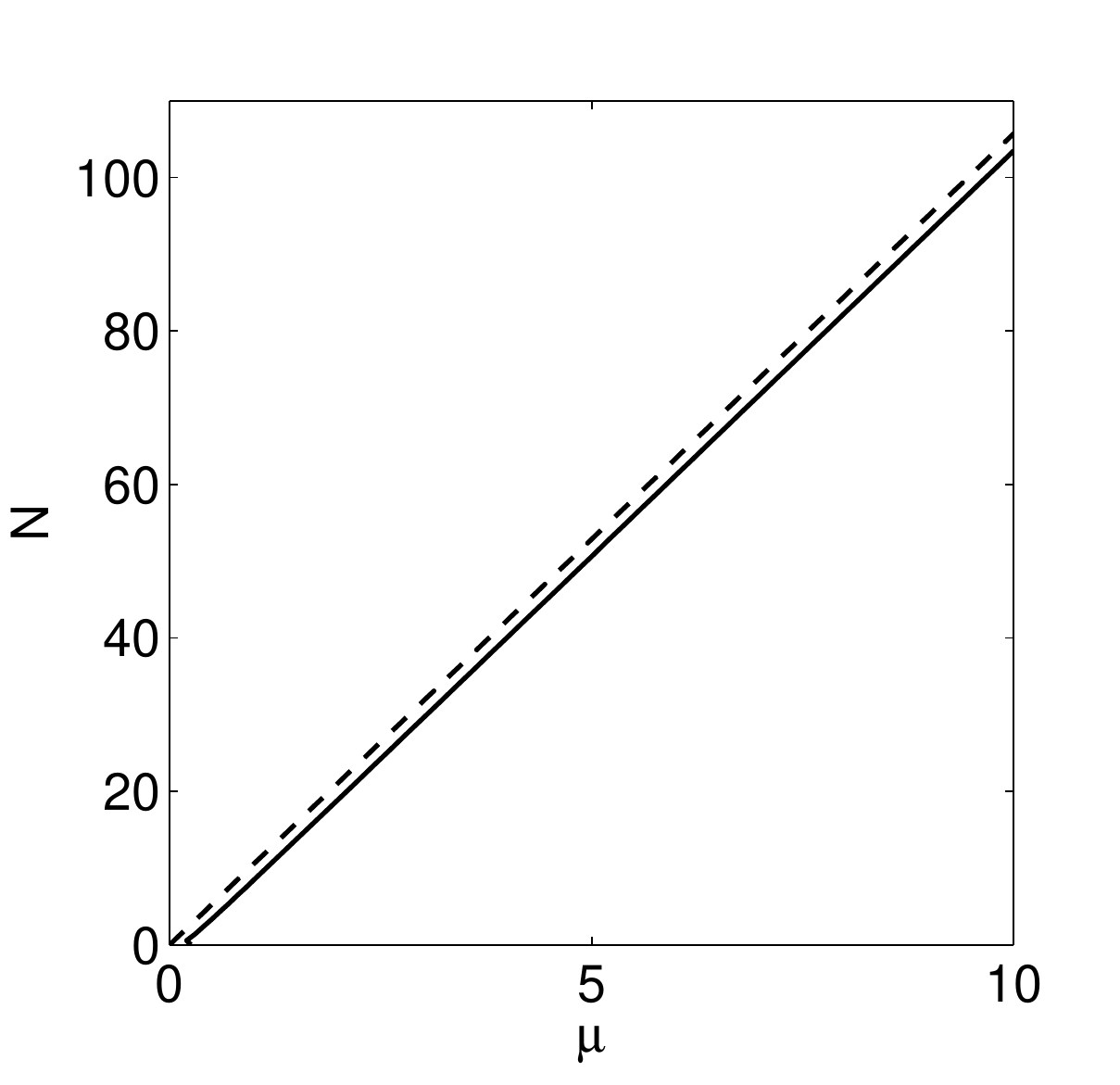}
\label{2WellSymmetricSolutionNvsMu}}
\caption{(Color online) (a) A typical example of the 2D stable symmetric fundamental state
in the model with the double-well nonlinearity profile [see Eq. (\protect\ref%
{simple})] for $\protect\mu =5$ $\protect\alpha =0.5$ and $x_{0}=1$. (b) The
cross-section profiles of the same solution, drawn through $x=0$ and $y=0$.
The solid and dashed lines display, respectively, the numerically found
solution and its counterpart produced by the TFA (Thomas-Fermi
approximation) based on Eq. (\protect\ref{2DTFA}). (c) The respective $N(%
\protect\mu )$ curve of stable symmetric solutions for $\protect\alpha =0.5$
and $x_{0}=1$, with the dashed line representing the respective TFA given by
Eq. (\protect\ref{2DNTFA}).}
\label{2WellSymmetricSolution}
\end{figure}

The most essential object of the 2D analysis is the model with the DW
nonlinearity profile, represented by Eq. (\ref{simple}), focusing on SSB
scenarios in this setting (cf. a 2D DW profile based on the spatial
modulation of the self-focusing nonlinearity, which was introduced in Ref.
\cite{2D-Thawatchai}). The respective fundamental symmetric state is
illustrated by Fig. \ref{2WellSymmetricSolutionMu5}, for $\mu =5,\alpha =0.5$
and $x_{0}=1$. Direct simulations have shown that this state is stable for
all values of $x_{0}$, $\mu $ and $\alpha $, with the respective $N(\mu )$
curve for $\alpha =0.5$ and $x_{0}=1$ displayed in Fig. \ref%
{2WellSymmetricSolutionNvsMu}).

This 2D symmetric state can be easily found in the approximate form in the
framework of the TFA, cf. its 1D version (\ref{1DTFA}):%
\begin{equation}
\phi _{\mathrm{TFA}}(x,y)=\sqrt{\mu }\exp \left\{ -\left( \alpha /2\right) %
\left[ \left( |x|-x_{0}\right) ^{2}+y^{2}\right] \right\} ,  \label{2DTFA}
\end{equation}%
the corresponding approximation for the norm being [cf. the 1D result (\ref%
{1DNTFA})]:%
\begin{equation}
N_{\mathrm{TFA}}(\mu )=\left( \pi /\alpha \right) \mu \left[ 1+\mathrm{erf}%
\left( \sqrt{\alpha }x_{0}\right) \right] .  \label{2DNTFA}
\end{equation}%
The latter result readily explains the nearly linear numerically found
dependence $N(\mu )$ displayed in Fig. \ref{2WellSymmetricSolutionNvsMu}(b).

Antisymmetric zero-vorticity states, as well as asymmetric ones bifurcating
from them, have been found too, but they turn out to be completely unstable
(not shown here in detail), unlike their 1D counterparts, cf. Figs. \ref%
{1D2WellBasicNVsMu} and \ref{1D2WellAsymmetric}. In direct simulations,
unstable states of these type spontaneously rearrange into stable symmetric
modes.

\subsection{Semi-vortices trapped in the double-well potential}

\begin{figure}[tbp]
\subfigure[]{\includegraphics[width=2.25in]{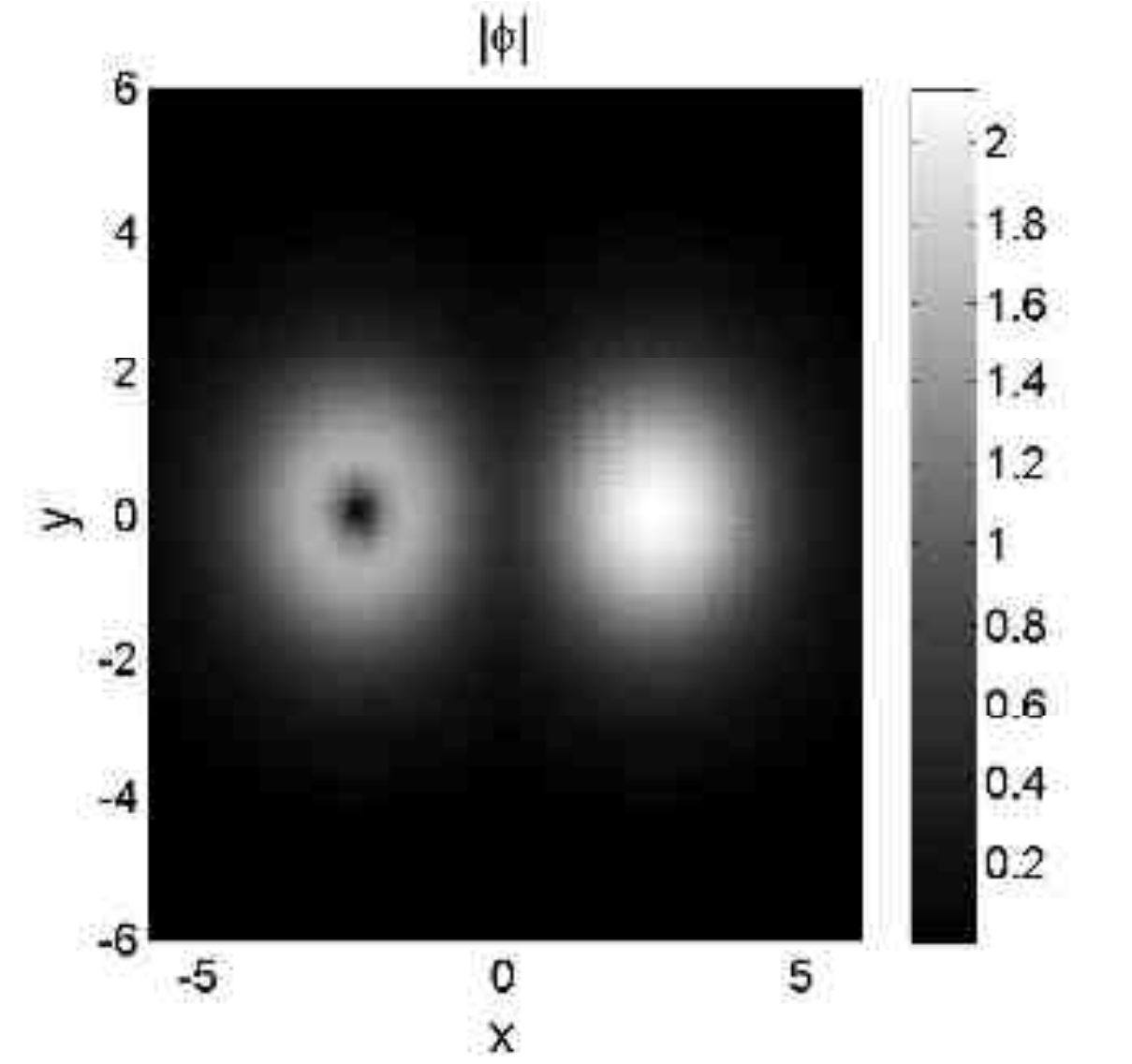}
\label{2WellFundM1VortexMu5Abs}}
\subfigure[]{\includegraphics[width=2.25in]{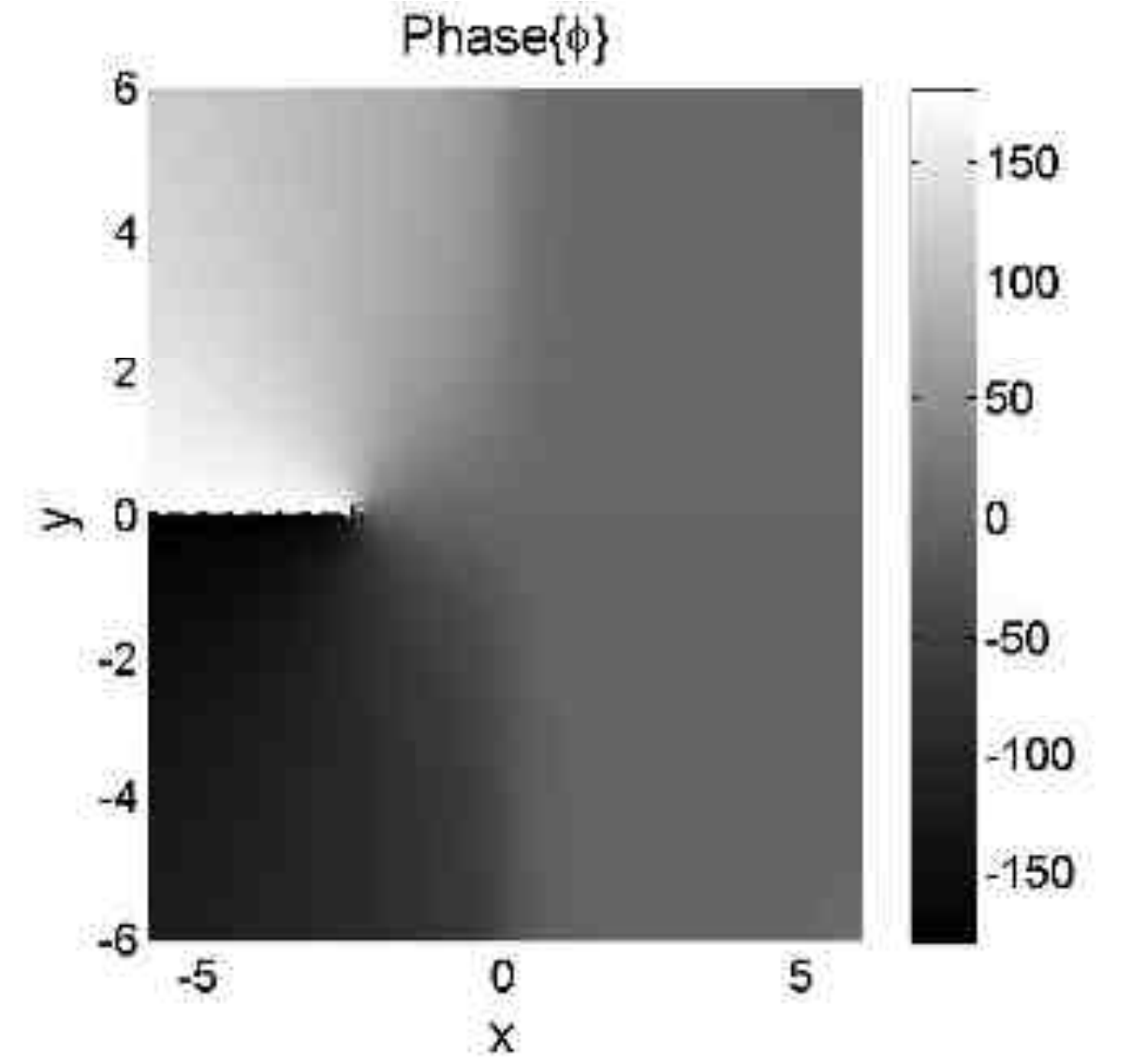}
\label{2WellFundM1VortexMu5Phase}}
\subfigure[]{\includegraphics[width=2.25in]{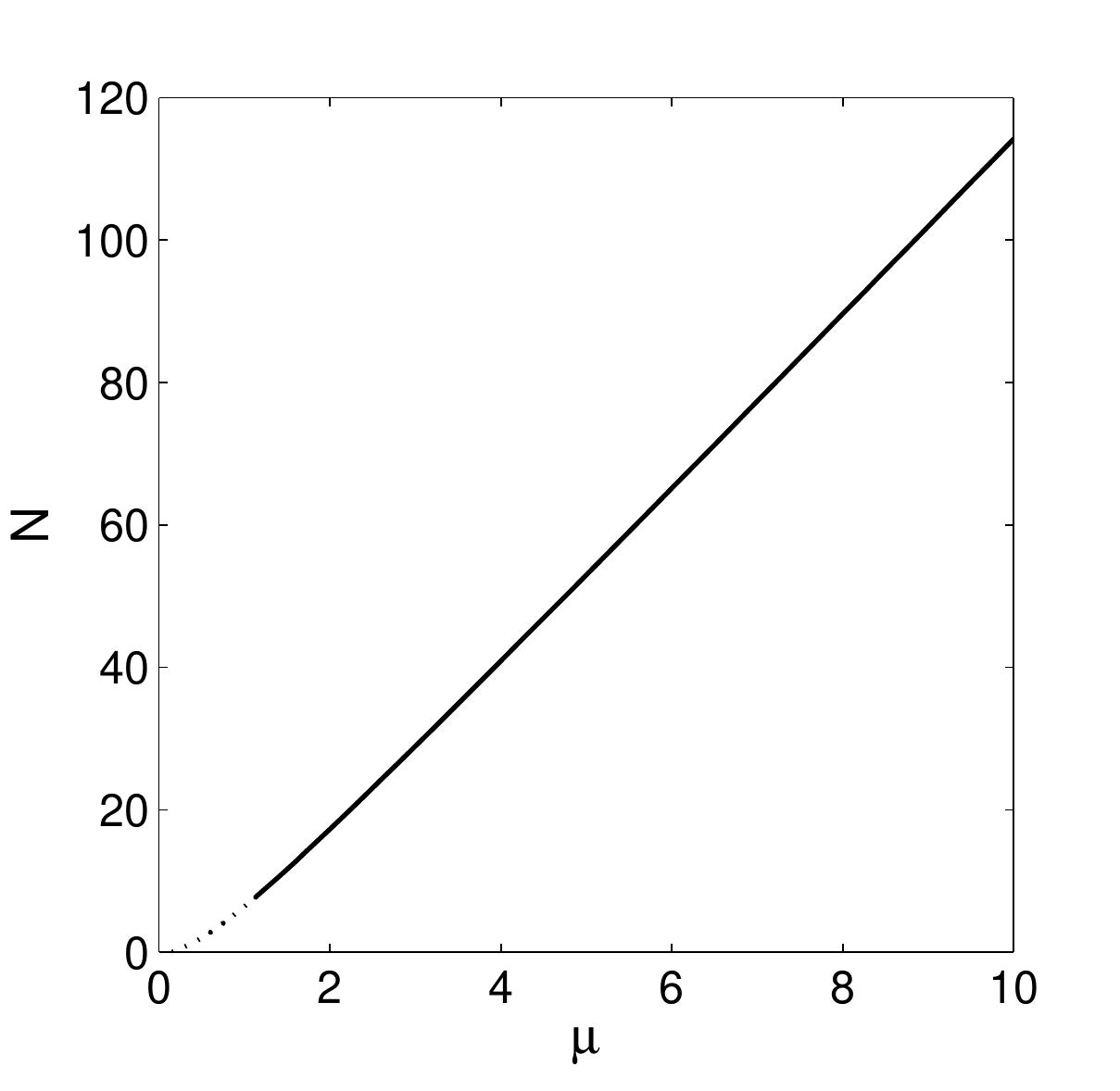}
\label{2WellFundM1VortexNvsMu}}
\subfigure[]{\includegraphics[width=2.25in]{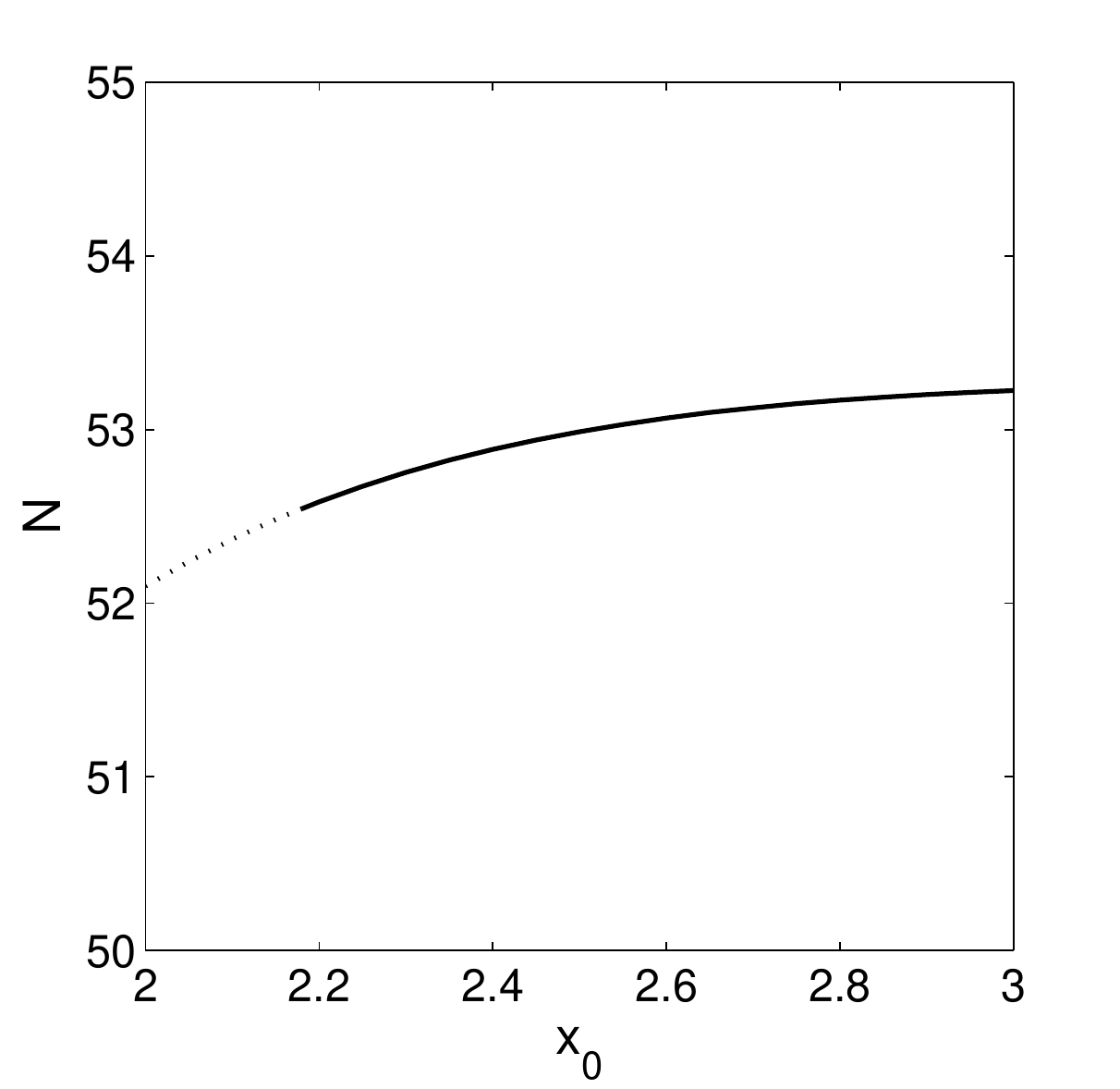}
\label{2WellFundM1VortexNvsX0}}
\caption{(Color online) (a,b) Amplitude and phase profiles of the semi-vortex composite
state, built of a vortex with topological charge $m=1$ trapped in the left
well, and a fundamental soliton trapped in the right well, for $x_{0}=2.5$, $%
\protect\mu =5$ and $\protect\alpha =0.5$. The respective $N(\protect\mu )$
and $N\left( x_{0}\right) $ curves are displayed in panels (c) and (d),
fixing $\protect\alpha =0.5$ and $x_{0}=0.5$ or $\protect\mu =5$,
respectively.}
\label{2WellFundM1Vortex}
\end{figure}

Stable \textit{semi-vortex} 2D states can be composed of fundamental and
vortex modes supported, severally, by the two nonlinear wells. A
representative example, built of the fundamental solution and the vortex
with topological charge $m=1$, is displayed in Fig. \ref{2WellFundM1Vortex}%
(a,b), for $x_{0}=2.5$, $\mu =5$ and $\alpha =0.5$. The respective stability
diagrams, produced by varying $\mu $ at fixed $x_{0}$, and varying $x_{0}$
at fixed $\mu $, are presented in Figs. \ref{2WellFundM1Vortex}(c,d). It is
seen that the composite state readily gets stabilized with the increase of $%
x_{0}$: at $\mu =5$, the solution is stable for $x_{0}>2.18$. For $x_{0}=2$,
the solution is unstable for all values of $\mu $ and $N$ (not shown here in
detail), while, at $x_{0}=2.5$, it is stable for $\mu >1.12$, $N>7.63$. When
the semi-vortex solutions are unstable, they evolve into stable symmetric
fundamental states.

\begin{figure}[tbp]
\subfigure[]{\includegraphics[width=2.25in]{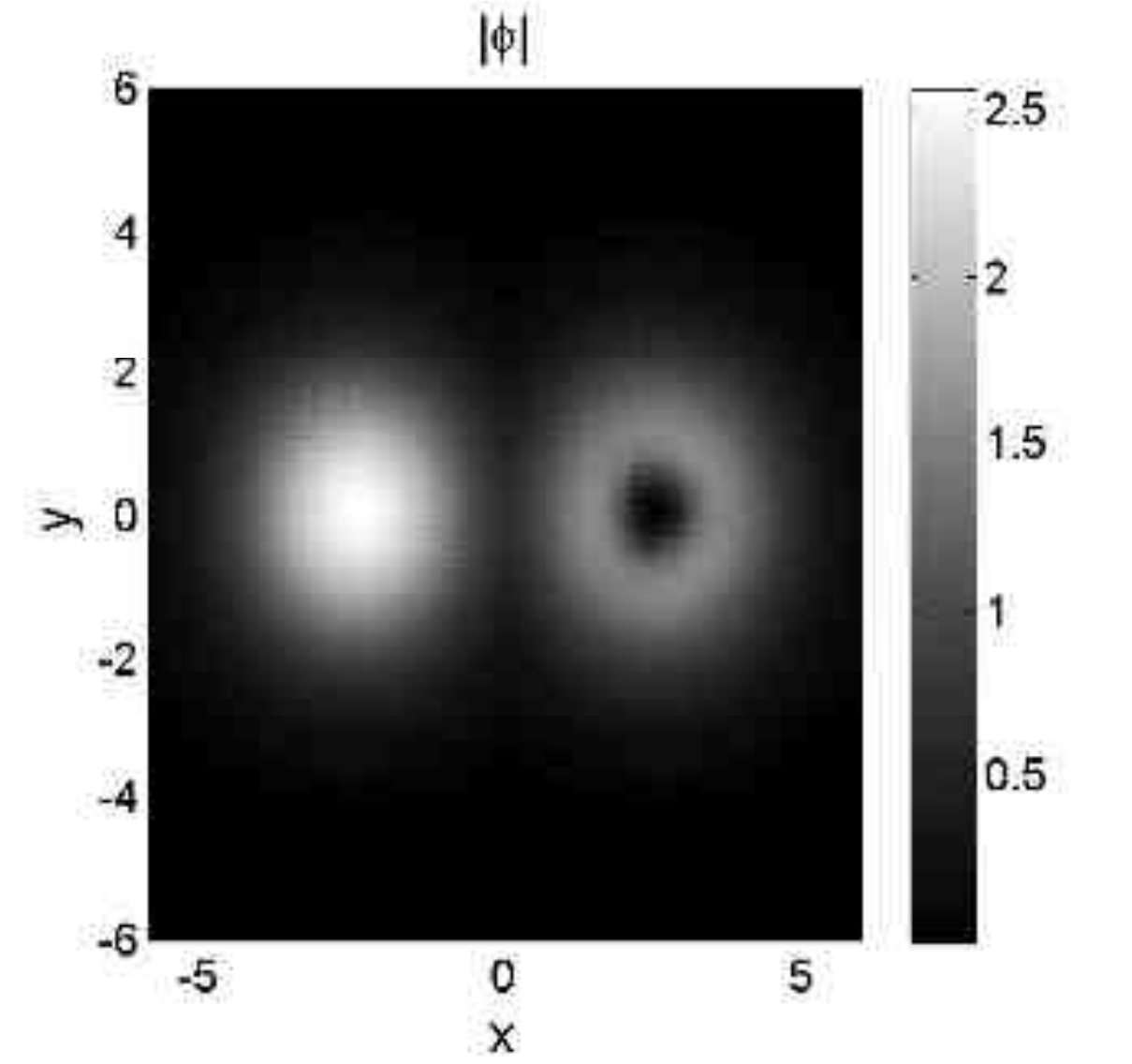}
\label{2WellFundM2VortexMu7Abs}}
\subfigure[]{\includegraphics[width=2.25in]{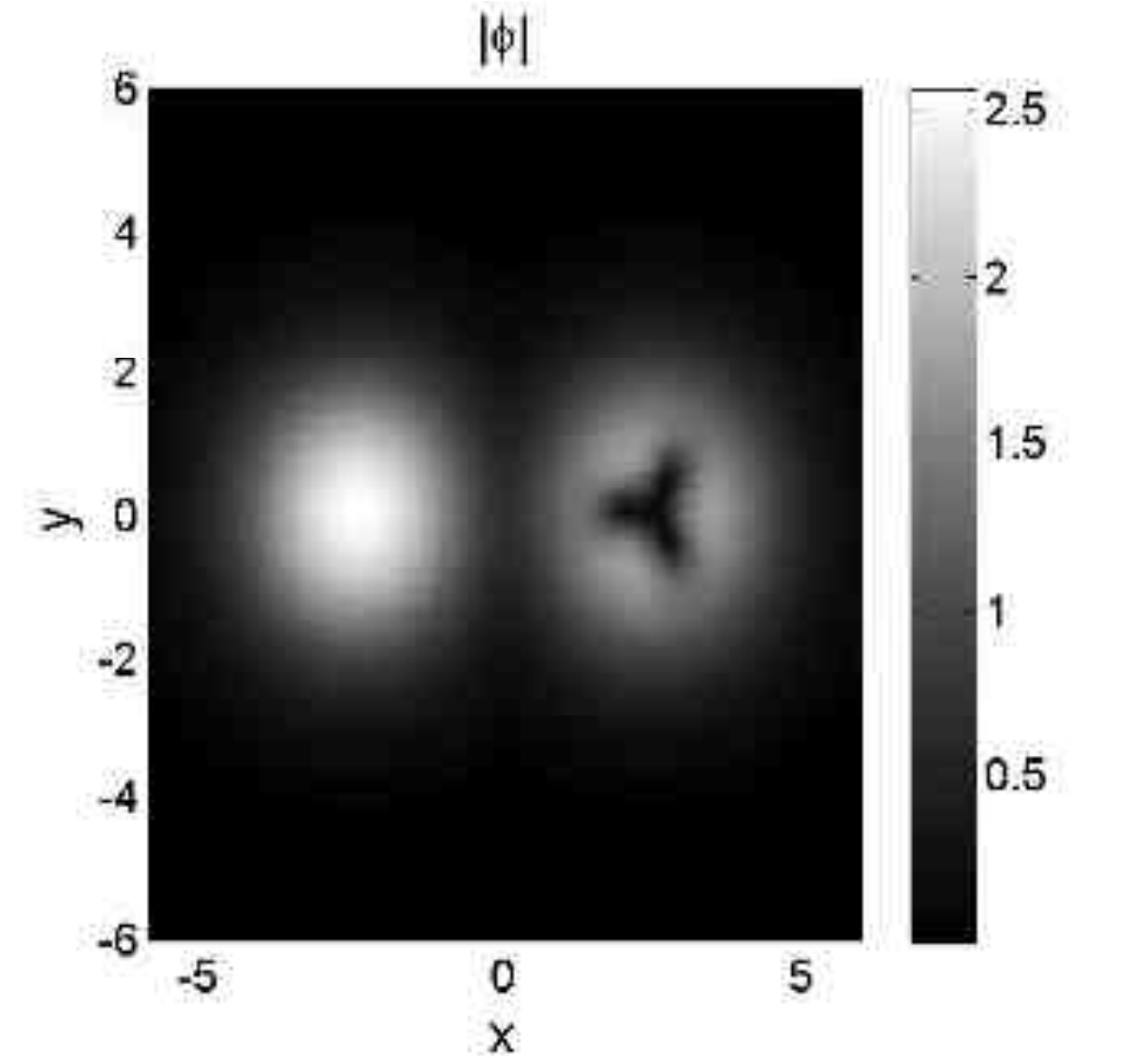}
\label{2WellFundTrioVortexMu7Abs}}
\subfigure[]{\includegraphics[width=2.25in]{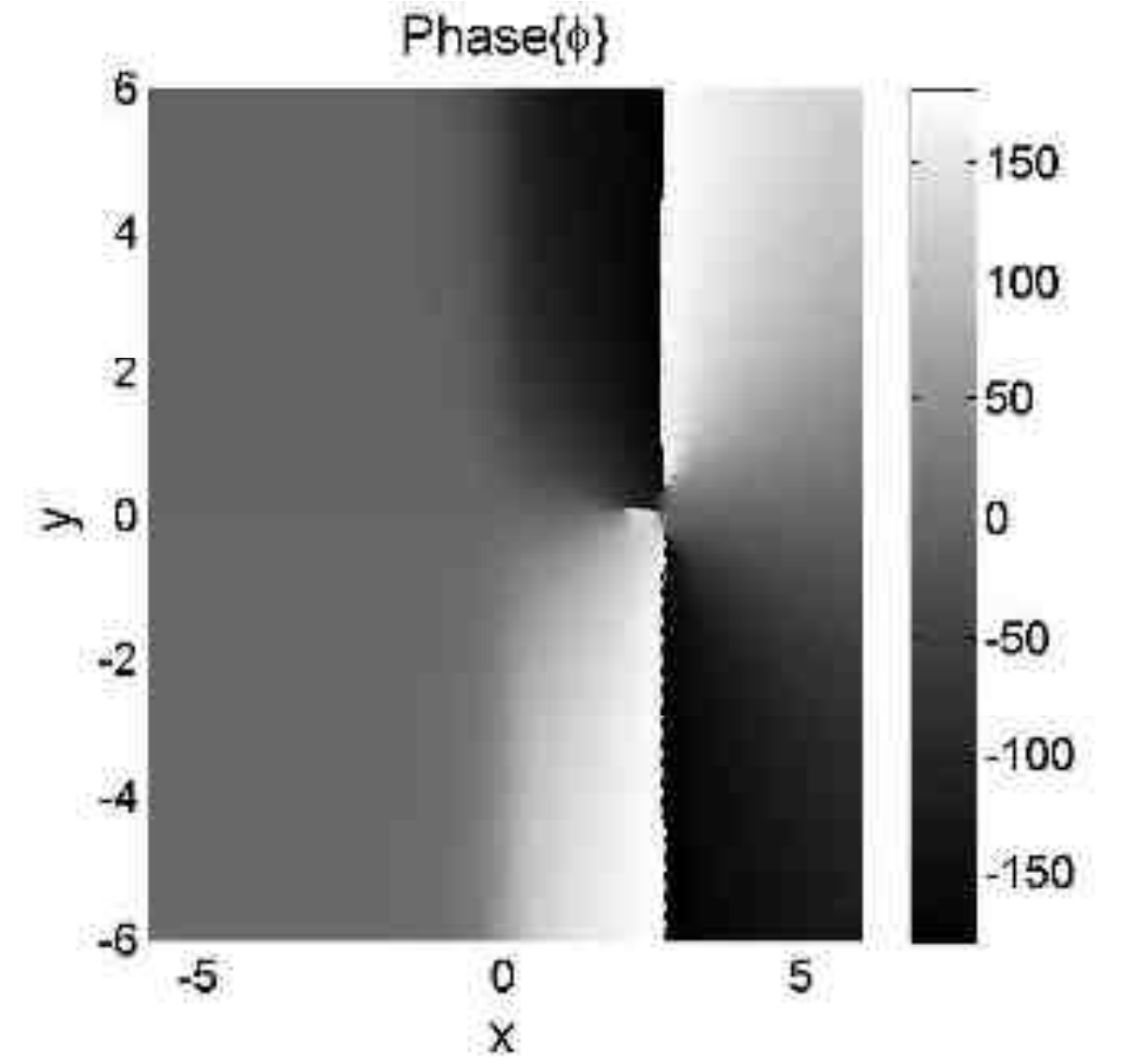}
\label{2WellFundTrioVortexMu7Phase}}
\subfigure[]{\includegraphics[width=2.25in]{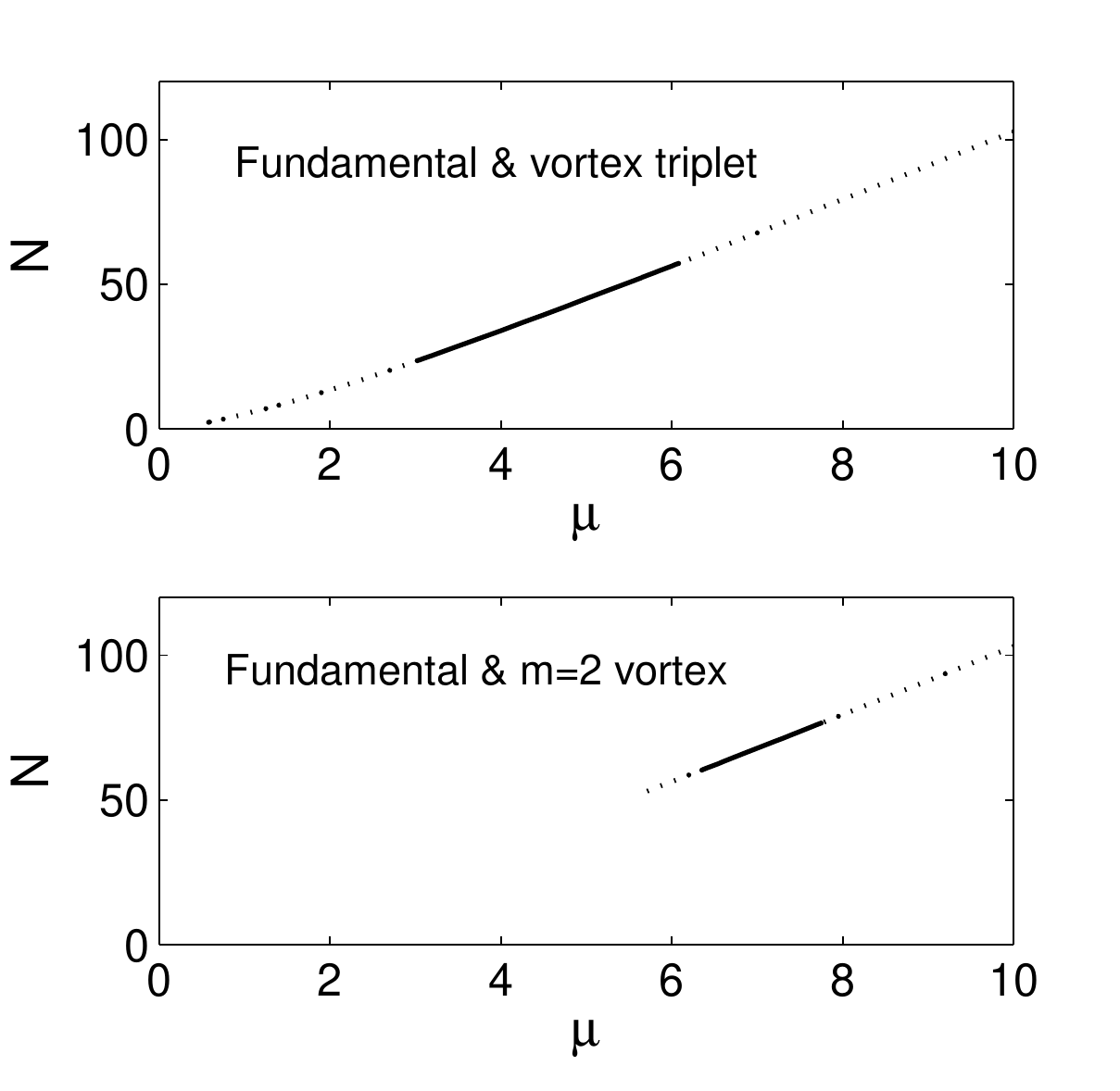}
\label{2WellFundM2VortexNvsMu}}
\subfigure[]{\includegraphics[width=2.25in]{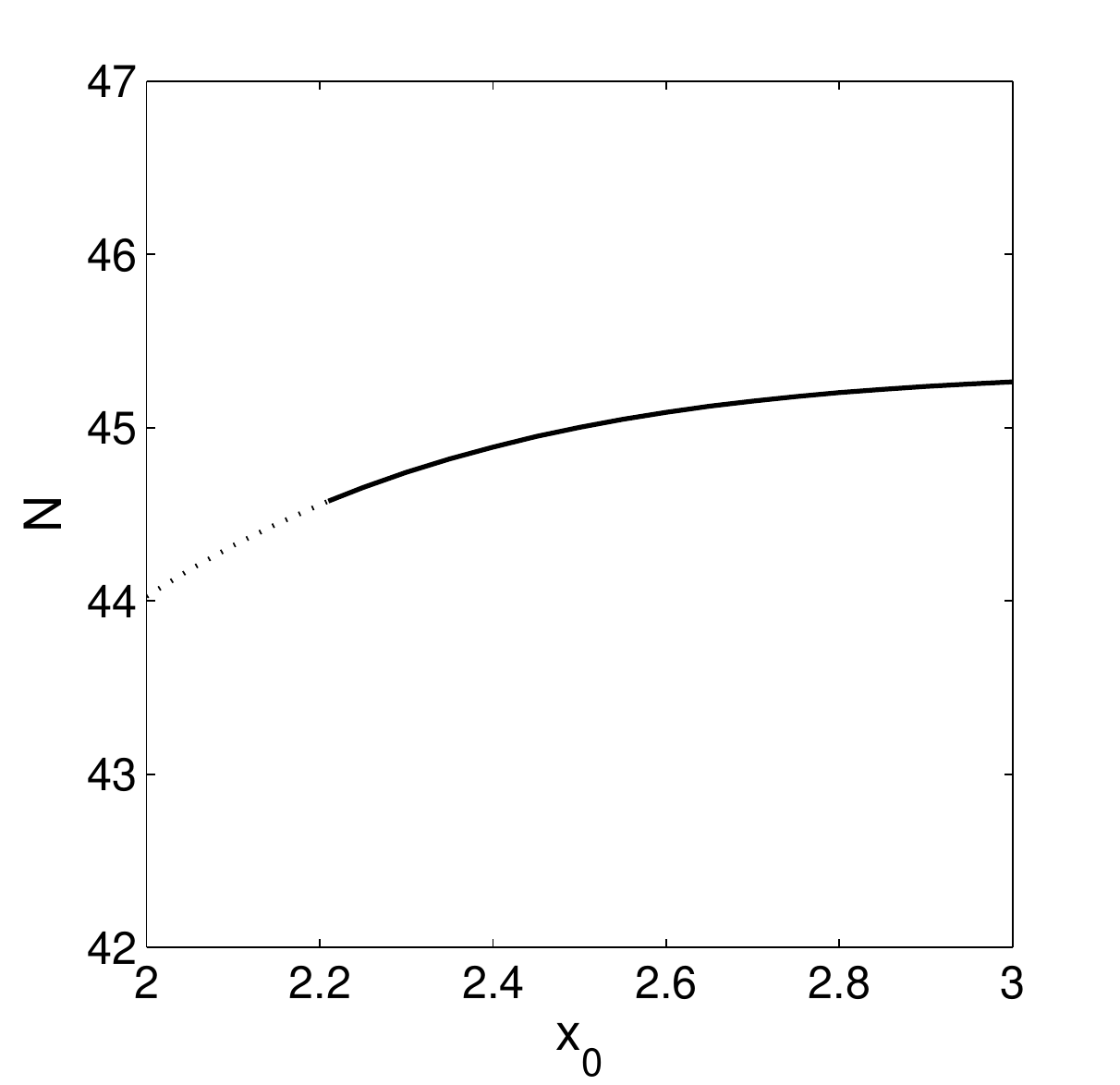}
\label{2WellFundM2VortexNvsX0}}
\caption{(Color online) (a)-(c) Amplitude profiles and a common phase pattern of 2D
composite states built of a single-well fundamental mode in the left well,
and a vortex with topological charge $m=2$, or a vortex triangle,
bifurcating from it, in the right well, for $x_{0}=2.5$, $\protect\mu =7$
and $\protect\alpha =0.5$. (d) $N(\protect\mu )$ curves for both species, at
$\protect\alpha =0.5$, $x_{0}=2.5$ (because of overlapping, the branches are
shown separately). (e) The $N\left( x_{0}\right) $ plane, for $\protect\mu %
=5 $, below the bifurcation, where only the soliton-vortex complex exists.}
\label{2WellFundM2Vortex}
\end{figure}

For the semi-vortical complex built of the fundamental soliton and vortex
with topological charge $m=2$, similar results were obtained, with an
essential addition: the bifurcation of the VT occurs from its vortex
component. Examples of composite modes of both types, including either the
vortex component with $m=2$ or the VT generated by it, are displayed in
panels (a)-(c) of Fig. \ref{2WellFundM2Vortex} for $\mu =7$, the phase
pattern being common to both. The stability of these complexes is presented
in panels (d,e) of Fig. \ref{2WellFundM2Vortex}.

\begin{figure}[tbp]
\subfigure[]{\includegraphics[width=2.25in]{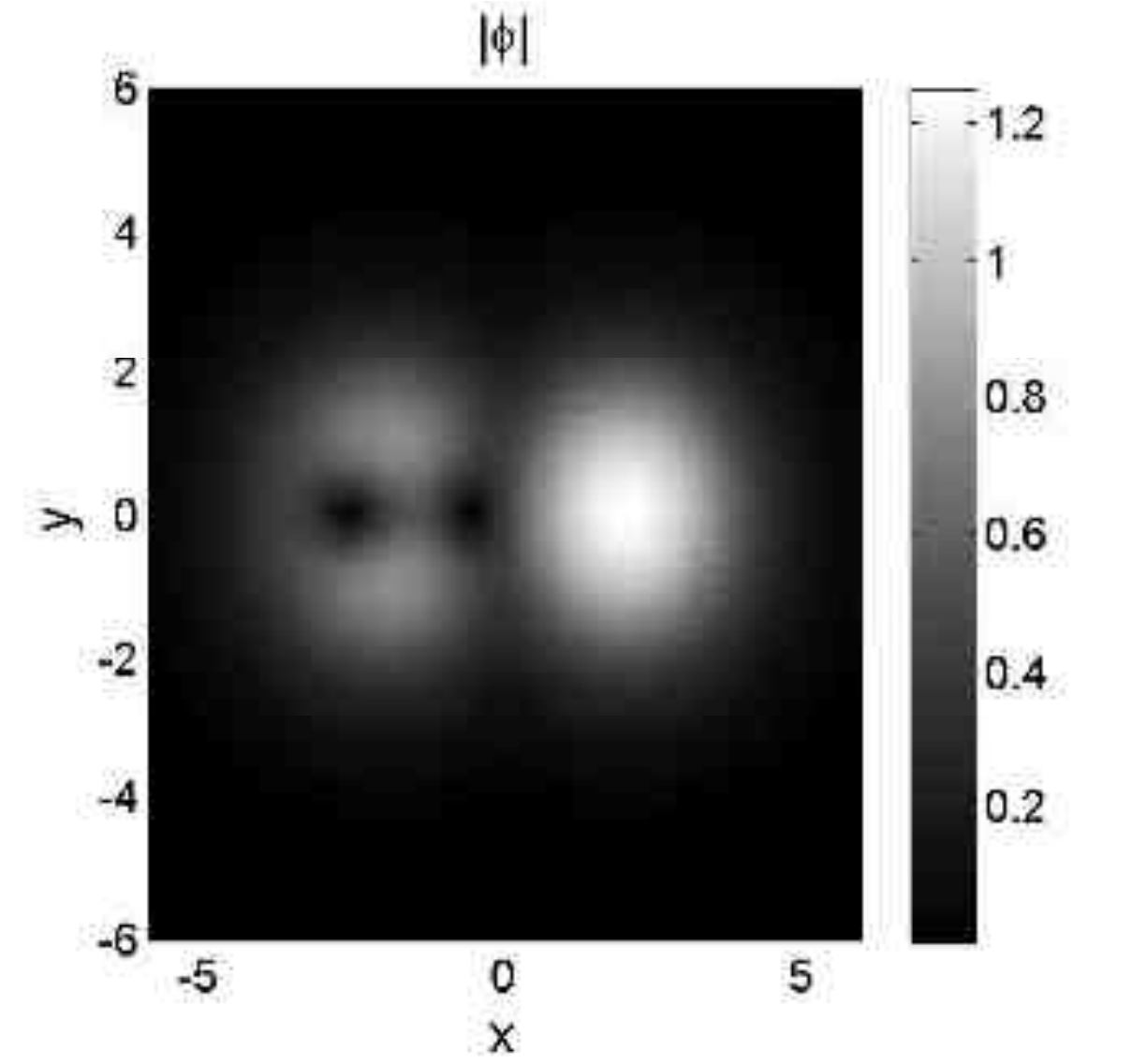}
\label{2WellFundPairMu5Abs1}}
\subfigure[]{\includegraphics[width=2.25in]{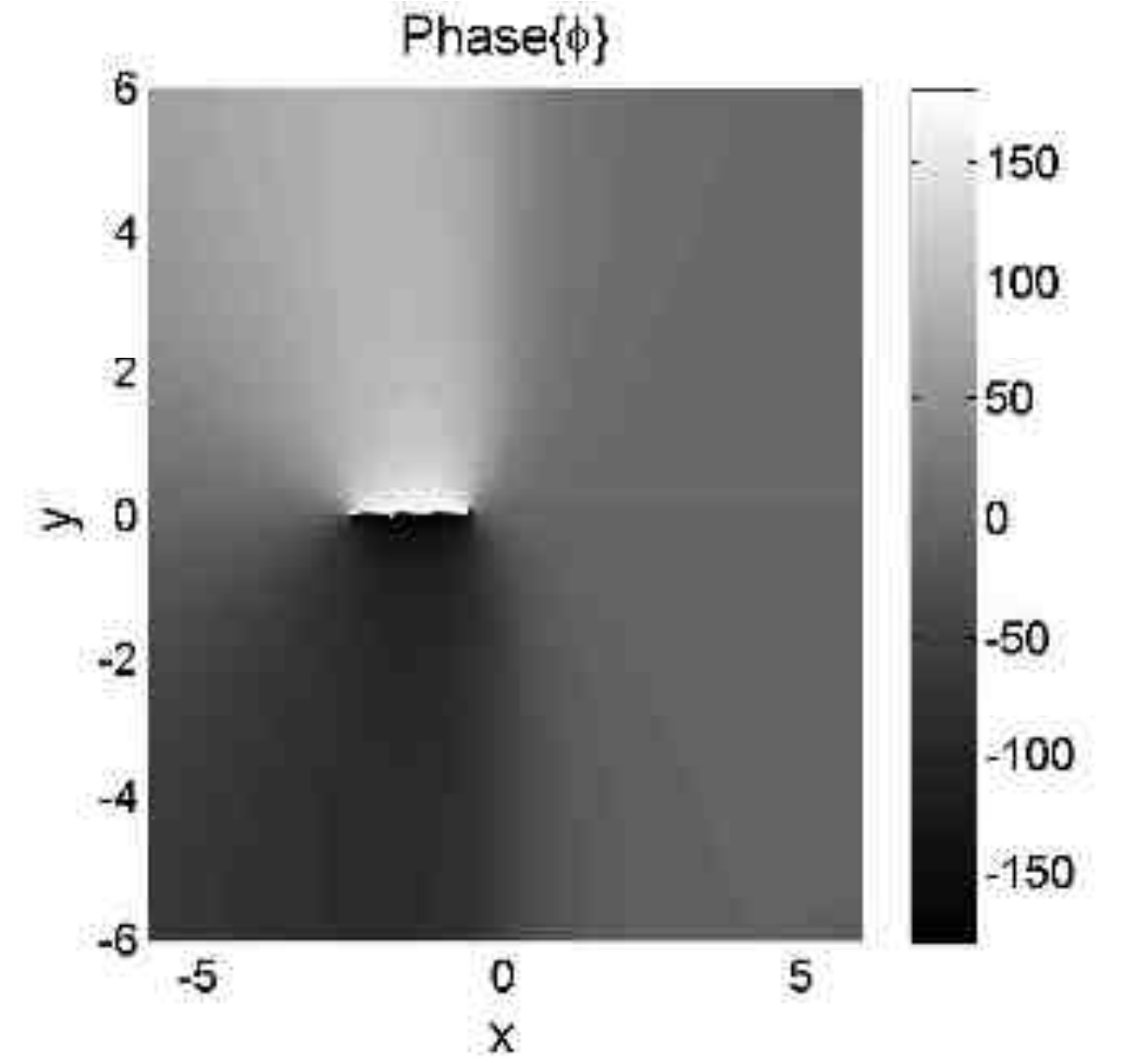}
\label{2WellFundPairMu5Phase1}}
\subfigure[]{\includegraphics[width=2.25in]{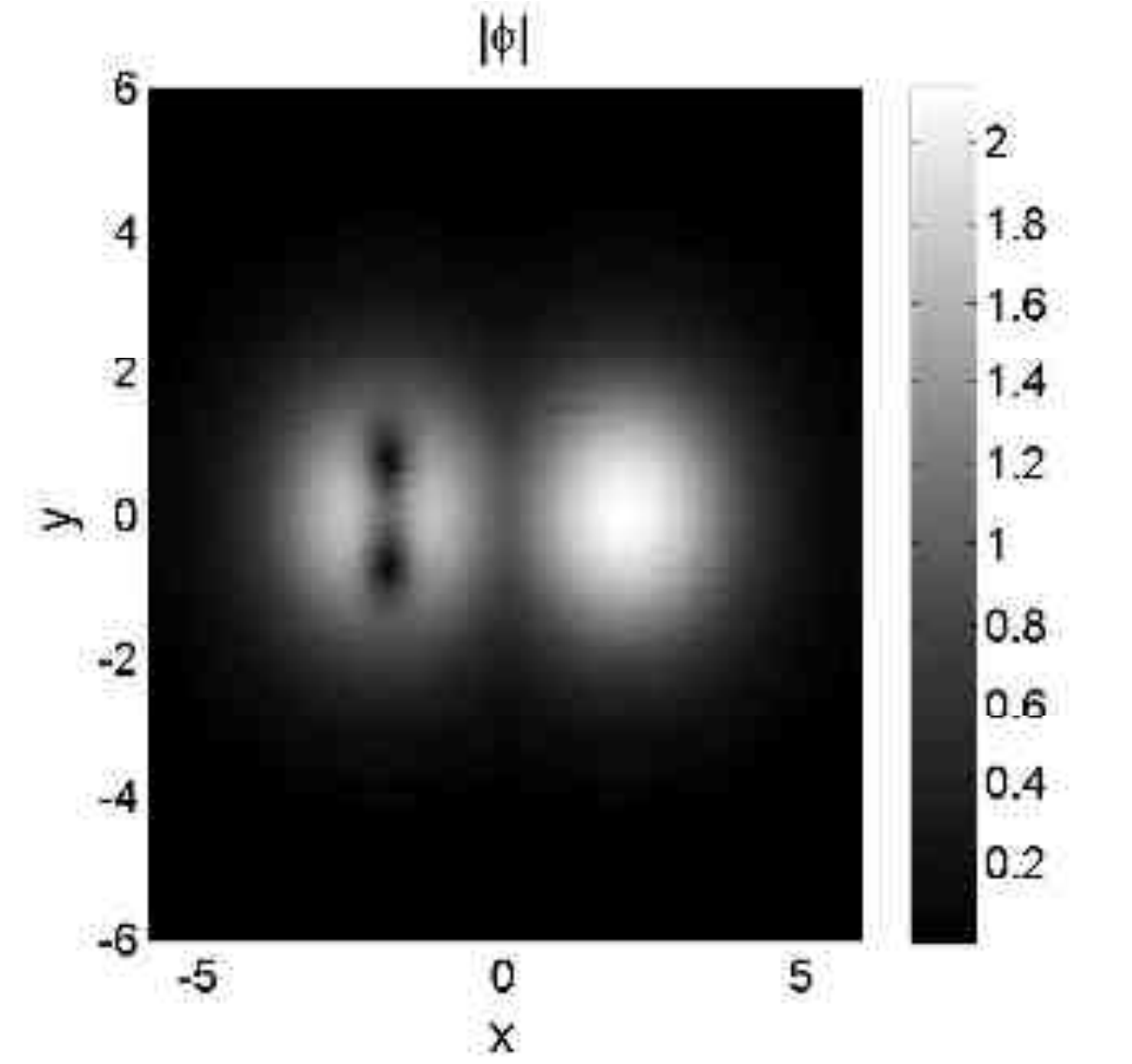}
\label{2WellFundPairMu5Abs2}}
\subfigure[]{\includegraphics[width=2.25in]{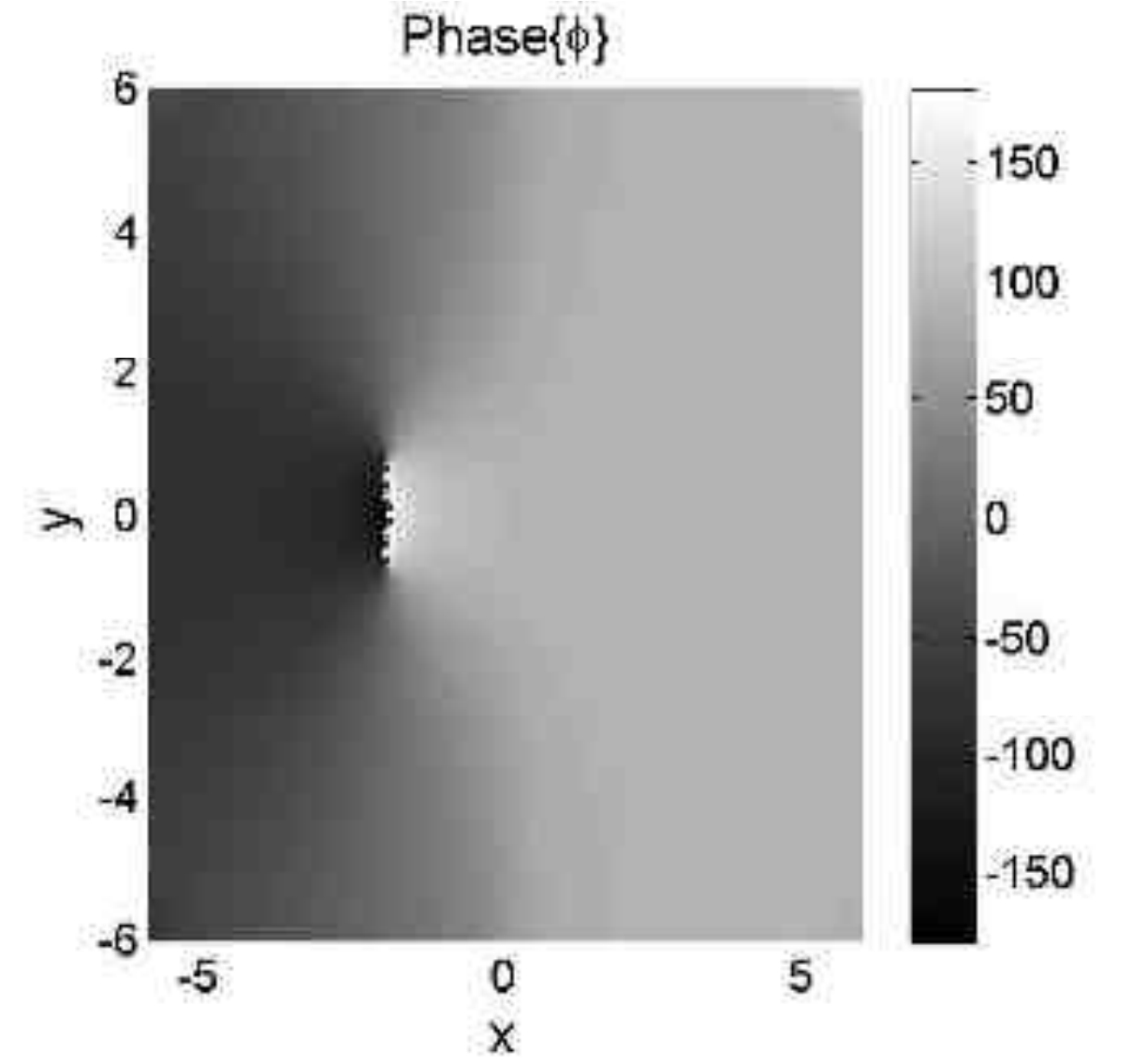}
\label{2WellFundPairMu5Phase2}}
\subfigure[]{\includegraphics[width=2.25in]{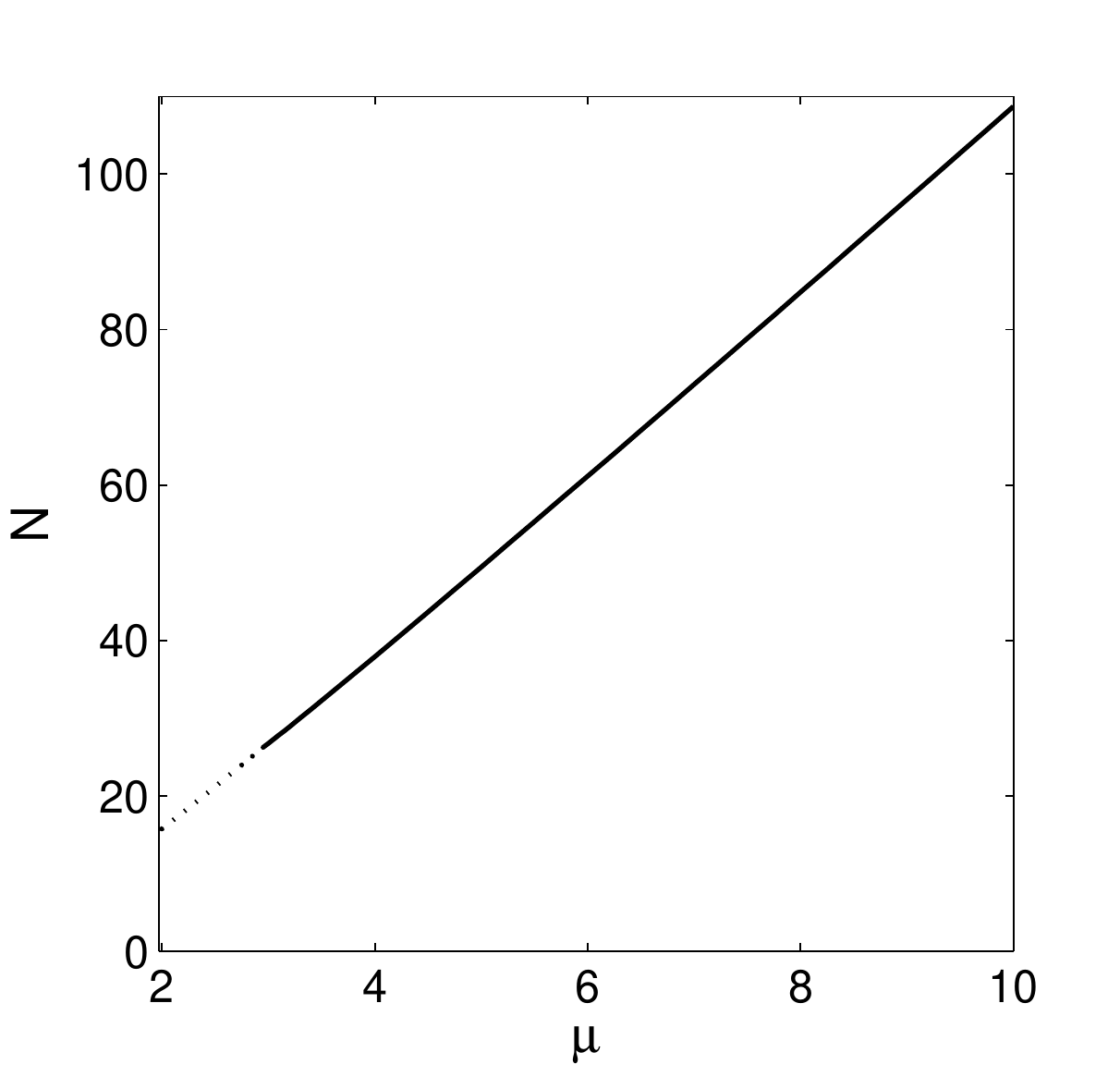}
\label{2WellFundPairNvsMu}}
\subfigure[]{\includegraphics[width=2.25in]{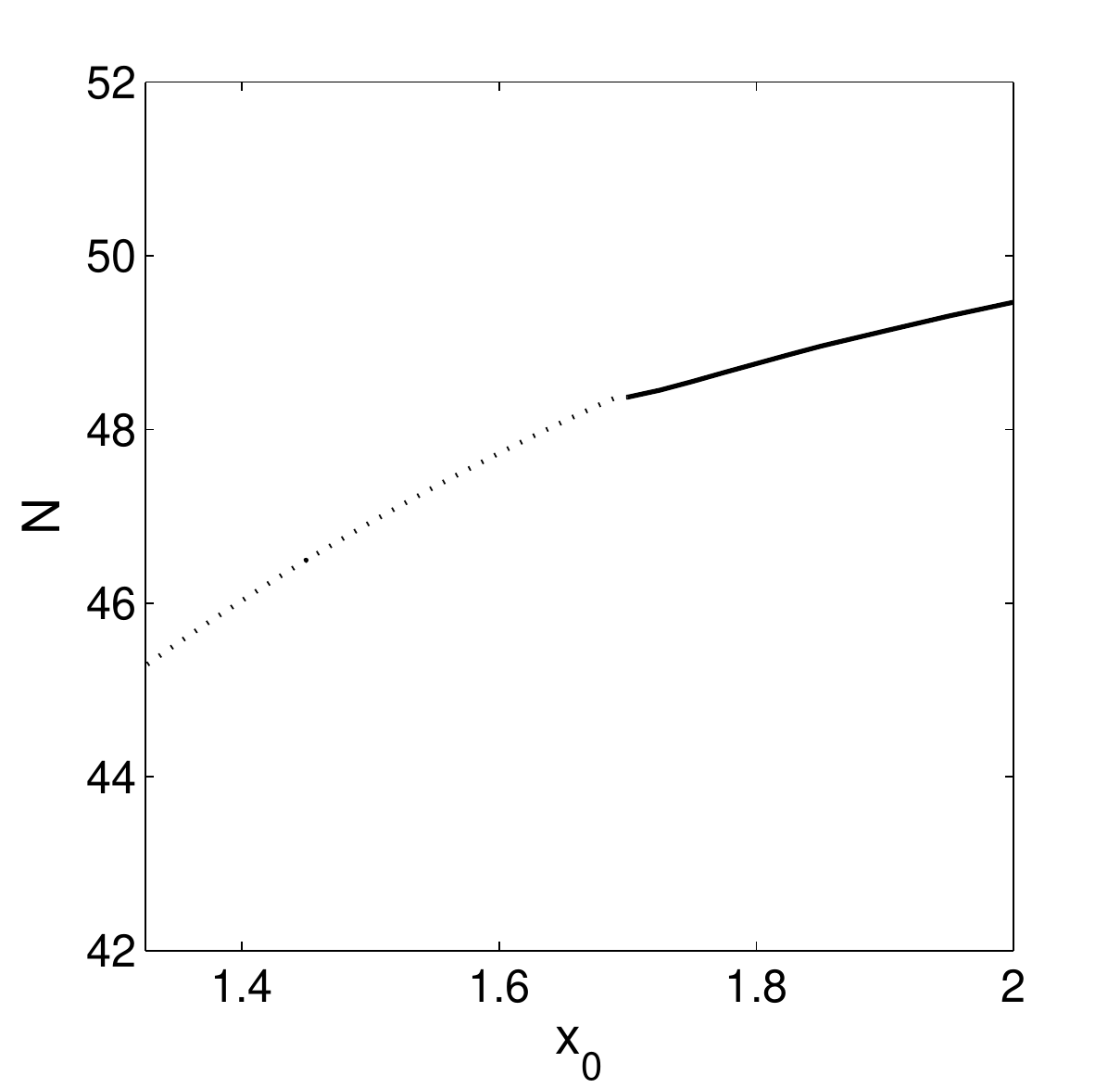}
\label{2WellFundPairNvsX0}}
\caption{(Color online) Examples of semi-vortex complexes built of the single-well
fundamental soliton in the right well, and a horizontal (a,b) or vertical
(c,d) vortex-antivortex dipole in the left one, for $x_{0}=2$, $\protect\mu %
=5$ and $\protect\alpha =0.5$. Complexes of the vertical type are completely
unstable. (e) The $N(\protect\mu )$ curve for complexes of the horizontal
type with $\protect\alpha =0.5$ and $x_{0}=2$. (f) The corresponding $%
N\left( x_{0}\right) $ curve, for $\protect\mu =5$.}
\label{2WellFundPair}
\end{figure}

In direct simulations, those semi-vortices with $m=2$ which are unstable
turn into a stable complex built of a single-well fundamental state in one
well and a VAD supported by the other. Following this observation, two
different combinations of the fundamental soliton and VAD, supported by the
individual wells, were investigated in detail: horizontal and vertical ones,
with the line connecting centers of the two vortices (which built the VAD)
oriented, respectively, parallel or perpendicular to the axis of the
two-well configuration, see Figs. \ref{2WellFundPair}(a,b) and (c,d),
respectively. The former species is partially stable: for instance, at $\mu
=5$, it is stable for $x_{0}>1.70$, see Fig. \ref{2WellFundPairNvsMu}. When
fixing $x_{0}=2$, the stability region is $\mu >2.96$, $N>26.38$ (Fig. \ref%
{2WellFundPairNvsX0}), while for $x_{0}=1.5$ the solution is entirely
unstable. Complexes of the latter (perpendicular) type are always unstable,
which can be understood, as only a horizontally aligned dipole may realize
an energy minimum in the present setting. In both cases, unstable complexes
converge to the ground-state symmetric modes.

\subsection{Dual-vortex configurations supported by the double-well potential%
}

\begin{figure}[tbp]
\subfigure[]{\includegraphics[width=2.25in]{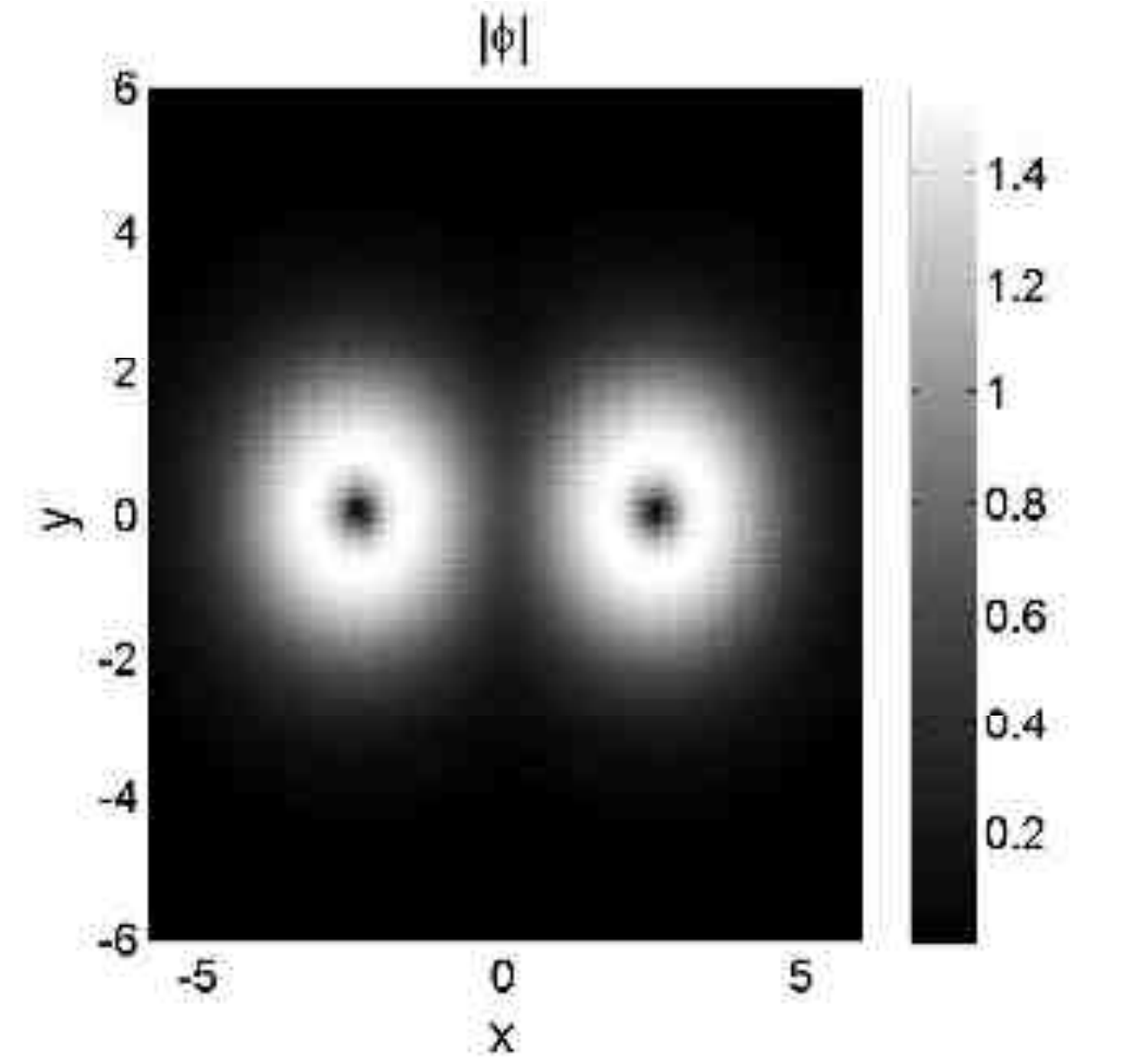}
\label{2WellM1M1Mu5Abs1}}
\subfigure[]{\includegraphics[width=2.25in]{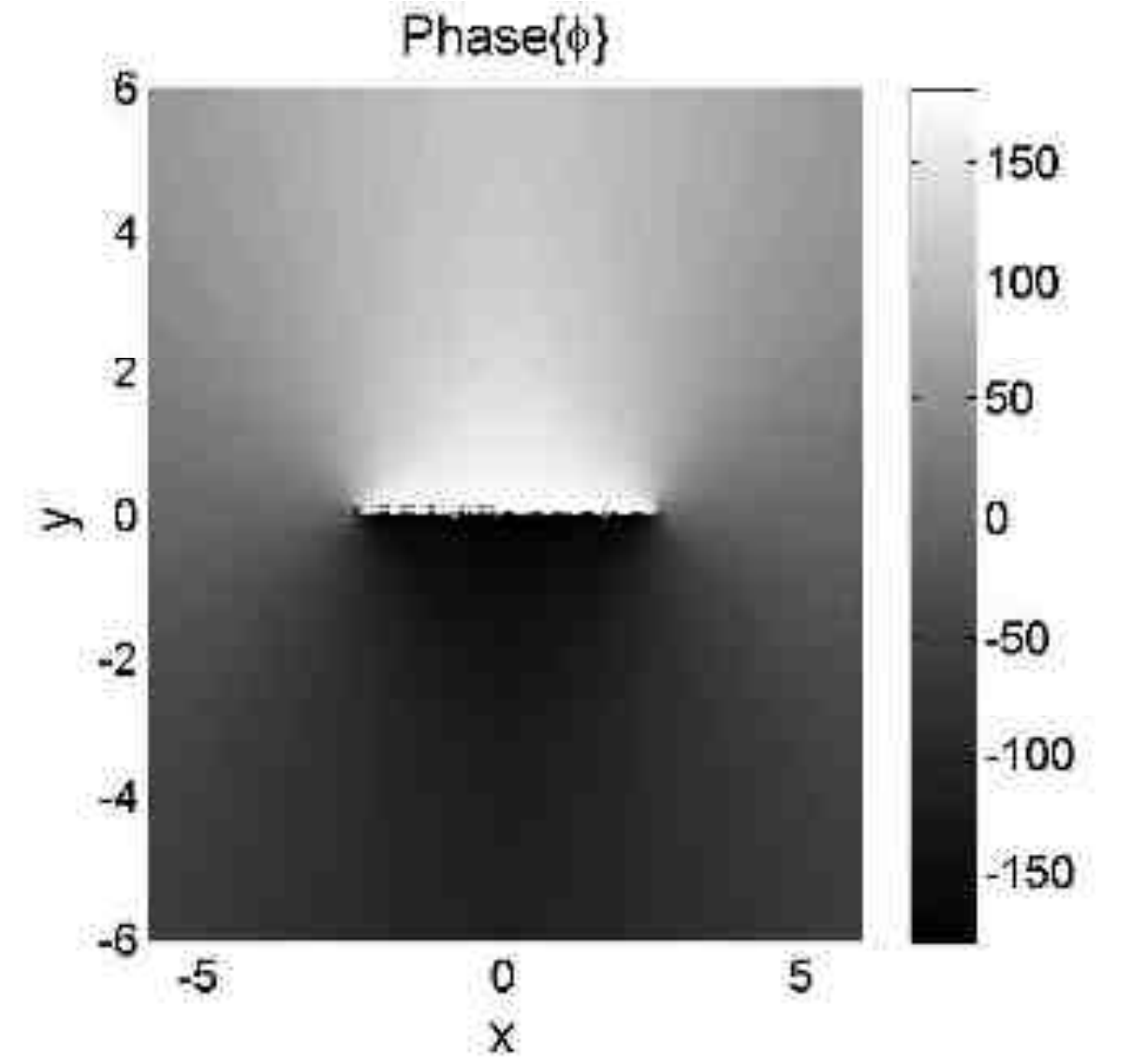}
\label{2WellM1M1Mu5Phase1}}
\subfigure[]{\includegraphics[width=2.25in]{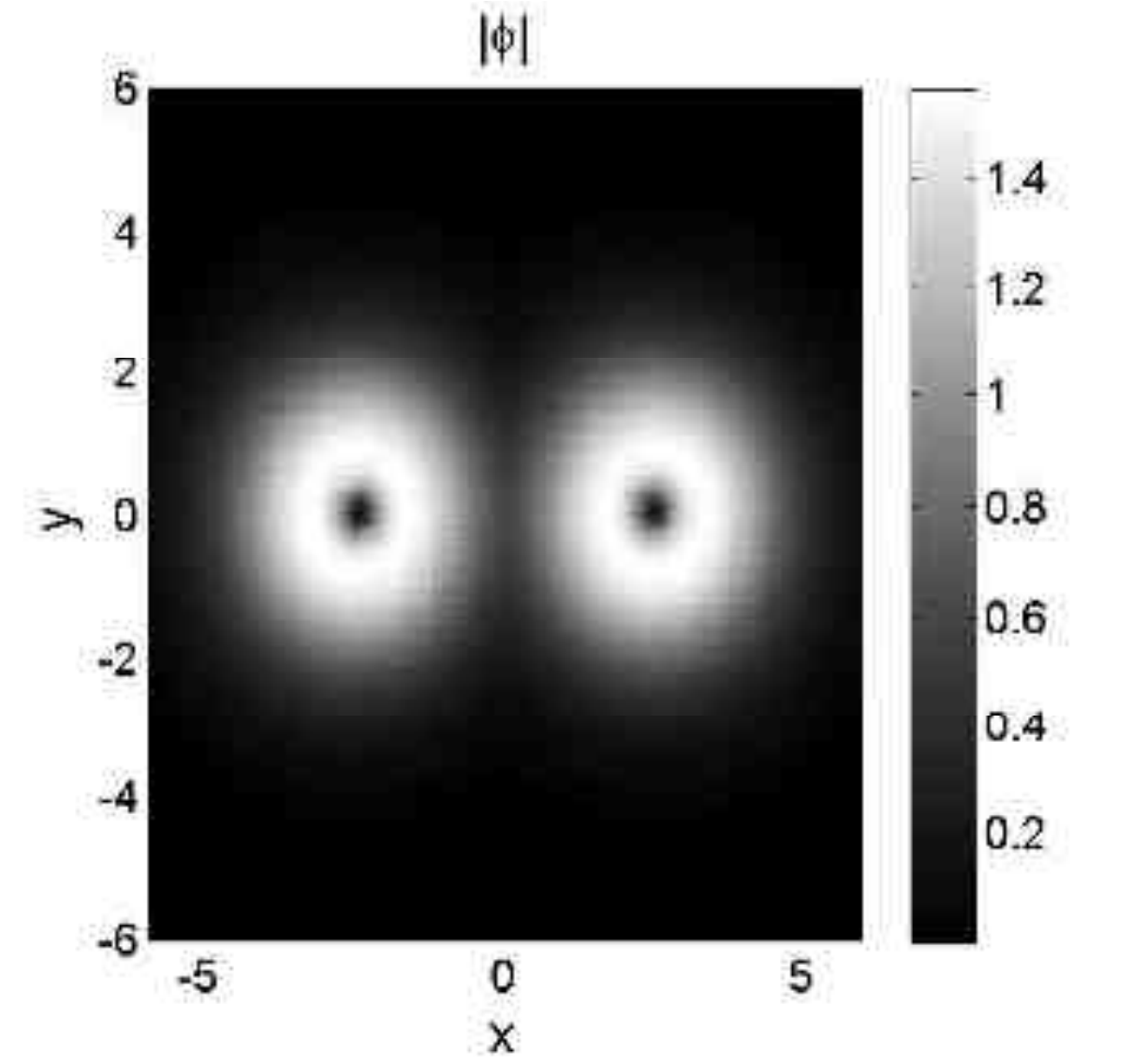}
\label{2WellM1M1Mu5Abs2}}
\subfigure[]{\includegraphics[width=2.25in]{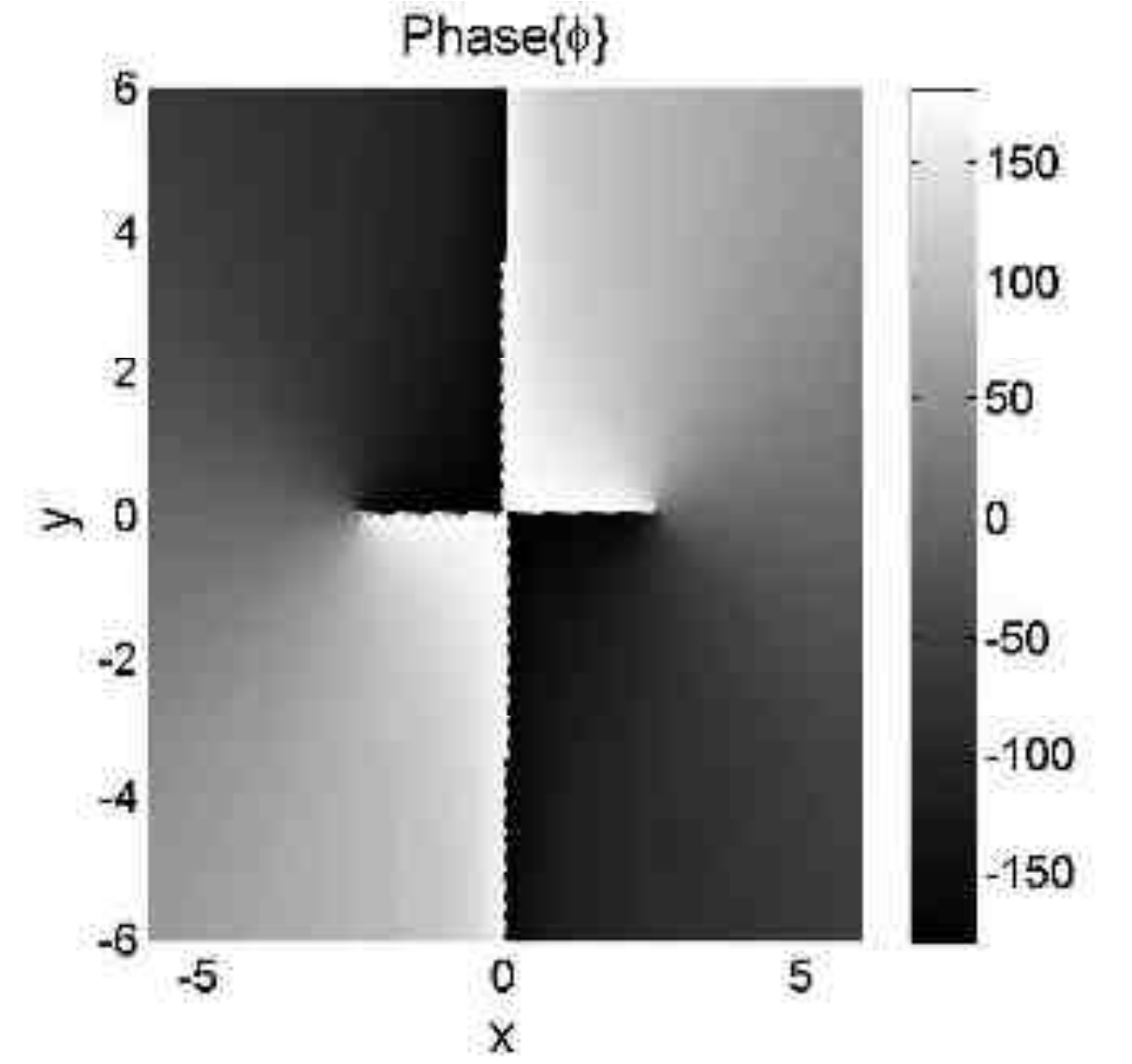}
\label{2WellM1M1Mu5Phase2}}
\subfigure[]{\includegraphics[width=2.25in]{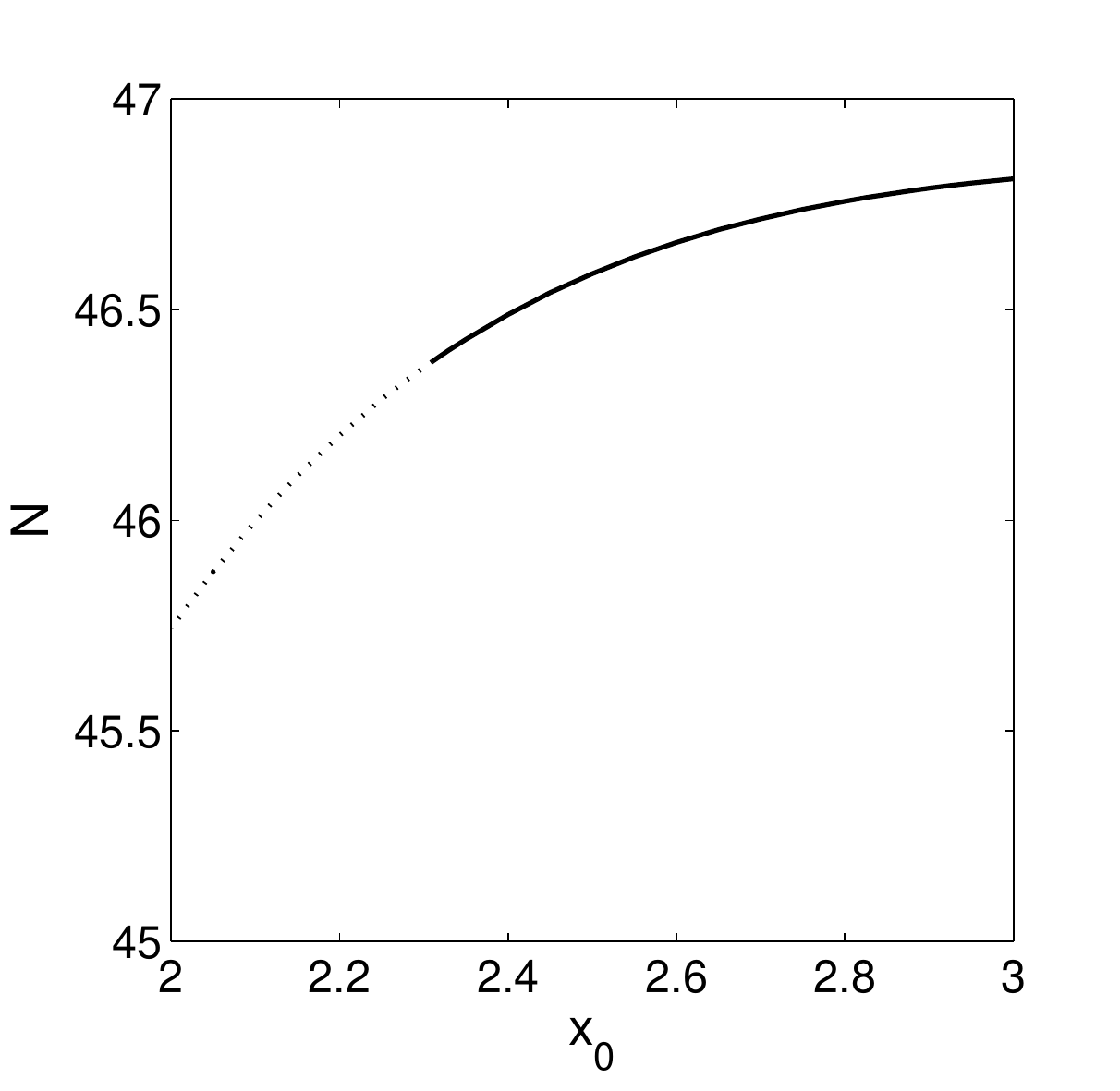}
\label{2WellM1M1NvsX01}}
\subfigure[]{\includegraphics[width=2.25in]{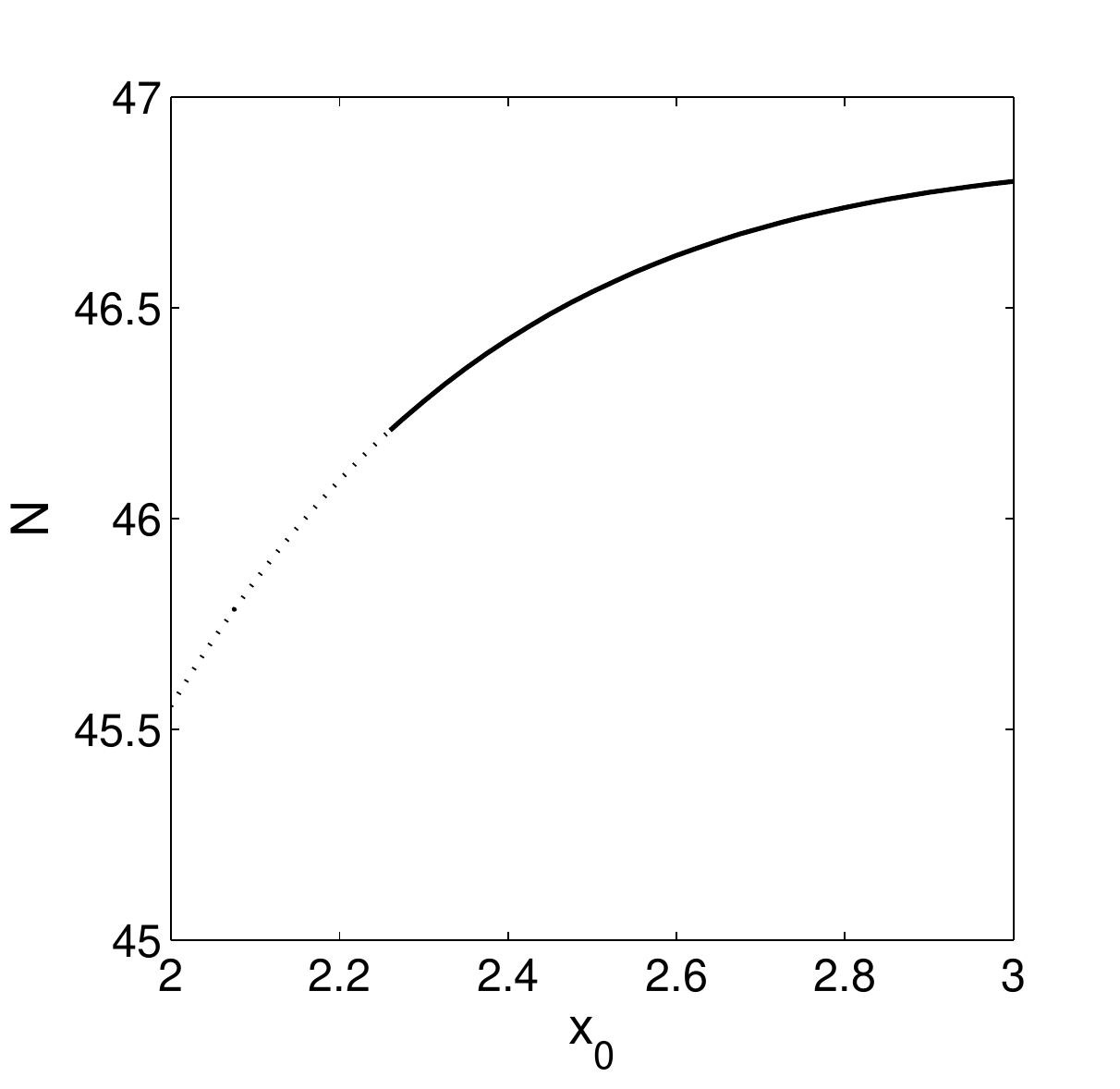}
\label{2WellM1M1NvsX02}}
\caption{(Color online) Examples of dual-vortex complexes with topological-charge sets $%
\left( +1,-1\right) $ (a,b) and $\left( +1,+1\right) $ (c,d), for $x_{0}=2.5$%
, $\protect\mu =5$ and $\protect\alpha =0.5$. The $N(x_{0})$ curves, with $%
\protect\mu =5$, for the complexes of these two types are displayed in
panels (e) and (f), respectively.}
\label{2WellM1M1}
\end{figure}

\textit{Dual-vortex} complexes, composed of vortices with $|m|=1$ trapped in
the two wells, were constructed too. Figures \ref{2WellM1M1}(a,b) and (c,d)
show examples of such complexes, built of individual vortices with
topological charges $\left( m=+1,m=-1\right) $ or $\left( m=+1,m=+1\right) $%
, so that the respective total charges are $0$ or $2$. In both cases, the
results are similar: the solutions are unstable at $x_{0}=2$, and stable at $%
x_{0}=2.5$. Fixing $\mu =5$, the $\left( +1,-1\right) $ complex is stable
for $x_{0}>2.31$, $N>46.38$ (Fig. \ref{2WellM1M1NvsX01}), while its
counterpart of the $\left( +1,+1\right) $ type is stable at $\ x_{0}>2.26$, $%
N>46.21$ (Fig. \ref{2WellM1M1NvsX02}). In this case too, unstable solutions
transform themselves into symmetric ground-state modes.

\begin{figure}[tbp]
\subfigure[]{\includegraphics[width=2.25in]{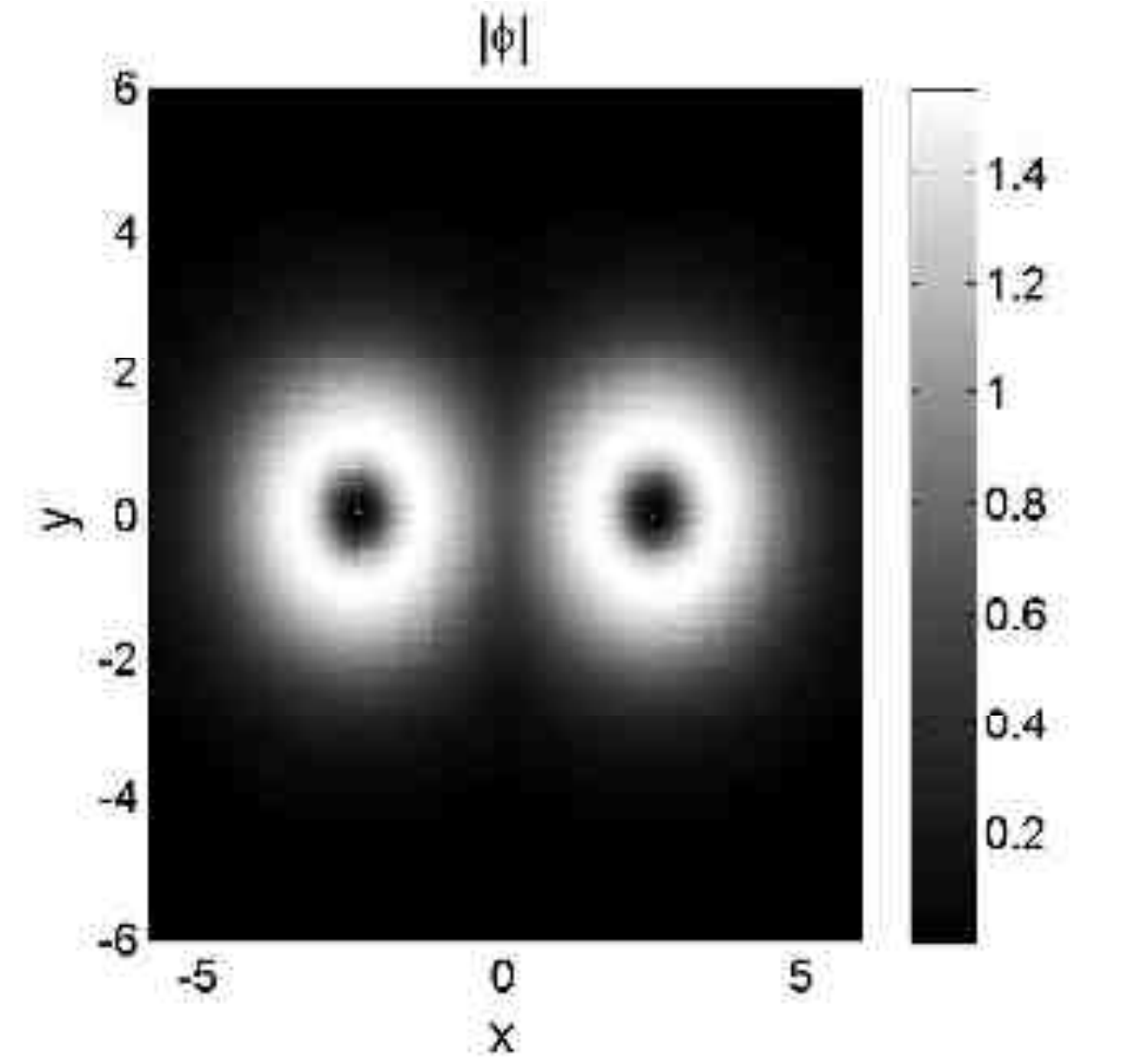}
\label{2WellM2M2Mu7Abs1}}
\subfigure[]{\includegraphics[width=2.25in]{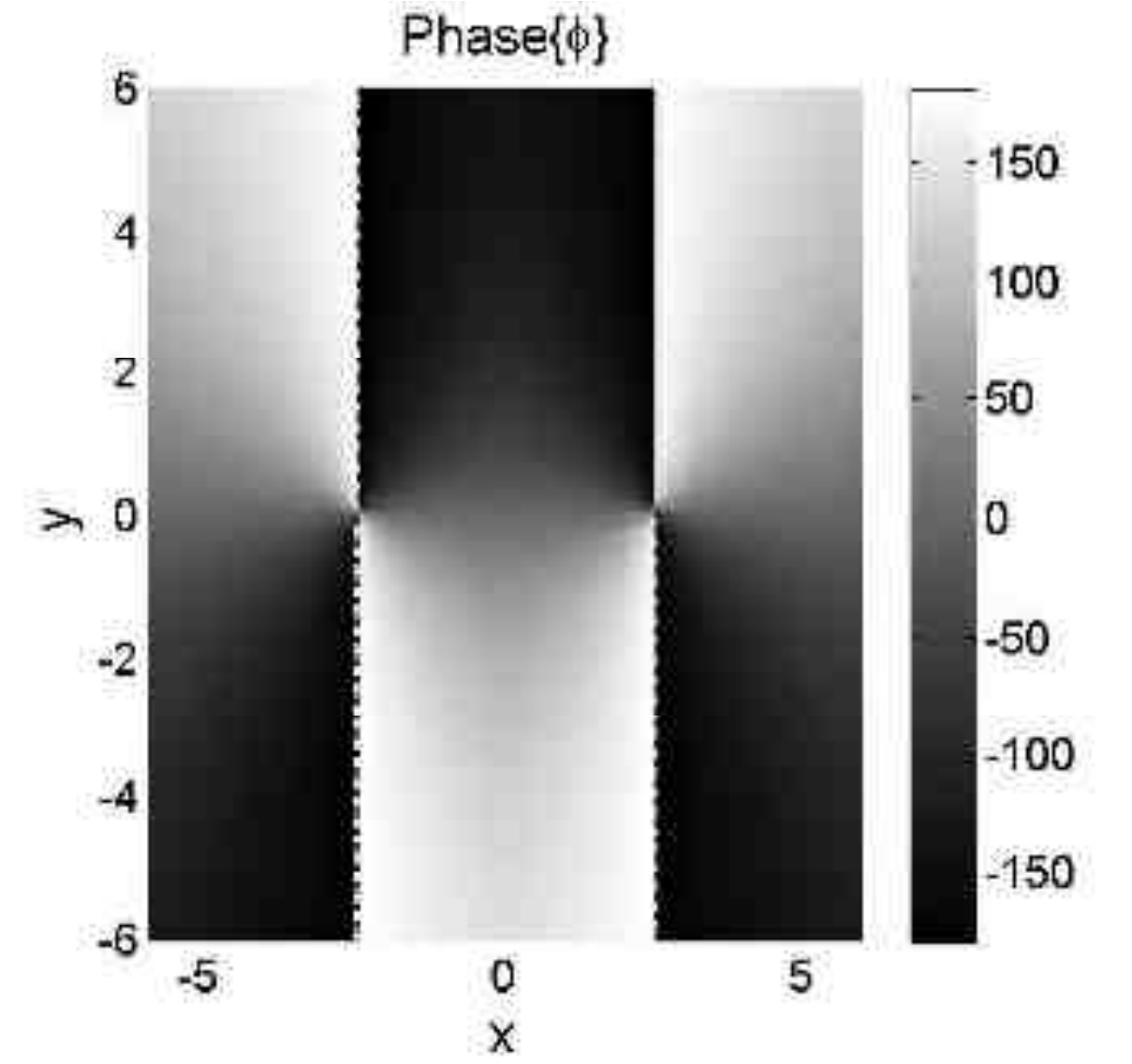}
\label{2WellM2M2Mu7Phase1}}
\subfigure[]{\includegraphics[width=2.25in]{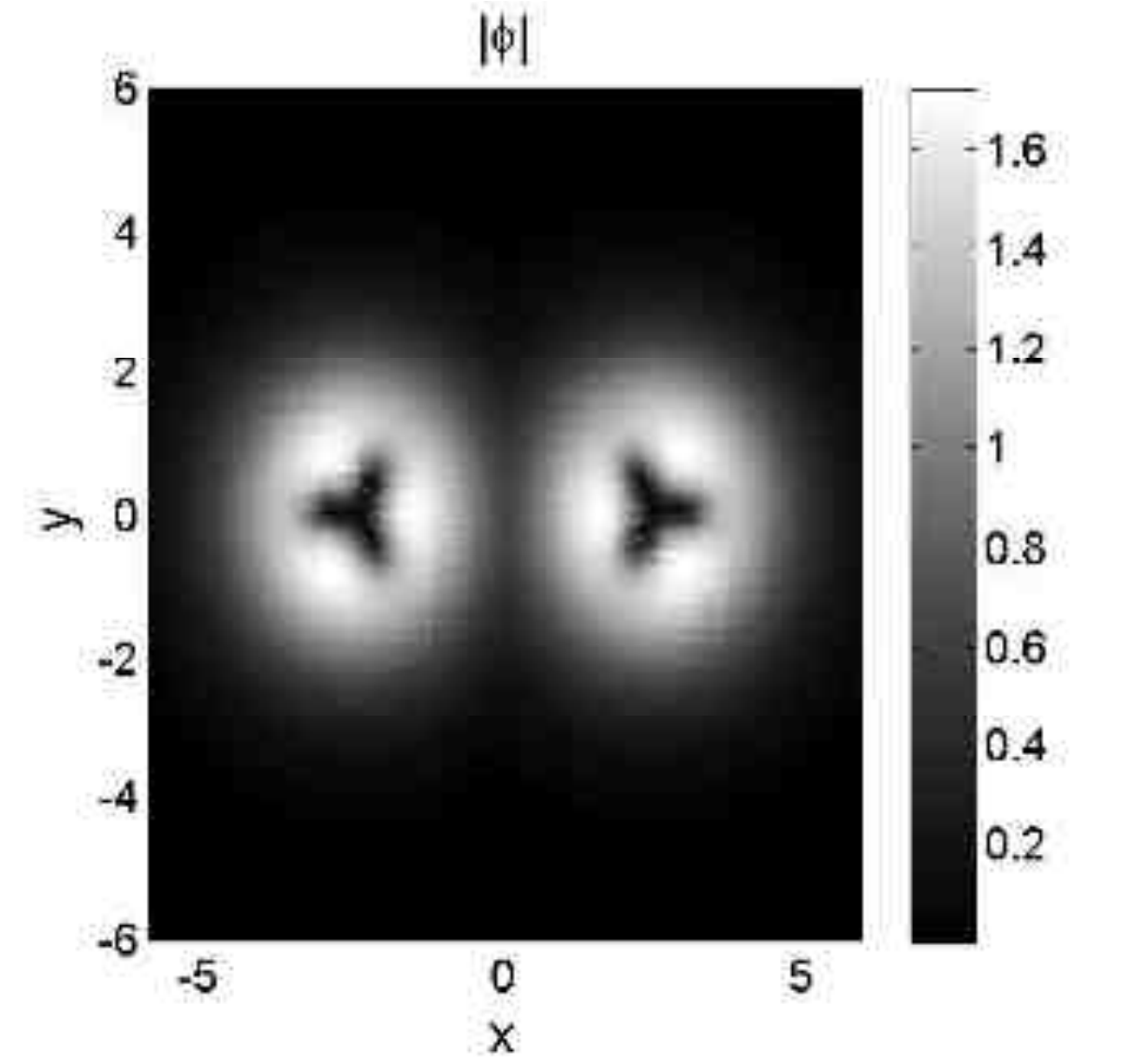}
\label{2WellTrioTrioMu7Abs1}}
\par
\subfigure[]{\includegraphics[width=2.25in]{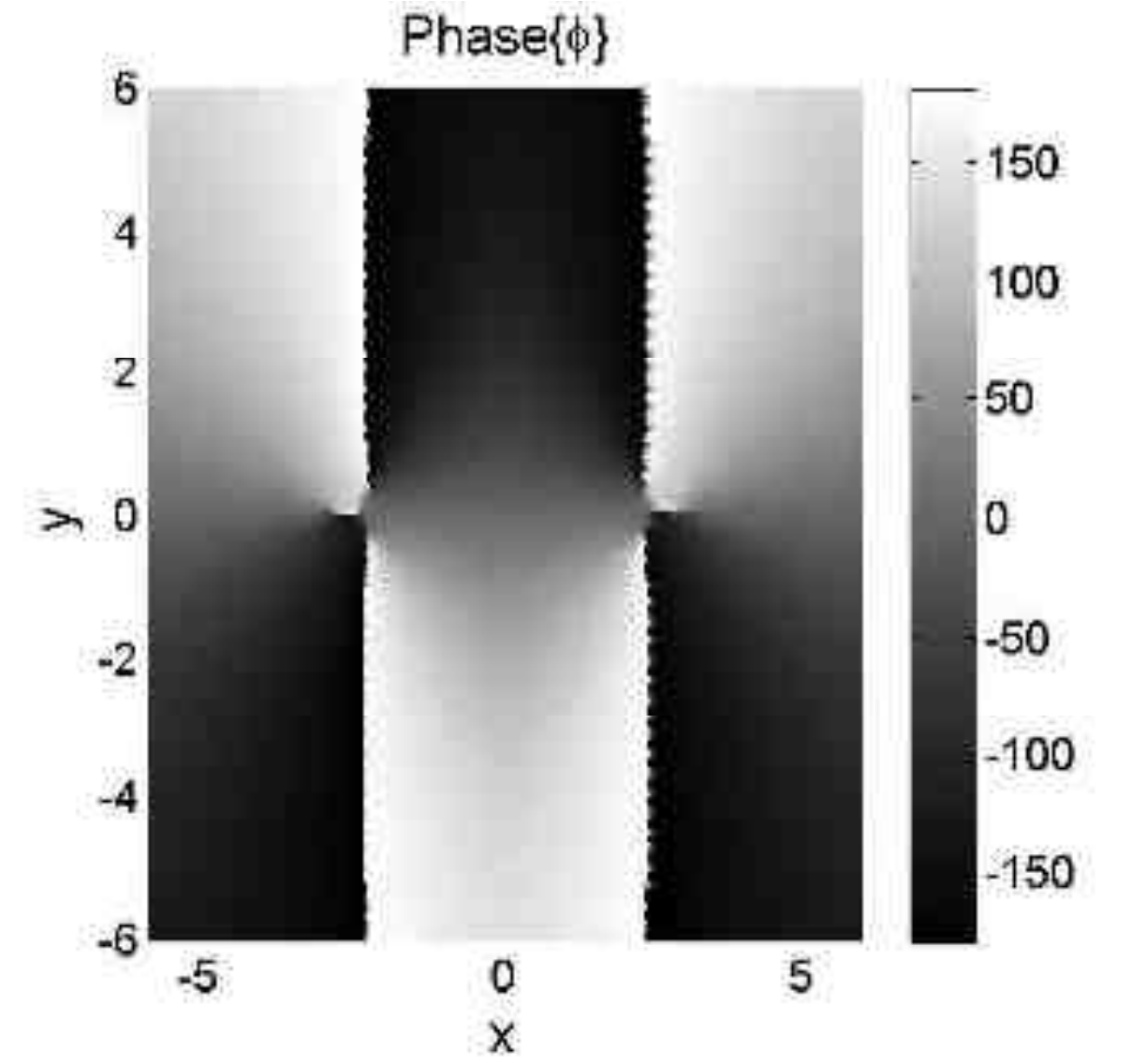}
\label{2WellTrioTrioMu7Phase1}}
\subfigure[]{\includegraphics[width=2.25in]{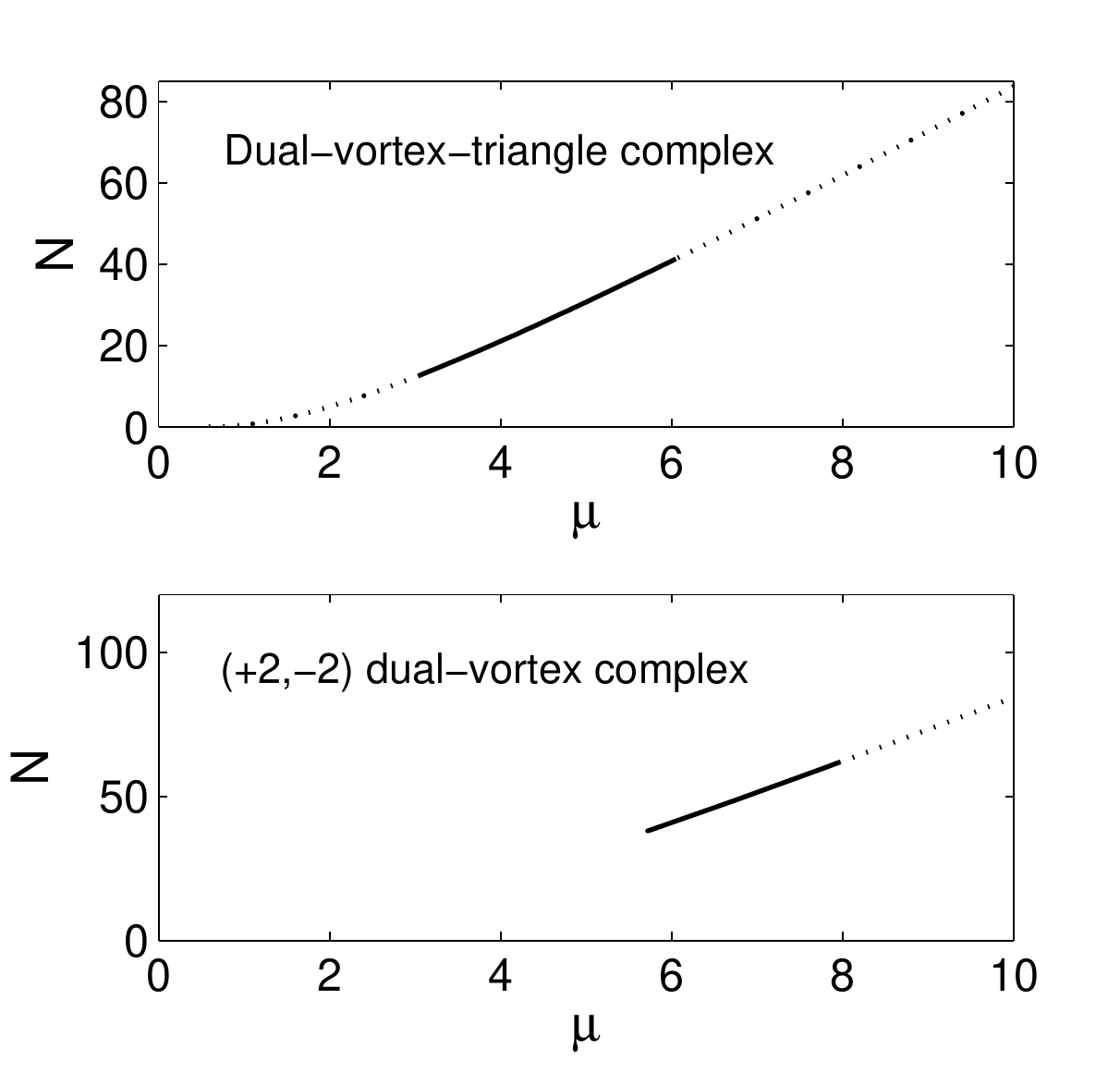}
\label{2WellTrioTrioNvsMu1}}
\subfigure[]{\includegraphics[width=2.25in]{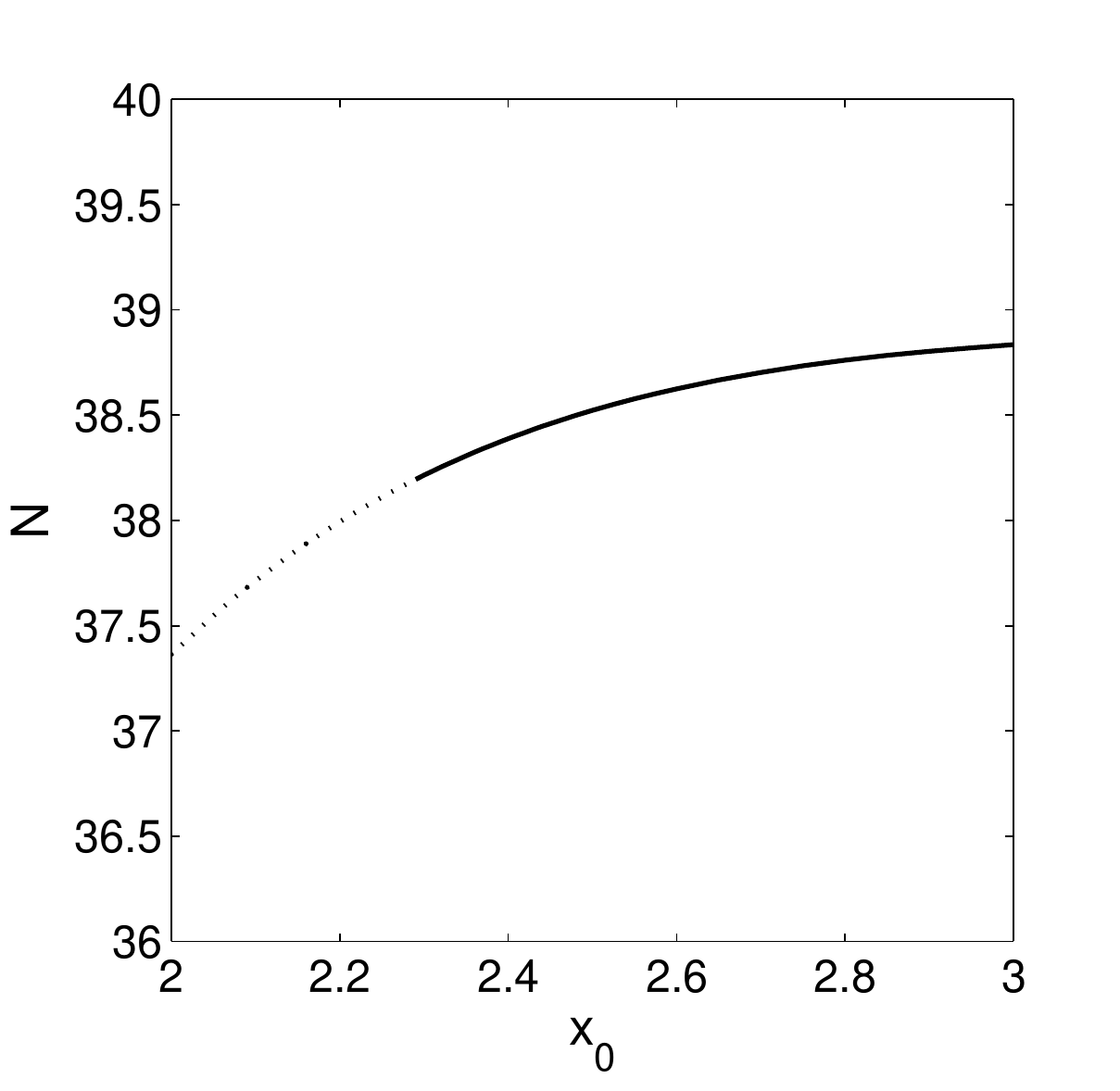}
\label{2WellTrioTrioNvsX01}}
\caption{(Color online) Dual-vortex complexes with topological charges $\left( +2,-2\right)
$, built of two single-well vortices (a,b), or vortex triangles (c,d), for $%
x_{0}=2.5$, $\protect\mu =7$ and $\protect\alpha =0.5$. (e) The respective $%
N(\protect\mu )$ curves, for $x_{0}=2.5$ and $\protect\alpha =0.5$. (f) The $%
N(x_{0})$ curve, for the dual-vortex complexes at fixed $\protect\mu =5$.}
\label{2WellTrioTrio1}
\end{figure}

Dual-vortex complexes built of vortices with $|m|=2$ were also addressed. In
both cases of topological-charge sets $\left( +2,-2\right) $ and $\left(
+2,+2\right) $, the transition (bifurcation) from the individual double
vortices to the VTs occurs, thus giving rise to additional \textit{dual-VT}
complexes. An example of the complex for the charge set $\left( +2,-2\right)
$ is displayed in Figs. \ref{2WellTrioTrio1}(a,b), with the respective
bifurcation diagram shown in Fig. \ref{2WellTrioTrioNvsMu1}. In this case,
the dual-vortex complexes are stable at $5.71<\mu <7.96$, i.e., $%
37.92<N<61.73$. Further, an example of the related dual-VT complex, with the
same charge set, $\left( +2,-2\right) $, and the same parameters is
presented in Fig. \ref{2WellTrioTrio1}(c,d). The dual-VT complexes of this
type are stable at $3.05<\mu <6.07$, i.e., $12.69<N<41.52$. For $\mu =5$ and
varying $x_{0}$, the dual-vortex complexes with topological charges $\left(
+2,-2\right) $ are stable at $x_{0}>2.40$ for fixed $\mu =0.5$, as shown in
Fig. \ref{2WellTrioTrioNvsX01} (in this case, VTs do not exist, as they are
generated by the bifurcation at larger values of $\mu $).

\begin{figure}[tbp]
\subfigure[]{\includegraphics[width=2.25in]{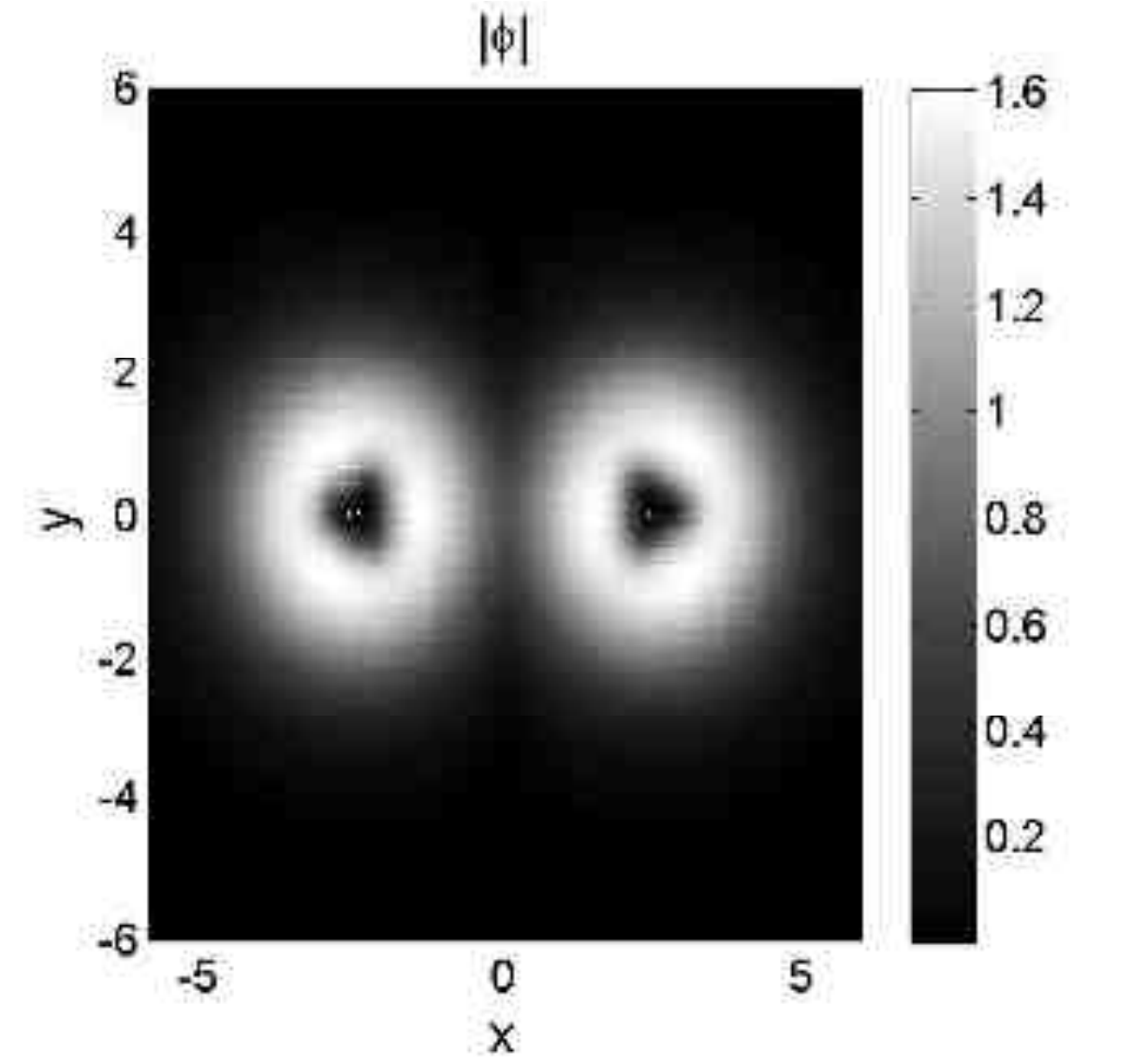}
\label{2WellM2M2Mu7Abs2}}
\subfigure[]{\includegraphics[width=2.25in]{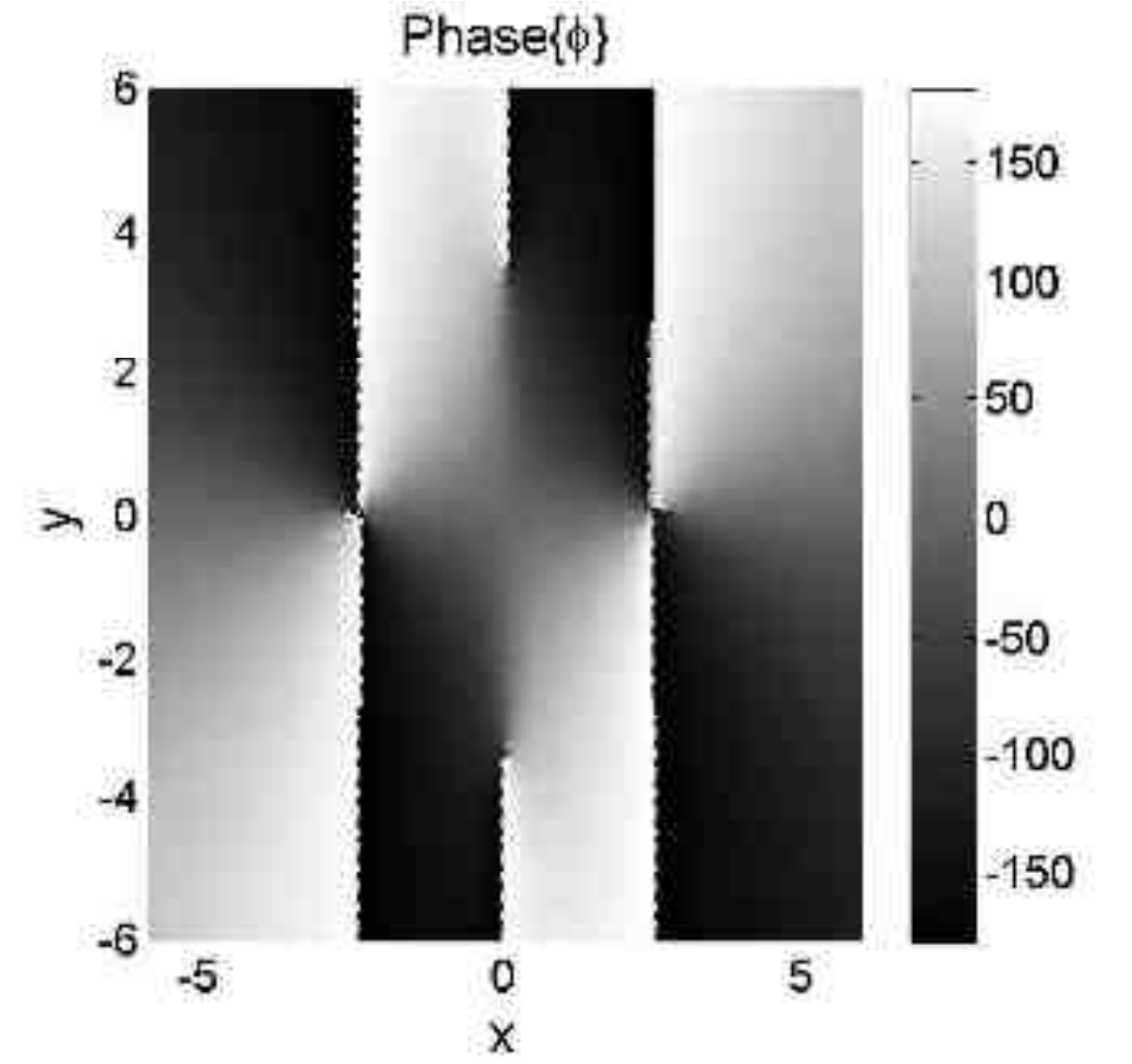}
\label{2WellM2M2Mu7Phase2}}
\subfigure[]{\includegraphics[width=2.25in]{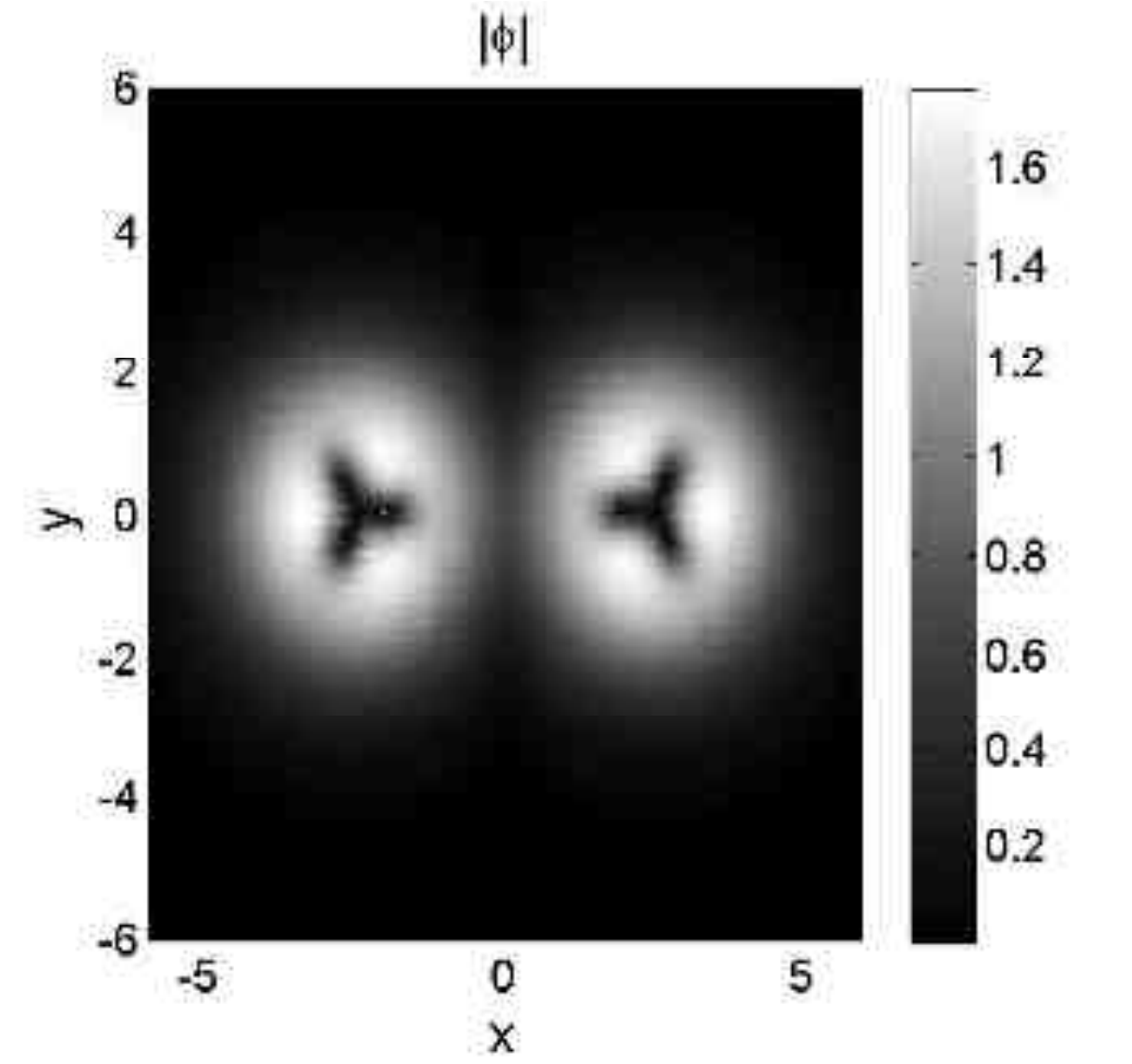}
\label{2WellTrioTrioMu7Abs2}}
\subfigure[]{\includegraphics[width=2.25in]{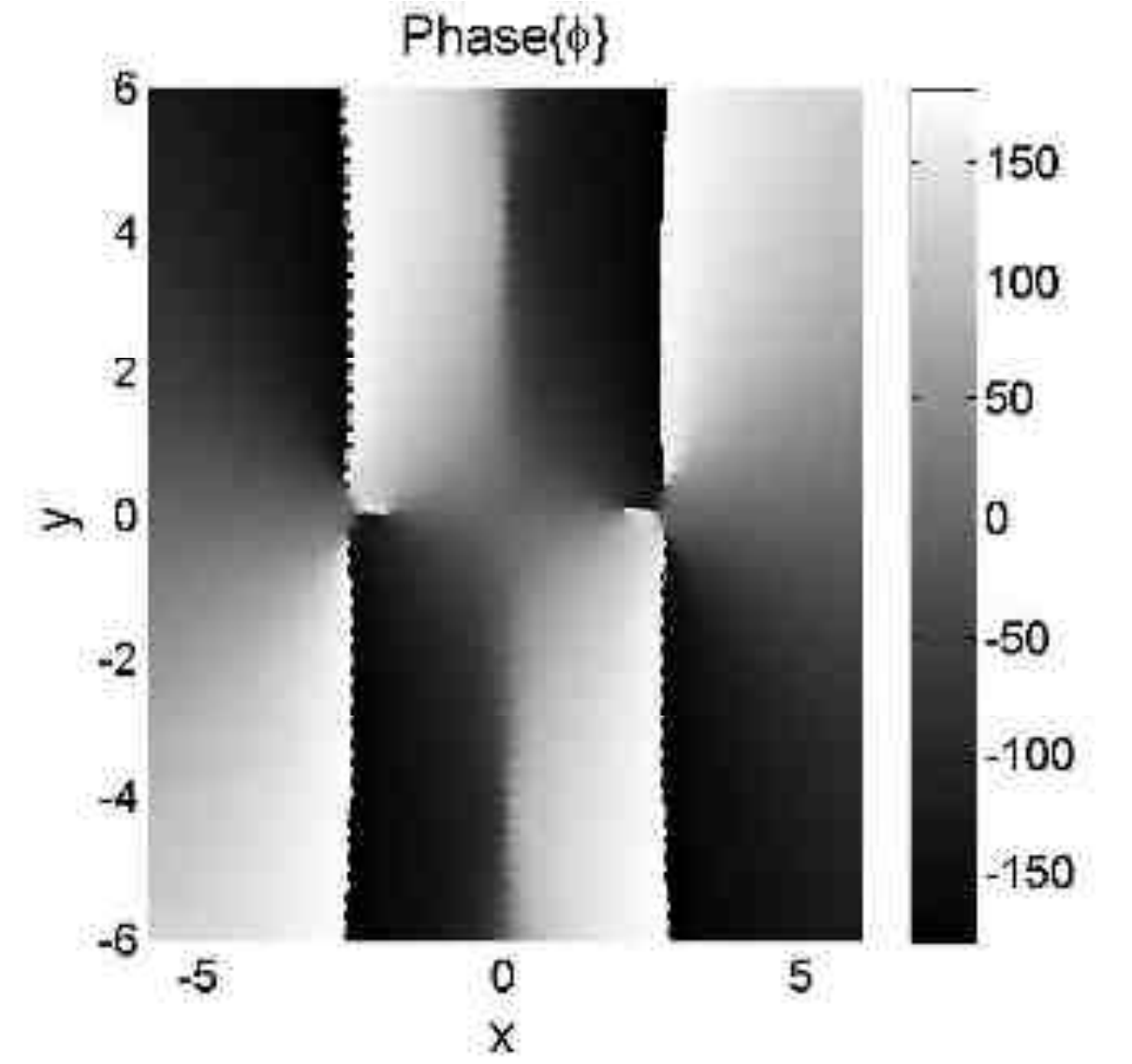}
\label{2WellTrioTrioMu7Phase2}}
\subfigure[]{\includegraphics[width=2.25in]{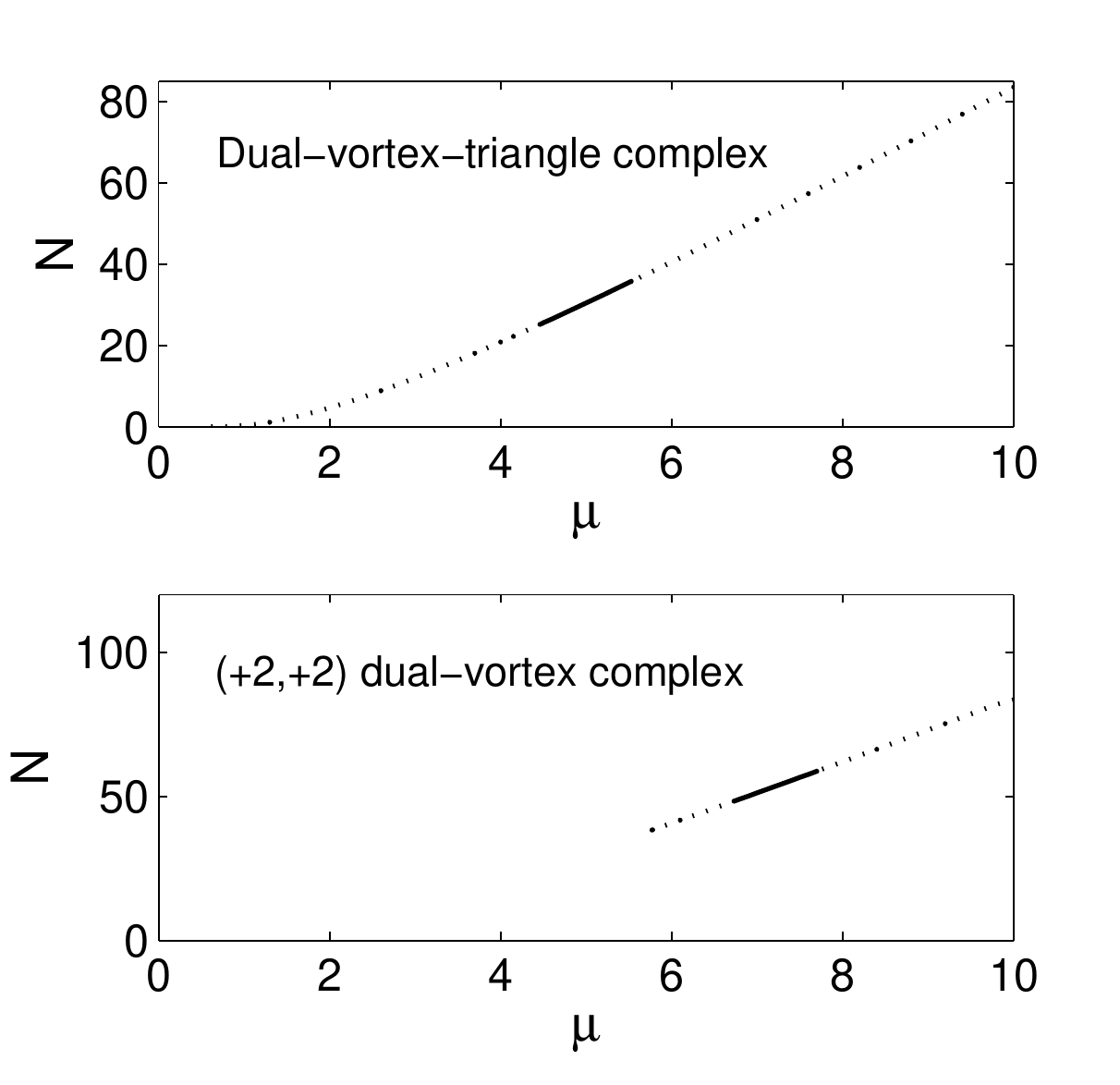}
\label{2WellTrioTrioNvsMu2}}
\subfigure[]{\includegraphics[width=2.25in]{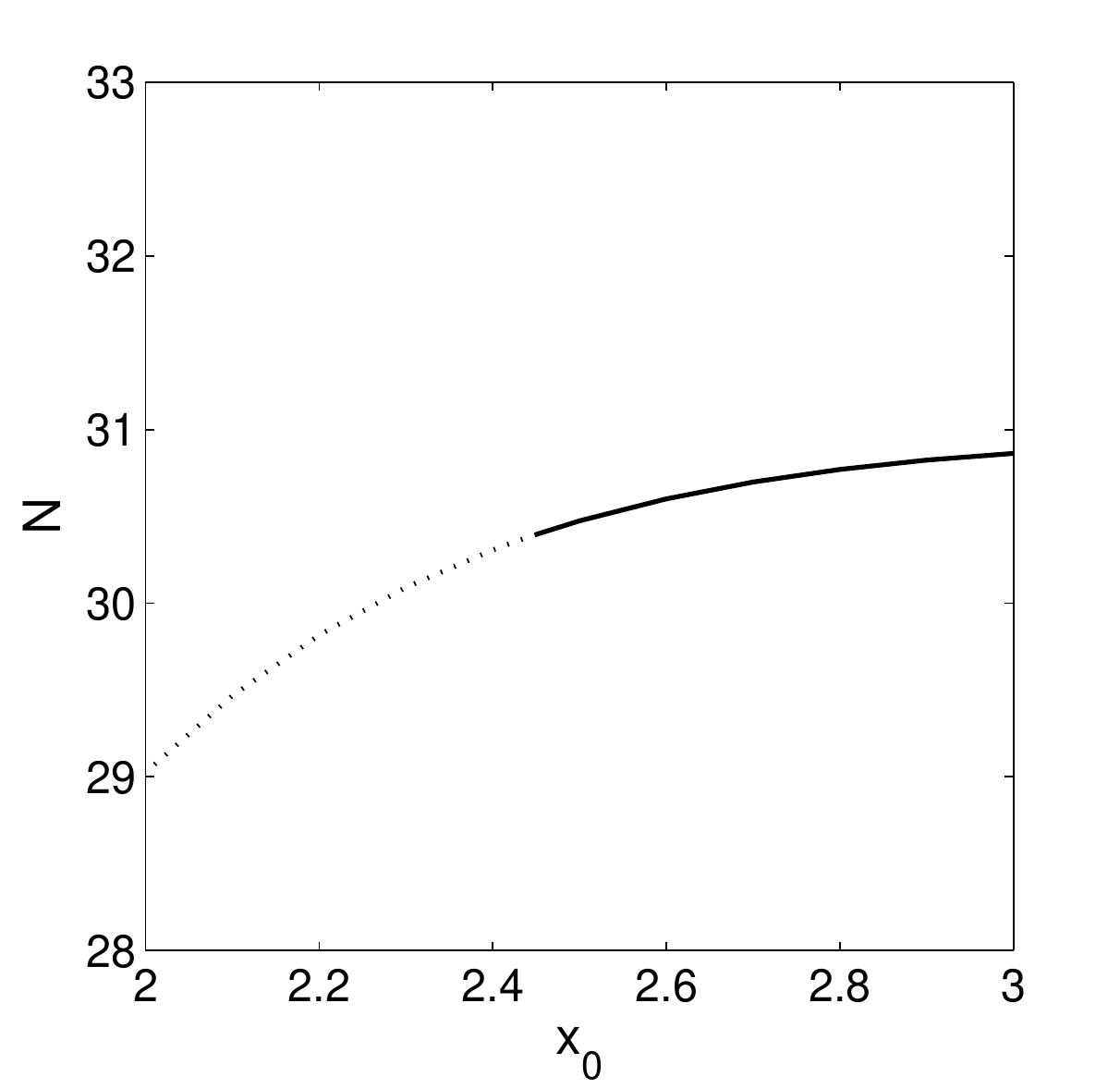}
\label{2WellTrioTrioNvsX02}}
\caption{(Color online) The same as in Fig. \protect\ref{2WellTrioTrio1}, but for the
vortex and vortex-triangle complexes with the topological-charge set $\left(
+2,+2\right) $.}
\label{2WellTrioTrio2}
\end{figure}

Similar results for the complexes pertaining to the topological-charge set $%
\left( +2,+2\right) $ are presented in Fig. \ref{2WellTrioTrio2}. Here, for $%
x_{0}=2.5$, the dual-vortex complexes, and ones built of the VTs, are stable
at $6.74<\mu <7.71$, i.e., $48.37<N<58.82$, and $4.47<\mu <5.54$, i.e., $%
25.30<N<35.87$, respectively. For $\mu =5$, the solutions are stable at $%
x_{0}>2.45$, as shown in Fig. \ref{2WellTrioTrioNvsX02}.

A conspicuous difference of the configuration with the charge set $(+2,+2)$
from the above one with charges $(+2,-2)$, observed in Fig. \ref%
{2WellTrioTrio2}(a) [cf. Fig. \ref{2WellTrioTrio1}(a)], is relatively strong
deformation of cores of both left and right vortices. This is explained by
attractive and destructive interference of overlapping fields of the
vortices in the gap between their cores, in the cases of opposite and
identical signs of the two vorticities, respectively. Indeed, in the latter
case the destructive interference removes the field from the gap, making
inner boundaries of the cores nearly flat, as seen in Fig. \ref%
{2WellTrioTrio2}(a). A similar argument helps to understand the difference
between the mutual orientations of the VTs, which is observed in Figs. \ref%
{2WellTrioTrio1}(c) and \ref{2WellTrioTrio2}(c): the interferometric removal
of the field from the inter-core gap in the configuration case makes it
possible to keep the empty corners inside the gap.

It is relevant to mention too that the stability regions are somewhat
smaller for the complexes with the charge set $(+2,+2)$ than for their
counterparts with charges $(+2,-2)$. Note that the difference between the
dual-vortex complexes corresponding to charge sets $\left( +m,-m\right) $
and $\left( +m,+m\right) $ is much smaller in the case of $m=1 $ than for $%
m=2$, cf. Fig. \ref{2WellM1M1}. This is explained by the fact that the size
of the vortex core is essentially smaller for $m=1$.

All the complexes with $|m|=2$ are completely unstable at $x_{0}=2$. Direct
simulations show that unstable solutions of these types transform themselves
into the ground-state symmetric state, or into symmetric complexes of VADs
(their description is following below). On the contrary to the VTs in the
single-well case (see Fig. \ref{1Well2DRotatingVortexTrioEvolution}), the
triangles forming stable dual-VT complexes in the DW potential do not
exhibit rotation when perturbations are added to the initial configuration.

\begin{figure}[tbp]
\subfigure[]{\includegraphics[width=2.25in]{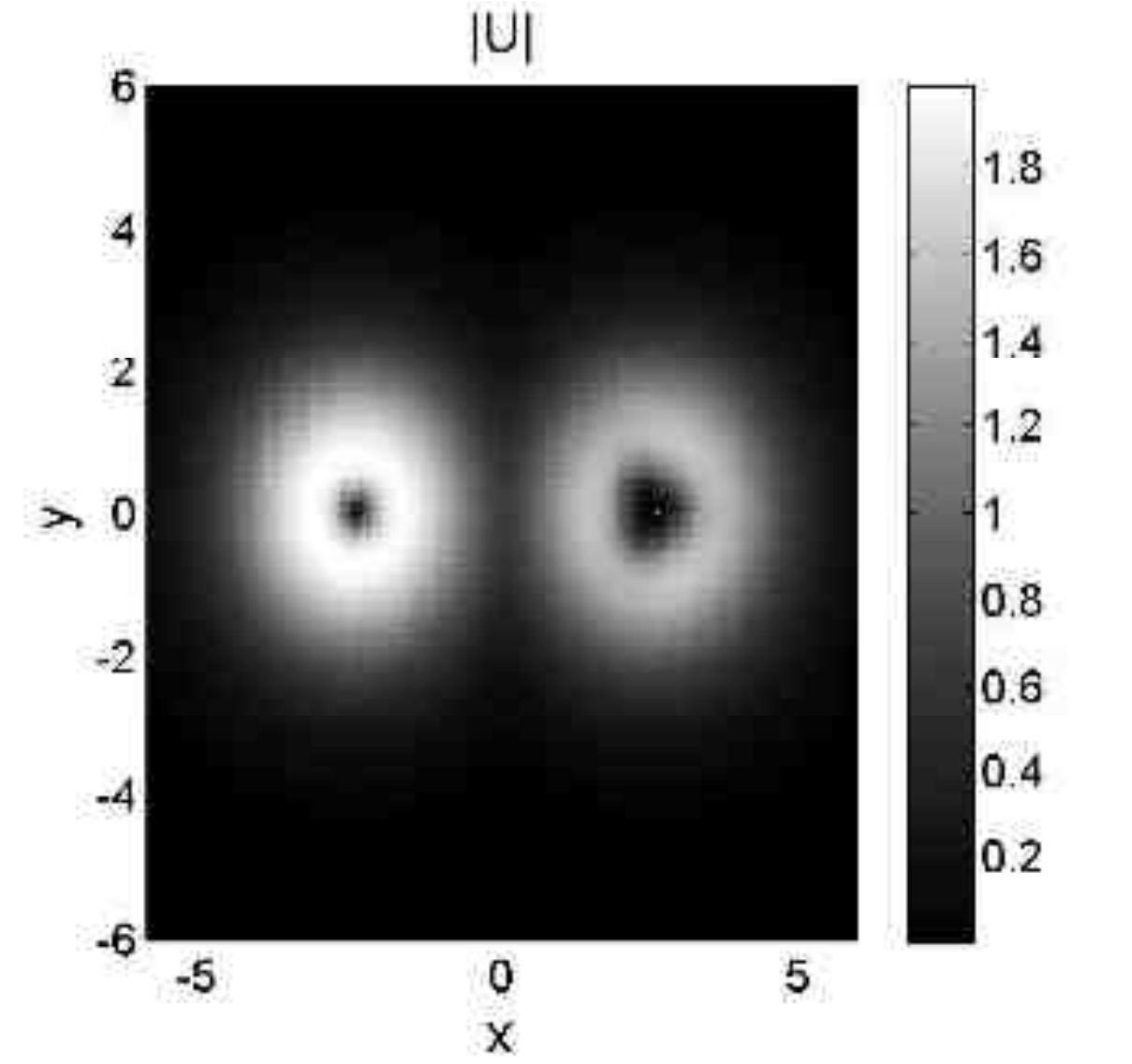}
\label{2WellM2M2Mu7Abs1}}
\subfigure[]{\includegraphics[width=2.25in]{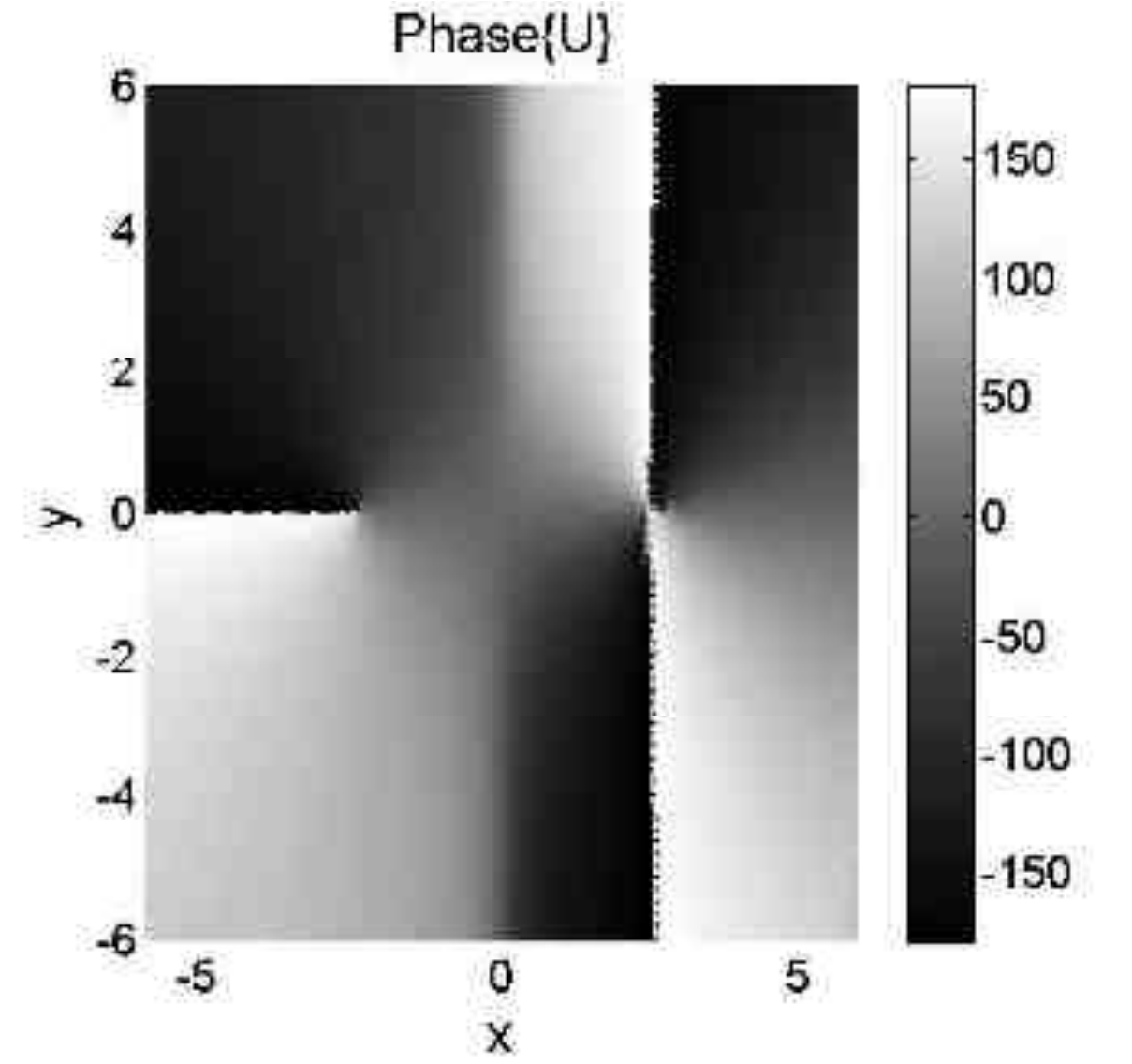}
\label{2WellM2M1Mu7Phase1}}
\subfigure[]{\includegraphics[width=2.25in]{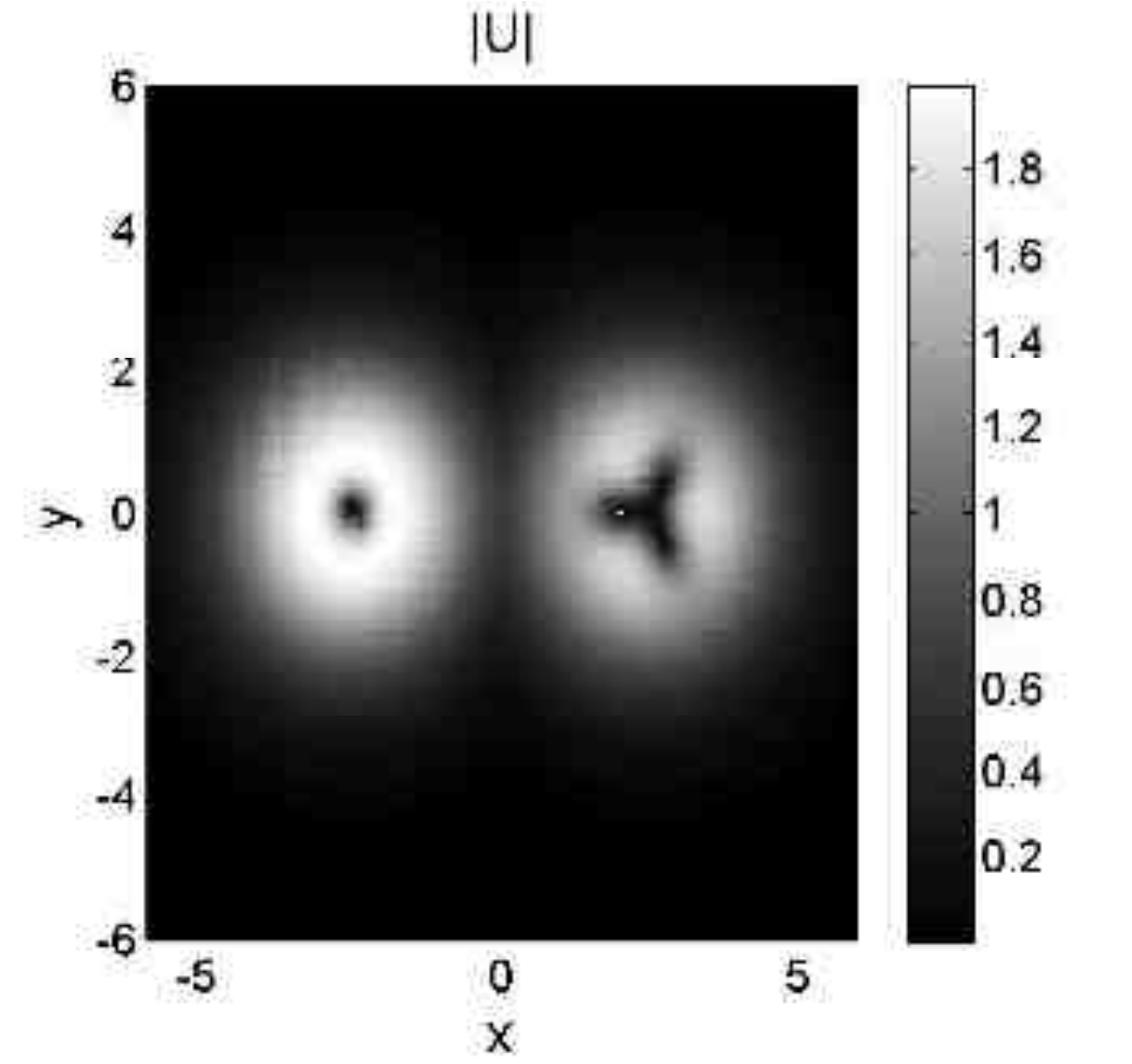}
\label{2WellTrioM1Mu7Abs1}}
\subfigure[]{\includegraphics[width=2.25in]{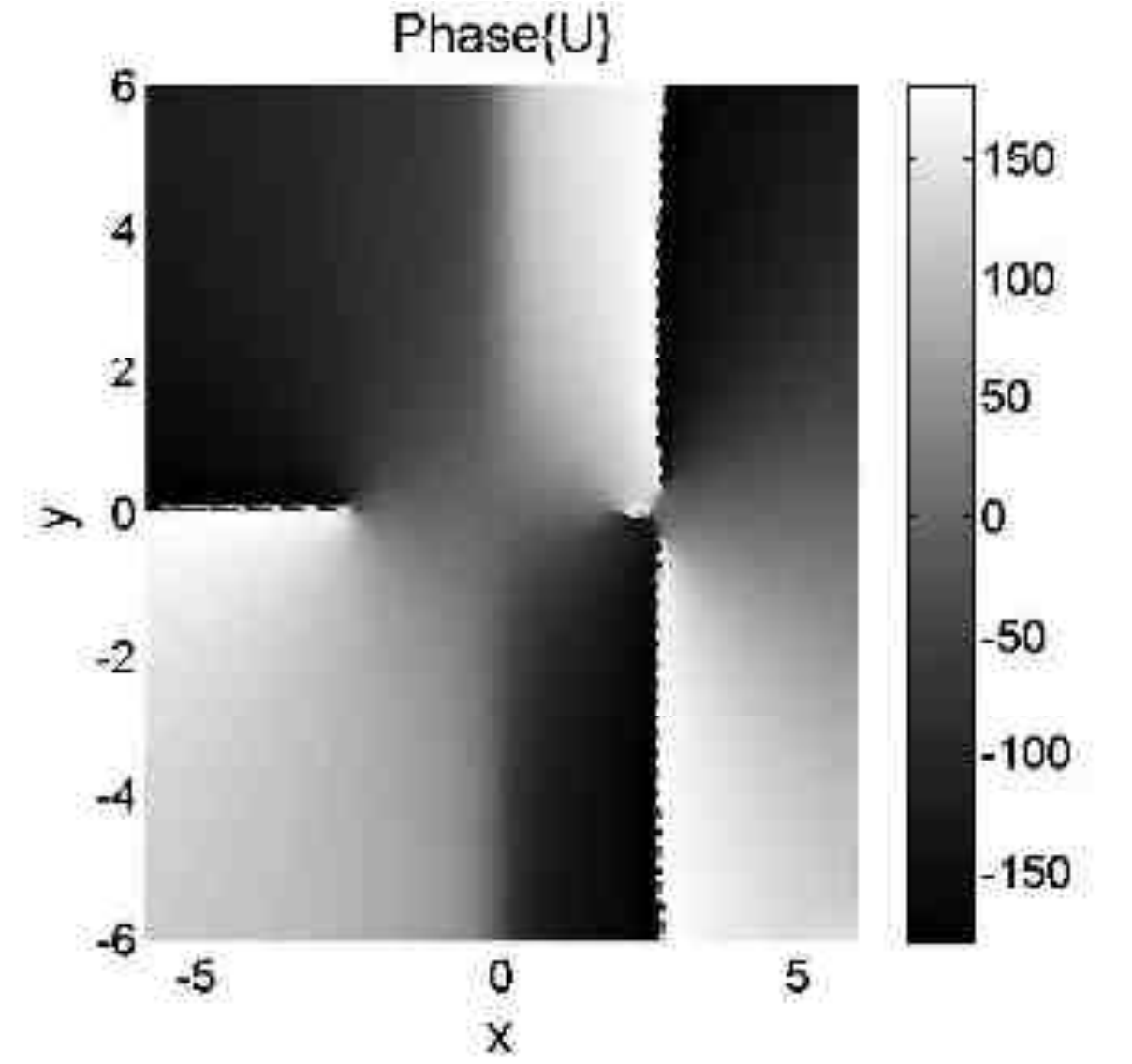}
\label{2WellTrioM1Mu7Phase1}}
\subfigure[]{\includegraphics[width=2.25in]{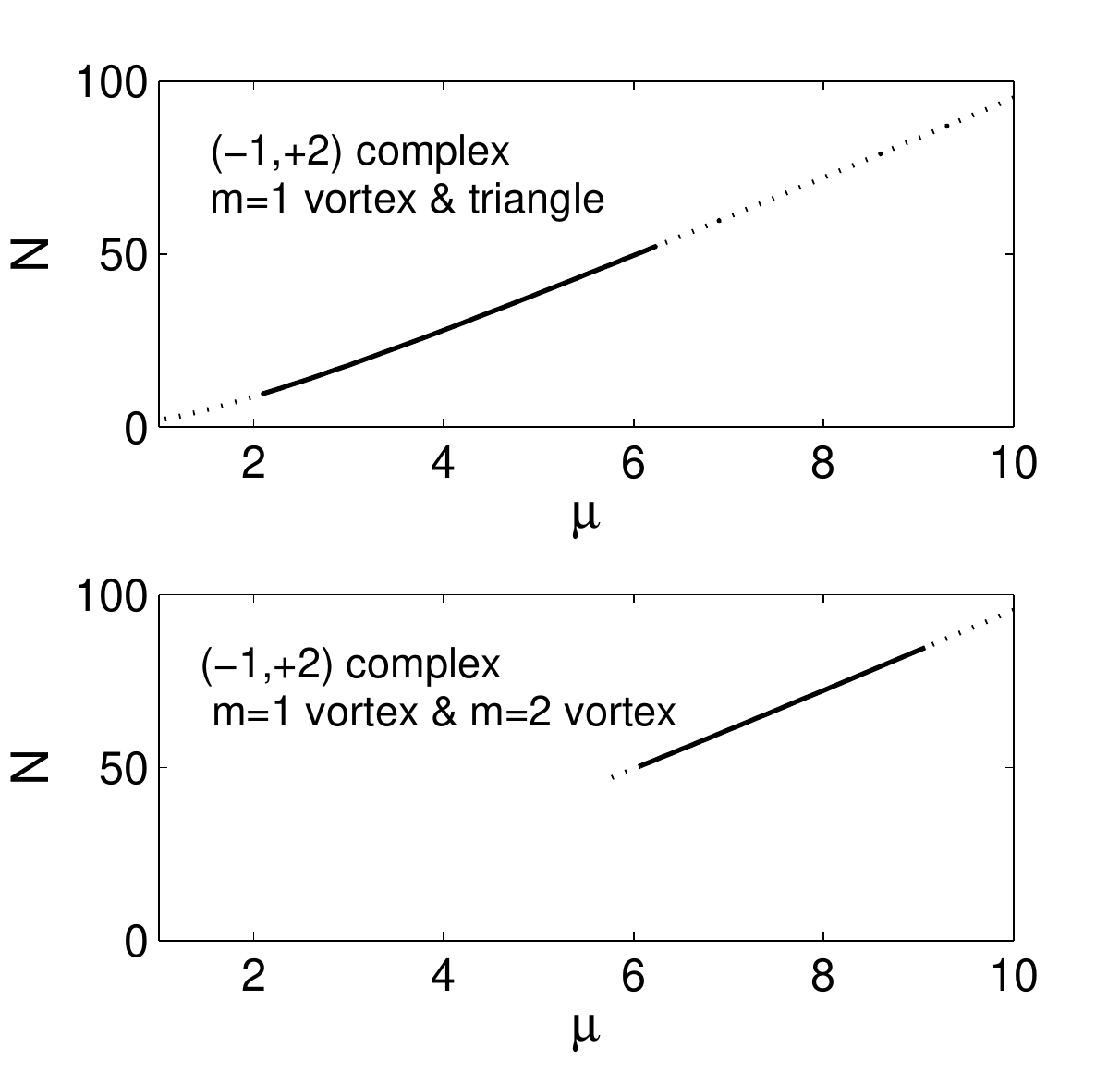}
\label{2WellM2M1NvsMu1}}
\subfigure[]{\includegraphics[width=2.25in]{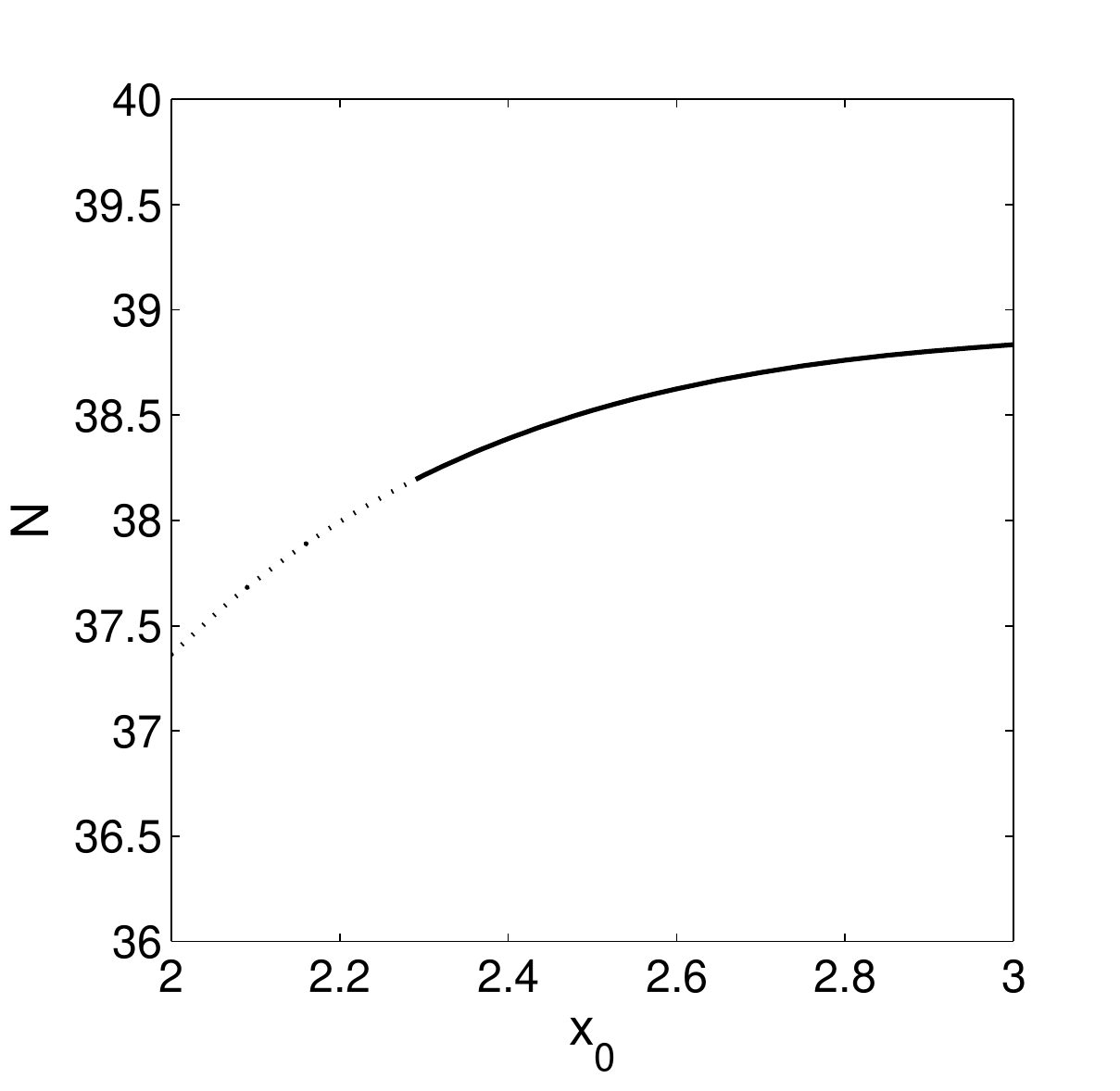}
\label{2WellM2M1NvsX01}}
\caption{(Color online) Examples of dual-vortex mixtures, with the topological-charge set $%
\left( +1,+2\right) $, composed of $m=1$ and $m=2$ vortices trapped in the
two wells (panels a and b), or an $m=1$ vortex and a VT (vortex triangle)
(panels c and d), both for $x_{0}=2.5$, $\protect\mu =7$. The $N(\protect\mu %
)$ curves for both families, are presented in panel (e), for fixed $%
x_{0}=2.5 $. Panel (f) shows the\ respective $N(x_{0})$ curve, for $\protect%
\mu =5$.}
\label{2WellM1M21}
\end{figure}

Dual-vortex composites, constructed of an $|m|=1$ vortex trapped in one
well, and an $|m|=2$ vortex in the other, were also investigated. In this
configuration, two families, with topological charges $\left( +1,+2\right) $
and $\left( -1,+2\right) $, were examined. Similar to the dual vortex
complexes with $|m|=2$ mentioned above, bifurcations of structures, with the
same charges, where VTs replace the vortex with $|m|=2$, are observed. An
example of the vortex-vortex complex with the charge set $\left(
+1,+2\right) $ is demonstrated in Fig. \ref{2WellM1M21}(a,b), for $\mu =7$
and $x_{0}=2.5$. For this charge set, the bifurcation diagram in the $\left(
N,\mu \right) $ plane is displayed in Fig. \ref{2WellM2M1NvsMu1} for $%
x_{0}=2.5$. In this case, the vortex-vortex complexes are stable at $%
6.58<\mu <7.84$, i.e., $56.02<N<70.36$. The corresponding vortex-VT
complexes with charges $\left( +1,+2\right) $, demonstrated in Fig. \ref%
{2WellM1M21}(c,d) for $\mu =7$ and $x_{0}=2.5$, are stable at $3.01<\mu
<5.91 $, i.e., $17.85<N<48.50$. When fixing $\mu =5$ and varying $x_{0}$,
the complexes with topological charges $\left( +1,+2\right) $ are stable at $%
x_{0}>2.29$.

\begin{figure}[tbp]
\subfigure[]{\includegraphics[width=2.25in]{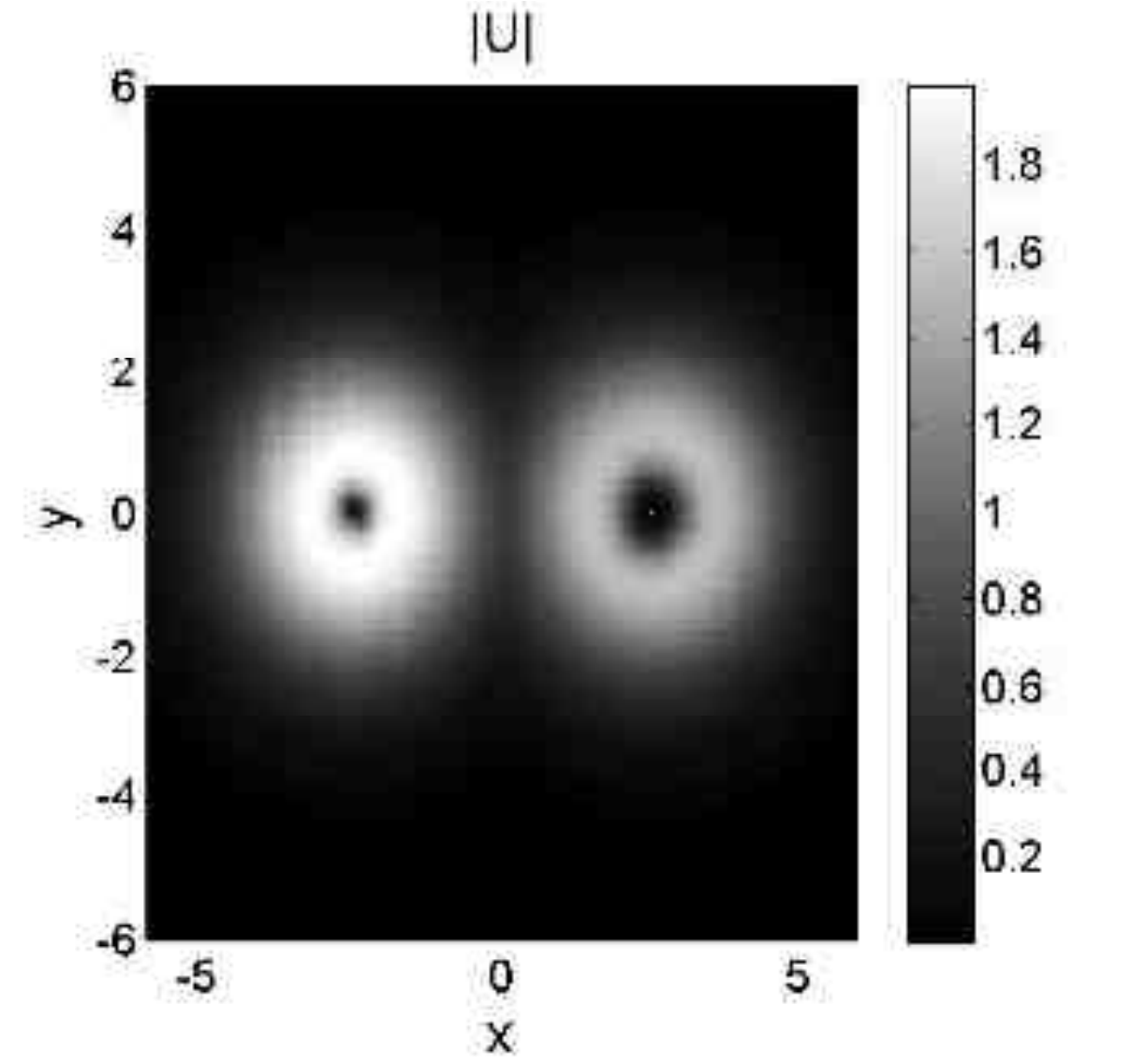}
\label{2WellM2M2Mu7Abs2}}
\subfigure[]{\includegraphics[width=2.25in]{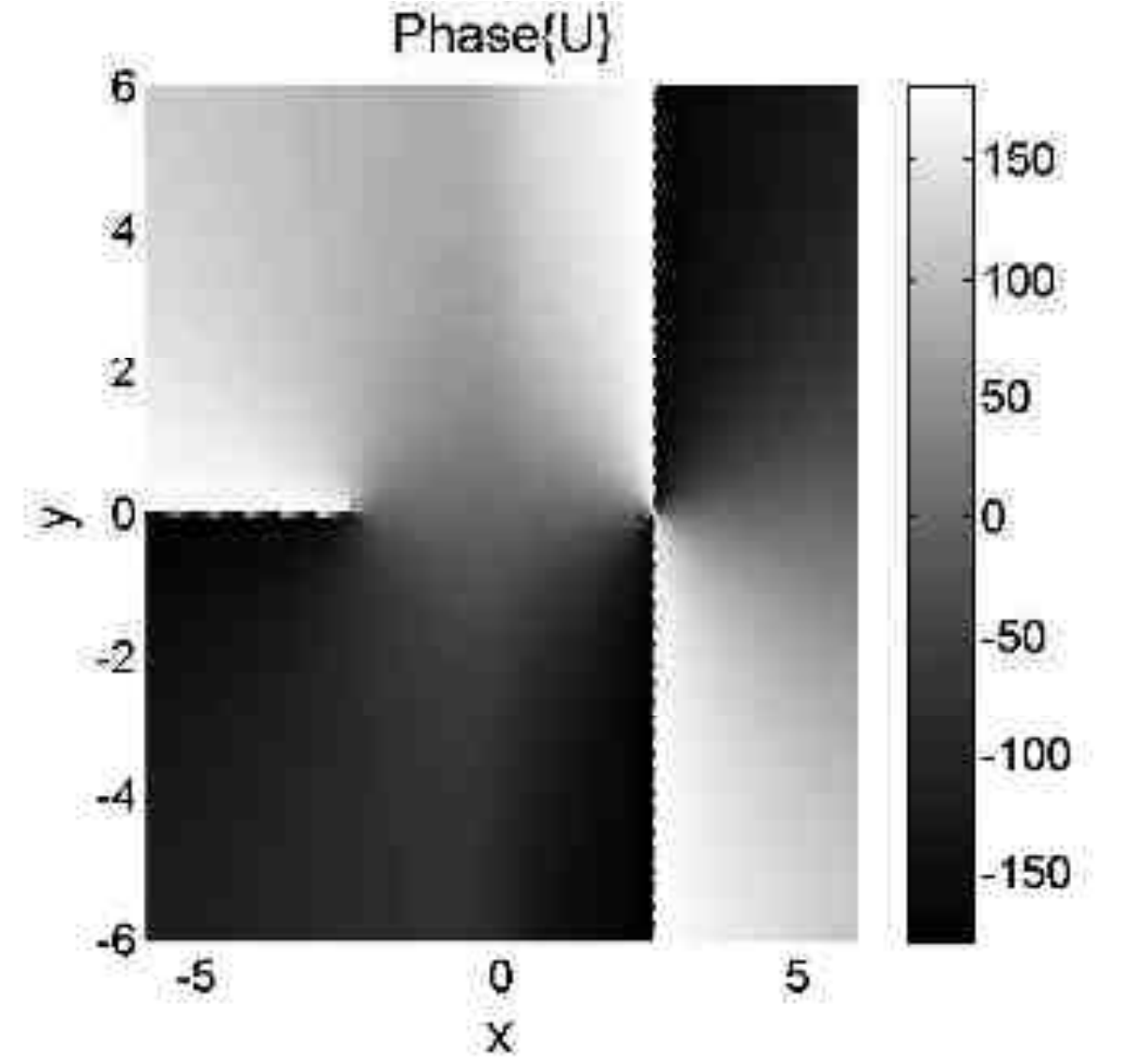}
\label{2WellM2M1Mu7Phase2}}
\subfigure[]{\includegraphics[width=2.25in]{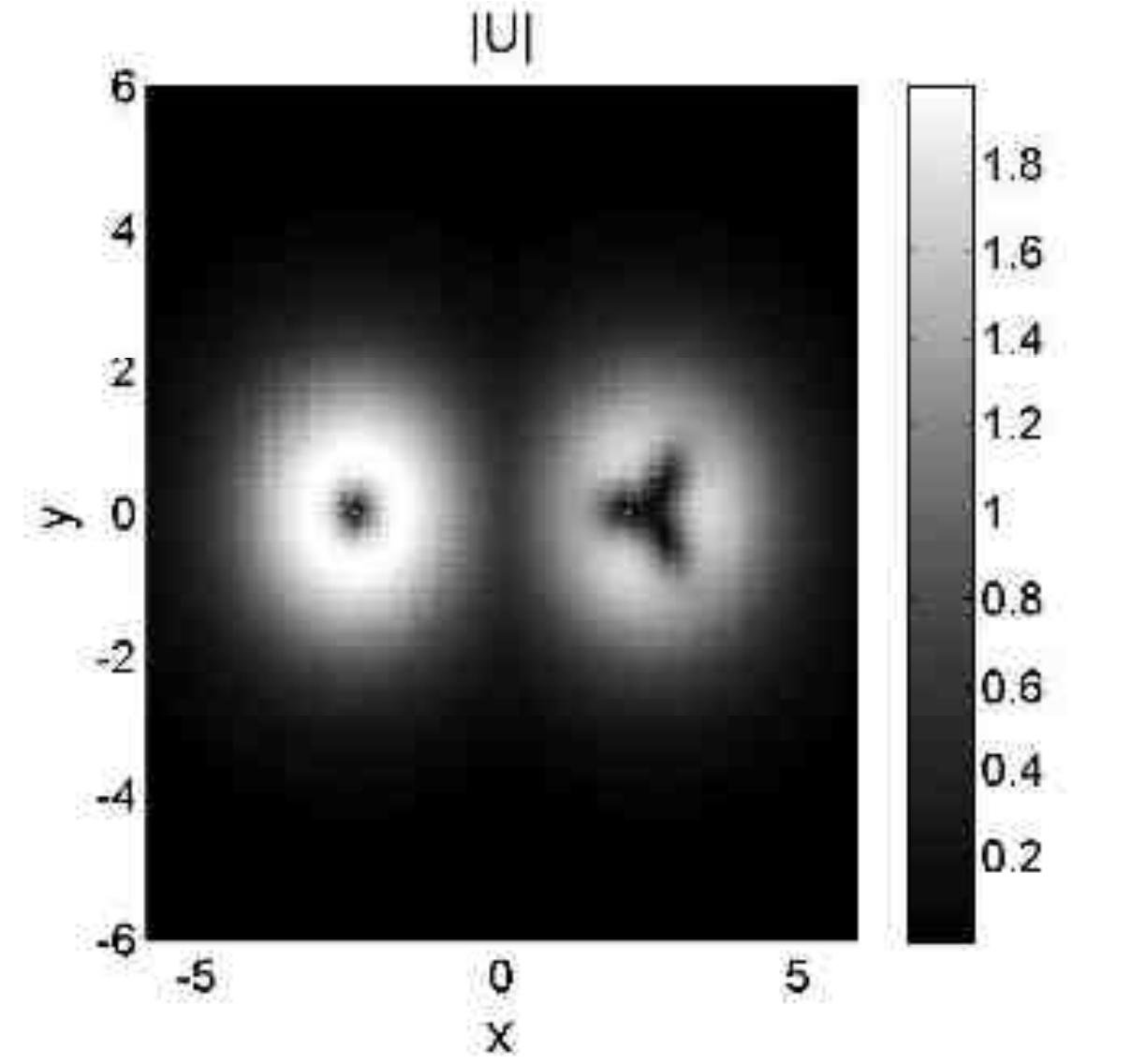}
\label{2WellTrioM1Mu7Abs2}}
\subfigure[]{\includegraphics[width=2.25in]{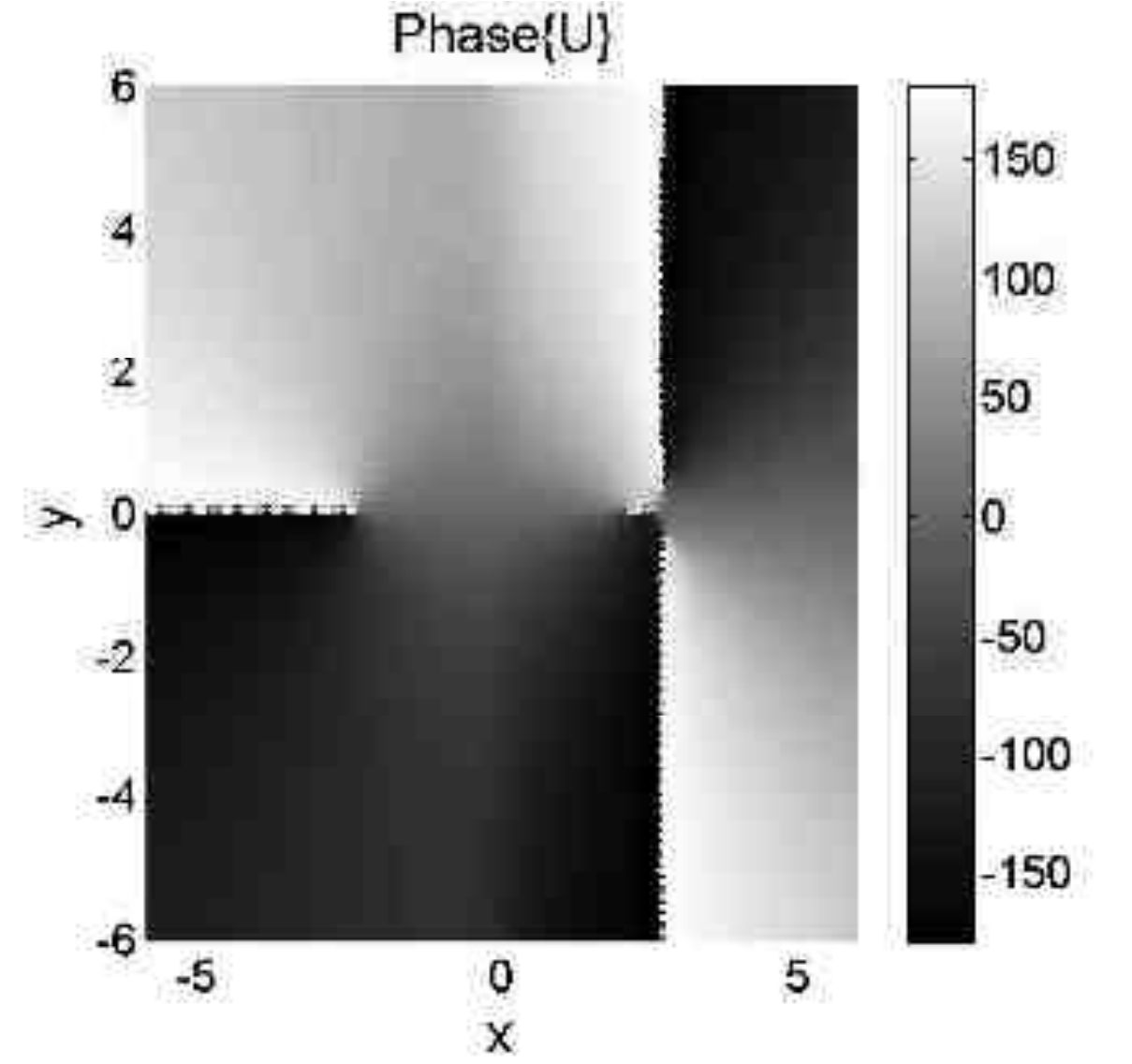}
\label{2WellTrioM1Mu7Phase2}}
\subfigure[]{\includegraphics[width=2.25in]{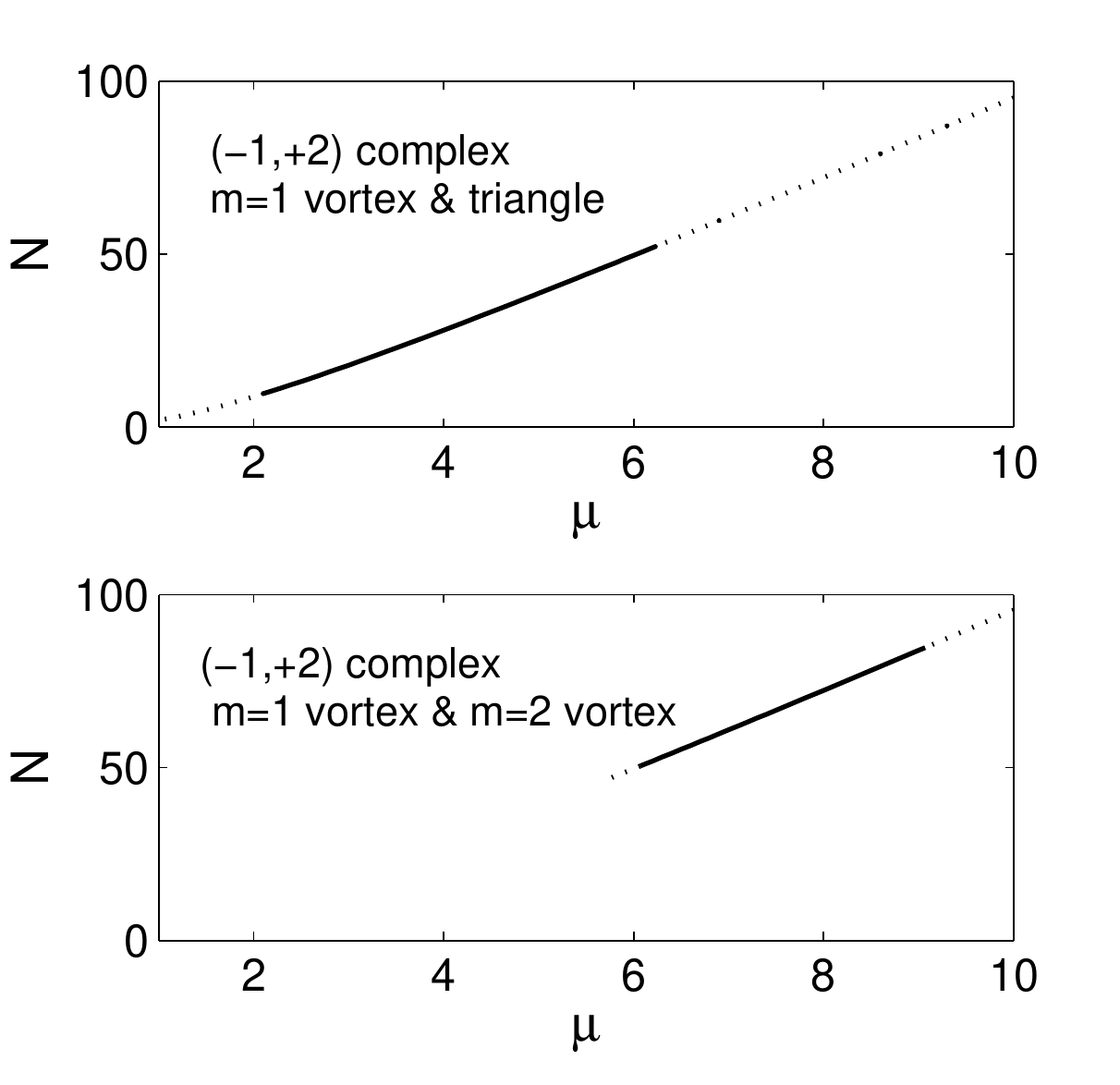}
\label{2WellM2M1NvsMu2}}
\subfigure[]{\includegraphics[width=2.25in]{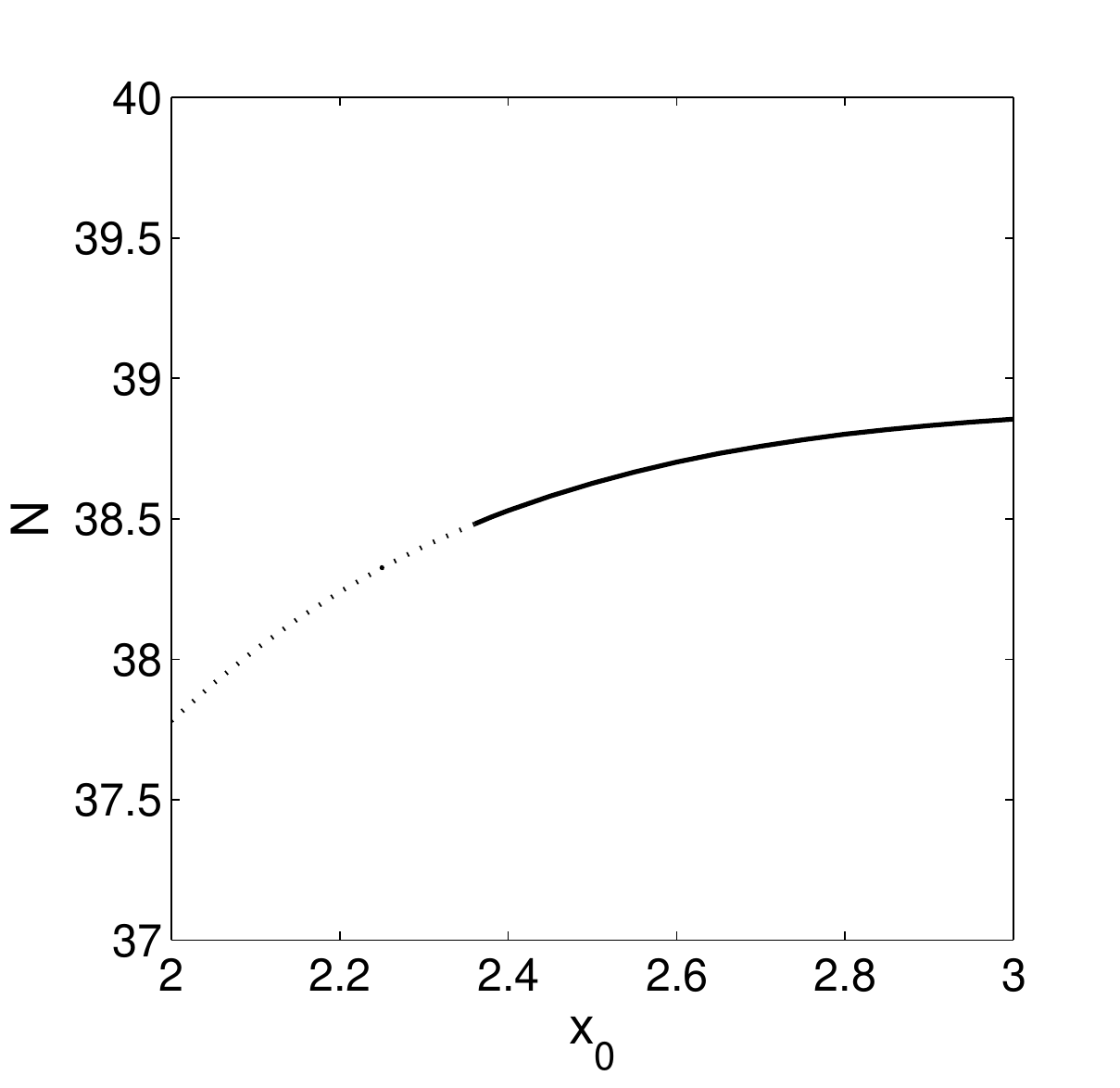}
\label{2WellM2M1NvsX02}}
\caption{(Color online) The same as in Fig. \protect\ref{2WellM1M21}, but for the
vortex-vortex and vortex-VT complexes with the topological-charge set $%
\left( -1,+2\right) $.}
\label{2WellM1M22}
\end{figure}

Equivalent results were also observed for the other charge set, $\left(
-1,+2\right) $, see Fig. \ref{2WellM1M22}). Examples of the vortex-vortex
and vortex-VT structures, in this topological setting, are displayed in
Figs. \ref{2WellM1M21}(a,b) and Figs. \ref{2WellM1M21}(c,d), respectively,
for $\mu =7$ and $x_{0}=2.5$. For $x_{0}=2.5$, the vortex-vortex complexes
are stable at $6.07<\mu <9.07$, i.e., $50.41<N<84.65$, and the vortex-VT
ones are stable at $2.11<\mu <6.24$, i.e., $9.69<N<52.27$ (Fig. \ref%
{2WellM2M1NvsMu2}). As seen in Fig. \ref{2WellM2M1NvsX02}, for fixed $\mu =5$%
, the stability holds at $x_{0}>2.36$.

\begin{figure}[tbp]
\subfigure[]{\includegraphics[width=1.6in]{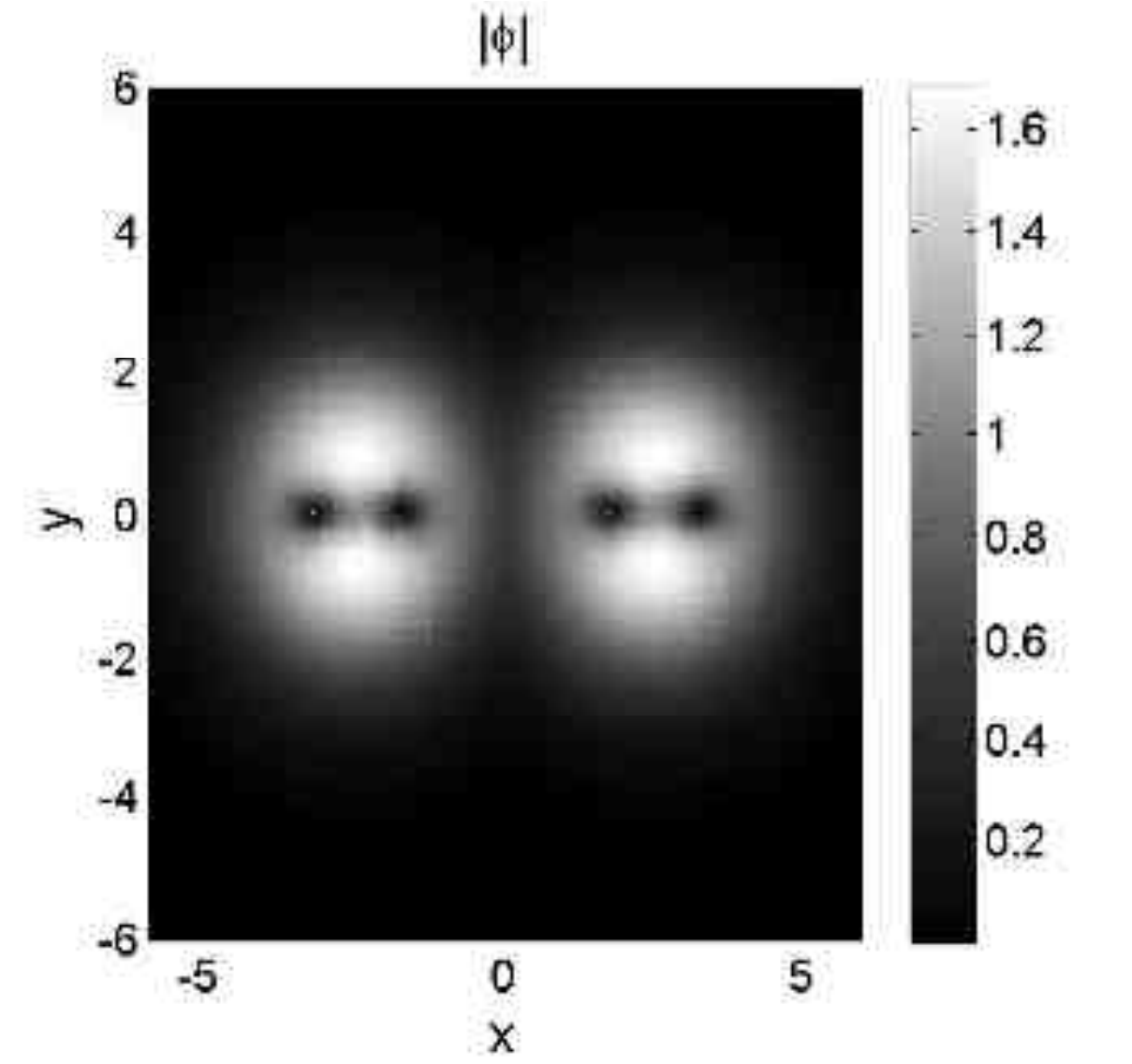}
\label{2WellPairPairMu5Abs2}}
\subfigure[]{\includegraphics[width=1.6in]{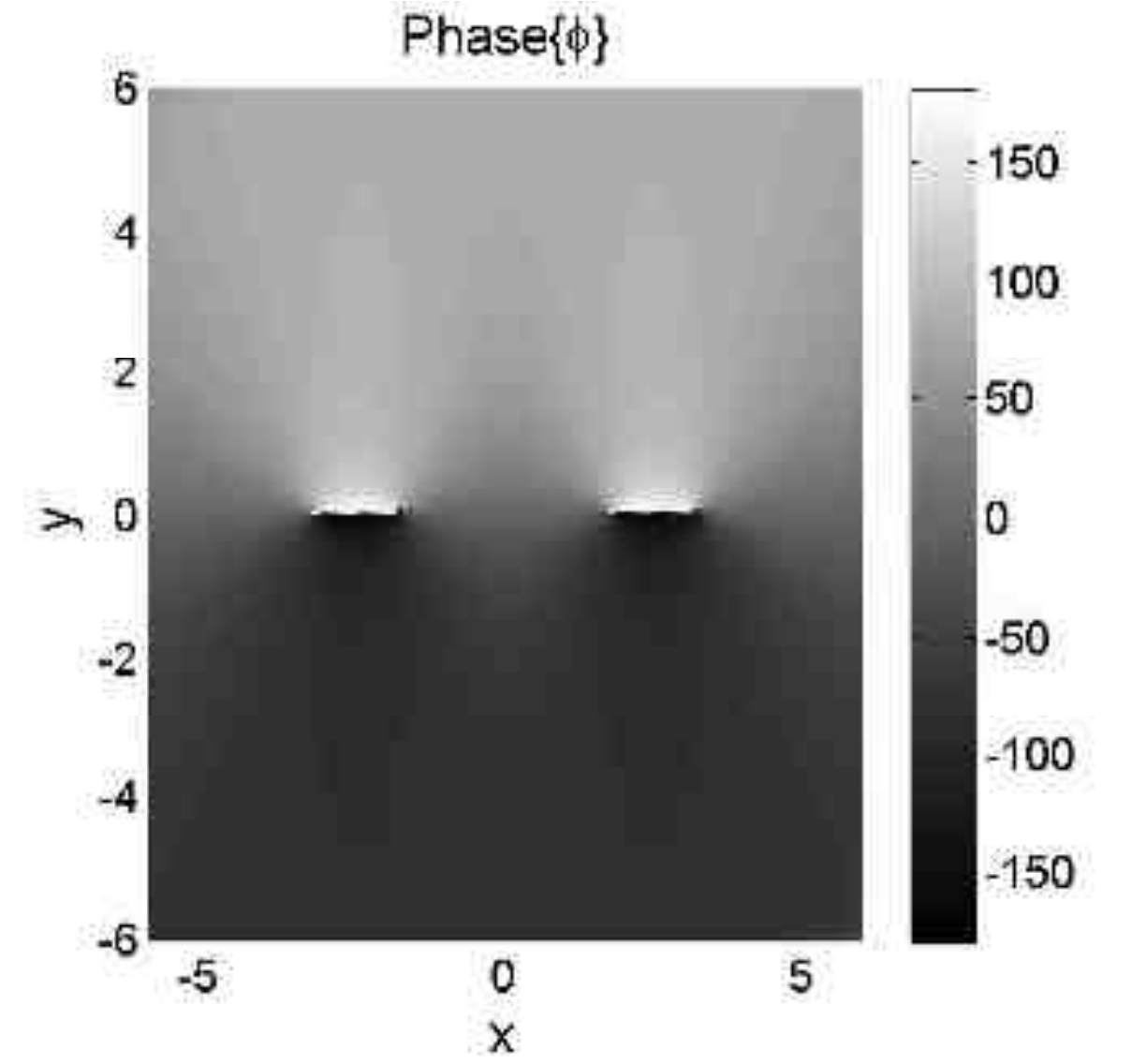}
\label{2WellPairPairMu5Phase2}}
\subfigure[]{\includegraphics[width=1.6in]{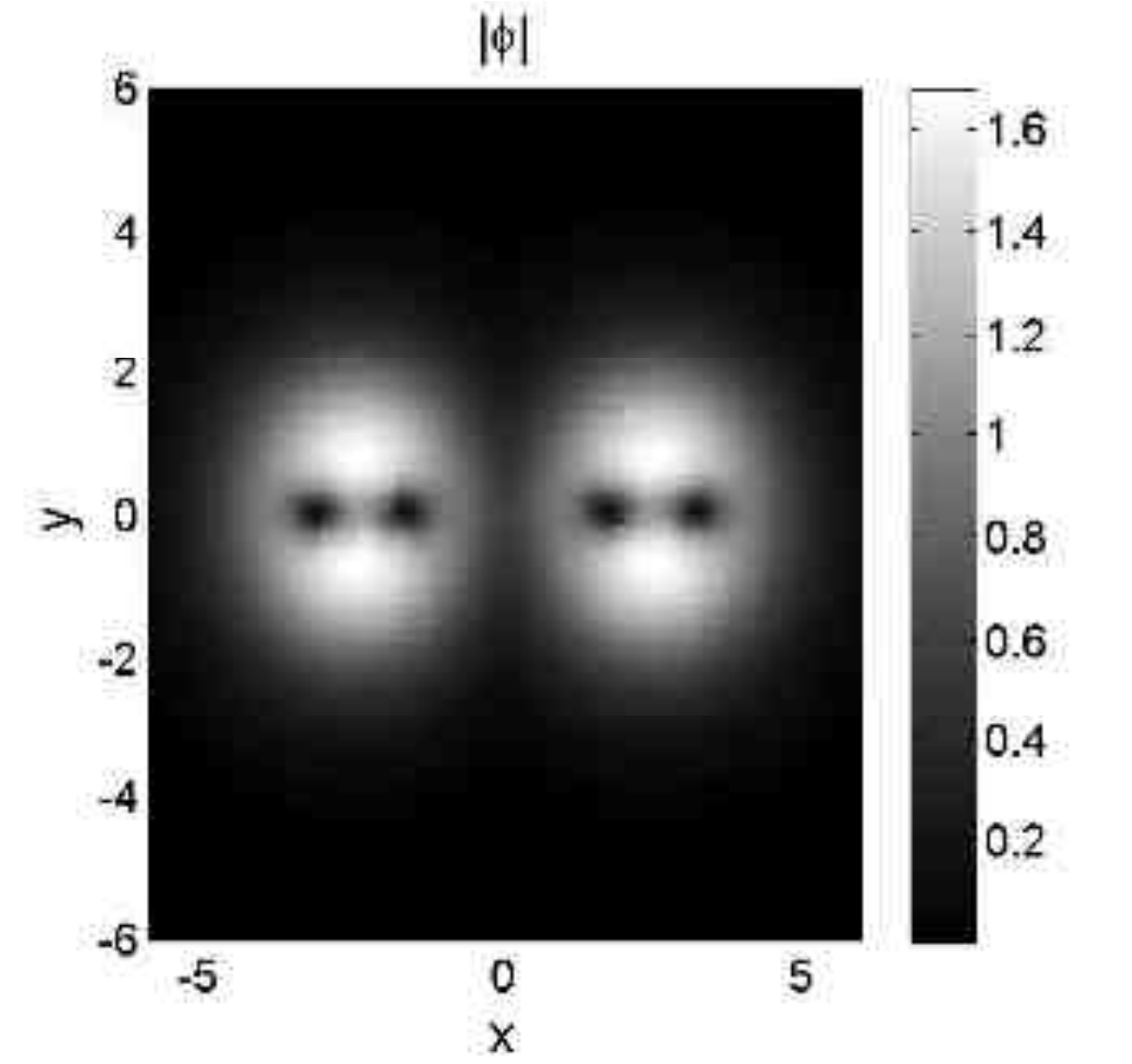}
\label{2WellPairPairMu5Abs1}}
\subfigure[]{\includegraphics[width=1.6in]{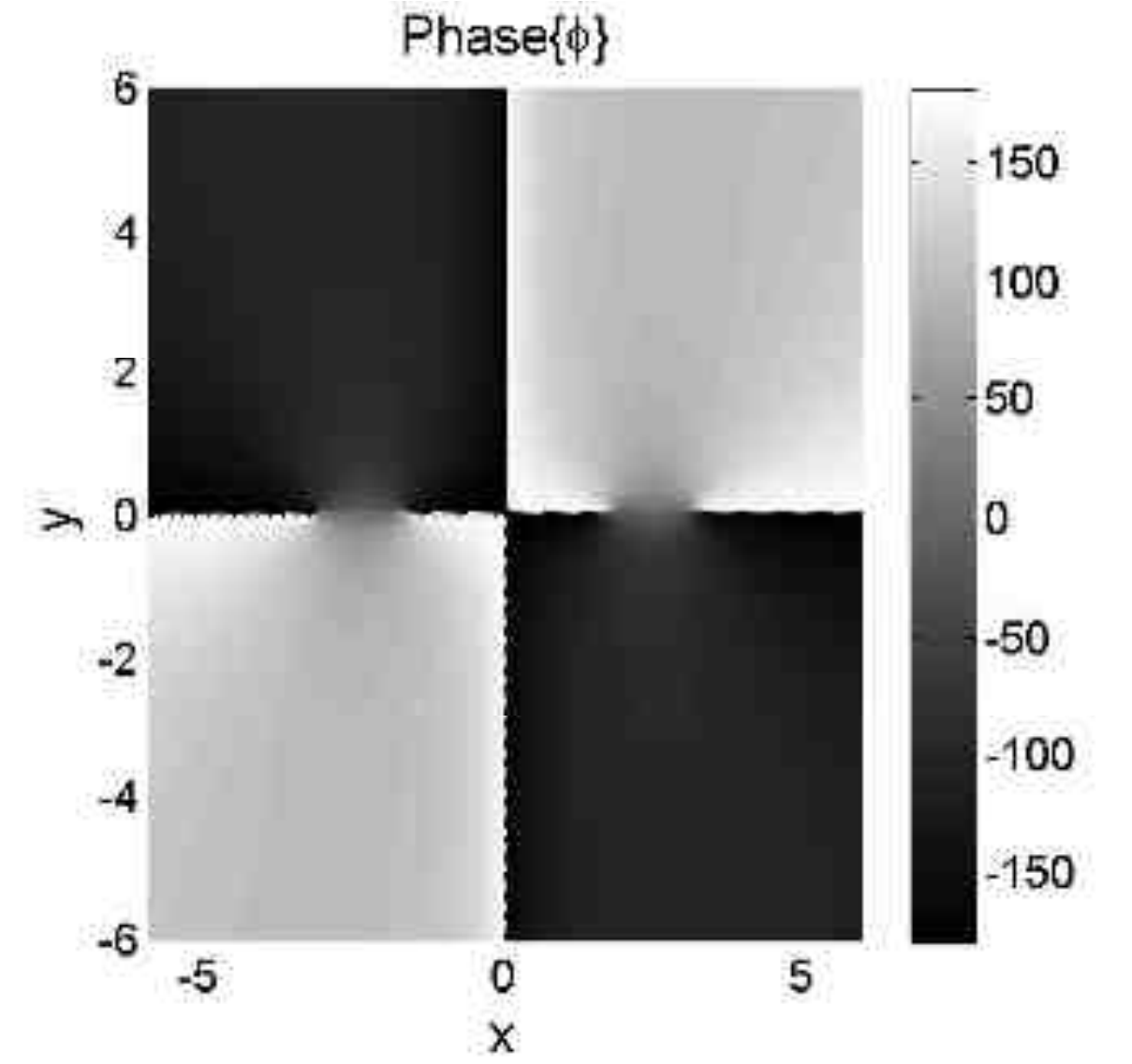}
\label{2WellPairPairMu5Phase1}} \newline
\subfigure[]{\includegraphics[width=1.6in]{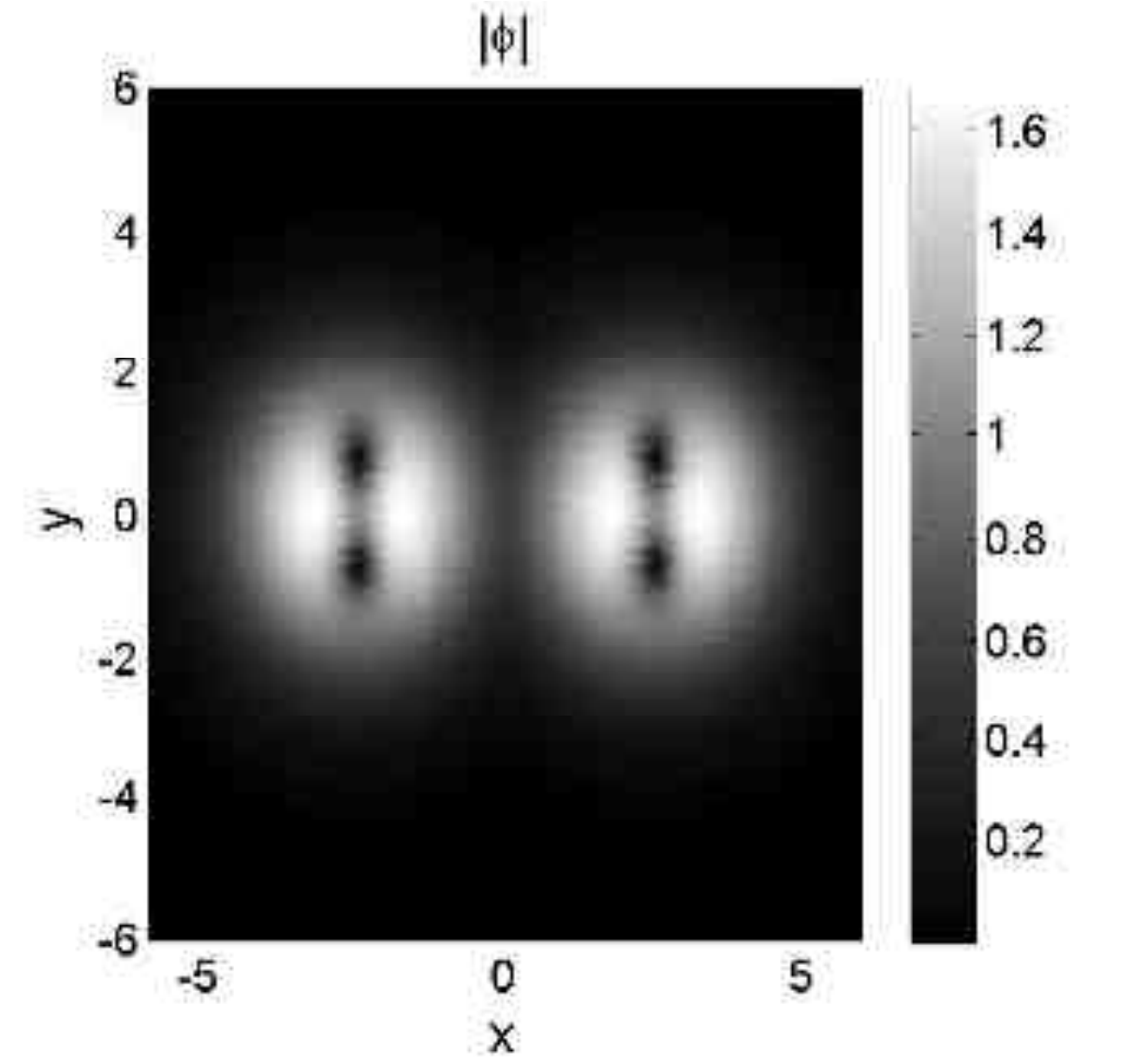}
\label{2WellPairPairMu5Abs4}}
\subfigure[]{\includegraphics[width=1.6in]{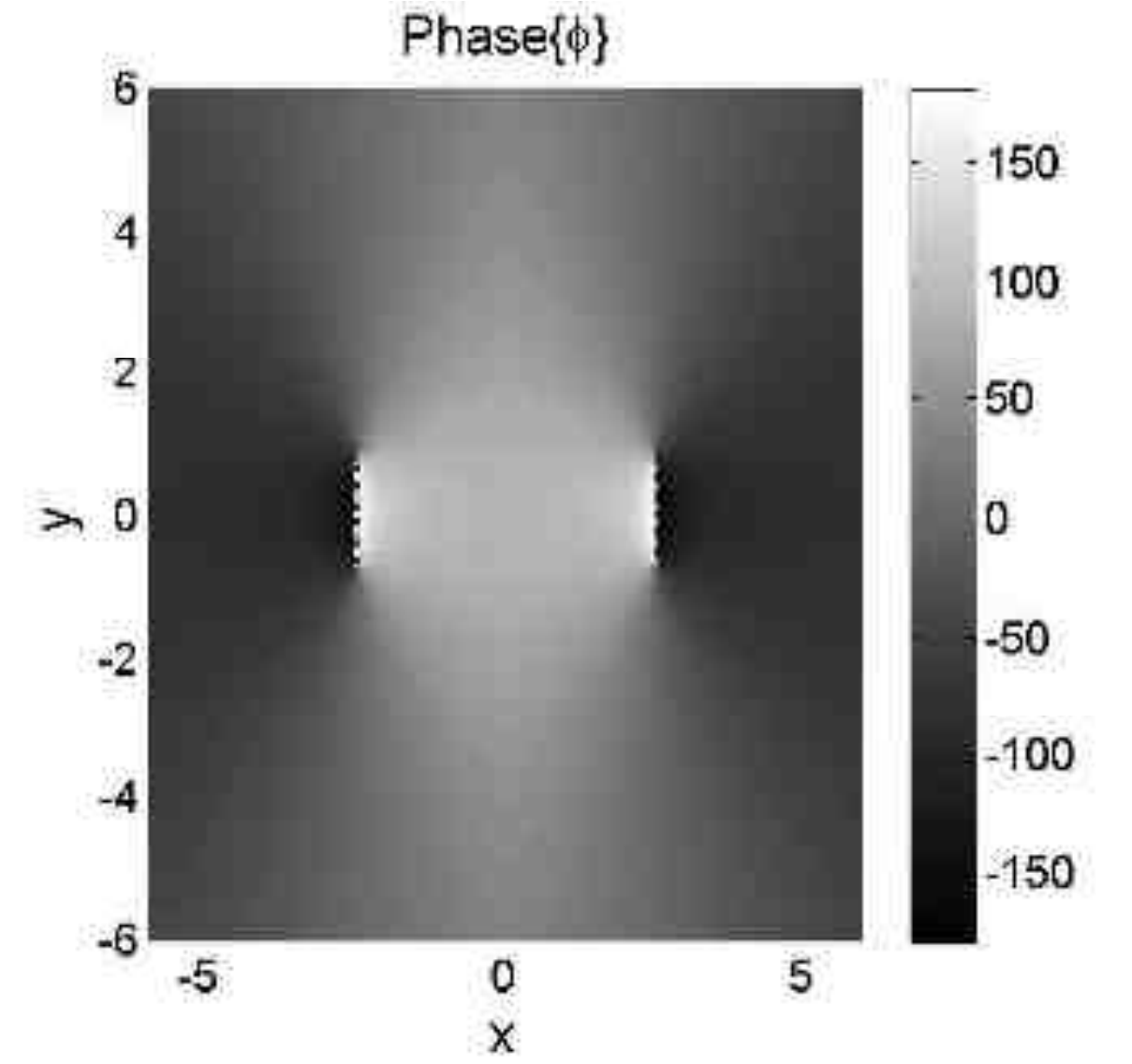}
\label{2WellPairPairMu5Phase4}}
\subfigure[]{\includegraphics[width=1.6in]{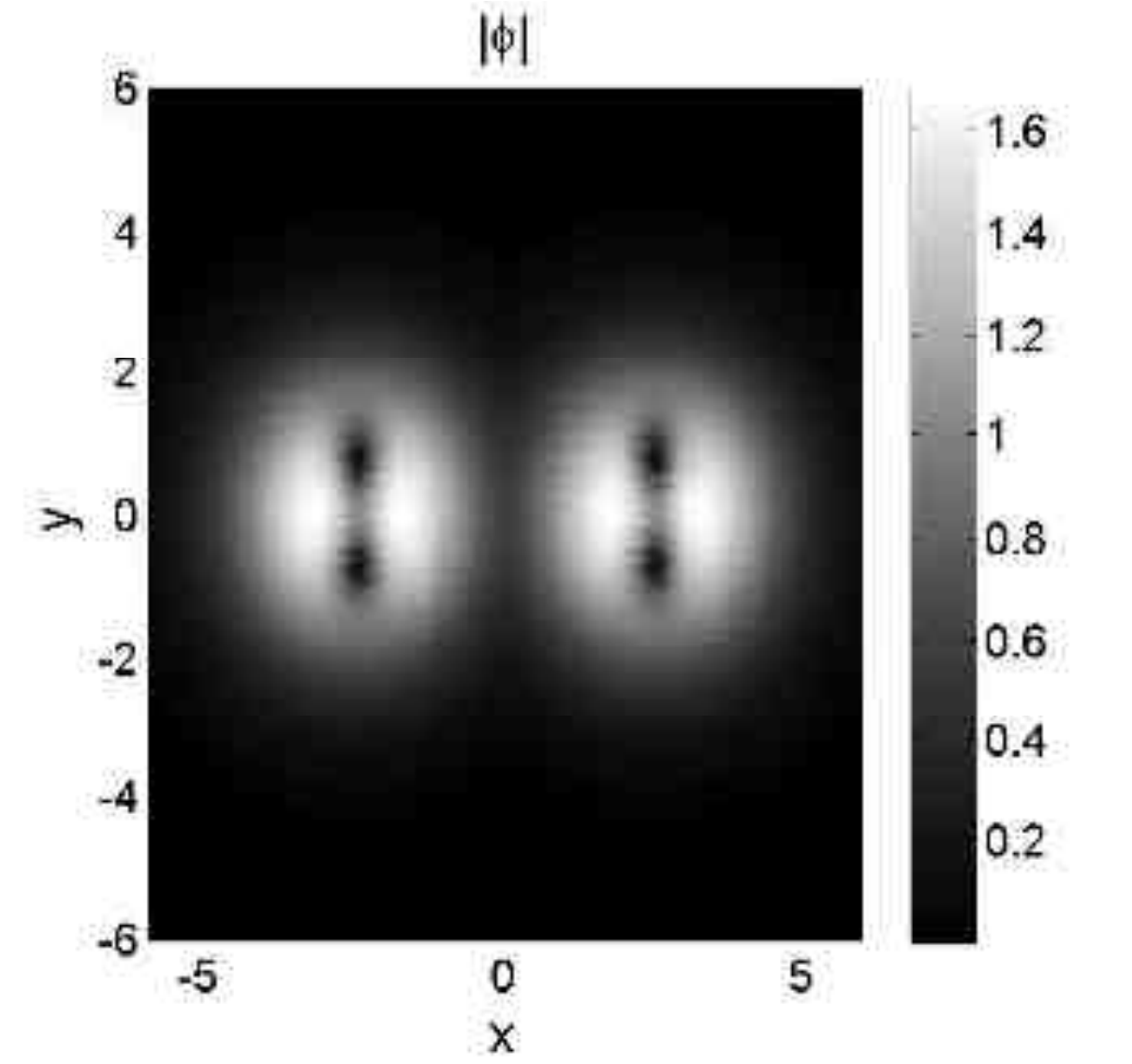}
\label{2WellPairPairMu5Abs3}}
\subfigure[]{\includegraphics[width=1.6in]{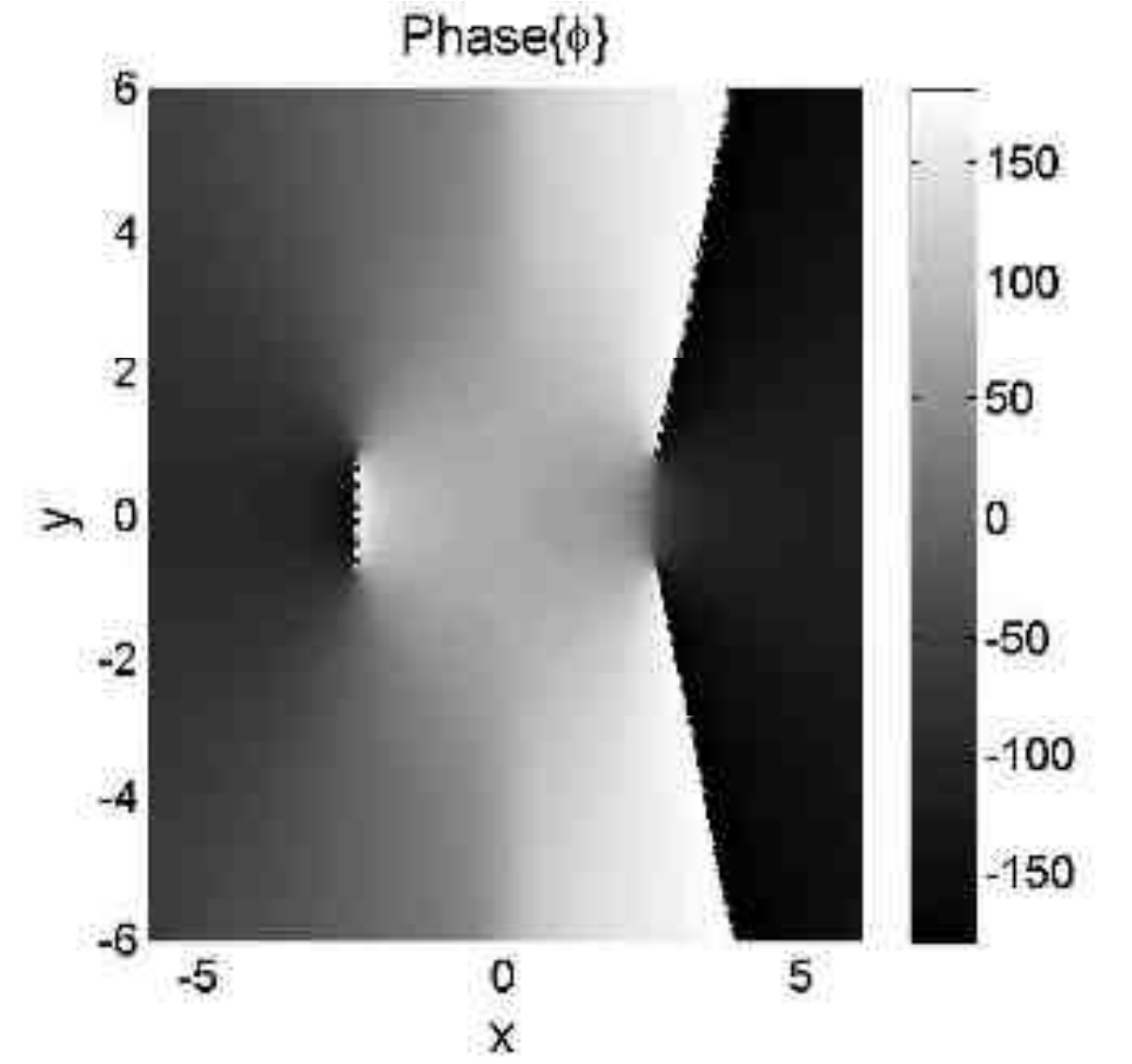}
\label{2WellPairPairMu5Phase3}} \newline
\subfigure[]{\includegraphics[width=2.25in]{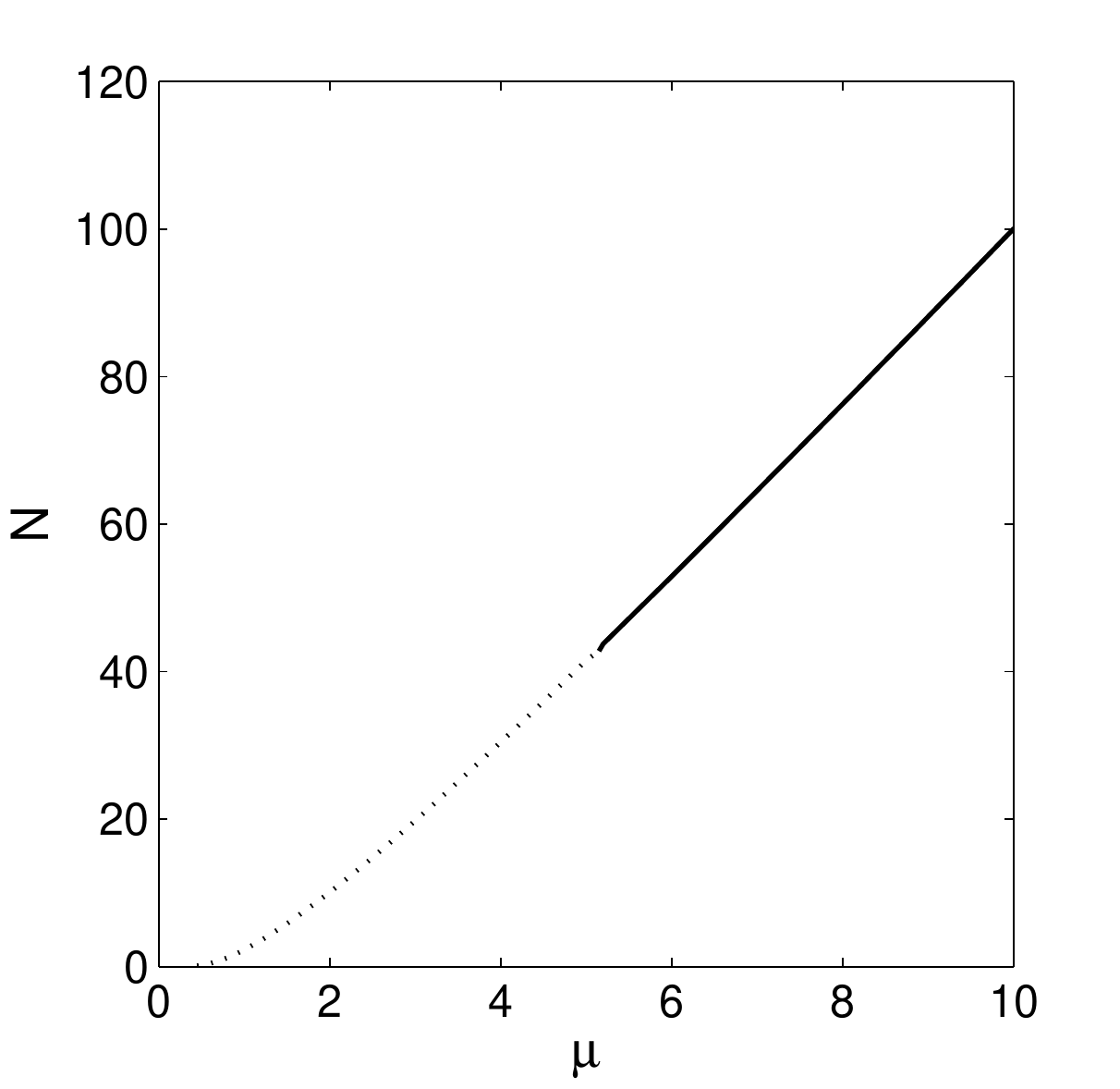}
\label{2WellPairPairNvsMu1}}
\subfigure[]{\includegraphics[width=2.25in]{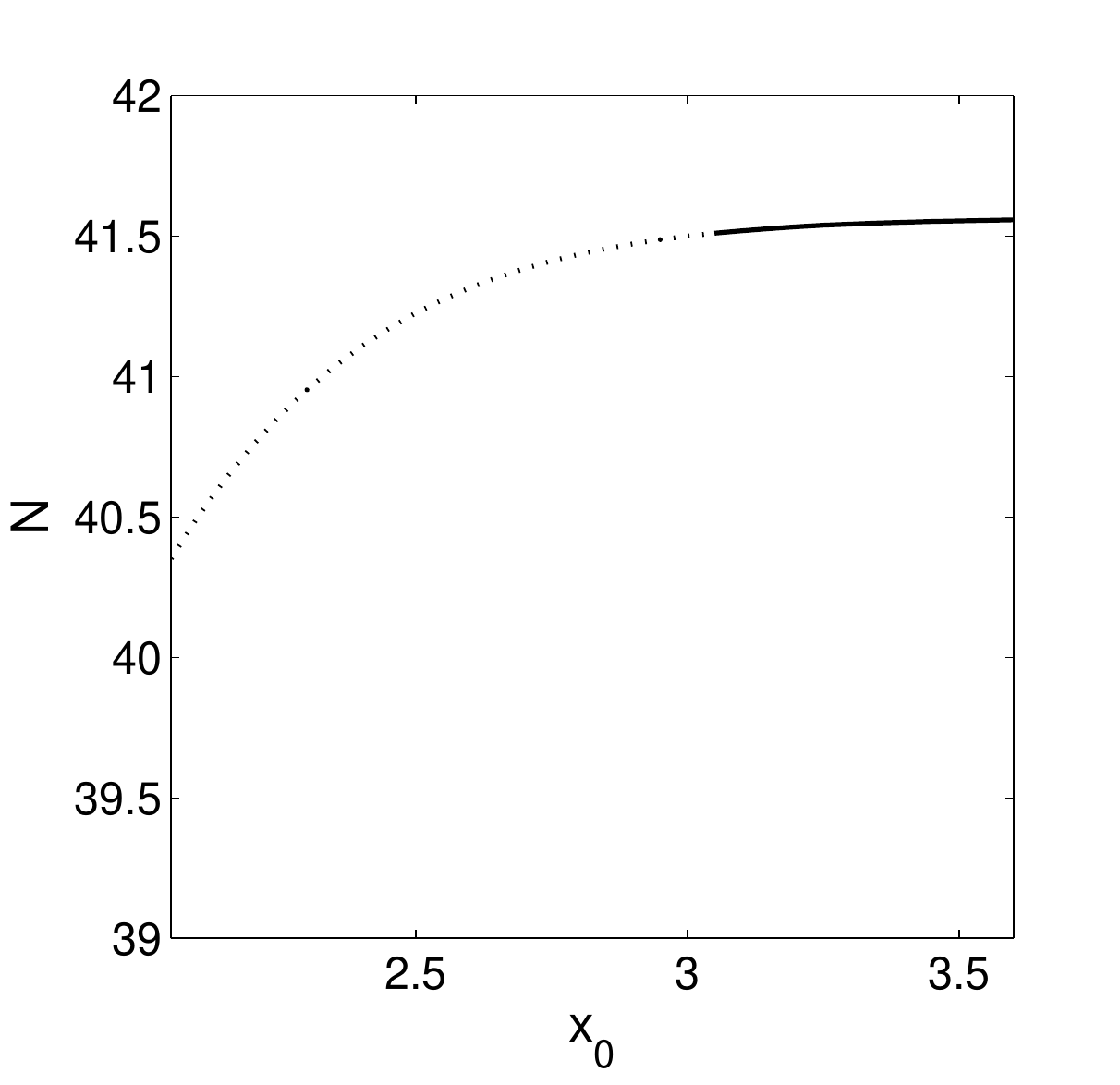}
\label{2WellPairPairNvsX01}}
\caption{(Color online) (a)-(h) Four different combinations of two single-well VADS
(vortex-antivortex dipoles), for $x_{0}=2.5$, $\protect\mu =5$ and $\protect%
\alpha =0.5$. (i) The $N(\protect\mu )$ curve, for the single (partly)
stable combination, shown in (c,d), for $x_{0}=3$ and $\protect\alpha =0.5$.
(j) The $N(x_{0})$ curve for the same family, at $\protect\mu =5$.}
\label{2WellPairPair}
\end{figure}

Lastly, four combinations of VADs, of the horizontal and vertical types, as
well as with identical or opposite orientations of the left and right
dipoles, were constructed and analyzed too, see Fig. \ref{2WellPairPair}%
(a-h). A single species among them which was found to be partly stable
corresponds to the horizontal structure shown in Fig. \ref{2WellPairPair}%
(c,d), in which the left and right components have opposite signs (recall
that the semi-vortex complex with a VAD component may also be stable solely
in the case when this component is horizontal, see Fig. \ref{2WellFundPair};
an explanation for the feasible stability of the horizontal structure in the
present case is essentially the same, as only the horizontal orientation of
the dipoles may realize an energy minimum). For this solution, the $N(\mu )$
curve, with fixed $x_{0}=3$ [Fig. \ref{2WellPairPairNvsMu1}], and the $%
N(x_{0})$ one, with fixed $\mu =5$ [Fig. \ref{2WellPairPairNvsX01}], exhibit
stability regions at $\mu >5.15$ and $x_{0}>3.05$, respectively. The
unstable solutions evolve into the symmetric ground-state mode, or, in some
cases, into stable combinations of VADs. In addition to these four complexes
built of parallel VADs, another one, composed of a vertical VAD in one well
and a horizontal VAD in the other, was also constructed. This state (which
is not shown here), is entirely unstable

To complete the study of dual-vortex complexes, it is relevant to mention
that ones of both $\left( +m,-m\right) $ and $\left( +m,+m\right) $ types,
composed of vortices proper (with $m=1$ and $2$) or VTs, as well as the VAD
complexes with identical or opposite orientations, do not demonstrate
spontaneous breaking of their symmetry or antisymmetry with respect to the
underlying DW structure, in the entire parameter region explored in our
analysis. Recall that, as mentioned above, antisymmetric complexes composed
of 2D fundamental modes feature spontaneous breaking of the antisymmetry,
but both the resulting asymmetric complexes and the antisymmetric parent
ones are completely unstable.

\section{Conclusion}

The purpose of this work is to extend the analysis of the recently
introduced class of models which admit self-trapping of stable solitons,
vortices, and more complex topologically structured modes, with the help the
defocusing cubic nonlinearity whose local strength grows fast enough from
the center to periphery. In addition to the previously studied models with
the single-well profile of the nonlinearity modulation, we have introduced
the DW (double-well) settings in the 1D and 2D geometry, which may be
realized in nonlinear optics and BEC. In both cases of the single- and
double-well profiles, we have investigated various scenarios of the
spontaneous formation of self-trapped modes, both nontopological and
topological ones, whose symmetry may be lower than that of the underlying
modulation pattern, due to the occurrence of the SSB (spontaneous symmetry
breaking) in these systems. In particular, in the 2D single-well setting we
have found unstable tripole and quadrupole modes and, on the other hand,
partly stable ordinary (fundamental) dipoles, VADs (vortex-antivortex
dipoles) and VTs (vortex triangles). The most essential results are reported
for the DW settings, in 1D and 2D alike. Due to the repulsive sign of the
nonlinearity, symmetric modes are always stable, realizing the system's
ground state. However, the antisymmetric (dipole) states are subject to the
SSB (more accurately speaking, this is spontaneous breaking of the
antisymmetry). The resulting asymmetric states are partly stable in 1D, but
unstable in 2D. While most results have been obtained in a numerical form,
the1D and 2D symmetric states were analyzed by means of the TFA
(Thomas-Fermi approximation). Stability domains have been identified for
diverse 2D semi-vortex and dual-vortex configurations, built of vortices
with topological charges $m=1$ and $2$, VTs and VADs, trapped in each
individual nonlinear potential well of the DW structure.

In the framework of the 1D system, we have also considered the model with
the rocking single well. If the rocking period is large enough, the
originally trapped dipole mode features Rabi oscillations between the
fundamental (single-peak) and dipole shapes.

In the 2D model, it may be interesting to introduce a system of three
(rather than two) nonlinear-potential wells, which form an equilateral
triangle, as such a configuration realizes the most fundamental 2D setting
(simplex) \cite{Pfau}. On the other hand, in terms of the BEC model, a
challenging problem is to consider 3D configurations generalizing their 2D
counterparts \cite{gyroscope}-\cite{Yasha}.


\newpage

\begin{thebibliography}{99}
\bibitem{Agrawal} Y. S. Kivshar and G. P. Agrawal, \textit{Optical Solitons:
From Fibers to Photonic Crystals} (Academic Press: San Diego, 2003).

\bibitem{Talanov} V. I. Talanov, Pis'ma Zh. Eksp. Teor. Fiz. \textbf{11},
303 (1970) [JETP Lett. \textbf{11}, 199 (1970)].

\bibitem{Townes} R. Y. Chiao, E. Garmire, and C. H. Townes, Phys. Rev. Lett
. \textbf{13}, 479 (1964).

\bibitem{Berge'} L. Berg\'{e}, Phys. Rep. \textbf{303}, 259 (1998); E. A.
Kuznetsov and F. Dias, \textit{ibid}. \textbf{507}, 43 (2011).

\bibitem{Mario} F. Kh. Abdullaev and M. Salerno, Phys. Rev. A \textbf{72},
033617 (2005).

\bibitem{pseudo} W. A. Harrison, \textit{Pseudopotentials in the Theory of
Metals} (Benjamin: New York, 1966).

\bibitem{BEC} S. Giorgini, L. P. Pitaevskii, and S. Stringari, Rev. Mod.
Phys. \textbf{80}, 1215 (2008); H. T. C. Stoof, K. B. Gubbels, and D. B. M.
Dickrsheid, \textit{Ultracold Quantum Fields} (Springer: Dordrecht, 2009).

\bibitem{RMP} Y. V. Kartashov, B. A. Malomed, and L. Torner, Rev. Mod. Phys.
\textbf{83}, 247 (2011).

\bibitem{FR-Randy} S. E. Pollack, D. Dries, M. Junker, Y. P. Chen, T. A.
Corcovilos, and R. G. Hulet, Phys. Rev. Lett. \textbf{102}, 090402 (2009).

\bibitem{FR-review} C. C. Chin, R. Grimm, P. Julienne, and E. Tsienga, Rev.
Mod. Phys. \textbf{82}, 1225 (2010).

\bibitem{FR-Tom} D. M. Bauer, M. Lettner, C. Vo, G. Rempe, and S. D\"{u}rr,
Nature Phys. \textbf{5}, 339 (2009); M. Yan, B. J. DeSalvo, B.
Ramachandhran, H. Pu, and T. C. Killian, Phys. Rev. Lett. \textbf{110},
123201 (2013).

\bibitem{experiment-inhom-Feshbach} R. Yamazaki, S. Taie, S. Sugawa, and Y.
Takahashi, Phys. Rev. Lett. \textbf{105}, 050405 (2010).

\bibitem{painting} K. Henderson, C. Ryu, C. MacCormick, and M. G. Boshier,
New J. Phys. \textbf{11}, 043030 (2009).

\bibitem{Cooper} N. R. Cooper, Phys. Rev. Lett. \textbf{106}, 175301 (2011).

\bibitem{magn-latt} H. S. Ghanbari, T. D. Kieu, A. Sidorov, and P.
Hannaford, J. Phys. B: At. Mol. Opt. Phys. \textbf{39}, 847 (2006); O.
Romero-Isart, C. Navau, A. Sanchez, P. Zoller, and J. I. Cirac, Phys. Rev.
Lett. \textbf{111}, 145304 (2013); S. Ghanbari, A. Abdalrahman, A. Sidorov,
and P. Hannaford, J. Phys. B: At. Mol. Opt. Phys. \textbf{47}, 115301
(2014); S. Jose, P. Surendran, Y. Wang, I. Herrera, L. Krzemien, S.
Whitlock, R. McLean, A. Sidorov, and P. Hannaford, Phys. Rev. A \textbf{89},
051602 (2014).

\bibitem{concentrator} C. Navau, J. Prat-Camps, and A. Sanchez, Phys. Rev.
Lett. \textbf{109}, 263903 (2012).

\bibitem{Kip} J. Hukriede, D. Runde, and D. Kip, J. Phys. D \textbf{36}, R1
(2003).

\bibitem{1D-attr} G. Theocharis, P. Schmelcher, P. G. Kevrekidis, and D. J.
Frantzeskakis, Phys. Rev. A \textbf{72}, 033614 (2005); H. Sakaguchi and B.
A. Malomed, Phys. Rev. E \textbf{72}, 046610 (2005); F. K. Abdullaev and J.
Garnier, Phys. Rev. A \textbf{72}, 061605(R) (2005); G. Dong, B. Hu, and W.
Lu, \textit{ibid.} \textbf{74}, 063601 (2006); G. Fibich, Y. Sivan, and M.
I. Weinstein, Physica D 217, 31 (2006); D. A. Zezyulin, G. L. Alfimov, V. V.
Konotop, and V. M. P\'{e}rez-Garc\'{\i}a, Phys. Rev. A \textbf{76}, 013621
(2007); F. Kh. Abdullaev, A. Gammal, M. Salerno, and L. Tomio, \textit{ibid.
}\textbf{77}, 023615 (2008); L. C. Qian, M. L. Wall, S. Zhang, Z. Zhou, and
H. Pu, \textit{ibid.} \textbf{77}, 013611 (2008); A. S. Rodrigues, P. G.
Kevrekidis, M. A. Porter, D. J. Frantzeskakis, P. Schmelcher, and A. R.
Bishop, \textit{ibid.} \textbf{78}, 013611 (2008); Y. Kominis and K.
Hizanidis, Opt. Exp. \textbf{16}, 12124 (2008); Y. V. Kartashov, V. A.
Vysloukh, and L. Torner, Opt. Lett. \textbf{33}, 1747 (2008); \textit{ibid.}
\textbf{33}, 2173 (2008); F. Kh. Abdullaev, R. M. Galimzyanov, M. Brtka, and
L. Tomio, Phys. Rev. E \textbf{79}, 056220 (2009); V. M. P\'{e}rez-Garc\'{\i}%
a and R. Pardo, Physica D \textbf{238}, 1352 (2009); A. V. Yulin, Yu. V.
Bludov, V. V. Konotop, V. Kuzmiak, and M. Salerno, Phys. Rev. A \textbf{84},
063638 (2011); J. Belmonte-Beitia, V. M. P\'{e}rez-Garc\'{\i}a, and V.
Brazhnyi, Commun. Nonlinear Sci. Numer. Simulat. \textbf{16}, 158 (2011); H.
J. Shin, R. Radha, and V. Ramesh Kumar, Phys. Lett. A \textbf{375}, 2519
(2011); X.-F. Zhou, S.-L. Zhang, Z.-W. Zhou, B. A. Malomed, and H. Pu, Phys.
Rev. A \textbf{85}, 033603 (2012); Y. Kominis, \textit{ibid}. \textbf{87},
063849 (2013); T. Wasak, V. V. Konotop, and M. Trippenbach, EPL \textbf{105}%
, 64002 (2014).

\bibitem{Thawatchai} T. Mayteevarunyoo, B. A. Malomed, and G. Dong, Phys.
Rev. A \textbf{78}, 053601 (2008); A. Acus, B. A. Malomed, and Y. Shnir,
Physica D \textbf{241}, 987 (2012).

\bibitem{2D-attr} Y. Sivan, G. Fibich, and M. I. Weinstein, Phys. Rev. Lett.
\textbf{97}, 193902 (2006); H. Sakaguchi and B. A. Malomed, Phys. Rev. E
\textbf{73}, 026601 (2006); Y. V. Kartashov, B. A. Malomed, V. A. Vysloukh,
and L. Torner, Opt. Lett. \textbf{34}, 770 (2009); O. V. Borovkova, Y. V.
Kartashov, and L. Torner, Phys. Rev. A \textbf{81}, 063806 (2010).

\bibitem{2D-Thawatchai} T. Mayteevarunyoo, B. A. Malomed, and A. Reoksabutr,
J. Mod. Opt. \textbf{58}, 1977 (2011).

\bibitem{Barcelona} O. V. Borovkova, Y. V. Kartashov, B. A. Malomed, and L.
Torner, Opt. Lett. \textbf{36}, 3088 (2011).

\bibitem{Barcelona2} O. V. Borovkova, Y. V. Kartashov, B. A. Malomed, and L.
Torner, Phys. Rev. E \textbf{84}, 035602 (R) (2011).

\bibitem{others} Q. Tian, L. Wu, Y. Zhang, and J.-F. Zhang, Phys. Rev. E
\textbf{85}, 056603 (2012); Y. Wu, Q. Xie, H. Zhong, L. Wen, and W. Hai,
Phys. Rev. A \textbf{87}, 055801 (2013).

\bibitem{gyroscope} R. Driben, Y. V. Kartashov, B. A. Malomed, T. Meier, and
L. Torner, Phys. Rev. Lett. \textbf{112}, 020404 (2014).

\bibitem{hybrid} R. Driben, Y. Kartashov, B. A. Malomed, T. Meier, and L.
Torner, New J. Phys. \textbf{16}, 063035 (2014).

\bibitem{Yasha} Y. V. Kartashov, B. A. Malomed, Y. Shnir, and L. Torner,
Phys. Rev. Lett. \textbf{113}, 264101 (2014).

\bibitem{discrete} G. Gligori\'{c}, A. Maluckov, L. Hadzievski, and B. A.
Malomed, Phys. Rev. E \textbf{88}, 032905 (2013).

\bibitem{LucaLuca} L. Barbiero, B. A. Malomed, and L. Salasnich, Phys. Rev.
A \textbf{90}, 063611 (2014).

\bibitem{Raymond} Y. Li, J. Liu, W. Pang, and B. A. Malomed, Phys. Rev. A
\textbf{88}, 053630 (2013).

\bibitem{China-DW} Q. Xie, L. Wang, Y. Wang, Z. Shen, and J. Fu, Phys. Rev.
E \textbf{90}, 063204 (2014).

\bibitem{SR} R. Driben, T. Meier, and B. A. Malomed, Sci. Rep. \textbf{5},
9420 (2015).

\bibitem{LL} D. Landau and E. M. Lifshitz, \textit{Quantum Mechanics}
(Moscow: Nauka Publishers, 1974).

\bibitem{book} \textit{Spontaneous Symmetry Breaking, Self-Trapping, and
Josephson Oscillations}, B. A. Malomed, editor (\noindent Springer-Verlag:
Berlin and Heidelberg, 2013).

\bibitem{NewsViews} B. A. Malomed, Nature Photonics \textbf{9}, 287-289
(2015).

\bibitem{Davies} E. B. Davies, Commun. Math. Phys. \textbf{64}, 191 (1979);
J. C. Eilbeck, P. S. Lomdahl, and A. C. Scott, Physica D \textbf{16}, 318
(1985).

\bibitem{Snyder} A. W. Snyder, D. J. Mitchell, L. Poladian, D. R. Rowland,
and Y. Chen, J. Opt. Soc. Am. B \textbf{8}, 2101 (1991).

\bibitem{bif} G. Iooss and D. D. Joseph, \textit{Elementary Stability
Bifurcation Theory} (Springer-Verlag: New York, 1980).

\bibitem{Wabnitz} E. M. Wright, G. I. Stegeman, and S. Wabnitz, Phys. Rev. A
\textbf{40}, 4455 (1989).

\bibitem{Pare} C. Par\'{e} and M. F\l orja\'{n}czyk, Phys. Rev. A \textbf{41}%
, 6287 (1990); A. I. Maimistov, Kvant. Elektron. \textbf{18}, 758 [Sov. J.
Quantum Electron. \textbf{21}, 687 (1991)].

\bibitem{Akhmed} N. Akhmediev and A. Ankiewicz, Phys. Rev. Lett. \textbf{70}%
, 2395 (1993).

\bibitem{Pak} P. L. Chu, B. A. Malomed, and G. D. Peng, J. Opt. Soc. Am. B
10, 1379 (1993).

\bibitem{tight} G. L. Alfimov, P. G. Kevrekidis, V. V. Konotop, and M.
Salerno, Phys. Rev. E \textbf{66}, 046608 (2002).

\bibitem{Milburn} G. J. Milburn, J. Corney, E. M. Wright, and D. F. Walls,
``Quantum dynamics of an atomic Bose-Einstein condensate in a double-well
potential", Phys. Rev. A \textbf{55}, 4318-4324 (1997); A. Smerzi, S.
Fantoni, S. Giovanazzi, and S. R. Shenoy, Phys. Rev. Lett. \textbf{79}, 4950
(1997).

\bibitem{junction} S. Raghavan, A. Smerzi, S. Fantoni, and S. R. Shenoy,
Phys. Rev. A \textbf{59}, 620-633 (1999); A. Smerzi and S. Raghavan, \textit{%
ibid}. \textbf{61}, 063601 (2000).

\bibitem{junction2} K. Sakmann, A. I. Streltsov, O. E. Alon, and L. S.
Cederbaum, Phys. Rev. Lett. \textbf{103}, 220601 (2009); M. Chuchem, K.
Smith-Mannschott, M. Hiller, T. Kottos, A. Vardi, and D. Cohen, Phys. Rev. A
\textbf{82}, 053617 (2010).

\bibitem{junction-reviews} R. Gatti and M. K. Oberthaler, J. Phys. B: At.
Mol. Opt. Phys. \textbf{40}, R61 (2007); M. A. Cazalilla, R. Citro, T.
Giamarchi, E. Orignac, and M. Rigol, Rev. Mod. Phys. \textbf{83}, 1405
(2011).

\bibitem{Warsaw} M. Matuszewski, B. A. Malomed, and M. Trippenbach, Phys.
Rev. A \textbf{75}, 063621 (2007).

\bibitem{Markus} M. Albiez, R. Gati, J. F\"{o}lling, S. Hunsmann, M.
Cristiani, and M. K. Oberthaler, Phys. Rev. Lett. \textbf{95}, 010402 (2005).

\bibitem{photo} P. G. Kevrekidis, Z. Chen, B. A. Malomed, D. J.
Frantzeskakis, and M. I. Weinstein, Phys. Lett. A \textbf{340}, 275 (2005).

\bibitem{lasers} T. Heil, I. Fischer, W. Els\"{a}sser, J. Mulet, and C. R.
Mirasso, Phys. Rev. Lett. \textbf{86}, 795 (2001); P. Hamel, S. Haddadi, F.
Raineri, P. Monnier, G. Beaudoin, I. Sagnes, A. Levenson, and A. M.
Yacomotti, Nature Photonics \textbf{9}, 311 (2015).

\bibitem{azi} A. S. Desyatnikov, A. A. Sukhorukov, and Y. S. Kivshar, Phys.
Rev. Lett. \textbf{95}, 203904 (2005).

\bibitem{Yang} J. Yang, \textit{Nonlinear Waves in Integrable and
Nonintegrable Systems} (SIAM: Philadelphia, 2010).

\bibitem{Lakoba} T. I. Lakoba and J. Yang, J. Comput. Phys. \textbf{226},
1668 (2007); T. I. Lakoba and J. Yang, Stud. Appl. Math. \textbf{118}, 153
(2007).

\bibitem{rocking-optical} I. L. Garanovich, S. Longhi, A. A. Sukhorukov, and
Y. S. Kivshar, Phys. Rep. \textbf{518}, 1 (2012).

\bibitem{Rabi} A. Gubeskys, B. A. Malomed, and I. M. Merhasin, Stud. Appl.
Math. \textbf{115}, 255 (2005); Y. V. Kartashov, V. A. Vysloukh, and L.
Torner, Phys. Rev. Lett. \textbf{99}, 233903 (2007); K. G. Makris, D. N.
Christodoulides, O. Peleg, M. Segev, and D. Kip, Opt. Exp. \textbf{16},
10309 (2008); F. Dreisow, A. Szameit, M. Heinrich, T. Pertsch, S. Nolte, A. T%
\"{u}nnermann, and S. Longhi, Phys. Rev. Lett. \textbf{102}, 076802 (2009);
T. Kanna, R. B. Mareeswaran, F. Tsitoura, H. E. Nistazakis, and D. J.
Frantzeskakis, J. Phys. A: Math. Theor. \textbf{46}, 475201 (2013).

\bibitem{Radik} R. Driben, N. Dror, B. A. Malomed, and T. Meier, to be
published.

\bibitem{Pfau} T. Lahaye, T. Pfau, and L. Santos, Phys. Rev. Lett. \textbf{%
104}, 170404 (2010).
\end{thebibliography}
\end{document}